\documentclass[11pt,letterpaper]{article}

\usepackage[T1]{fontenc}
\usepackage[utf8]{inputenc}
\usepackage{lmodern}
\usepackage[margin=1in]{geometry}
\usepackage{amsmath,amssymb,amsthm,mathtools}
\usepackage{graphicx}
\usepackage{tikz}
\usetikzlibrary{arrows.meta,positioning}
\definecolor{cssOI1}{HTML}{0072B2}
\definecolor{cssOI2}{HTML}{E69F00}
\definecolor{cssOI3}{HTML}{009E73}
\definecolor{cssOI4}{HTML}{CC79A7}
\definecolor{cssOI5}{HTML}{56B4E9}
\definecolor{cssOI6}{HTML}{F0E442}
\definecolor{cssOI7}{HTML}{D55E00}
\definecolor{cssOI8}{HTML}{000000}
\definecolor{cssGrey}{HTML}{BBBBBB}
\colorlet{cssR}{cssOI1}
\colorlet{cssV}{cssOI3}
\colorlet{cssT}{cssOI7}
\tikzset{
  cssfill/.style={fill=#1},%
  cssfig/.style={font=\footnotesize, >={Stealth[length=1.7mm,width=1.3mm]}, line join=round},
  cssnode/.style={circle, minimum size=4.5mm, inner sep=0pt, line width=0.5pt, draw=black!85, text=black, font=\footnotesize},
  cssdep/.style={->, >={Stealth[length=1.7mm,width=1.3mm]}, line width=0.45pt, draw=black!80},
  csscont/.style={line width=2.2pt, draw=black!22},
  cssproc/.style={rounded corners=2mm, line width=0.4pt, draw=black!55, fill=black!13},
  cssRbox/.style={rounded corners=2mm, line width=1.1pt, draw=cssR, fill=cssR!14},
  cssVbox/.style={rounded corners=2mm, line width=1.1pt, draw=cssV, fill=cssV!14},
  cssTbox/.style={rounded corners=2mm, line width=1.1pt, draw=cssT, fill=cssT!14},
  cssadded/.style={densely dashed, line width=0.9pt},
  csstag/.style={font=\footnotesize\itshape, text=black!75, inner sep=1pt},
  csswset/.style={rounded corners=4mm, line width=1.2pt, draw=black!85},
  cssaxis/.style={->, >={Stealth[length=2.4mm,width=1.9mm]}, line width=0.8pt, draw=black},
  csstube/.style={line width=0.5pt, draw=black, line cap=round},
  cssflow/.style={->, >={Stealth[length=2.2mm,width=1.8mm]}, line width=0.7pt, draw=black},
  cssmap/.style={->, >={Straight Barb[length=1.8mm,width=2.4mm]}, line width=0.4pt, densely dashed, draw=black},
  cssinner/.style={rounded corners=1.2mm, line width=0.4pt, draw=black!80},
}

\usepackage{array,booktabs,tabularx}
\usepackage{placeins}%
\usepackage[font=small]{caption}%
\usepackage{enumitem}
\usepackage[round]{natbib}
\usepackage[titletoc,title]{appendix}
\usepackage[hidelinks]{hyperref}

\usepackage{xcolor}
\usepackage[normalem]{ulem}
\newif\ifmarkchanges
\markchangesfalse%
\ifmarkchanges
  \DeclareRobustCommand{\new}[1]{\ifmmode{\color{red}#1}\else\textcolor{red}{#1}\fi}
  \DeclareRobustCommand{\del}[1]{\textcolor{red}{\sout{#1}}}
  \DeclareRobustCommand{\delmark}{\ifmmode{\color{red}{}^{\times}}\else\textcolor{red}{\textsuperscript{$\times$}}\fi}
\else
  \DeclareRobustCommand{\new}[1]{#1}
  \DeclareRobustCommand{\del}[1]{}
  \DeclareRobustCommand{\delmark}{}
\fi
\DeclareRobustCommand{\revnote}[1]{\ifmarkchanges\textcolor{red}{\textbf{#1}}\fi}
\newenvironment{newtext}{\ifmarkchanges\color{red}\fi\ignorespaces}{\ignorespacesafterend}
\DeclareRobustCommand{\prelim}[1]{\ifmmode\colorbox{yellow}{$#1$}\else\colorbox{yellow}{#1}\fi}
\theoremstyle{definition}
\newtheorem{definition}{Definition}
\newtheorem{proposition}{Proposition}
\newtheorem{corollary}{Corollary}
\newtheorem{openproblem}{Open Problem}
\newtheorem*{conjecture}{Conjecture}

\renewcommand{\bibpreamble}{Citations are given as (Author, year). Bibliographic details
  for entries newly added in this version have not been machine-verified.\par\medskip}

\title{Which Structures Carry Information?\\[8pt]%
  {\large The Inverse Problem of Heredity and a Computational Test in Digital Evolution\par}}%
\author{{\small Wouter van der Wijngaart}\\[2pt]%
  \scriptsize Department of Micro and Nanosystems, School of Electrical Engineering and
  Computer Science,\\ \scriptsize KTH Royal Institute of Technology, Stockholm, Sweden}
\date{}
\makeatletter\g@addto@macro\abstract{\footnotesize}\makeatother%

\begin{document}

\maketitle

\begin{abstract}
The laws of physics do not distinguish a gene from another molecule, or computer code from another physical pattern.
Yet genes and computer code are copied and direct the construction of other structures, because they act as \textit{information carriers}.
Among physical structures, can we identify which are information carriers without predefining them?

Because information carriers both enable cumulative heredity and emerge from it, we hypothesise that information carriers can be identified as the hereditary \new{information }carriers in systems undergoing cumulative heredity (as in Universal Darwinian Evolution).
We therefore call the query for defining information carriers \textit{the inverse problem of heredity}.
To investigate this question, we introduce the \textit{Causal Structure Set} (CSS): a model of an observed world episode in the form of a directed acyclic graph (DAG) of persistent physical structures and their causal ancestry, in which hereditary roles are not \textit{designated} in advance but \textit{inferred} from the causal organisation.
As a first criterion for cumulative hereditary organisation, we propose that structure replication, variation, variant-dependent translation, and continuing lineages occur jointly within a single causally connected family of information carrier variants.
\new{This paper develops the theory, i.e., the CSS, the criterion and their formal properties, and demonstrates its application in a preliminary form in two domains: the cultural transmission of cooking recipes, and digital evolution.} %

\new{For culture, a reconstruction of cooks who teach, modify and prepare recipes, under stated physical hypotheses, shows that the criterion can identify recipe knowledge as a family of information carrier variants.} %
\new{In an Avida-like simulator of digital evolution, in which self-copying programs mutate and multiply, we build CSSs of 848,000–884,000 nodes from complete execution histories, with the designated roles, i.e., those that the simulator's design fixes, removed, and test each causal link by replay.
In the development runs, the inferred organisms match the designated ones, the time to first division follows the inserted program in 95.9\% of 4,314 program swaps, and a restricted role-blind search finds a candidate information family of 4–6 instruction types in each run.
Frozen and tested on 40 unseen random seeds, the analysis finds a candidate family in all 39 runs with births, and the time to first division follows the inserted program in 96.2\% of 30,422 program swaps.
The preregistered five-part test nevertheless fails, in two runs, because of a choice we made in defining the designated family, not because of an error in the inference.} %

The \new{theory and its preliminary demonstrations} have several potentially impactful consequences. 
First, the CSS offers a route to identifying hereditary organisation without starting from designated genes, organisms or programs. In the longer term, such analysis could help specify the physical conditions under which evolvable, information-bearing or living systems emerge.
Second, changes in how hereditary information is physically stored, copied and used, from nucleic acids to writing and computer code, have accompanied the emergence of new levels of organisation, including biology, social organisation, civilisational institutions, and software ecosystems.
If information carriers can be identified from causal histories alone, the same analysis could find and compare hereditary systems across these domains, including forms not yet recognised as hereditary, and identify the causal signatures that make such systems persist, become fragile or collapse.

\end{abstract}

\begin{newtext}
\section*{Overview: identifying information carriers from causal histories}
\addcontentsline{toc}{section}{Overview: identifying information carriers from causal histories}

\subsection*{The inverse problem of heredity}

A molecule can participate in heredity, a neural configuration can preserve a learned skill, and a pattern in computer memory can direct the production of further structures. Their informational roles are not additional physical substances or forces. They depend on how ordinary physical configurations are copied, transformed and used within larger causal organisations. Can those roles be identified without first being told which structures are genes, memories or programs?

This paper formulates that question as the \emph{inverse problem of heredity}. Rather than beginning with recognised hereditary entities and analysing their evolution, it asks which structures in an observed causal history occupy hereditary roles. We introduce a representation for this inference, propose an initial criterion, and implement the resulting procedure in a digital-evolution simulator (Figure~\ref{fig:overview}). The computational demonstration turns the formal proposal into a method whose constructions, detections and failures can be examined directly.

\begin{figure}[htbp]
\centering
\begin{tikzpicture}[cssfig,
  ovn/.style={circle, minimum size=2.8mm, inner sep=0pt, line width=0.4pt, draw=black!85},
  ovcar/.style={circle, minimum size=4.4mm, inner sep=0pt, line width=0.5pt, draw=black!85, font=\scriptsize},
  ovprod/.style={rectangle, minimum size=2.8mm, inner sep=0pt, line width=0.4pt, draw=black!85},
  ovbadge/.style={rectangle, rounded corners=0.6mm, minimum size=3.4mm, inner sep=0pt, line width=0.6pt, font=\scriptsize\bfseries},
  ovtext/.style={text width=3.2cm, align=center, font=\scriptsize, inner sep=0pt},
  ovtag/.style={font=\scriptsize\itshape, text=black!70, inner sep=0pt}]

\def\ovbox#1#2#3{%
  \begin{scope}
    \clip[rounded corners=1.5mm] (#1,0) rectangle ++(3.5,4.4);
    \fill[#2] (#1,3.58) rectangle ++(3.5,0.82);
  \end{scope}
  \draw[line width=0.6pt, draw=black!60, rounded corners=1.5mm] (#1,0) rectangle ++(3.5,4.4);
  \draw[line width=0.4pt, draw=black!40] (#1,3.58) -- ++(3.5,0);
  \node[font=\footnotesize\bfseries, align=center, inner sep=0pt] at (#1+1.75,3.99) {#3};
}

\ovbox{0}{black!12}{Observed episode}
\node[ovtag] at (1.75,3.3) {no types, no roles yet};
\draw[->, line width=0.5pt, draw=black!70] (0.42,1.24) -- (0.42,3.1);
\draw[->, line width=0.5pt, draw=black!70] (0.36,1.3) -- (3.25,1.3);
\node[font=\scriptsize, rotate=90] at (0.22,2.2) {time};%
\node[font=\scriptsize, anchor=north, inner sep=1pt] at (1.8,1.26) {space};%
\draw[csstube] (1.0,1.4) to[out=90,in=-80] (1.1,2.05);
\draw[csstube] (1.1,2.05) to[out=100,in=-90] (0.85,3.0);
\draw[csstube] (1.1,2.05) to[out=75,in=-95] (1.5,3.0);
\draw[csstube] (2.0,1.4) to[out=85,in=-125] (2.4,2.35);
\draw[csstube] (2.85,1.4) to[out=95,in=-55] (2.4,2.35);
\draw[csstube] (2.4,2.35) to[out=105,in=-90] (2.25,3.0);
\draw[csstube] (2.4,2.35) to[out=60,in=-95] (2.9,3.0);
\node[ovtext] at (1.75,0.6) {a bounded spacetime\\ episode};

\ovbox{4}{cssOI5!35}{Causal Structure Set}
\node[ovtag] at (5.75,3.3) {stages 1--5};
\node[ovn, fill=cssOI5] (a) at (5.0,1.45) {};
\node[ovn, fill=cssOI2] (b) at (6.5,1.45) {};
\node[ovn, fill=cssOI5] (c) at (4.75,2.15) {};
\node[ovn, fill=cssOI5] (d) at (5.65,2.2) {};
\node[ovn, fill=cssOI4] (e) at (6.65,2.15) {};
\node[ovn, fill=cssOI2] (f) at (5.15,2.9) {};
\node[ovn, fill=cssOI4] (g) at (6.3,2.9) {};
\foreach \src/\dst in {a/c, a/d, b/d, b/e, c/f, d/f, d/g, e/g}{
  \draw[cssdep] (\src) -- (\dst);
}
\node[ovtext] at (5.75,0.6) {instances $W$, labels $\ell$,\\ causal order $\prec$,\\ under perspective $P$};

\ovbox{8}{cssOI2!35}{Causal motif\\ identification}
\node[ovtag] at (9.75,3.3) {stage 6: blind search};
\node[font=\scriptsize, text=black!80] at (9.75,2.97) {initial example: RVT};
\node[ovbadge, draw=cssR, fill=cssR!14] at (8.55,2.6) {R};
\node[anchor=west, font=\scriptsize] at (8.85,2.6) {copy};
\node[ovbadge, draw=cssV, fill=cssV!14] at (8.55,2.2) {V};
\node[anchor=west, font=\scriptsize] at (8.85,2.2) {vary};
\node[ovbadge, draw=cssT, fill=cssT!14] at (8.55,1.8) {T};
\node[anchor=west, font=\scriptsize] at (8.85,1.8) {different outputs};
\node[ovbadge, draw=black!55, fill=black!8] at (8.55,1.4) {L};
\node[anchor=west, font=\scriptsize] at (8.85,1.4) {continued descent};
\node[ovtext] at (9.75,0.6) {all four, jointly, for\\ one family of variants:\\ \textit{a hypothesis under test}};

\ovbox{12}{cssOI2!35}{Inferred information\\ family}
\node[ovtag] at (13.75,3.3) {stage 6: roles assigned};
\node[ovcar, fill=cssOI5!80] (A1) at (12.7,1.35) {$A$};
\node[ovcar, fill=cssOI6!80] (A2) at (13.75,1.7) {$A'$};
\node[ovcar, fill=cssOI4!80] (A3) at (14.8,2.05) {$A''$};
\node[ovprod, fill=cssOI2] (P1) at (12.7,2.2) {};
\node[ovprod, fill=cssOI1!70] (P2) at (13.75,2.55) {};
\node[ovprod, fill=cssOI3!80] (P3) at (14.8,2.9) {};
\draw[cssdep, draw=cssV] (A1) -- (A2);
\draw[cssdep, draw=cssV] (A2) -- (A3);
\foreach \src/\dst in {A1/P1, A2/P2, A3/P3}{
  \draw[cssdep, draw=cssT] (\src) -- (\dst);
}
\node[ovtext] at (13.75,0.58) {information carrier\\ variants $A$, $A'$, $A''$\\ and their products};%

\foreach \xa in {3.5, 7.5, 11.5}{
  \draw[cssflow] (\xa+0.05,2.1) -- (\xa+0.45,2.1);
}

\tikzset{ovlow/.style={draw=black!60, rounded corners=1.5mm, line width=0.6pt, align=center,
  font=\scriptsize, minimum width=3.5cm, minimum height=1.5cm, inner sep=2pt}}
\draw[densely dashed, draw=black!50, line width=0.6pt, rounded corners=2mm] (-0.2,-5.75) rectangle (15.7,-0.65);
\node[anchor=north west, font=\footnotesize\bfseries, inner sep=0pt] at (0.0,-0.82)
  {Validation in digital evolution (Sect.~\ref{sec:comp})};
\node[ovlow, fill=black!12] (S) at (1.75,-2.05)
  {{\footnotesize\bfseries Stripped run}\\ the logged run with its\\ designated roles removed};
\node[ovlow, fill=cssOI5!35, font=\footnotesize\bfseries] (K) at (5.75,-2.05) {Causal Structure Set};
\node[ovlow, fill=cssOI2!35, font=\footnotesize\bfseries] (M) at (9.75,-2.05) {Causal motif\\ identification};
\node[ovlow, fill=cssOI2!35] (I) at (13.75,-2.05)
  {{\footnotesize\bfseries Inferred roles}\\ information carrier\\ variants, operators,\\ products};
\foreach \xa in {3.5, 7.5, 11.5}{
  \draw[cssflow] (\xa+0.05,-2.05) -- (\xa+0.45,-2.05);
}
\node[draw=black!60, rounded corners=1.5mm, fill=black!5, line width=0.6pt, align=center, font=\scriptsize,
  minimum width=4.5cm, minimum height=1.5cm, inner sep=2pt] (D) at (4.25,-4.7)
  {{\footnotesize\bfseries Digital evolution}\\ simulated run with its designated\\ roles: organisms, ancestry,\\ copies, mutations};
\draw[draw=black!75, fill=black!5, line width=0.9pt] (13.75,-3.95) -- (14.65,-4.7) -- (13.75,-5.45) -- (12.85,-4.7) -- cycle;
\node[font=\LARGE] at (13.75,-4.7) {$\stackrel{?}{=}$};
\draw[cssflow] (2.75,-3.95) -- (2.75,-2.8)
  node[midway, right=2pt, font=\scriptsize\itshape, text=black!70] {designated roles removed};
\draw[cssflow] (D.east) -- (12.85,-4.7)
  node[midway, above=1pt, font=\scriptsize\itshape, text=black!70] {designated roles, withheld};
\draw[cssflow] (I.south) -- (13.75,-3.95)
  node[midway, left=2pt, font=\scriptsize\itshape, text=black!70] {inferred roles};%

\end{tikzpicture}
\caption{\textbf{\new{Conceptual overview of the six-stage pipeline} that derives hereditary roles from an observed episode, and \new{of }its validation against designated roles withheld from the search.} Top row: a bounded spacetime episode, drawn here as world-tubes of persisting structures, enters the analysis without structural types and without designated roles. Stages 1--5 of Figure~\ref{fig:pipeline} build from it, under a declared perspective $P$, the CSS: structure instances $W$, structural labels $\ell$ and causal order $\prec$. In stage 6, a blind search identifies causal motifs, i.e., patterns of several structures and processes; the initial example tested here is the RVT criterion: replication (R), variation (V), translation (T) and lineage continuation (L), jointly for one family of variants. Where the criterion holds, the search assigns the inferred roles of an information family: \new{information }carriers $A$, $A'$ and $A''$, the operators that copy and translate them, and their products. Lower panel: the validation, which requires a record of designated roles. In Section~\ref{sec:comp}, the episode is a run of an Avida-like digital-evolution simulator, whose design fixes and whose records document the designated roles, i.e., organisms, ancestry, copies and mutations (Section~\ref{sec:comp.4}). The designated roles are removed from the logged run, the stages of the top row run on the stripped run, and the decision diamond ($\stackrel{?}{=}$) compares the inferred roles with the designated roles, which the search never reads. Section~\ref{sec:cultural} runs the stages of the top row on stipulated histories, which have no designated roles and hence permit no such validation. Box headers are coloured as in Figure~\ref{fig:pipeline}: blue for the construction of the CSS, orange for inference; node colours give structural labels; time runs upward. The figure was inspired by discussions with Priyanka Mehra (ORCID 0000-0001-9523-6689) and \v{S}ar\={u}nas Girdzijauskas (ORCID 0000-0003-4516-7317).}
\label{fig:overview}
\end{figure}

The distinction matters most where the hereditary entities are unknown, contested or themselves emerging. A framework that begins with genes cannot independently discover that a previously unrecognised structure plays their role. Nor does statistical dependence alone establish such a role: two physical quantities can be correlated without either configuration being copied and used to produce differentiated outcomes. CSS does not replace measures of statistical, functional or semantic information. It addresses the prior identification of the physical structures to which a hereditary interpretation might apply.

The guiding conjecture is that information-carrier functions arise within hereditary organisation, or through arrangements established by hereditary processes. This motivates identifying \new{information }carriers and their organisation jointly. The conjecture is not proved by defining a criterion that assigns them jointly; its general adequacy remains a scientific question.

\subsection*{A physical description before a hereditary interpretation}

The Causal Structure Set (CSS) separates what is supplied by observation from what is to be inferred. An analysis begins with a bounded spacetime episode and a declared observational \emph{perspective}: the available physical or recorded variables, their resolution, the dynamics and interventions available, and the rules for recognising persistent structures and structural types. A description is therefore not ontology-free. What must remain absent from its construction is the hereditary-role assignment that the analysis is intended to discover.

For a perspective \(P\), a CSS is written

\[
C_{P,\ell}=(W,\prec,\ell).
\]

Here, \(W\) is a finite set of persistent physical structure instances; \(\ell\) assigns each instance a structural type; and \(\prec\) records realised causal ancestry. Instances remain distinct even when they share a type. This distinguishes a succession of actual descendants from repeated independent occurrences of similar structures. A structure is represented over an interval of its existence, rather than as an instantaneous event.

Structural types must be established independently of hereditary roles. Almost-invariant regions---regions of a structure's state space with slow escape relative to internal relaxation---provide one possible basis for persistent typing. They are not compulsory. Alternative procedures are permitted when they satisfy the same requirements of physical grounding, persistence, complete and exclusive typing, and hereditary-role independence. The computational example uses declared value classes rather than an automatically constructed almost-invariant-region partition.

Causal dependencies concern the establishment or maintenance of persistent structural types. Suppressing, replacing or perturbing a source must affect whether the target type is realised, under the declared intervention rules. With known dynamics, these dependencies can be evaluated by replay. Chains of supported dependencies define causal ancestry: an ancestry relation does not mean that its endpoints have separately passed a direct intervention test.

Figure~\ref{fig:pipeline} gives the single six-stage overview of construction and inference. The first five stages construct the CSS; the sixth searches it for candidate hereditary organisation. Sections~\ref{sec:3}--\ref{sec:5} supply the definitions and Appendix~\ref{app:A} the admissibility requirements. The novelty is not the use of an acyclic graph itself, but the discipline separating physical identity, structural classification and causal ancestry from the hereditary roles subsequently assigned.

\subsection*{The candidate criterion: copying, consequential differences and continuation}

Hereditary roles are sought in patterns involving several structures and processes, not in isolated nodes. Such patterns are called causal motifs. A process is a \emph{causally convex} part of the CSS: if two of its structures have a represented causal intermediary between them, that intermediary is also included. Its earliest and latest elements define input and output boundaries. A structural type satisfies the \emph{operator condition} when its positive multiplicity is conserved across those boundaries and an input instance contributes causally to an output of another type. This identifies possible process-mediating structures without yet calling them hereditary machinery.

The initial criterion combines replication, variation and translation, abbreviated RVT, with lineage continuation. \textbf{Replication} requires a net increase in instances of a structural type, with each output instance causally descended from an input of that type. \textbf{Variation} connects distinct candidate types through realised causal processes; its definition does not presuppose mutation, editing or another domain-specific mechanism. \textbf{Translation} requires alternative variants, acting through corresponding operator organisation, to produce distinguishable output types absent from their respective input boundaries. The output distinction must depend causally on the variant distinction. Additional copies alone do not establish translation.

\textbf{Lineage continuation} requires an additional instance produced by replication to enter a subsequent replication process and itself be replicated. The minimum witness therefore contains two successive replication transitions. Longer observed lineages measure how extensively the organisation continues; they do not compensate for a missing component of the criterion.

All these relations must concern one coherent family of variants. An unrelated copying process, a transformation elsewhere and a separate production process cannot simply be combined into heredity. Only after the joint criterion holds are information-carrier, replication-operator, translation-operator and realised-product roles assigned. The resulting organisation is called an \emph{information family}. Roles are relational: a structure can participate in different roles in different processes.

The representation and this particular criterion must be distinguished. CSS specifies how to pose the inference problem; RVT with lineage continuation is an initial hypothesis about the organisation to identify. The formal development gives sufficient conditions for preserving its witnesses under admissible changes of representation and a way to compare process-boundary organisation across substrates. It does not establish invariance under arbitrary changes of resolution or demonstrate empirical transfer between domains.

\subsection*{What the conceptual examples establish}

Section~\ref{sec:6} separates the components of the criterion through synthetic positive and negative cases. A catalyst can satisfy the operator condition without belonging to a hereditary family; copying can occur without translation; and differentiated production can occur without continuing replication. A complete synthetic configuration shows how the joint criterion assigns roles only after all its conditions are supplied.

Section~\ref{sec:cultural} applies the construction to two hypothetical histories of recipe transmission. In the first, teaching establishes some copying and continuation, but the proposed two-recipe family lacks required witnesses. In the second, added modification, teaching and preparation events, together with a stipulated causal comparison, complete the family. Both use the same structural types and physical hypotheses. Their different verdicts follow from their specified histories, not from labelling a recipe as information in advance. These are conditional demonstrations of the criterion, not empirical evidence about human capability structures.

The distinction also clarifies the limits of observation. Failure to identify a role in a bounded episode does not establish that the structure lacks it. An isolated transmission may omit the wider organisation needed to recognise its information-bearing role. A larger, causally connected observation may supply that evidence, but unobserved events cannot be inserted merely because the system is familiar.

\subsection*{From formal specification to executable tests}

Section~\ref{sec:comp} implements the framework in an Avida-like digital-evolution simulator. Self-copying programs occupy a grid of 36 cells. Five development seeds were analysed; four produced births and one ended without offspring. The simulator records execution-level state changes separately from its descriptions of organisms, ancestry and mutations. These \new{designated roles} are withheld during reconstruction and search and consulted afterwards for comparison.

At instruction level, the representation consists of time-bounded variable-value sections. Memory sites retain their instruction types, while other variables are classified by kind and by whether their values are zero or non-zero. One-step interventions replay execution with a changed input and controlled random draws to establish causal dependencies.

For seed 1, 386,860,320 variable-step pairs are represented by \textbf{847,911 CSS nodes}. The four reproducing runs yield approximately 848,000--884,000 nodes each. Construction for seed 1 took 48 seconds with four worker processes on an Apple M2; this is the construction time, not the runtime of the complete motif search. The result demonstrates that the proposed instance-level representation can be built for a substantial recorded history.

The instruction-level search tests replication, lineage, variation and translation without instruction names or \new{designated roles}. It searches declared process classes, and \new{information }carrier candidates are restricted to memory labels after the initial replication pass. Under the primary causal order, it finds information families containing \textbf{5, 4, 6 and 4 instruction types}, with the same four ancestral types in every family. \new{An exploratory null that reassigns the labels at random while keeping the causal history shows that the labels align with replication and continuing lineages more than by chance: the structural part of the search finds 2--5 instruction types after reassignment, against 10--14 in the real CSS of the four runs, and the swap tests then confirm only 5 of the 13 of seed 1.} Post-inference comparison shows that 17,131 of the 17,133 processes detected as replicative in seed 1 are copy steps in the withheld record. This checks replication-event correspondence; it is not a precision estimate for complete information families.

At organism level, the analysis uses declared rules for identifying births, lifetimes, memory content at birth and time to first division. Reconstructed records agree with the withheld simulator records, including all 706 organisms in seed 1. The memory-content and division-time descriptions are supplied by these rules, not discovered without an observational specification. The inferred families contain 13, 21, 14 and 17 memory-content types across the four reproducing runs.

Memory-content swaps test the causal relationship between these descriptions. An organism receives another candidate's memory content at birth, and execution is replayed with the recorded schedule and keyed random draws. In \textbf{4,135 of 4,314 interventions (95.9\%)}, the time to first division matches the characteristic value of the inserted memory-content type. Of the remaining interventions, 166 are censored before division and 13 end without division; the latter occur in replay contexts containing copying errors. These results establish a tested causal dependence under the declared description, rather than unrestricted discovery of a genotype--phenotype relation.

\subsection*{What has been achieved, and what remains open}

The principal advance is an executable route from recorded dynamics to candidate hereditary roles, with comparisons against withheld records and intervention-based tests. The experiments also reveal where the initial formulation is restrictive or permissive.

\new{A control run without mutations has a single memory-content type and hence no information family at organism level, as expected. At instruction level it still yields a family of three instruction types, because mutation is only one source of variation: the ancestral program already contains distinct instruction types, and causal links between them satisfy V. This control therefore tests the criterion at organism level only. At instruction level, V links nearly every pair of candidate instruction types and so does not discriminate between them; specificity there remains to be shown, since the family-level null of Section~\ref{sec:comp.4} tests only whether the labels align with replication and lineage.} Conversely, the copying instruction itself is not recovered as \del{a}\new{an} \new{information }carrier in the principal instruction-level families\new{ of the development runs, although it is a member in 17 of the 39 held-out families}. In the full causal order, enclosing one executed step in a causally convex process can include later activity, changing its boundaries and preventing some expected replication witnesses. Processes spanning a complete copy loop have not yet been searched blind at instruction level.

Structural typing, process selection, operator correspondence and observational scale consequently affect what is identified. These findings delimit the current implementation; they neither establish the general adequacy of RVT nor invalidate CSS as a representation. The paper separately discusses hereditary systems for which translation, as presently formulated, may be absent or ambiguous.

The development runs establish feasibility, and held-out runs on 40 new seeds test reproducibility\new{ under a frozen preregistration: the information family recurs in all 39 held-out runs with births, but the preregistered test fails in 2 runs, because of a choice we made in defining the designated family, not because of an error in the inference (Appendix~\ref{app:H.8})}. Neither\new{ set of runs} tests whether the criterion is specific, which needs experiments with independently characterised positive and negative cases (Section~\ref{sec:8.world}). The results concern this Avida-like simulator, not Avida itself (Section~\ref{sec:comp}). Appendix~\ref{app:H} documents the computational methods; Appendices~\ref{app:B}--\ref{app:C} develop the formal results, and Appendices~\ref{app:F}--\ref{app:G} address incomplete observation and conditional reconstruction.

Selection, cumulative evolutionary accumulation and hereditary stability remain downstream questions. The longer-term opportunity is to investigate them in systems whose relevant hereditary structures have first been inferred, rather than assumed. \textbf{The contribution of this paper is to make that prior identification problem explicit, computationally accessible and open to causal tests.}
\end{newtext}

\clearpage
\tableofcontents
\clearpage

\noindent\textbf{Note on this manuscript version:} \del{The conceptual framework in this manuscript was developed across versions 1 to 5 of this preprint. This Version 6 turns it into an explicit formal argument. Three changes are substantive. The joint identification of information carriers and hereditary organisation is now stated as an operational criterion rather than as a conceptual claim. The relation to prior work on biological information, individuality inference and substrate-independent formulations of replication is set out explicitly, including where that work already makes moves the present framework builds on. And the advance over it is argued directly rather than asserted. The framework itself is substantially unchanged since version 1; what has changed is its formulation and its positioning.}
\new{The conceptual framework in this manuscript was developed across versions 1 to 5 of this preprint, and from version 6 on it is stated as an explicit formal argument.
Version 8 makes three changes.
First, it adds a computational test, Section~\ref{sec:comp}: a CSS is constructed from the complete causal record of an Avida-like digital-evolution simulator and searched blind for an information family, and the result is checked against the simulator's withheld record; Figures~\ref{fig:pipeline}, \ref{fig:pipeline-digital}, \ref{fig:css-L1}, \ref{fig:css-L2}, \ref{fig:family}, \ref{fig:zooms} and~\ref{fig:validation} are new, and Appendix~\ref{app:H} gives the methods.
Second, it reworks Section~\ref{sec:6}, with new Figures~\ref{fig:ex-6.1}--\ref{fig:ex-6.5}, and gives the reconstruction of recipe transmission, Section 6.6 of version 7, a section of its own, Section~\ref{sec:cultural}: a cultural episode, which now pairs an incomplete and a completed conditional reconstruction, Case A and Case B, with new Figures~\ref{fig:pipeline-cultural} and~\ref{fig:recipe} and the full reconstruction in the new Appendix~\ref{app:G}.
Third, it corrects errors of notation and typography and makes three conditions explicit: Section~\ref{sec:5} now requires an observed lineage depth $L \geq 2$, in line with Section~\ref{sec:4.7}; Section~\ref{sec:4.6} requires the re-entering descendant to be replicated again, and; Section~\ref{sec:4.4} states the dependence of the product distinction on the \new{information }carrier distinction, which Section~\ref{sec:5} and Appendix~\ref{app:B} already required of T.
Apart from these corrections, the representation and the criterion are unchanged.
The computational results come from development runs on seeds 1--5 and from a held-out test on seeds 101--140, run under a preregistration frozen before the runs; the held-out test checks reproducibility, not specificity (Section~\ref{sec:8.world})\new{. The information family recurs in all 39 held-out runs with births, but under the frozen rule the held-out test fails on one of its five predictions, a check of the organism-level reading (Appendix~\ref{app:H.8})}.} %

\section{Introduction}\label{sec:1}

Persistent organisation is widespread in the physical world. Atoms, crystals, stars, vortices and reaction--diffusion patterns recur because the physical dynamics repeatedly generate or maintain them. Other forms of organisation display a different kind of persistence. Biological structures regenerate through inherited molecular organisation; learned behaviours propagate through neural and material structures; institutions persist through records, people and practices \citep{Acemoglu2021}; and software organisation propagates through copied and executed machine-readable structures \citep{Lehman1980}.

These cases raise a general physical question: \textbf{when does recurrent organisation constitute hereditary organisation rather than repeated spontaneous organisation?}
Fundamental physics describes the behaviour of matter and spacetime, but it does not designate particular physical structures as genes, memories or programs. Such roles depend on how structures interact within larger organisations. In some of these organisations, persistent configurations are copied, and differences between them cause different downstream structures to be produced through corresponding physical machinery. This is the kind of information-bearing organisation investigated here: not an additional physical force or substance, but a particular causal organisation realised through ordinary physical processes.

The problem has a long history. Schrödinger emphasised the physical problem of stable hereditary order \citep{Schrodinger1992}, and von Neumann showed that self-reproduction requires an organisation in which descriptions and constructive machinery occupy distinct but interacting roles \citep{vonNeumann1966}. Modern information theory, evolutionary theory, universal biology, constructor theory, theories of individuality, autopoiesis and autocatalytic closure each illuminate parts of the same problem. Innovations in how hereditary information is stored, transmitted and interpreted have repeatedly underpinned the emergence of new organisational levels \new{\citep{Szathmary1995,Jablonka2006}}, and different inheritance systems can coexist and coevolve. Yet a fundamental identification difficulty remains.

Most theories start from entities whose relevance is already known. Population genetics begins with organisms, genes or alleles. Cultural evolution begins with transmissible traits or socially learned behaviours \new{\citep{Mesoudi2017,Beach2003}}. Software evolution begins with programs, repositories or versions. Shannon information requires specified random variables \citep{Shannon1948}; functional information requires a specified function; semantic information requires specified system and viability variables; and physical theories of replication generally begin by specifying the information medium or replicator.

These frameworks answer important questions \emph{about already identified entities}. They do not generally answer the prior inverse question:

\textbf{Given an observed physical episode and its realised causal organisation, can we infer which structures occupy hereditary roles without first specifying genes, replicators, information carriers, programs or other domain-specific hereditary entities?}

This is not merely a philosophical concern. A framework that prespecifies the hereditary entities can recognise only hereditary organisations compatible with that supplied ontology. The problem is particularly acute for origin-of-life questions, unfamiliar artificial systems, major evolutionary transitions and comparisons across physical substrates, where the relevant hereditary entities may themselves be emergent \citep{GodfreySmith2009}. Existing work explicitly recognises both the lack of a general physical measure of biological information \new{\citep{Wagner2010,vanBaalen2013}} and the difficulty of prespecifying evolutionary individuals or transmission units.

It is reasonable to ask why corresponding causal organisation should be expected across domains as different as biology, culture and software. Two considerations make the apparent recurrence unsurprising. First, physically different systems may operate under analogous organisational constraints and therefore repeatedly realise corresponding causal architectures; on this interpretation, the recurrence of concepts such as replication, maintenance, parasitism, recovery and collapse reflects genuine regularities of causal organisation. Second, scientific models are constructed by human observers using recurrent strategies of abstraction, compression and explanation, so similar conceptual structures may also be imposed on physically different phenomena. These explanations are not mutually exclusive and may apply differently to different concepts and domains. Cross-domain similarity should therefore not itself be taken as evidence of a universal causal structure. The relevant question is whether an apparent correspondence survives role-neutral reconstruction of the underlying systems and can be expressed as an explicit role-preserving equivalence between independently inferred causal organisations. The framework developed here provides a means to pose that question without assuming its answer. Which explanation dominates in any particular case is left open.

We call the task of recovering hereditary roles from realised causal organisation the \textbf{inverse-heredity problem}.

Hereditary roles are treated here as relational properties of realised causal organisation. Whether a physical structure carries hereditary information, helps reproduce it, or acquires its form through its effects depends on the causal relations in which its instances participate.

This premise is not new. The replicator/interactor distinction \citep{Hull1980} and developmental systems accounts of information \new{\citep{Oyama1985,Griffiths1994}} have long treated hereditary roles as functional rather than as kinds of structure. The central proposal of the present work is to recover those roles from a realised causal history in which none has been labelled.

To make this inference possible, we introduce the \textbf{Causal Structure Set (CSS)} as an intermediate representation between observation and hereditary interpretation. Mathematically, a CSS is a finite labelled strict partial order, representable as a \textbf{directed acyclic graph (DAG)}. Its nodes represent persistent physical structure instances, its labels describe their structural types, and its order records their realised causal ancestry.

\delmark\new{Figure~\ref{fig:pipeline} summarises how a CSS is constructed and how hereditary roles are inferred from it; Figures~\ref{fig:pipeline-cultural} and~\ref{fig:pipeline-digital} implement the same six stages for a cultural episode and for digital-evolution episodes. Unlike approaches that begin with predefined hereditary entities, CSS assigns candidate hereditary roles only after a role-neutral representation of physical structures and their causal organisation has been constructed.}

Structural labels may be visualised as node colours. These colours distinguish physical types; they do not designate hereditary functions. Different instances can therefore share a structural colour while retaining their individual identities and ancestry. Keeping structural identity separate from hereditary function is the methodological reason for introducing CSS.

Partial orders and causal DAGs are not themselves novel. The novelty is the combination of constraints on what constitutes a node, what information may enter its structural label, what causal relation is represented, how physical processes are extracted from the order, and when hereditary roles may subsequently be assigned.

Processes are described through their input and output structures. A structural type satisfies the \textbf{operator condition} when it has the same positive multiplicity at input and output and at least one of its input instances causally contributes to an output of another type. This identifies a possible process-mediating role without yet assigning any hereditary function.

As an initial operational criterion for hereditary organisation, we consider coherent \textbf{replication--variation--translation (RVT)} organisation together with continuing causal lineage structure. Replication identifies causal copying of structural variants. Variation connects different variants through realised transformations. Translation requires alternative variants, acting through shared operator organisation, to cause distinguishable downstream structural types absent from their respective process inputs. Additional copies of an input type alone do not satisfy this requirement. Descendants that subsequently re-enter the replication organisation establish increasingly deep observed hereditary lineages. Section~\ref{sec:4} develops these component conditions, and Section~\ref{sec:5} states the joint criterion under which hereditary roles are assigned.

\del{The present work is deliberately foundational and definitional. It aims to specify the object and the inference problem sufficiently precisely that algorithms, proofs, counterexamples and empirical tests can subsequently be developed.} \new{The present work makes two contributions. The first is foundational and definitional: it formalises the inverse-heredity problem and its CSS representation precisely enough that algorithms, proofs, counterexamples and empirical tests can be developed. The second is an initial executable implementation of the framework and a computational demonstration on an Avida-like digital-evolution simulator (Section~\ref{sec:comp}). The demonstration establishes feasibility, i.e., that a CSS can be constructed from a complete causal record and searched for RVT organisation without designated roles (Section~\ref{sec:comp.4}). It does not establish generality: the computational results are formal RVT detections on development runs of one small simulated world, not verified heredity.} \del{It}\new{The present work} does \textbf{not} claim that RVT is necessarily a complete criterion for every possible form of heredity. It does not define selection, cumulative evolutionary accumulation or hereditary stability. Those become distinct downstream questions once the relevant hereditary organisation has been identified.

\del{The paper makes five main contributions:}\new{The formal contribution has five main parts:}
\begin{itemize}
  \item it defines a physical, instance-level CSS representation derived from bounded observations;
  \item it introduces a constrained theory of admissible spatial and temporal coarse-grainings of persistent world-tubes;
  \item it formalises processes, operators and coherent RVT organisation within CSS;
  \item it treats lineage continuation as graded, imposing only the minimal closure requirement that one descendant re-enter replication, and treating observed depth beyond that as a measure of evidence rather than of class membership; and
  \item it defines cross-substrate hereditary equivalence through post-inference causal boundary organisation rather than material similarity.
\end{itemize}

\del{The main text develops the representation, the hereditary criterion and its discriminating cases.} \new{Section~\ref{sec:2} relates the framework to existing approaches, and Sections~\ref{sec:3} to~\ref{sec:5} develop the representation and the hereditary criterion. Three sections then show the criterion at work. Section~\ref{sec:6} gives synthetic examples and counterexamples, which start from a stipulated CSS and show what the criterion accepts and rejects. Sections~\ref{sec:cultural} and~\ref{sec:comp} run all six stages of Figure~\ref{fig:pipeline}, each in one domain and at a different level of evidence: Section~\ref{sec:cultural} reconstructs a cultural episode, recipe transmission, under stated hypotheses, and Section~\ref{sec:comp} constructs a CSS from the recorded history of digital-evolution episodes and gives computational evidence. Sections~\ref{sec:7}, \ref{sec:8} and~\ref{sec:9} state what the framework does and does not claim, discuss open problems, and conclude.} Formal material that supports but does not carry the central argument is placed in appendices: admissible perspectives (Appendix~\ref{app:A}), coarse-graining and the physical-witness results (Appendix~\ref{app:B}), cross-substrate equivalence (Appendix~\ref{app:C}), computational search (Appendix~\ref{app:D}), an extended comparison with prior frameworks (Appendix~\ref{app:E}), and composition, identifiability and quantitative extensions (Appendix~\ref{app:F}). \new{Appendix~\ref{app:G} gives the full conditional reconstruction of recipe transmission behind Section~\ref{sec:cultural}, and Appendix~\ref{app:H} the methods of Section~\ref{sec:comp}.}

\subsection{What an inverse formulation would make askable}\label{sec:1.1}

The \new{formal} contribution of this paper is a representation and a criterion. Its motivation is the class of questions that become askable once hereditary organisation can be identified rather than assumed. \del{None is addressed here; each is stated as a question the framework would open.} \new{Section~\ref{sec:comp} addresses one precondition shared by these questions, i.e., the computational feasibility of identifying hereditary organisation from a recorded causal history, and only for one small simulated system. The evolutionary questions themselves are not addressed here; each is stated as a question the framework would open.}

\textbf{Where heredity begins.} A theory that starts from genomes or organisms cannot use those categories to identify the transition at which genomes or organisms first appeared. An inverse criterion can in principle be applied to a prebiotic episode and asked when coherent consequential replication first occurs, without the answer being presupposed by the question.

\textbf{Unfamiliar and engineered systems.} The same applies wherever the hereditary entities are unknown or contested: artificial chemistries, synthetic and minimal cells, self-replicating machines, and evolving software ecosystems. In each case the substantive question is whether hereditary organisation is present at all, which prespecified frameworks cannot ask.

\textbf{Transitions in hereditary architecture.} Major transitions repeatedly change how hereditary information is stored, transmitted and interpreted \new{\citep{Szathmary1995,Jablonka2006}}. Because CSS assigns roles relationally and permits nesting, a transition appears as a change in which structures occupy which roles, rather than requiring a new ontology at each level.

\textbf{Stability, fragility and collapse.} Generic indicators of approaching critical transitions have been proposed across ecosystems, climate and physiology \citep{Scheffer2009}, and low-dimensional resilience relations can emerge across physically different networks \citep{Gao2016}. These methods require a specified state variable, network or dynamics. Predictors derived from inferred hereditary organisation would not, and could therefore in principle apply to systems whose relevant variables have not yet been identified.

\textbf{Parasitism and exploitation.} Parasitism is described independently in biology, culture, institutions and software, but each description uses domain-specific entities. In CSS terms a candidate formulation becomes available: \del{a}\new{an} \new{information }carrier family that propagates through the replication and translation operators of another family without contributing to that family's continued propagation. Whether this captures what the separate domain accounts describe is an empirical question, and one that cannot be posed without first identifying the families and operators.

\textbf{Transferable intervention principles.} If two physically different systems instantiate equivalent hereditary organisation under the criterion of Appendix~\ref{app:C}, then interventions effective in one may become candidate interventions in the other.

The common feature is that each question requires the hereditary organisation to be identified before it can be posed. That identification is the problem addressed here.

\subsection{What the framework offers established fields}\label{sec:1.2}

The inverse formulation stands in different relations to different domains, and the distinction is not between fields but between situations within them.

\textbf{Where hereditary entities are independently characterised, the framework can be validated.} Digital-evolution platforms record genomes, replication events and lineages by construction \new{\citep{Ofria2005,Lenski2003}}. Recovering those entities is of no practical value there, but that is what makes such platforms suitable as reference environments: \del{independently established annotations}\new{designated roles} can be withheld during inference and used afterwards to assess what the framework recovers and what it misses. \new{Section~\ref{sec:comp} uses an Avida-like simulator in this way: the simulator's record of organisms, memory contents and mutations is withheld from the inference and used afterwards to check what a blind search of the CSS recovers.
In the development runs, the organisms found in the CSS match that record in every run, e.g., 706 of 706 in seed 1 (Section~\ref{sec:comp.2}).} %
The same logic applies wherever conventional entities are well established. Asking whether independently reconstructed causal organisation recovers them, whether alternative role assignments are admissible, and how one system's organisation corresponds to another's, remains meaningful even where nothing is in doubt.

\textbf{Where hereditary status is uncertain, the framework offers candidate discovery.} In origin-of-life research, a central question is when hereditary organisation first arises. Approaches using genome or organism concepts can propose candidate entities and test when their defining properties emerge; the specific difficulty is assuming the hereditary status that the analysis is intended to establish. What CSS attempts is to identify the relevant participants without requiring their hereditary roles to be designated beforehand. Related motivations underlie autocatalytic and assembly-theoretic approaches \new{\citep{Hordijk2018,Huson2024,Sharma2023}}. Cultural and institutional inheritance presents the same difficulty, since transmission units depend on domain-specific entities and analytical granularity \citep{OBrien2010}, though the observational requirements are considerably harder to meet.

Software ecosystems are of interest for a different reason. Versioned artefacts, build environments and deployments leave unusually complete records, and reproducible builds and continuous integration permit some dependencies to be re-evaluated rather than only observed. These records are evidence from which a CSS might be reconstructed; they are not themselves the required causal genealogy, and structures such as compilers and build systems are candidate physical implementations of CSS-derived hereditary \emph{operators} (cf\del{section}\new{.\ Section}~\ref{sec:3.9}) rather than operators by designation. Whether they satisfy the operator condition depends on the represented process boundaries, structural labels and causal relations.

\textbf{Four uses of the inferred representation.} Once hereditary roles have been inferred, the same representation supports several further operations.
\begin{enumerate}[label=\arabic*.]
  \item Explicit grounds become available for rejecting a candidate role assignment: a specified candidate can be shown to fail a stated RVT condition.
  \item Two systems can be compared through the boundary equivalence of Appendix~\ref{app:C}, which asks whether they instantiate corresponding hereditary organisation rather than whether both are described using the same vocabulary.
  \item Distinctions that informal usage combines can be examined, such as the difference between propagation through another family's operators and propagation using its resources (Section~\ref{sec:8.5} shows how this is a proposed application).
  \item And observed lineage depth provides a specified quantitative descriptor of inferred lineage continuation, comparable across substrates because it is defined relative to the same criterion in each.
\end{enumerate}

None of these operations is individually unavailable to frameworks that prespecify hereditary entities. What may be distinctive is performing them from a single role-neutral representation, so that the entities being compared, rejected or measured are themselves outputs of the analysis rather than inputs to it.

Section~\ref{sec:2} and Appendix~\ref{app:E} provide more details about the current work and its relation to the state of the art.

\begin{newtext}
\subsection{A computational test on a digital-evolution simulator}\label{sec:1.3}

One run of the Avida-like simulator of Section~\ref{sec:comp}, 36 cells over 100 updates, holds 4,752 state variables at each of 81,410 executed steps, i.e., $3.9 \times 10^{8}$ variable-step pairs, and its record states the causes of every change among them. %
Section~\ref{sec:comp} implements the six-stage procedure of Figure~\ref{fig:pipeline} on such records (Figure~\ref{fig:pipeline-digital}).
The recorded execution history provides the observed episode; persistent variable sections are represented as nodes with declared structural labels; causal dependencies are reconstructed by one-step intervention replay (Section~\ref{sec:3.5}), and; the resulting CSS is searched blind, i.e., without instruction names or \del{simulator annotations}\new{designated roles}, for candidate information families.
The inferred organisations are then compared with the withheld simulator record and tested by interventions.
These comparisons are post-inference validation, not additional stages of CSS construction.

The test gives three main results. First, at the level of organisms, the organisms, their births, their memory contents at birth, G, and their times to first division, P, found blind in the CSS match the simulator's withheld record in every run\new{, development and held-out}; P follows the inserted memory content in 808 of 824 memory-content swaps in seed 1, and none gives another value of P. %
Second, at the level of instructions, a role-blind search in the order of the CSS as defined finds one information family in each of the four development runs with births, with 5, 4, 6 and 4 instruction labels and the same 4 ancestral instructions, \texttt{nop-A}, \texttt{nop-B}, \texttt{nop-C} and \texttt{h-alloc}, in every family; the copy instruction \texttt{h-copy} is a member of none, because in that order a single executed instruction is not a causally closed process (Section~\ref{sec:comp.3}).\new{ This development observation does not carry over: in the held-out runs, \texttt{h-copy} is a member in 17 of 39 (Appendix~\ref{app:H.8}).} %
Third, the CSS of seed 1 condenses these variable-step pairs into 847,911 nodes, a 456-fold reduction, and forms a strict partial order. %
The significance of the test lies in that the hereditary roles, i.e., which instructions act as \new{information }carriers, which CPU state copies them and which products their execution yields, are outputs of the search rather than inputs to it; the organisms are reconstructed blind from the CSS, whereas G and P follow rules we declared (Section~\ref{sec:comp.2}).

The scope of the test is narrow. The development runs on seeds 1--5 establish feasibility, and held-out runs on seeds 101--140 test reproducibility, not specificity (Section~\ref{sec:8.world})\new{. In the held-out runs, the information family recurs in 39 of 39 runs with births, but under the frozen rule the test fails on its check of the organism-level reading, in 2 runs, because the reference definition gives one heritable genotype no phenotype (Appendix~\ref{app:H.8})}. %
The simulator reconstructs Avida's semantics but is not Avida itself, and its world holds only 36 cells. %
\new{A control run without mutations has no information family at the level of organisms, as expected, because all organisms then carry the same program; at the level of instructions it still has one, because the instruction types of the ancestral program differ without any mutation, so this control tests the criterion at the level of organisms only (Section~\ref{sec:comp.4}).} \new{An exploratory family-level null, which reassigns the labels at random while keeping the causal history, shows that the labels align with replication and continuing lineages more than by chance (Section~\ref{sec:comp.4}); it does not test the specificity of the full criterion.} %

\begin{figure}[htbp]
\begin{newtext}
\centering
\begin{tikzpicture}[cssfig,
  pipenum/.style={circle, minimum size=4.6mm, inner sep=0pt, line width=0.5pt,
    draw=cssOI5!70!black, fill=cssOI5!35, font=\footnotesize\bfseries},
  pipeinf/.style={circle, minimum size=4.6mm, inner sep=0pt, line width=0.5pt,
    draw=cssOI2!70!black, fill=cssOI2!35, font=\footnotesize\bfseries}]

\node[anchor=north west, inner sep=0pt] (h) at (0,0) {%
\begin{tabular}{@{\hspace{0.6cm}}>{\raggedright}p{2.85cm}@{\hspace{0.12cm}}>{\raggedright}p{11.62cm}@{}}
\rule{0pt}{0.3cm}\textbf{Stage}
&
\rule{0pt}{0.3cm}\textbf{General symbolic operation}
\end{tabular}};
\draw[line width=0.5pt] ([yshift=-0.03cm]h.south west) -- ([yshift=-0.03cm]h.south east);

\node[anchor=north west, inner sep=0pt] (r1) at ([yshift=-0.08cm]h.south west) {%
\begin{tabular}{@{\hspace{0.6cm}}>{\raggedright}p{2.85cm}@{\hspace{0.12cm}}>{\raggedright}p{11.62cm}@{}}
\rule{0pt}{0.62cm}\textbf{Perspective}
&
\rule{0pt}{0.62cm}$(\mathcal{R},P)\xrightarrow{\pi_P,\,\rho_P}(X_P,\,x_P(t))$\\
$P=(\pi_P,D_P,J_P,\rho_P,\Sigma_P)$\par\vspace{1pt}
{\scriptsize\color{black!70}$\pi_P$ state map, $D_P$ dynamics, $J_P$ interventions, $\rho_P$ resolution, $\Sigma_P$ individuation\par}
\end{tabular}};

\node[anchor=north west, inner sep=0pt] (r2) at ([yshift=-0.08cm]r1.south west) {%
\begin{tabular}{@{\hspace{0.6cm}}>{\raggedright}p{2.85cm}@{\hspace{0.12cm}}>{\raggedright}p{11.62cm}@{}}
\rule{0pt}{0.45cm}\textbf{Structures}
&
\rule{0pt}{0.45cm}$(\pi_P,\Sigma_P,\rho_P)\to Y_P$\\
$S_j\subset\mathcal{R},\quad z_{S_j}(t)\in Y_P$\par\vspace{1pt}
{\scriptsize\color{black!70}Candidates: world-tube sections in $Y_P$.\par}
\end{tabular}};

\node[anchor=north west, inner sep=0pt] (r3) at ([yshift=-0.08cm]r2.south west) {%
\begin{tabular}{@{\hspace{0.6cm}}>{\raggedright}p{2.85cm}@{\hspace{0.12cm}}>{\raggedright}p{11.62cm}@{}}
\rule{0pt}{0.45cm}\textbf{Persistence and types}
&
\rule{0pt}{0.45cm}$(\Sigma_P,\rho_P,\mathcal{R})\to\mathcal{S},\quad(\Sigma_P,Y_P)\to\mathcal{L}_P$\par\vspace{1pt}
{\scriptsize\color{black!70}\textit{(optional: AIR partition)} $\mathcal{A}(P)=\{A_c : c\in\mathcal{L}_P\}$\par}
\vspace{1pt}
{\scriptsize\color{black!70}Structural types carry no hereditary role.\par}
\end{tabular}};

\node[anchor=north west, inner sep=0pt] (r4) at ([yshift=-0.08cm]r3.south west) {%
\begin{tabular}{@{\hspace{0.6cm}}>{\raggedright}p{2.85cm}@{\hspace{0.12cm}}>{\raggedright}p{11.62cm}@{}}
\rule{0pt}{0.45cm}\textbf{Nodes and labels}
&
\rule{0pt}{0.45cm}$M_P\colon\mathcal{S}\to W,\quad M_P(S_j)=s_j$\\
$\ell\colon W\to\mathcal{L}_P$\par\vspace{2pt}
\begin{tabular}[t]{@{}c@{\hspace{2pt}}c@{\hspace{2pt}}c@{\hspace{2pt}}c@{\hspace{2pt}}c@{}}
$S_j$ & $\to$ & $s_j$ & $\to$ & $\ell(s_j)$\\[-1pt]
{\scriptsize\itshape\color{black!70}physical section} & & {\scriptsize\itshape\color{black!70}CSS node} & & {\scriptsize\itshape\color{black!70}structural type}
\end{tabular}
\end{tabular}};

\node[anchor=north west, inner sep=0pt] (r5) at ([yshift=-0.08cm]r4.south west) {%
\begin{tabular}{@{\hspace{0.6cm}}>{\raggedright}p{2.85cm}@{\hspace{0.12cm}}>{\raggedright}p{11.62cm}@{}}
\rule{0pt}{0.45cm}\textbf{Causal order and CSS}
&
\rule{0pt}{0.45cm}$(W,D_P,J_P,X_P,\ell)\to D$\\
$\prec\,=D^{+},\quad C_{P,\ell}=(W,\prec,\ell)$\par\vspace{1pt}
{\scriptsize\color{black!70}$D$: direct, supported dependencies; $\prec$: ancestry by transitive closure.\par}
\end{tabular}};

\node[anchor=north west, inner sep=0pt] (r6) at ([yshift=-0.08cm]r5.south west) {%
\begin{tabular}{@{\hspace{0.6cm}}>{\raggedright}p{2.85cm}@{\hspace{0.12cm}}>{\raggedright}p{11.62cm}@{}}
\rule{0pt}{0.45cm}\textbf{Processes, motifs, roles}
&
\rule{0pt}{0.45cm}$C_{P,\ell}\to\{Q\},\quad Q\to I(Q),\,O(Q),\,\mathrm{Op}(Q)$\\
RVT motifs $+$ lineage\\
$\to$ candidate hereditary roles\par\vspace{1pt}
{\scriptsize\color{black!70}Roles only after the joint criterion; a conserved label is first an operator candidate.\par}
\end{tabular}};

\foreach \rowname in {r1, r2, r3, r4}{
  \draw[black!30, line width=0.3pt] ([yshift=-0.04cm]\rowname.south west) -- ([yshift=-0.04cm]\rowname.south east);
}
\draw[black!70, line width=0.7pt] ([yshift=-0.04cm]r5.south west) -- ([yshift=-0.04cm]r5.south east);

\node[pipenum] (c1) at ([xshift=0.3cm, yshift=-0.49cm]r1.north west) {1};
\node[pipenum] (c2) at ([xshift=0.3cm, yshift=-0.32cm]r2.north west) {2};
\node[pipenum] (c3) at ([xshift=0.3cm, yshift=-0.32cm]r3.north west) {3};
\node[pipenum] (c4) at ([xshift=0.3cm, yshift=-0.32cm]r4.north west) {4};
\node[pipenum] (c5) at ([xshift=0.3cm, yshift=-0.32cm]r5.north west) {5};
\node[pipeinf] (c6) at ([xshift=0.3cm, yshift=-0.32cm]r6.north west) {6};
\foreach \cfrom/\cto in {c1/c2, c2/c3, c3/c4, c4/c5, c5/c6}{
  \draw[cssdep, draw=black!55] (\cfrom) -- (\cto);
}

\draw[line width=0.9pt, cssOI5!70!black] ([xshift=-0.06cm, yshift=-0.03cm]r1.north west) -- ([xshift=-0.17cm, yshift=-0.03cm]r1.north west)
  -- ([xshift=-0.17cm, yshift=0.03cm]r5.south west) -- ([xshift=-0.06cm, yshift=0.03cm]r5.south west);
\path (r1.north west) -- (r5.south west) coordinate[pos=0.5] (m1);
\node[rotate=90, anchor=south, text=cssOI5!45!black, inner sep=1pt] at ([xshift=-0.24cm]m1) {construction of the CSS};
\draw[line width=0.9pt, cssOI2!70!black] ([xshift=-0.06cm, yshift=-0.03cm]r6.north west) -- ([xshift=-0.17cm, yshift=-0.03cm]r6.north west)
  -- ([xshift=-0.17cm, yshift=0.03cm]r6.south west) -- ([xshift=-0.06cm, yshift=0.03cm]r6.south west);
\path (r6.north west) -- (r6.south west) coordinate[pos=0.5] (m2);
\node[rotate=90, anchor=south, text=cssOI2!55!black, inner sep=1pt] at ([xshift=-0.24cm]m2) {inference};

\node[anchor=north west, draw=black!60, densely dashed, line width=0.6pt, rounded corners=1mm,
  inner xsep=4pt, inner ysep=7pt, text width=14.86cm, align=left, font=\scriptsize] (ft) at ([yshift=-0.4cm]r6.south west)
  {\textbf{Post-inference validation:} withheld designated roles, interventions and controls, as far as the domain provides them.};
\node[csstag, fill=white, anchor=east, inner sep=2pt] at ([xshift=-0.35cm]ft.north east) {assesses the output of stage 6};

\node[anchor=north west, draw=black!60, line width=0.6pt, rounded corners=1mm,
  inner xsep=4pt, inner ysep=5pt, text width=14.86cm, align=left, font=\scriptsize] (dv) at ([yshift=-0.25cm]ft.south west)
  {\textbf{Domain versions:} Figure~\ref{fig:pipeline-cultural}, a cultural episode reconstructed under stipulated hypotheses (Section~\ref{sec:cultural}), and Figure~\ref{fig:pipeline-digital}, digital-evolution episodes constructed from a recorded history (Section~\ref{sec:comp}).};

\end{tikzpicture}
\caption{\textbf{Six-stage construction of a Causal Structure Set and inference of candidate hereditary roles.} Stages 1--5 transform a bounded episode $\mathcal{R}$, observed under a declared perspective $P$, into the labelled causal order $C_{P,\ell} = (W, \prec, \ell)$ of CSS nodes $W$, causal ancestry $\prec$ and structural labels $\ell$. Stage 6 identifies causally convex processes and evaluates coherent RVT-and-lineage motifs before it assigns hereditary roles. Structural labels carry no predefined hereditary meaning. Structural typing by almost-invariant regions (AIRs, Section~\ref{sec:3.3}) is one optional route. Figures~\ref{fig:pipeline-cultural} and~\ref{fig:pipeline-digital} give the two domain versions: a cultural episode reconstructed under stipulated hypotheses (Section~\ref{sec:cultural}) and digital-evolution episodes constructed from a recorded history (Section~\ref{sec:comp}). Validation, e.g., against withheld designated roles or by interventions, assesses the output of stage 6 and is performed separately, after inference. Arrows denote operations relative to a chosen perspective, not unique or automatically valid representations.}
\label{fig:pipeline}
\end{newtext}
\end{figure}
\end{newtext}

\section{Relation to existing approaches}\label{sec:2}

\subsection{Information theories address complementary questions}\label{sec:2.1}

Shannon information quantifies uncertainty and statistical dependence between specified variables \citep{Shannon1948}. Landauer established the physical cost of information processing \citep{Landauer1991}. Functional information quantifies the fraction of configurations that achieve a specified function \citep{Hazen2007}. Semantic-information approaches ask which correlations are causally necessary for a specified system to maintain viability \citep{Kolchinsky2018}.

CSS does not replace these concepts. Its target occurs one inferential step earlier.

If a DNA sequence is given as a variable, Shannon theory can quantify its uncertainty. If a biological function is specified, functional information can quantify configurations achieving it. If an organism and viability criterion have been specified, semantic information can identify correlations that matter to viability.

CSS asks instead:

\textbf{Which realised structures, if any, should first be treated as the hereditary variables?}

The distinction is between assigning an information measure to a chosen physical description and inferring from causal organisation which physical structures occupy an information-bearing hereditary role.

\textbf{Information and heredity presuppose one another}

How do physical structures first acquire the ability to carry information that can be copied and used to produce other structures? We conjecture that the answer is inseparable from the emergence of hereditary organisation.

Shannon's theory specifies a source, transmitter, channel, receiver and destination, and establishes quantitative limits on communication \citep{Shannon1948}. His statement that semantic aspects are ``irrelevant to the engineering problem'' makes its scope explicit: the theory characterises communication within a specified system; it does not explain how the physical mechanisms implementing that system first arise.

In contrast, biological accounts connect information to heredity and adaptation, including the acquisition of information about an environment \citep{Kimura1961,MaynardSmith2000,Adami2012}. They typically begin with specified genomes, populations and environments. Constructor Theory instead characterises information media through physically possible and impossible transformations \citep{Deutsch2015}. Characterising such capabilities and identifying how \new{information }carrier functions arise in a realised physical history are distinct explanatory tasks.

It is worth separating two layers of Shannon's framework, since they behave very differently here, and conflating them has caused sustained confusion.

The first layer is a family of quantities defined over probability distributions: entropy, conditional entropy and mutual information (MI). These require only random variables with a joint distribution. There is no source, no encoder, no channel and no receiver. MI is symmetric, so it does not describe transmission in any direction. Wherever two physical quantities are statistically dependent, \del{a}\new{an} MI can be computed: {\new{$I(X;Y)$}} is the reduction in the description length of {\new{$X$}} afforded by knowing {\new{$Y$}}. MI says how much more economically one variable can be described once the other is known. But that is a property of descriptions, not of the things described. \textbf{Mutual information attributes nothing informational to the physical structures the variables stand for.} The ice does not become an information carrier because the description of past climate shortens when the ice is examined, any more than a shadow becomes a description of the object casting it. The word \emph{information} in \emph{mutual information} names a property of representations; it does not locate anything in either variable's referent.

This has been a persistent source of confusion, and it is worth stating plainly. From the fact that {\new{$I(X;Y) > 0$}} it follows only that {\new{$X$}} and {\new{$Y$}} are statistically dependent. It does not follow that {\new{$X$}} encodes {\new{$Y$}}, that {\new{$X$}} is about {\new{$Y$}}, that {\new{$X$}} carries a message, or that any structure in the world is doing anything informational at all. Ordinary physical processes generate such dependence everywhere: between crater sizes and impact energies, between tree rings and rainfall, between rock strata and depositional conditions. The universe is saturated with mutual information and, on the account developed here, contained no information carriers whatever for most of its history.

The second layer in Shannon theory is the communication setting: a source selecting among alternative messages, an encoder producing a signal, a channel, and a receiver decoding into distinguishable outputs. Shannon's principal results belong here, and channel capacity is defined as the maximum of the mutual information over input distributions, so the second layer is built from the first without being implied by it. A dependence can be computed anywhere; a communication architecture is a physical arrangement that is either present or absent.

Described physically, that architecture is \del{a}\new{an} \new{information }carrier architecture in the sense used here: alternative configurations at one location are copied into another medium and produce distinguishable further states through shared decoding machinery. Shannon's framework presupposes such an arrangement and quantifies what it can achieve. The question pursued here is the prior one of whether such an arrangement is present in a physical episode at all, and which structures constitute it. These are two questions about the same phenomenon, not two kinds of information.

\textbf{An example: the ice core.} Consider four stages. Atmospheric conditions at some past time determine the composition of the snow that falls and is buried. The snow compacts into ice, whose isotopic and gas composition varies with depth according to the conditions under which each layer formed. Much later, a core is drilled and measured, and the measured values are written down. Those records are then copied, circulated, entered into analyses and used to produce figures, models and reports, which are themselves copied, corrected and revised.

\emph{The first two stages support the first layer (mutual information) only.} Because the ice composition depends on the past atmospheric state, {\new{$I(\mathrm{ice};\mathrm{climate})$}} is well posed and positive. Nothing further follows. There is no source selecting among alternatives, no encoder, no signal and no receiver. The ice is a persistent physical structure with a composition; the criterion assigns it no hereditary role, and neither does the communication layer of Shannon's framework.

\emph{The recording practice.} The fourth stage is different in every respect. Measured values are written down and the records are copied, so distinguishable configurations are replicated. Different cores yield different values, so alternative configurations exist and propagate. And different recorded values cause different downstream structures to be produced through the same analytical apparatus: different figures, different models, different conclusions. Replication, variation and variant-dependent production are all present here and none is present in the ice. The core participates as an input to measurement and acquires no role from doing so.

\emph{What the example shows.} The dependence between ice and climate obtains whether or not anyone examines it. The quantity representing that dependence, however, is computed relative to a chosen variable set and an assumed distribution, and is instantiated as a written figure or a remembered value. It is therefore instantiated in exactly the kind of structure the criterion identifies as \del{a}\new{an} \new{information }carrier candidate, and in none of the ice, the snow or the atmosphere. The measure of the dependence is carried; the dependence itself is not. Which structures qualify follows from the causal organisation in which they participate, not from what they are made of and not from what they depend on.

\textbf{A second illustration.} A written recipe can be copied, and differences between recipes cause differences between the dishes prepared. The same three features are present as in the recording practice above. \new{The illustration is intuitive only: copying a recipe and cooking from it does not by itself satisfy the RVT-and-lineage criterion of Section~\ref{sec:5}, which also requires variation connecting the recipes, corresponding operators, a causal comparison showing that the dish depends on the recipe distinction, and a copy that is copied again. Section~\ref{sec:cultural} applies the criterion to recipe transmission between people in two conditional reconstructions, one that lacks these relations and one that supplies them by stipulation.}

We use \emph{carrier information} for this second sense. Candidates can be identified without initially assuming heredity: distinguishable physical configurations are copied, and their differences cause different further structures to be produced through a common physical process. ``Interpreted'' denotes this physical consequence, and requires neither meaning nor intention. Nothing in it requires an observer either: the same three features appear in the copying of nucleic acid sequences by a polymerase and their differentiated expression through shared cellular machinery, with no observer anywhere in the process. These observations do not by themselves establish continuing heredity: a program can be copied and used once, with no descendant subsequently copied again. A Shannon communication system is a further instance of this: messages are encoded, transmitted and decoded, but the received message does not re-enter transmission, so replication and interpretation occur without hereditary continuation.

\textbf{The asymmetry.} This is the framework's starting point. Physical and chemical processes generate statistical dependence everywhere and copy nothing. Carrier information appears only with organisation that replicates configurations and makes their differences consequential, and each subsequent innovation in how such configurations are stored, transmitted and interpreted has opened a further organisational level: nucleic acids, transmissible neural patterns, durable records, machine-readable code \citep{Szathmary1995,Jablonka2006}.

\begin{conjecture}
Hereditary organisation requires such physical \new{information }carriers. Conversely, \new{information }carrier functions arise only within hereditary organisation, or through arrangements established by hereditary processes.
\end{conjecture}

\del{A}\new{An} \new{information }carrier can continue to function even when transmission ends. A plant that leaves no offspring still uses its inherited genome in development and maintenance. Likewise, a copied program can direct a machine without ever being copied again. The conjecture therefore concerns both present hereditary participation and the hereditary processes that established the \new{information }carrier function. Manufactured devices are one instance of this broader distinction: a machine's memory carries information because human hereditary organisation established the practices that built it, not because the machine reproduces. Establishing such a historical connection requires tracing the relevant inherited configurations and their causal contribution to that function; the material object's ancestry alone is insufficient.

The conjecture permits \new{information }carriers and hereditary organisation to emerge together from non-hereditary precursors. It does not require an earlier hereditary system in every case.

The conjecture motivates identifying \new{information }carriers and hereditary organisation jointly. The Causal Structure Set framework provides a representation and an initial RVT-and-lineage criterion for that inference. Within the framework, their identification is joint by construction; whether the proposed interdependence holds generally in physical systems remains an empirical question.

\subsection{Physical theories of replication}\label{sec:2.2}

Constructor Theory of Information describes information media through physically possible copying and permutation tasks \citep{Deutsch2015}, and Constructor Theory of Life develops substrate-independent physical requirements for replication and self-reproduction \citep{Marletto2015}.

These theories are particularly important because they demonstrate that replication and information processing can be formulated without reference to particular terrestrial biochemistry.

CSS approaches the problem from the opposite direction. Constructor theory primarily asks whether specified transformations are \emph{possible}. CSS asks what hereditary organisation is \emph{realised in an observed episode}.

Thus the question is not ``Could this structure replicate?'' but ``Did realised instances of this structural type occupy a replication or \new{information }carrier role in the observed causal history?''\del{.}

This distinction becomes important below. CSS contains no unobserved capacity relation.

\subsection{Closure and self-maintenance}\label{sec:2.3}

Autopoiesis \citep{Maturana1980} and autocatalytic and RAF theories \citep{Hordijk2018,Huson2024} characterise aspects of self-production and organisational closure. They show how relationships among components can support their collective regeneration. Work on autocatalytic sets also investigates conditions for heredity and evolvability \del{(Hordijk and Steel, 2014)}\new{\citep{Hordijk2014}}.
Self-production also leaves open how hereditary distinctions are preserved despite copying errors. Eigen's analysis relates the amount of sequence information maintainable under selection to copying fidelity \citep{Eigen1971}; these constraints remain central to models of prebiotic replicator organisation \citep{Szilagyi2017}. They motivate examining whether the products generated by hereditary variants can influence the accuracy of subsequent replication, a possibility considered in Section~\ref{sec:4.5}.

CSS offers a complementary question: \textbf{which structures within such a system, if any, function as information carriers?} Autocatalytic closure alone does not determine whether some or all components occupy \new{information }carrier roles, or whether those roles arise at the level of persistent assemblies. Given an admissible representation of a sufficiently observed episode, CSS provides a means to test these possibilities without assigning \new{information }carrier status beforehand.

The distinction is between establishing that components collectively support their production and establishing that particular structural variants are propagated, generate differentiated downstream structures and continue through causal lineages. The same physical system may exhibit both forms of organisation. \del{A}\new{An} \new{information }carrier may also depend on a wider self-maintaining system without itself constituting that system.

Autocatalytic closure therefore provides a setting in which the hereditary criterion can be investigated. CSS aims to identify where \new{information }carrier roles occur within that organisation and how they relate to the processes sustaining it.

\subsection{Individuality, historical organisation and universal biology}\label{sec:2.4}

Work on the information theory of individuality has demonstrated that individuals may sometimes be inferred from dynamical organisation rather than assumed a priori \citep{Krakauer2020}. Universal-biology approaches seek principles of living organisation that are less dependent on terrestrial biochemical details \citep{Goldenfeld2017}. Assembly Theory identifies historical-selection signatures associated with the pathways required to construct complex objects \citep{Sharma2023}.

These developments establish that domain-independent abstractions of biological organisation are possible.

The specific inverse problem addressed by CSS differs in that CSS preserves individual realised structure instances and causal ancestry so that \textbf{hereditary role itself} can subsequently be inferred. The relevant question is not simply whether a structure has a complicated history or belongs to an individual, but whether a family of structural variants participates in a particular causal organisation of replication, variation, differentiated realisation and lineage continuation.

\subsection{Causal graphs and causal sets}\label{sec:2.5}

CSS is representable as a directed acyclic graph (DAG). Partial orders and directed graphs are established representations of causal structure in at least two distinct settings. In causal discovery, DAGs relate statistical variables, and the task is to infer edges among a specified variable set from observational or interventional data \new{\citep{Spirtes2001,Vowels2022}}. In causal set theory, a locally finite partial order is proposed as the fundamental structure of spacetime, its elements being discrete events without internal structure or persistence \new{\citep{Sorkin1991,Surya2019}}.

We therefore do not claim that CSS is novel because its graph is acyclic or because causality can be expressed as an order. Neither is any single one of the following properties unique to it. The methodological distinction lies in their combination.

First, CSS nodes are persistent realised physical structure instances, not statistical variables or discrete spacetime events.

Second, the structural label attached to a node is fixed before hereditary inference and contains no hereditary role.

Third, the causal order records the realised ancestry required to distinguish genealogical continuation from mere recurrence of structurally similar states.

Fourth, physical processes are represented as causally convex subsets with explicit input and output boundaries.

Fifth, hereditary roles are not carried by the representation but assigned through joint motif identification within it.

The comparison with causal set theory is instructive on a further point. Its order relation is causal precedence: whether one event lies in the causal past of another. The CSS relation is stronger, requiring that suppressing or perturbing the source reduce the probability that the target's structural type is realised (Section~\ref{sec:3.5}). Two structures may therefore stand in causal precedence without standing in $\prec$. The frameworks also differ in purpose: causal set theory proposes a fundamental structure for spacetime, whereas CSS represents an observed organisational history at a declared resolution.

The comparison with causal discovery is different again. In standard formulations the variable set is supplied and the edges are inferred. CSS likewise depends on a declared observational perspective, which supplies primitives, resolution and structure-detection rules (Appendix~\ref{app:A}); what differs is the inference target. Causal discovery infers relations among given variables; CSS infers which structures occupy hereditary roles within a reconstructed history.

Thus the CSS contribution should not be understood as a new general graph data structure. It is a representational discipline for posing an inverse organisational inference problem.

\subsection{The specific boundary against the closest precedents}\label{sec:2.6}

CSS is not the first framework to \emph{infer} higher-level organisation rather than \emph{prespecify} it. Several existing frameworks move substantially beyond theories that simply assume genes, organisms or other evolutionary units, and each removes an important piece of domain-specific ontology. The distinction from CSS is not that these approaches are domain specific while CSS is generic. Rather, each solves a different part of the inverse problem and stops before the \textbf{joint identification of hereditary \new{information }carriers and hereditary process roles from a realised causal history}. Appendix~\ref{app:E} develops each comparison in detail; the essential boundaries are as follows.

\textbf{Krakauer et al.} \citep{Krakauer2020} show that the \emph{individual} can be inferred from temporal informational organisation rather than prespecified. Their inference target is a system--environment boundary. CSS does not primarily seek such a boundary, and a CSS may contain no uniquely privileged individual at all; what CSS aims to recover is a relational division of hereditary labour among \new{information }carrier variants, operators and realised outputs.

\textbf{Walker and Davies} \citep{Walker2013} argue that the \emph{informational causal architecture of life} should be identified as an emergent transition in causal architecture rather than equated with replication alone, and their separation of stored instructions from implementing machinery is a direct conceptual antecedent to T. They do not, however, provide an instance-level inverse procedure that begins with hereditary-role-neutral persistent structures and returns \new{information }carrier variants, replication operators, translation operators, realised products and genealogical relations.

\textbf{Assembly Theory} \citep{Sharma2023} shows that \emph{historical selection} can be inferred from properties of observed object ensembles without prespecifying conventional units of selection. It does not attempt to recover the hereditary causal role decomposition responsible for those copies.

\textbf{Constructor Theory} \citep{Deutsch2015,Marletto2015} gives substrate-independent physical definitions of \emph{information media, copying and replication} in terms of possible and impossible transformations. It characterises capability rather than inferring which structures occupied hereditary roles in a realised episode.

CSS builds directly at the intersection of these moves. Its additional step is to infer the hereditary ontology itself from realised causal organisation. The claim is not merely that the \new{information }carrier can be identified independently. \del{Carrier}\new{Information carrier} status is defined only through its participation in the same causal organisation that establishes heredity; conversely, the hereditary process is not recognised until its \new{information }carrier variants and corresponding replication and translation operators can be identified. The inference is therefore \textbf{joint}:

\new{information }carrier variants $\Longleftrightarrow$ R/V/T organisation + genealogy.

Neither side is supplied independently as ground truth.

\begin{table}[htbp]
\centering
\small
\begin{tabularx}{\textwidth}{@{}>{\raggedright\arraybackslash}p{0.12\textwidth}>{\raggedright\arraybackslash}X>{\raggedright\arraybackslash}X>{\raggedright\arraybackslash}X@{}}
\textbf{Framework} & \textbf{What is inferred rather than assumed?} & \textbf{Input abstraction it still requires} & \textbf{Where it stops relative to CSS} \\
\midrule
Krakauer et al. & Individual/system boundary and degree/type of individuality & Time-indexed microvariables and candidate coarse-grainings & Does not jointly identify \new{information }carrier variants, R/T operators and realised hereditary genealogy \\
Walker \& Davies & Transition to life as emergence of information control and context-dependent/top-down causation & System-level dynamical/informational organisation & Does not provide a role-neutral instance-level inverse algorithm for identifying the physical bearers of \new{information }carrier/operator roles \\
Assembly Theory & Historical selection/contingency from object assembly and copy number; no conventional unit of selection required & Distinguishable objects, building blocks/assembly operations and copy counts & Does not recover the causal \new{information }carrier--operator architecture that transmitted the variants \\
Constructor Theory & Physical information media and replication in terms of possible/impossible tasks & Candidate substrates, attributes/variables and tasks & Characterises capability; does not infer which structures occupied hereditary roles in a realised causal episode \\
\textbf{CSS} & \textbf{\del{Carrier}\new{Information carrier} variants, R/T operators, realised structures and genealogy jointly} & \textbf{Role-neutral persistent structures and realised causal history} & --- \\
\end{tabularx}
\label{tab:c1-1}
\end{table}

\section{The Causal Structure Set (CSS) framework}\label{sec:3}

\new{Figure~\ref{fig:pipeline} summarises the six-stage procedure. Sections~\ref{sec:3.1} to~\ref{sec:3.6} formalise the construction of a CSS, and Sections~\ref{sec:3.7} to~\ref{sec:3.9} develop the process and operator concepts used in stage 6. Sections~\ref{sec:4} and~\ref{sec:5} then define the candidate hereditary motifs and their joint role assignment. Appendix~\ref{app:A} specifies the admissibility conditions on the observational perspective.}

\subsection{A bounded observed spacetime episode}\label{sec:3.1}

\delmark\new{Consider a bounded spacetime episode $\mathcal{R}$ observed under a declared perspective $P = (\pi_P, D_P, J_P, \rho_P, \Sigma_P)$. In this tuple, $\pi_P$ maps the episode to its physical or recorded state-space description; $D_P$ specifies the available dynamical or transition description; $J_P$ specifies the admissible interventions or counterfactual perturbations; $\rho_P$ declares the spatial, temporal and structural resolution, and; $\Sigma_P$ specifies the rule by which persistent structure instances and their structural types are identified. The tuple is schematic, i.e., it does not require every application to use the same mathematical representation of dynamics. Under this perspective, the episode is represented by a state space $X_P$ and an observed trajectory $x_P(t)$. The perspective must be established independently of the hereditary roles subsequently inferred from the CSS.}

CSS is therefore explicitly \textbf{perspective-dependent}.

Independence from prespecified hereditary roles does not mean independence from observation. No empirical theory can identify physical structure without specifying what is observed and at what resolution. What is required is that structural description and hereditary interpretation remain distinct: physical primitives and structural types may be supplied by the observation perspective, while genome, organism, program, lineage and related hereditary semantics are withheld until after CSS construction. Appendix~\ref{app:A} \del{states}\new{specifies} the admissibility conditions that make this separation precise.

\delmark\new{Here, ``emergent'' has a deliberately limited meaning. A hereditary role is not an intrinsic property of a physical structure or its structural label; it depends on the realised causal organisation in which its instances participate. The framework seeks to infer such roles from a declared physical description, not to derive that description from fundamental physical laws.}

\new{Establishing that an observational perspective is genuinely free of prespecified hereditary roles is not always straightforward. Perspectives constructed for cultural, institutional or technological episodes may incorporate categories that already imply hereditary functions. For example, ``skilled practitioner'' or ``instruction'' may encode distinctions that the analysis is supposed to discover. Definition~\ref{def:1} (Appendix~\ref{app:A}) states the admissibility conditions, but satisfying them for a particular perspective requires separate justification. Where the necessary physical structures or causal relations remain hypothetical, the resulting role assignments must be reported conditionally, as explained in Appendix~\ref{app:F.3}.}

\subsection{Primitive structures as world-tubes}\label{sec:3.2}

At the finest declared observational level, let
\[
S_0 = \{S_j\} \text{ with } j=1\ldots m
\]
denote the complete set of persistent \textbf{primitive structures} $S_j$ available under $P$.

A physical structure is not represented as an instantaneous point. It occupies spatial extent and persists over an interval of time. We therefore represent a realised structure geometrically as a finite \textbf{world-tube}
\[
S_j \subset \mathcal{R}.
\]

For point-like idealisations this reduces to a worldline; for extended structures, world-tube terminology is more appropriate.

Examples of primitive structures may be, depending on perspective:
\begin{itemize}
  \item a persistent excitation of a quantum field;
  \item a molecule;
  \item a cell;
  \item a tracked physical artefact;
  \item a person over a bounded interval;
  \item a versioned software artefact;
  \item a memory-state region;
  \item another directly individuated persistent object.
\end{itemize}

An emergent structure may instead be an organised aggregate of primitive structures whose collective state itself persists.

Only persistent structures enter the representation. Transient coincidences do not constitute CSS nodes. Accepted structure instances are finite worldline sections, and the accepted set must satisfy completeness and non-overlap relative to the declared observation.

\subsection{Persistent structural types from almost-invariant regions}\label{sec:3.3}

We require a hereditary role-neutral way to decide when realised structures count as the same structural type. Two requirements pull in different directions. The classification must be broad enough to apply to any dynamics, since the framework is not restricted to a substrate, and it must nonetheless separate structures that persist from configurations that merely occur. Almost-invariant regions (AIR) meet both.\footnote{The literature also uses the term \emph{almost-invariant sets}.} They are defined by a ratio of timescales, internal relaxation against escape, which is available wherever a dynamics is defined and which appeals to no property of the particular system \citep{Froyland2014}. A configuration that arises and disperses fails the criterion whatever it is made of; one that the dynamics holds together satisfies it. AIRs therefore offer the most general basis we know of for distinguishing persistence from coincidence. A generic implementation of the structural classification accordingly uses AIRs of the dynamics. \new{Any other typing rule that meets the conditions stated at the end of this section also qualifies; Section~\ref{sec:comp} uses such a rule.}

Structural typing applies to individual structures, so it is stated in a local state space rather than in the episode-wide space $X_P$. For a candidate structure, meaning a world-tube section $S_j \subset \mathcal{R}$ admitted under the individuation and persistence rules of Section~\ref{sec:3.2}, let $Y_P$ denote the state space in which that structure's own configuration is described at the declared resolution, and write its reduced trajectory as $z_{S_j}(t) \in Y_P$. The notation $Y_P$ here refers to the shared space in which all admitted structures are described.

The reduced trajectory $z_{S_j}(t)$ is the projection of $x_P(t)$ onto the degrees of freedom constituting $S_j$. It describes the configuration of that structure alone, while $x_P(t)$ continues to describe the episode as a whole. A region of $Y_P$ is therefore a set of configurations the structure may occupy, and not a spatial region it occupies.

AIRs identify regions with low escape probability over a declared timescale. For an AIR candidate region $A \subseteq Y_P$, let $\tau_{\mathrm{mix}}(A)$ denote the characteristic time for internal relaxation and $\tau_{\mathrm{esc}}(A)$ the characteristic time for escape. In metastable regimes with rapid internal relaxation, a characteristic separation is $\tau_{\mathrm{mix}}(A) \ll \tau_{\mathrm{esc}}(A)$. The configuration thus relaxes within the region much faster than it leaves it.\footnote{The criterion as stated presumes a dynamics under which a region's mixing and escape timescales are well defined. Where the dynamics is itself time-dependent, related finite-time coherent-set methods apply, but structural typing then additionally requires a declared correspondence of types across time, since the regions themselves change.}

Let $Y_P^{\mathrm{acc}} \subseteq Y_P$ denote the domain of structural states admitted by the declared individuation and persistence rules. Let $\mathcal{L}_P$ be the set of structural labels, with each label $c \in \mathcal{L}_P$ identifying a non-empty region $A_c$. An AIR-based implementation requires these regions to form a qualifying partition $\mathcal{A}(P) = \{A_c : c \in \mathcal{L}_P\}$, such that
\[
\bigcup_{c \in {\new{\mathcal{L}}}_P} A_c = Y_P^{\mathrm{acc}},\ A_c \cap A_{c'} = \emptyset\ (c \neq c').
\]

Thus the regions cover the admitted domain without overlap.

For each accepted structure section $S_j$, its structural label is written $\ell(S_j) = c$. The label $c$ identifies the unique region $A_c$ to which the section's state trajectory $z_{S_j}(t)$ is assigned under the declared persistence rule. This assignment concerns the represented section, rather than membership at a single instant. A valid implementation must establish a unique assignment for every accepted section.

These are conditions on a valid implementation, not a guarantee that every dynamical system admits such an AIR partition. Resolving overlapping candidate regions into disjoint cells does not automatically preserve their persistence properties; those properties must be checked for the resulting cells.

Completeness concerns all instances qualifying under the declared structural rule. It does not require every state in $Y_P$, including transient configurations, to belong to an AIR. Transient configurations failing the persistence criterion do not supply nodes. A qualifying persistent instance whose type remains unresolved is a gap in the reconstruction; it cannot simply be omitted while completeness is claimed.

Hereditary-role independence follows from 1) the physically grounded $Y_P$ and 2) the AIR condition itself.
\begin{enumerate}[label=\arabic*), leftmargin=*]
  \item An AIR is defined by the behaviour of trajectories within and at the boundary of the region: fast internal relaxation and slow escape. These are intrinsic properties of the region at the time in question, and make no reference to what its occupants later produce. A class defined by downstream consequence is therefore generally not an AIR, and splitting an existing AIR according to the outputs its occupants later generate yields subregions that fail the condition, since internal relaxation within the parent region mixes them.
  \item A coordinate of the structural state space $Y_P$ that records which downstream structure a configuration will later produce is not a degree of freedom of that structure at the time in question, and its inclusion would violate the physical grounding condition of Definition~\ref{def:1}.
\end{enumerate}

Thus, given a physically grounded $Y_P$, an AIR partition may be computed automatically without further concern about hereditary contamination.

Selection among partitions that are each independently admissible is unrestricted. Admissibility is established for each member of the family separately, so choosing which member to examine adds no hereditary content to any of them, and the existential formulation of Appendix~\ref{app:B.11} presupposes this freedom.

AIRs remain a possible implementation, not a requirement of CSS. Alternative procedures are admissible when they provide persistent individuation, complete exclusive typing and hereditary-role independence; for these, role independence must be established rather than inherited from the construction. \new{AIR-based typing is therefore one route to structural types, not the only one. Any typing rule within $\Sigma_P$ qualifies if it yields a complete, mutually exclusive and hereditary-role-neutral classification of the admitted structural states $Y_P^{\mathrm{acc}}$ and satisfies Definition~\ref{def:1}. Section~\ref{sec:comp} uses declared value classes rather than an AIR partition.}

\subsection{Mapping realised structures to CSS instances}\label{sec:3.4}

Let ${\new{\mathcal{S}}} = \{S_1, \ldots, S_n\}$ be an admissible collection of persistent world-tube sections in $\mathcal{R}$.

Define a mapping\new{ onto the set $W$ of CSS instances,}
\[
M_P \colon {\new{\mathcal{S}}} \to W \colon M_P(S_j) = s_j.
\]

The distinction
\[
S_j \longrightarrow s_j \longrightarrow \ell(s_j)
\]
is fundamental.

$S_j$ is the physical world-tube section.

$s_j$ is its CSS instance.

$\ell(s_j)$ is its structural type.

Different physical instances may carry the same structural label while remaining distinct nodes with distinct ancestry.

\subsection{Realised structure-creating causality}\label{sec:3.5}

For temporally ordered physical structures $s_i$ and $s_j$, a realised structure-creating causal dependency exists when the presence or state of {\new{$s_i$}} contributes causally to the realisation of $s_j$'s persistent structural type.

Operationally, this can be tested through suppression, replacement or perturbation of the source while holding other admissible conditions fixed\new{, evaluated on the state description $X_P$ under the dynamics $D_P$ and the admissible interventions $J_P$ of the perspective (Section~\ref{sec:3.1})}. If such an intervention reduces the probability that the target realises structural type $\ell(s_j)$ according to the declared persistence criterion, the source is causally relevant to the target's realisation.

Thus the target must not merely exhibit a transient response; the causal outcome is the establishment or maintenance of an identified persistent structure. This distinguishes structure creation from arbitrary microscopic influence: a relevant episode establishes or maintains an identified persistent structural type, while perturbing or suppressing the source decreases the probability of that persistent realisation.

This is an interventionist reading of causal dependence in the sense of \citep{Woodward2000}. In systems with known dynamics, the relevant intervention need not be experimentally performed. It may be evaluated by simulation, replay, modification of initial or boundary conditions, or another counterfactual evaluation grounded in the dynamical model.

\paragraph{Dependence on the structural classification.} The causal dependence defined here is relative to the structural classification. Its target is the realisation of a persistent structural type, not every physical detail of the target instance. Two representations of the same episode may therefore contain exactly the same realised world-tube sections and yet support different causal relations, if they classify those structures differently. Merely renaming structural labels changes nothing; merging or splitting structural classes changes the outcome whose dependence is evaluated.

For example, consider a training episode in which one person causes another to acquire a persistent skill. Under a classification that distinguishes skill configurations, suppressing the training may prevent the target's observed structural type from being realised. Under a classification that distinguishes only persons independent of skills, the same intervention leaves the target's structural type unchanged. The training supports a causal dependence at the finer classification that need not be supported at the coarser one.

Structural classification $\ell$ and causal reconstruction are therefore coupled. The classification is fixed independently of hereditary-role inference, and is then used in evaluating the structure-creating dependencies from which the causal order is constructed. This extends to causal relations the perspective dependence already stated for structures in Appendix~\ref{app:A.2}: the physical history is fixed, its admissible representations need not be unique.

\paragraph{From intervention-supported dependencies to the causal order.} Let $D$ denote the relation of structure-creating dependencies established by the intervention criterion above. The temporal segmentation must make $D$ acyclic and consistent with temporal precedence. The CSS order $\prec$ is defined as the transitive closure of $D$ and is consequently a strict partial order. A pair related by $\prec$ is connected by a path of intervention-supported dependencies but need not itself satisfy the intervention criterion; $\prec$ therefore records causal ancestry through such dependencies. $D$ and its transitive closure are distinguished wherever a statement requires intervention-supported dependence rather than ancestry alone. $D$ is not in general recoverable from $\prec$, and is retained as part of the construction record where later analysis requires it.

\subsection{Definition of a CSS}\label{sec:3.6}

A \textbf{Causal Structure Set} is a triple
\[
C_{P,\ell} = (W, \prec, \ell)
\]
where:
\begin{itemize}
  \item $W$ is a finite admissible set of persistent realised structure instances;
  \item $\ell \colon W \to \mathcal{L}_P$ is a hereditary-role-neutral structural labelling;
  \item $\prec$ is a realised structure-creating strict causal order.
\end{itemize}

Throughout what follows, the motif predicates are evaluated on $\prec$ rather than on $D$. Where a process condition states that one instance causally contributes to another, the requirement is ancestry through intervention-supported dependencies within that process, not a direct intervention-supported dependency between the two instances. This choice is deliberate: requiring $D$ would make motif detection unstable under refinement, since resolving an intermediate converts a direct dependency into a path. The process boundary bounds the relation, so ancestry within a causally convex subset is considerably more restrictive than ancestry in the CSS as a whole.

The relation satisfies $s \nprec s$ (antireflexivity) and $s_i \prec s_j$, $s_j \prec s_k \Longrightarrow s_i \prec s_k$ (transitivity) and is therefore a strict partial order and can be represented by a DAG or by its Hasse diagram.

The finite partial-order construction intentionally retains:
\begin{itemize}
  \item physical instance identity;
  \item temporal precedence;
  \item causal ancestry;
  \item structural type;
\end{itemize}
while abstracting away substrate-specific dynamical details not required for the organisational question. This is the defining role of CSS: a finite labelled strict partial order preserving instance-level causal ancestry while separating neutral structural identity from hereditary function.

\begin{figure}[htbp]
  \centering
  \input{figures/fig-worldtubes-css}
  \caption{\textbf{From observed world-tubes to CSS.} \new{A worked illustration of stages 2--5 of Figure~\ref{fig:pipeline}, from world-tube sections $S_j$ to the CSS $(W, \prec, \ell)$.}}
  \label{fig:1}
\end{figure}

\subsection{Causally convex processes}\label{sec:3.7}

CSS represents a complete causal episode at a chosen resolution. Hereditary motifs are identified not from arbitrary isolated nodes but from structured \textbf{processes} within this order.

A subset
\[
Q \subseteq W
\]
is a process if it is causally convex:
\[
x,z \in Q,\ x \prec y \prec z \Longrightarrow y \in Q.
\]

Thus if a represented causal intermediary lies between two structures included in the process, it cannot be omitted.

\subsection{Input and output boundaries}\label{sec:3.8}

Define the input boundary
\[
I(Q) = \{\, x \in Q : \nexists\, y \in Q \text{ such that } y \prec x \,\},
\]
and output boundary
\[
O(Q) = \{\, x \in Q : \nexists\, y \in Q \text{ such that } x \prec y \,\}.
\]

For structural label $c$, define the boundary multiplicities
\begin{align*}
n^{-}_{Q}(c) &= |\{\, x \in I(Q) : \ell(x) = c \,\}|, \\
n^{+}_{Q}(c) &= |\{\, x \in O(Q) : \ell(x) = c \,\}|,
\end{align*}
where $|G|$ denotes the cardinality of a set $G$.

These quantities describe the causal transformation performed by $Q$ without giving its structures any hereditary meaning.

\subsection[Operators]{Operators\footnote{An earlier formulation of this framework called the corresponding structures \textbf{templates} and gave essentially this boundary-based formal definition. We use \emph{operator} here because the term avoids conflict with translation T and emphasises causal role rather than resemblance or copying.}}\label{sec:3.9}

Hereditary organisation depends on the physical arrangements through which structures are copied and their variants produce differentiated outputs. These arrangements constrain the underlying dynamics, enabling complex organisation to be generated repeatedly when the necessary inputs are supplied\del{, and}\new{; this} is what makes successive replication events instances of a recurrent arrangement rather than coincidences. Identifying them therefore complements identifying \new{information }carriers. Their continued availability can involve persistence of existing instances or production of replacements. The operator condition captures conserved contributing structural types across a process.

A structural label $o$ satisfies the \textbf{operator condition} in process $Q$ when
\[
n^{-}_{Q}(o) = n^{+}_{Q}(o) > 0,
\]
and at least one input instance with label $o$ causally contributes to an output instance with another label. The set of all operator labels for $Q$ is thus
\[
\mathrm{Op}(Q) = \left\{ o \in \mathcal{L}_P :
\begin{array}{c}
n^{-}_{Q}(o) = n^{+}_{Q}(o) > 0, \\
\exists x \in I(Q),\ y \in O(Q) : \\
\ell(x) = o,\ \ell(y) \neq o,\ x \prec y
\end{array}
\right\}.
\]

The causal relation used here is ancestry ($\prec$), as defined in Section~\ref{sec:3.6}.

Conservation is assessed across the chosen process, not necessarily at every internal step. Consider as an example a master mould used to manufacture a disposable mould, which in turn shapes a casting and is then destroyed. In the disposable mould casting process alone, the disposable mould is consumed and fails the operator condition. An expanded causally convex process including the disposable mould manufacture can instead have the master-mould type on both boundaries with equal positive multiplicity. If the master mould satisfies the causal condition, it qualifies as an operator; the disposable mould is an internal structure, present on neither boundary. Such process expansion requires the generating process to occur within the recorded episode.

Examples of operators include, after interpretation:
\begin{itemize}
  \item catalysts in chemical processes;
  \item replication or translation machinery in cells;
  \item tools or machinery in manufacturing;
  \item computers in digital information processing.
\end{itemize}

The operator condition depends on the declared individuation rule, structural classification and process boundary. For example, a whole-cell representation includes internal water within the cell node, whereas a constituent representation may retain separate water instances. The latter must satisfy persistence and non-overlap, and cannot simultaneously include overlapping whole-cell nodes. Where both constructions are admissible, their boundary multiplicities and operator assignments may differ. The same physical history can therefore support different operator assignments under different admissible representations. This does not imply that consumed structures cannot shape production.

Satisfying the operator condition is a necessary condition for the later assignment of a replication-operator or translation-operator role, not that assignment itself; roles are assigned only after coherent RVT organisation has been identified (Section~\ref{sec:5}).

Note that the operator condition is also satisfied by ambient structures conserved across a process and contributing causally to its outputs, so $\mathrm{Op}(Q)$ is typically not small. What the criterion requires downstream is correspondence of selected operator organisation across events (Appendix~\ref{app:C.1}), not that the operator set be unique or minimal.

The operator condition is not intrinsic to a structural label: the same structural type may satisfy it in one process and not in another.

\section{Causal motifs in the CSS}\label{sec:4}

A CSS contains no hereditary roles. Hereditary organisation, if present, must therefore emerge as a pattern in the causal order: a recurrent arrangement of processes, structural labels and boundary incidences that no single node or edge exhibits on its own. We call such a pattern a \textbf{causal motif}, and the identification problem is to find motifs whose occurrence warrants assigning hereditary roles to the structures participating in them.

The framework does not determine which motifs those are. That is an empirical question, and Section~\ref{sec:7.1} discusses systems that press on the answer. What the framework supplies is a representation in which candidate motifs can be stated precisely enough to be tested and, where necessary, replaced.

We adopt as an initial candidate the combination of replication (R), variation (V) and translation (T) together with lineage continuation. Each component is motivated independently. Replication and variation follow the substrate-independent treatment of copying in constructor theory \citep{Deutsch2015,Marletto2015}. Translation follows the observation that copying alone does not distinguish hereditary organisation from the propagation of inert patterns \citep{Walker2013}, and supplies the outward causal branch on which environmental coupling depends \citep{Day2003}. Lineage continuation distinguishes realised descent from recurrence of similar structures.

The combination is a hypothesis about which causal organisation constitutes heredity, not a definition of it. Sections~\ref{sec:6} and~\ref{sec:7.1} examine where it discriminates and where it may not.

\paragraph{Coherence.} Let
\[
I = \{i_1, \ldots, {\new{i_n}}\} \subseteq \mathcal{L}_P
\]
be a candidate family of alternative structural labels. R, V and T must be satisfied coherently by the same family. An unrelated replication process in one region of the CSS, an unrelated variation elsewhere and an unrelated translation-like process elsewhere do not together constitute the motif. This joint requirement is central to the criterion and is stated in full in Section~\ref{sec:5}.

\subsection{Replication: R}\label{sec:4.1}

Let $Q \subseteq W$ be a process, with boundaries $I(Q)$ and $O(Q)$. For a structural label $i \in \mathcal{L}_P$ write
\begin{align*}
I_i(Q) &= \{\, x \in I(Q) : \ell(x) = i \,\}, \\
O_i(Q) &= \{\, y \in O(Q) : \ell(y) = i \,\},
\end{align*}
so that $n^{-}_{Q}(i) = |I_i(Q)|$ and $n^{+}_{Q}(i) = |O_i(Q)|$.

$Q$ is an R candidate for $i$ when
\begin{enumerate}[label=\arabic*.]
  \item $n^{+}_{Q}(i) > n^{-}_{Q}(i) \geq 1$, and
  \item $\forall\, y \in O_i(Q),\ \exists\, x \in I_i(Q) : x \prec y$.
\end{enumerate}

Condition 1 requires a net increase in the multiplicity of $i$ across the process. Condition 2 requires every output instance of type $i$ to lie in the realised causal future of some input instance of the same type, so that the increase is produced from the inputs rather than arising independently within $Q$. As established in Section~\ref{sec:3.6}, the relation used here is $\prec$, that is, ancestry through intervention-supported dependencies within $Q$, not direct dependence.

Schematically,
\[
R_i : \{ r, \ldots, i, \ldots \} \longrightarrow \{ r, \ldots, i, i, \ldots \}
\]
where $r$ denotes one or more labels satisfying the operator condition of Section~\ref{sec:3.9}.

Replication is therefore not an increase in the number of structurally similar objects. The additional instances must belong to the realised causal future of the input \new{information }carriers.

For the family $I$, the R processes must satisfy the joint requirements of Section~\ref{sec:5}, including non-empty corresponding operator organisation under Appendix~\ref{app:C.1}. Physical operator instances may differ between events; the selected operator labels, multiplicities and boundary incidences must correspond.

\subsection{Variation: V}\label{sec:4.2}

A process $Q$ is a V candidate from $i_j$ to $i_k$ when
\[
i_j \neq i_k
\]
and there exist
\[
x \in I{\new{(}}Q{\new{)}},\ y \in O{\new{(}}Q{\new{)}}
\]
such that
\[
\ell(x) = i_j,\ {\new{\ell}}(y) = i_k,\ x \prec y.
\]

Schematically,
\[
V \colon i_j \longrightarrow i_k,\ i_j \neq i_k.
\]

Variation is therefore defined structurally and causally.

Terms such as \emph{mutation}, \emph{editing}, \emph{learning} or \emph{software update} may describe familiar domain interpretations, but they are not part of the CSS definition.

\subsection{Why R and V are insufficient}\label{sec:4.3}

R and V identify the propagation of distinguishable structural states. They do not establish that those distinctions are consequential.

A copying system could faithfully reproduce arbitrary inert patterns. A crystal could generate structurally variant descendants. A storage device could duplicate multiple bit patterns that never influence anything except later copying.

Such systems may exhibit copying and variation without a causal relation between the variant distinction and differentiated downstream organisation.

An additional condition is therefore required. That replication alone is insufficient for the organisation characteristic of living systems has been argued independently, in the distinction between trivial and non-trivial replication and in the proposal that the relevant transition concerns the emergence of informational control over matter rather than copying as such \citep{Walker2013}. The condition developed below is an attempt to make a comparable requirement detectable within a realised causal history rather than to characterise the transition itself.

\subsection{Translation: T}\label{sec:4.4}

Consider two alternative candidate \new{information }carrier labels
\[
i_1, i_2 \in I,\ i_1 \neq i_2.
\]

Processes
\[
Q_1, Q_2
\]
form a \textbf{translation motif} when:
\begin{itemize}
  \item $i_1$ participates as an input in $Q_1$;
  \item $i_2$ participates as an input in $Q_2$;
  \item the two processes share (non-empty) corresponding operator organisation;
  \item the \new{information }carrier instances causally contribute to downstream outputs;
  \item each designated downstream product label is absent from the corresponding process's input boundary;
  \item the corresponding downstream structural labels are distinguishable\del{.}\new{;}
  \item \new{the product distinction depends on the \new{information }carrier distinction: replacing $i_1$ by $i_2$ in $Q_1$, with the operator organisation and every other input at its realised type, changes the designated product from $y_1$ to $y_2$, evaluated by intervention replay or by declared dynamics as in Section~\ref{sec:3.5}.}
\end{itemize}

Schematically,
\begin{align*}
T_1 &\colon {\new{\{}} t, \ldots, i_1{\new{,}} \ldots {\new{\}}} \longrightarrow {\new{\{}} t, \ldots, y_1{\new{,}} \ldots {\new{\}}} \\
T_2 &\colon {\new{\{}} t, \ldots, i_2{\new{,}} \ldots {\new{\}}} \longrightarrow {\new{\{}} t, \ldots, y_2{\new{,}} \ldots {\new{\}}}
\end{align*}
where $y_1 \neq y_2$, and each translated product label is absent from its corresponding input boundary:
\[
n^{-}_{Q_k}(y_k) = 0,\ n^{+}_{Q_k}(y_k) > 0,\ k \in \{1, 2\}.
\]

Absence concerns structural type, not physical instance identity. Additional copies of \del{a}\new{an} \new{information }carrier type already present on the input boundary cannot serve as the translated product. The output multiplicity of the \new{information }carrier label $i_k$ remains unconstrained: translation may retain, consume or increase its instances, provided the other T conditions are satisfied.

A simple example is useful.

Copying a text file from one storage medium to another illustrates R.

Using two different files through the same printer to produce two different printed documents illustrates T.

Nothing about the file is intrinsically an information carrier. Its role follows only from the causal organisation of the realised episode.

\subsection{Why T matters}\label{sec:4.5}

T distinguishes \textbf{causally consequential variation} from causally inert copying.

It also has a deeper evolutionary role, and the requirement has a conceptual antecedent: the separation of stored instructions from the machinery that implements them has been argued to characterise the causal architecture of living organisation \citep{Walker2013}. T is an attempt to render a comparable distinction detectable within a realised causal history; Appendix~\ref{app:E.2} develops the comparison.

If alternative \new{information }carrier variants are ever to co-evolve with their environment, those alternatives must first be capable of influencing the environment differently.

T supplies the outward causal branch
\[
i_k \longrightarrow y_k.
\]

Thus T is proposed as a \textbf{necessary but not sufficient prerequisite for \new{information }carrier--environment co-evolution}.

Actual co-evolution additionally requires some downstream causal pathway
\[
y_k \text{ or modified environment} \longrightarrow \text{future information carrier dynamics.}
\]

Such feedback is central to niche-construction perspectives \citep{Day2003}, but the existence of that return pathway is not part of the minimal T criterion used here.

One particularly consequential return pathway concerns copying fidelity. A translated product may form or modify the machinery responsible for subsequent replication, including mechanisms that prevent or correct copying errors. \del{Carrier}\new{Information carrier} variation can therefore affect the processes through which later \new{information }carrier instances are produced:
\[
\text{information carrier variant} \ \overset{T}{\rightarrow} \ \text{translated product} \ \longrightarrow \ \text{fidelity of subsequent replication.}
\]

This connects T to Eigen's analysis of the constraints that imperfect copying places on the preservation of hereditary information \citep{Eigen1971}. Translation provides a route through which inherited configurations can influence those constraints by producing structures that improve copying accuracy. Whether such a pathway exists, improves fidelity, and remains effective across generations requires additional causal and dynamical evidence. Neither T nor lineage continuation alone establishes these properties. Other mechanisms may also preserve hereditary information, as studies of prebiotic replicator systems illustrate \citep{Szilagyi2017}.

T therefore identifies a causal branch through which \new{information }carrier variants may influence their environment or their subsequent replication. Co-evolution, error reduction and sustained hereditary stability require additional organisation and evidence beyond the minimal criterion.

\subsection{Causal re-entry into replication}\label{sec:4.6}

A single local RVT episode does not provide the same evidence for heredity as a causal organisation that recurs through hundreds of successive generations. Lineage continuation should therefore not be treated as simply present or absent\new{ beyond the minimal closure requirement of Section~\ref{sec:4.7}}.

Consider an R process $R_g$ with \new{information }carrier input $x_g$ and an additional, causally produced \new{information }carrier occurrence $x_{g+1}$ on its output boundary:
\[
x_g \in I(R_g),\ x_{g+1} \in O(R_g),\ x_g \prec x_{g+1}.
\]

Write
\[
x_g \rightsquigarrow x_{g+1}
\]
for this realised replication transition. Temporal continuation of the original occurrence alone does not constitute such a transition.

Lineage continuation occurs when the descendant re-enters a subsequent R process\new{ and is itself replicated there}:
\[
x_{g+1} \in O(R_g) \cap I(R_{g+1}),{\new{\qquad x_{g+1} \rightsquigarrow x_{g+2},}}
\]
\del{Where}\new{where} $R_g$ and $R_{g+1}$ have non-empty corresponding operator organisation under Appendix~\ref{app:C.1}.

The variant identity of the lineage must remain associated with its relevant T relations. Closure is genealogical and causal: output descendants re-enter replication through shared operator labels or types, and this is explicitly distinguished from mere repeated occurrence of similar labels.

\subsection{Observed lineage depth}\label{sec:4.7}

Let
\[
x_0 \rightsquigarrow x_1 \rightsquigarrow \cdots \rightsquigarrow x_L
\]
be a causally connected sequence of observed \new{information }carrier generations satisfying the corresponding R organisation.

Define the \textbf{observed lineage depth}
\[
L(C,I)
\]
as the greatest number of successive causal replication transitions supported in CSS $C$ for the coherent \new{information }carrier family $I$.

Each arrow counts as one replication transition. A single replication gives $L = 1$; one descendant re-entry requires two successive transitions and therefore $L \geq 2$.

The conceptual point is:
\[
L=2
\]
is weaker evidence of hereditary organisation than
\[
L=10,
\]
which is weaker than
\[
L=10^3.
\]

Beyond the minimal requirement that at least one descendant re-enters the corresponding replication organisation, there is no further generation threshold at which a system suddenly becomes ``hereditary.''

Membership in the class is therefore binary and minimal: closure requires one realised re-entry. The \textbf{evidence} that the organisation is sustained is graded, and $L$ measures it.

RVT identifies an organisation capable of propagating causally consequential structural variation; increasing $L$ supplies increasing evidence that this organisation is sustained as a hereditary lineage.

An application that requires a stronger classification may choose a preregistered minimum depth beyond $L \geq 2$, but any such higher threshold belongs to the application rather than to the foundational CSS definition.

\subsection{Accidental local motifs and lineage depth}\label{sec:4.8}

A finite non-hereditary history could nevertheless contain a short causal pattern satisfying some local pieces of R, V or T by accident. No representation theory can prevent physical coincidences that genuinely occurred.

This is a different problem from representational overfitting. The distinction is:

representational false positive

versus

physical but non-diagnostic motif.

The first is addressed by admissibility: a representation may not invent the required physical relations (Appendix~\ref{app:B}). The second concerns the \textbf{sufficiency of the hereditary criterion itself}.

This is one reason lineage continuation is graded\new{, beyond the minimal closure requirement of Section~\ref{sec:4.7},} rather than represented by a single local binary closure condition. A one-generation configuration containing local R, V and T provides much weaker evidence of sustained hereditary organisation than a causally connected lineage in which variants repeatedly propagate through the corresponding replication and translation organisation.

Thus increasing observed lineage depth $L$ does not compensate for a missing hereditary relation; rather, once the required organisation is present, it measures how extensively that organisation is realised in the observed history.

Whether RVT plus sufficiently deep lineage continuation is fully discriminative against every possible non-hereditary physical process remains an empirical and theoretical question. It should be tested through adversarial counterexamples and controlled systems rather than resolved by representational restrictions.

\subsection{Finite observations impose finite claims}\label{sec:4.9}

CSS describes only the realised observation.

Suppose a biological lineage has existed for a million generations but the observation contains only three. CSS contains evidence for three observed generations. The previous unobserved generations are not silently inserted.

Likewise, if a DNA molecule is observed for a short interval and no replication process occurs, biological knowledge that DNA \emph{can} replicate cannot be used to assign an R role in that CSS.

There is therefore no separate CSS concept of a \textbf{capacity for heredity}.

CSS asks exclusively:

What hereditary organisation is realised in the available causal observation?

Motif identification is necessarily episode-dependent in finite causal histories.

\new{Failure to identify an information-carrier role within a bounded CSS does not establish that the observed structure lacks such a role (cf.\ Case~A in Section~\ref{sec:cultural}). Hereditary roles are relational properties of causal organisation, and the organisation required to identify them may extend beyond the observed episode. For example, an isolated electromagnetic transmission may reveal variant-dependent effects at a receiver without exhibiting the replication and lineage continuation that the present criterion requires. A more extensive causal history, including repeated transmission, copying, modification and downstream interpretation, may establish that the same waveform structural types participate in a coherent information family. Enlarging the observation does not confer information-carrier status on the original signal; it may supply the evidence that was missing for identifying that status.}

\section{Information families and joint role assignment}\label{sec:5}

A candidate \textbf{information family} is not the union of independently detected R, V and T motifs. It is a coherent organisation in which the same family of structural variants satisfies the relevant relations.

Let
\[
I = {\new{\{}}i_1{\new{,}} \ldots{\new{,}} i_m{\new{\}}}.
\]
A candidate information family requires:
\begin{itemize}
  \item \textbf{variation connectivity:} members of $I$ are connected through observed V processes\new{, which may pass through instances of labels outside $I$};
  \item \textbf{replication:} members participate in corresponding R processes;
  \item \textbf{shared replication organisation:} the R processes use corresponding operator patterns;
  \item \textbf{translation:} alternative members participate in T processes;
  \item \textbf{shared translation organisation:} those T processes use corresponding operator patterns;
  \item \textbf{variant-dependent downstream realisation:} different \new{information }carrier variants generate distinguishable realised output structures;
  \item \textbf{observed lineage continuation:} at least one replicated descendant \new{of a member of $I$} re-enters the corresponding replication organisation ($L \geq {\new{2}}$)\new{; the condition applies to the family as a whole, not to each variant}. Beyond this minimal requirement, $L(C,I)$ reports the strength of the observed lineage and does not affect membership.
\end{itemize}

Only after this organisation has been identified are hereditary roles assigned.

The resulting relational roles are:
\begin{itemize}
  \item \textbf{information-carrier variants};
  \item \textbf{replication operators};
  \item \textbf{translation operators};
  \item \textbf{realised downstream structures}.
\end{itemize}

The union of these structures and relations forms the inferred information family.

Because the multiplicity of \del{a}\new{an} \new{information }carrier on the output boundary of a T process is unconstrained, \del{a}\new{an} \new{information }carrier may itself satisfy the operator condition of Section~\ref{sec:3.9} within that process. This is not a conflict: the operator condition is necessary but not sufficient for an operator role, and the joint criterion assigns the \new{information }carrier role to structures whose variants satisfy R, V and T coherently. Role assignment is made by the criterion as a whole, not by any single boundary condition.

This is the operational inversion central to CSS:

structural labels + causal organisation $\longrightarrow$ hereditary roles.

No label initially means ``information''.

\new{Section~\ref{sec:6} applies this criterion to a synthetic configuration that satisfies it (Section~\ref{sec:6.2}) and to configurations that each lack at least one of its components (Sections~\ref{sec:6.1} and~\ref{sec:6.3}--\ref{sec:6.5}). Section~\ref{sec:cultural} applies it to two conditional reconstructions of recipe transmission, one incomplete and one completed by stipulation, and Section~\ref{sec:comp} to the recorded history of a digital-evolution simulator.}

\subsection{Roles are relational}\label{sec:5.1}

The same structural type can occupy different roles in different processes. The same physical structure can even occupy different roles in different information families. For example, a printed book may be the downstream realised structure produced by a printing process and later become an information carrier in a copying or teaching process.

Thus, hereditary role is a property of causal organisation, not an intrinsic property of material type.

This naturally permits nested hereditary organisation. A structure may be a downstream product at one organisational level while acting as \del{a}\new{an} \new{information }carrier or operator at another.

Because roles are relational rather than material, they also become comparable across substrates: two physically unrelated processes may bear the same inferred roles in the same boundary organisation (Section~\ref{sec:8.4}, Appendix~\ref{app:C}).

\begin{figure}[htbp]
  \centering
  \begin{tikzpicture}[cssfig, cssnode/.append style={minimum size=3.2mm}]
\node at (3.599,11.73) {Causal structure set $(W,\prec,\ell)$};
\draw[csswset, rounded corners=4mm] (0.38,5.244) rectangle (6.934,11.396);
\node at (0.253,11.489) {\textbf{W}};
\draw[cssRbox] (0.805,9.073) rectangle (3.174,11.19);
\draw[cssinner] (0.943,9.729) rectangle (1.438,11.005);
\draw[cssinner] (2.484,9.258) rectangle (3.013,11.017);
\node at (1.196,9.947) {\textbf{...}};
\node at (2.748,9.441) {\textbf{...}};
\node at (1.92,9.269) {\textbf{...}};
\node[cssnode, cssfill=cssOI3] (m1) at (1.242,10.81) {};
\draw[cssdep] (0.655,10.81) -- (m1);
\draw[cssdep] (m1) -- (1.805,10.81);
\node[cssnode, cssfill=cssOI1] (m2) at (1.242,10.293) {};
\draw[cssdep] (0.655,10.293) -- (m2);
\draw[cssdep] (m2) -- (1.805,10.293);
\node[cssnode, cssfill=cssOI3] (m3) at (2.794,10.81) {};
\draw[cssdep] (2.231,10.81) -- (m3);
\draw[cssdep] (m3) -- (3.358,10.81);
\node[cssnode, cssfill=cssOI3] (m4) at (2.794,10.281) {};
\draw[cssdep] (2.231,10.281) -- (m4);
\draw[cssdep] (m4) -- (3.358,10.281);
\node[cssnode, cssfill=cssOI1] (m5) at (2.794,9.752) {};
\draw[cssdep] (2.231,9.752) -- (m5);
\draw[cssdep] (m5) -- (3.358,9.752);
\draw[cssRbox] (4.151,9.073) rectangle (6.521,11.19);
\draw[cssinner] (4.29,9.729) rectangle (4.784,11.005);
\draw[cssinner] (5.83,9.258) rectangle (6.359,11.017);
\node at (4.542,9.947) {\textbf{...}};
\node at (6.095,9.441) {\textbf{...}};
\node at (5.267,9.269) {\textbf{...}};
\node[cssnode, cssfill=cssOI6] (m6) at (4.588,10.81) {};
\draw[cssdep] (4.002,10.81) -- (m6);
\draw[cssdep] (m6) -- (5.152,10.81);
\node[cssnode, cssfill=cssOI1] (m7) at (4.588,10.293) {};
\draw[cssdep] (4.002,10.293) -- (m7);
\draw[cssdep] (m7) -- (5.152,10.293);
\node[cssnode, cssfill=cssOI6] (m8) at (6.141,10.81) {};
\draw[cssdep] (5.577,10.81) -- (m8);
\draw[cssdep] (m8) -- (6.704,10.81);
\node[cssnode, cssfill=cssOI6] (m9) at (6.141,10.281) {};
\draw[cssdep] (5.577,10.281) -- (m9);
\draw[cssdep] (m9) -- (6.704,10.281);
\node[cssnode, cssfill=cssOI1] (m10) at (6.141,9.752) {};
\draw[cssdep] (5.577,9.752) -- (m10);
\draw[cssdep] (m10) -- (6.704,9.752);
\draw[cssVbox] (0.84,7.763) rectangle (3.163,8.889);
\draw[cssinner] (0.978,7.958) rectangle (1.472,8.717);
\draw[cssinner] (2.518,7.958) rectangle (3.047,8.717);
\node at (1.23,8.13) {\textbf{...}};
\node at (2.783,8.153) {\textbf{...}};
\node at (1.955,8.038) {\textbf{...}};
\node[cssnode, cssfill=cssOI3] (m11) at (1.276,8.487) {};
\draw[cssdep] (0.69,8.487) -- (m11);
\draw[cssdep] (m11) -- (1.84,8.487);
\node[cssnode, cssfill=cssOI6] (m12) at (2.829,8.487) {};
\draw[cssdep] (2.265,8.487) -- (m12);
\draw[cssdep] (m12) -- (3.393,8.487);
\draw[cssTbox] (0.805,5.497) rectangle (3.174,7.59);
\draw[cssinner] (0.943,6.13) rectangle (1.438,7.406);
\draw[cssinner] (2.484,5.658) rectangle (3.013,7.417);
\node at (1.196,6.348) {\textbf{...}};
\node at (2.748,5.842) {\textbf{...}};
\node at (1.92,5.67) {\textbf{...}};
\node[cssnode, cssfill=cssOI3] (m13) at (1.242,7.21) {};
\draw[cssdep] (0.655,7.21) -- (m13);
\draw[cssdep] (m13) -- (1.805,7.21);
\node[cssnode, cssfill=cssOI7] (m14) at (1.242,6.693) {};
\draw[cssdep] (0.655,6.693) -- (m14);
\draw[cssdep] (m14) -- (1.805,6.693);
\node[cssnode, cssfill=cssOI3] (m15) at (2.794,7.21) {};
\draw[cssdep] (2.231,7.21) -- (m15);
\draw[cssdep] (m15) -- (3.358,7.21);
\node[cssnode, cssfill=cssOI4] (m16) at (2.794,6.681) {};
\draw[cssdep] (2.231,6.681) -- (m16);
\draw[cssdep] (m16) -- (3.358,6.681);
\node[cssnode, cssfill=cssOI7] (m17) at (2.794,6.152) {};
\draw[cssdep] (2.231,6.152) -- (m17);
\draw[cssdep] (m17) -- (3.358,6.152);
\draw[cssTbox] (4.151,5.497) rectangle (6.543,7.59);
\draw[cssinner] (4.29,6.13) rectangle (4.784,7.406);
\draw[cssinner] (5.83,5.658) rectangle (6.359,7.417);
\node at (4.542,6.348) {\textbf{...}};
\node at (6.095,5.842) {\textbf{...}};
\node at (5.267,5.67) {\textbf{...}};
\node[cssnode, cssfill=cssOI6] (m18) at (4.588,7.21) {};
\draw[cssdep] (4.002,7.21) -- (m18);
\draw[cssdep] (m18) -- (5.152,7.21);
\node[cssnode, cssfill=cssOI7] (m19) at (4.588,6.693) {};
\draw[cssdep] (4.002,6.693) -- (m19);
\draw[cssdep] (m19) -- (5.152,6.693);
\node[cssnode, cssfill=cssOI6] (m20) at (6.141,7.21) {};
\draw[cssdep] (5.577,7.21) -- (m20);
\draw[cssdep] (m20) -- (6.704,7.21);
\node[cssnode, cssfill=cssOI8] (m21) at (6.141,6.681) {};
\draw[cssdep] (5.577,6.681) -- (m21);
\draw[cssdep] (m21) -- (6.704,6.681);
\node[cssnode, cssfill=cssOI7] (m22) at (6.141,6.152) {};
\draw[cssdep] (5.577,6.152) -- (m22);
\draw[cssdep] (m22) -- (6.704,6.152);
\node at (4.37,8.372) {\textbf{...}};
\node[font=\large] at (7.4,8.34) {$\Longleftrightarrow$};
\node[cssVbox, minimum width=2.65cm, minimum height=0.75cm, inner sep=0pt] at (9.175,8.34) {Variation (\textbf{V})};
\node[cssRbox, minimum width=2.65cm, minimum height=0.75cm, inner sep=0pt] at (11.925,8.34) {Replication (\textbf{R})};
\node[cssTbox, minimum width=2.65cm, minimum height=0.75cm, inner sep=0pt] at (14.675,8.34) {Translation (\textbf{T})};
\node[cssnode, cssfill=cssOI3] at (8.165,7.66) {};
\node[cssnode, cssfill=cssOI6] at (8.685,7.66) {};
\node[cssnode, cssfill=cssOI1] at (10.675,7.66) {};
\node[cssnode, cssfill=cssOI7] at (12.925,7.66) {};
\node[cssnode, cssfill=cssOI4] at (14.915,7.66) {};
\node[cssnode, cssfill=cssOI8] at (15.435,7.66) {};
\node[align=center] at (8.425,7.04) {Information\\carriers};
\draw[line width=0.8pt, draw=black] (8.425,6.59) -- (8.425,4.49);
\node[align=center] at (10.675,7.04) {Replication\\operator};
\draw[line width=0.8pt, draw=black] (10.675,6.59) -- (10.675,4.49);
\node[align=center] at (12.925,7.04) {Translation\\operator};
\draw[line width=0.8pt, draw=black] (12.925,6.59) -- (12.925,4.49);
\node[align=center] at (15.175,7.04) {Realised\\structures};
\draw[line width=0.8pt, draw=black] (15.175,6.59) -- (15.175,4.49);
\node[anchor=north west, inner sep=0pt] (tab) at (4.95,4.441) {%
{\setlength{\tabcolsep}{0pt}\renewcommand{\arraystretch}{1.35}%
\begin{tabular}{@{\hspace{2mm}}>{\bfseries\raggedright\arraybackslash}m{2.15cm}*{4}{>{\centering\arraybackslash}m{2.25cm}}@{}}
\hline
Biology: & Genes in DNA & DNA replication\newline machinery & Cell\newline machinery & Phenotype \\
\hline
Culture: & Transferrable\newline ideas in neural\newline patterns & Biological\newline communication\newline system & Biological\newline motoric\newline system & Cultural\newline expression \\
\hline
Civilisation: & Instructions in\newline documents & Reproduction\newline technologies & Institutions & Civilisational\newline output \\
\hline
Cybernetics: & Code in\newline computer\newline memory & Computer & Robotic\newline machines & Workpieces \\
\hline
\end{tabular}}%
};
\node[rotate=90, anchor=south, align=center] at ([xshift=-1.5mm]tab.west) {Examples across\\layers of self-organization};
\end{tikzpicture}%
  \caption{\textbf{Input/output boundary motifs define R, V, T and information family.}}
  \label{fig:2}
\end{figure}

\section{Conceptual examples and counterexamples}\label{sec:6}

The discriminatory value of the framework becomes clearer through simple cases. %
{\delmark}
\begin{newtext}
Sections~\ref{sec:6.1} to~\ref{sec:6.5} are formal synthetic configurations, each evaluated by the criterion of Section~\ref{sec:5}. Each starts from a stipulated CSS, i.e., from the output of stage 5 of Figure~\ref{fig:pipeline}, and runs only stage 6. Sections~\ref{sec:cultural} and~\ref{sec:comp} then run all six stages, each in one domain: a cultural episode, recipe transmission, reconstructed under stated hypotheses, and digital-evolution episodes constructed from a recorded history. No case of this section is empirical evidence; each result has the conditional form of Appendix~\ref{app:F.3}.
\end{newtext}

\new{\textbf{Time in the figures.} A relation $x \prec y$ records realised causal influence, and causal precedence between events does not depend on the choice of reference frame (\citealp{Sorkin1991}; \citealp[Sect.~3.2]{Surya2019}). The node set, however, is declared by the perspective, which may cut world-tubes into sections using a time coordinate, so we claim no frame independence for a reconstructed CSS as a whole. In a non-relativistic setting, all observers share one global time, and every causal relation respects it, i.e., $x \prec y$ implies $t(x) < t(y)$. The Hasse diagrams of this section, of Sections~\ref{sec:cultural} and~\ref{sec:comp} and of Appendix~\ref{app:G} therefore draw time running from bottom to top, with every edge pointing upward; two instances at different heights need not be causally related.} %

\subsection{Catalytic chemistry: operator without heredity}\label{sec:6.1}

Consider \new{a} carbon oxidation process $Q$, mediated by a persistent platinum (Pt) catalyst \new{(Figure~\ref{fig:ex-6.1})}. The physical episode contains persistent structures corresponding to reactants, intermediate states, products and catalyst world-tube sections.
\new{The catalyst label is conserved across $Q$,}{\delmark}
\[ n^{-}_{Q}(\mathrm{Pt}) = n^{+}_{Q}(\mathrm{Pt}) > 0,{\delmark} \]
\new{and} its input instance causally contributes to output structures of other labels.
\new{Pt therefore satisfies the operator condition of Section~\ref{sec:3.9}.}{\delmark}

\begin{figure}[!htbp]
\centering
\begin{tikzpicture}[cssfig, x=1cm, y=1cm]
\path[cssproc, sharp corners, draw=none] (0.350,1.412) -- (0.357,1.347) -- (0.379,1.286) -- (0.413,1.231) -- (0.459,1.185) -- (0.514,1.150) -- (0.575,1.129) -- (0.640,1.122) -- (2.200,1.122) -- (2.265,1.129) -- (2.326,1.150) -- (2.381,1.185) -- (2.427,1.231) -- (2.461,1.286) -- (2.483,1.347) -- (2.490,1.412) -- (2.483,1.476) -- (2.054,2.916) -- (2.032,2.977) -- (1.998,3.032) -- (1.952,3.078) -- (1.897,3.113) -- (1.836,3.134) -- (1.771,3.142) -- (0.640,3.142) -- (0.575,3.134) -- (0.514,3.113) -- (0.459,3.078) -- (0.413,3.032) -- (0.379,2.977) -- (0.357,2.916) -- (0.350,2.852) -- cycle;
\path[cssproc, sharp corners, fill=none] (0.350,1.412) -- (0.357,1.347) -- (0.379,1.286) -- (0.413,1.231) -- (0.459,1.185) -- (0.514,1.150) -- (0.575,1.129) -- (0.640,1.122) -- (2.200,1.122) -- (2.265,1.129) -- (2.326,1.150) -- (2.381,1.185) -- (2.427,1.231) -- (2.461,1.286) -- (2.483,1.347) -- (2.490,1.412) -- (2.483,1.476) -- (2.054,2.916) -- (2.032,2.977) -- (1.998,3.032) -- (1.952,3.078) -- (1.897,3.113) -- (1.836,3.134) -- (1.771,3.142) -- (0.640,3.142) -- (0.575,3.134) -- (0.514,3.113) -- (0.459,3.078) -- (0.413,3.032) -- (0.379,2.977) -- (0.357,2.916) -- (0.350,2.852) -- cycle;
\node[cssnode, minimum size=5.4mm, font=\scriptsize, cssfill=cssOI1!55!white] (aP0) at (0.640,1.412) {$\mathrm{Pt}$};
\node[cssnode, minimum size=5.4mm, font=\scriptsize, cssfill=cssOI2!55!white] (aC0) at (2.200,1.412) {$\mathrm{C}$};
\node[cssnode, minimum size=5.4mm, font=\scriptsize, cssfill=cssGrey!75!white] (aX0) at (1.420,1.412) {$\mathrm{O}_2$};
\node[cssnode, minimum size=5.4mm, font=\scriptsize, cssfill=cssOI6!60!white] (aW1) at (1.225,2.132) {$\mathrm{O}^{*}$};
\node[cssnode, minimum size=5.4mm, font=\scriptsize, cssfill=cssOI1!55!white] (aP1) at (0.640,2.852) {$\mathrm{Pt}$};
\node[cssnode, minimum size=5.4mm, font=\scriptsize, cssfill=cssOI4!55!white] (aY2) at (1.771,2.852) {$\mathrm{CO}_2$};
\draw[cssdep] (aC0) -- (aY2);
\draw[csscont] (aP0) -- (aP1);
\draw[cssdep] (aP0) -- (aW1);
\draw[cssdep] (aW1) -- (aY2);
\draw[cssdep] (aX0) -- (aW1);
\node[csstag, font=\scriptsize, anchor=west] at (2.239,2.348) {$Q\colon\ \mathrm{Op}(Q) = \{\mathrm{Pt}\}$};
\node[anchor=north west, inner sep=0pt, align=left, font=\scriptsize, text=black] at (0.350,1.042) {no R: no label gains instances\\no T: no comparative rule; no lineage\\\textbf{no family}};
\draw[cssdep, draw=black!45] (0.130,1.412) -- (0.130,2.162);
\node[text=black!60, font=\scriptsize, anchor=south] at (0.130,2.162) {$t$};
\path[cssRbox, rounded corners=0.8mm] (5.350,2.912) rectangle (5.850,3.132);
\node[anchor=west, inner sep=0pt, font=\scriptsize, text=black] at (6.000,3.022) {replication (R)};
\path[cssVbox, rounded corners=0.8mm] (5.350,2.552) rectangle (5.850,2.772);
\node[anchor=west, inner sep=0pt, font=\scriptsize, text=black] at (6.000,2.662) {variation (V)};
\path[cssTbox, rounded corners=0.8mm] (5.350,2.192) rectangle (5.850,2.412);
\node[anchor=west, inner sep=0pt, font=\scriptsize, text=black] at (6.000,2.302) {translation (T)};
\path[cssproc, rounded corners=0.8mm] (5.350,1.832) rectangle (5.850,2.052);
\node[anchor=west, inner sep=0pt, font=\scriptsize, text=black] at (6.000,1.942) {process, no motif};
\draw[csscont] (5.350,1.582) -- (5.850,1.582);
\node[anchor=west, inner sep=0pt, font=\scriptsize, text=black] at (6.000,1.582) {persistence (one world-tube)};
\draw[cssdep] (5.350,1.222) -- (5.850,1.222);
\node[anchor=west, inner sep=0pt, font=\scriptsize, text=black] at (6.000,1.222) {Hasse edge of $\prec$};
\end{tikzpicture}%
\caption{\new{\textbf{Minimal stipulated CSS of Section~\ref{sec:6.1}, evaluated by the joint criterion.} Conventions of Figures~\ref{fig:ex-6.1} to~\ref{fig:ex-6.5}: each figure is the Hasse diagram of the complete CSS, i.e., the transitive reduction of $\prec$ over all instances, with precedence running upward. Colour gives the structural label of an instance within its figure only, never a hereditary role; grey bars join sections of one world-tube; each rounded shape encloses one declared process. A shape carries the colour of R, V or T only where the script identifies that motif for the candidate family $I$ of the figure, and is grey otherwise. The legend in this figure applies to all five. Text under each diagram gives the verdicts and names the missing components, which are annotations, not edges. Every boundary, multiplicity and verdict of Figures~\ref{fig:ex-6.1} to~\ref{fig:ex-6.5} was computed from the stipulated instance and dependency lists by a script that implements the definitions of Sections~\ref{sec:3}--\ref{sec:5}. Pt satisfies the operator condition in $Q$, but no label gains instances, so no R process exists for any candidate family; no comparative rule is stipulated, so no T; nothing is replicated, so no lineage continuation, and no family is assigned. Formally, $Q$ is also a V candidate (Section~\ref{sec:4.2}) from C, O$_2$ and Pt to CO$_2$, but no family containing these labels satisfies R.}}
\label{fig:ex-6.1}
\end{figure}

Nothing follows about heredity. \new{$Q$ formally matches the V pattern between labels, since instances of other labels cause the $\mathrm{CO}_2$ instance, but variation connectivity concerns members of a candidate family, and every family containing these labels fails R, so $Q$ supplies V for no information family.} \new{The operator condition is necessary for a replication-operator or translation-operator role but does not assign one, and it establishes none of R, V, T and lineage continuation.} If no candidate \new{information }carrier variants are replicated, translated into differentiated structures and continued genealogically, no information family is inferred\new{, and Pt receives no operator role}. This case illustrates that physical structures can be mapped into a CSS, and \del{operator roles assigned}\new{the operator condition evaluated}, entirely independently of hereditary motifs.

\subsection{Minimal hereditary organisation}\label{sec:6.2}

The case is stated abstractly, in the order the framework prescribes: structures and neutral labels first, causal relations next, motifs identified in the resulting CSS, roles assigned only afterwards. \new{As the positive control of this section, the configuration satisfies the complete criterion without any physical assumption about a particular substrate.}

\begin{figure}[!htbp]
\centering
\input{figures/examples-hasse-b}
\caption{\new{Conventions as in Figure~\ref{fig:ex-6.1}. The configuration of Section~\ref{sec:6.2} (Appendix~\ref{app:G.4}), $I = \{c_1, c_2\}$: $Q_1$, $Q_2$ and $Q_4$ are R processes, $Q_3$ is a V process, $Q_5$ and $Q_6$ satisfy T under the stipulated rule $F(c_1, c_4) = c_5$, $F(c_2, c_4) = c_6$, and $a_0 \rightsquigarrow a_2 \rightsquigarrow a_4$ gives $L = 2$; the family is assigned.}}
\label{fig:ex-6.2}
\end{figure}

Consider a synthetic history with six distinct structural labels
\[ \mathcal{L}_{P} = \{c_1, c_2, c_3, c_4, c_5, c_6\}. \]
These are neutral identifiers. All represented sections are stipulated to satisfy the declared persistence and non-overlap conditions.

\revnote{[v7 instance list, dependency relation, process boundaries, comparative rule and detailed joint evaluation moved unchanged to Appendix~\ref{app:G.4}.]}

\begin{newtext}
\textbf{Organisation and joint evaluation.} The stipulated history has 17 instances and 22 direct dependencies, on which six causally convex processes, $Q_1$ to $Q_6$, are declared (Appendix~\ref{app:G.4}\new{; Figure~\ref{fig:ex-6.2}}). $Q_1$ and $Q_2$ satisfy R for $c_1$, and $Q_4$ satisfies R for $c_2$: in each, a $c_3$ instance and \del{a}\new{an} \new{information }carrier instance produce a continuing $c_3$ section, a continuing \new{information }carrier section and one additional \new{information }carrier instance, i.e., \new{information }carrier multiplicity $1 \to 2$, and $c_3$ satisfies the operator condition with corresponding boundary patterns in all three. $Q_3$ is a V process from $c_1$ to $c_2$. $Q_5$ and $Q_6$ satisfy T: a $c_4$ instance and a $c_1$ or $c_2$ instance produce a continuing $c_4$ section and a $c_5$ or $c_6$ instance, and a stipulated product-type rule, $F(c_1, c_4) = c_5$ and $F(c_2, c_4) = c_6$ with other conditions fixed, supplies the variant-dependent comparison that the ancestry relations alone would not establish. The additional $c_1$ instance produced by $Q_1$ re-enters $Q_2$, which gives a lineage of two replication transitions, $L = 2$. The script used for Appendix~\ref{app:G} reproduces every boundary, multiplicity and verdict of this example.
\end{newtext}

\textbf{Role assignment.} The same family satisfies R, V, T and lineage continuation. Only now are roles assigned: $c_1$ and $c_2$ as information-carrier variants, $c_3$ as their replication operator, $c_4$ as their translation operator, and $c_5$ and $c_6$ as the corresponding realised products. None was supplied before motif detection. This is a conditional result for the stipulated history, not empirical validation of the criterion.

\subsection{Autocatalytic growth without T}\label{sec:6.3}

Consider
\new{$r, a \to r, a, a$.}{\delmark}
Suppose \new{an} occasional variant $a'$ also occurs and is copied \new{(Figure~\ref{fig:ex-6.3})}. \del{If $a$ and $a'$ do not generate distinguishable downstream structures through corresponding shared operators, the system lacks T.}\new{$a'$ instances produce nothing but further $a'$ instances, whose label is already on the input boundary, and no stipulated rule makes any product depend on the $a$/$a'$ distinction, so the system lacks T.} It therefore contains copying and variation but does not satisfy the initial candidate information-family criterion. This provides a distinction between hereditary organisation and generic autocatalytic proliferation\new{: replication of alternative structures does not imply variant-dependent downstream production through corresponding operators}. \new{Such a system may nonetheless be hereditary under other legitimate definitions (Section~\ref{sec:7.1}). Failing the RVT criterion shows that the system lacks the organisation this criterion requires, not that it lacks every broader form of heredity.}

\begin{figure}[!htbp]
\centering
\begin{tikzpicture}[cssfig, x=1cm, y=1cm]
\path[cssRbox, sharp corners, draw=none] (0.350,1.412) -- (0.357,1.347) -- (0.379,1.286) -- (0.413,1.231) -- (0.459,1.185) -- (0.514,1.150) -- (0.575,1.129) -- (0.640,1.122) -- (1.360,1.122) -- (1.425,1.129) -- (1.486,1.150) -- (1.541,1.185) -- (2.405,1.905) -- (2.451,1.951) -- (2.485,2.006) -- (2.507,2.067) -- (2.514,2.132) -- (2.507,2.196) -- (2.485,2.257) -- (2.451,2.312) -- (2.405,2.358) -- (2.350,2.393) -- (2.289,2.414) -- (2.224,2.422) -- (0.640,2.422) -- (0.575,2.414) -- (0.514,2.393) -- (0.459,2.358) -- (0.413,2.312) -- (0.379,2.257) -- (0.357,2.196) -- (0.350,2.132) -- cycle;
\path[cssRbox, sharp corners, draw=none] (1.070,2.132) -- (1.077,2.067) -- (1.099,2.006) -- (1.133,1.951) -- (1.179,1.905) -- (1.234,1.870) -- (1.295,1.849) -- (1.360,1.842) -- (2.224,1.842) -- (2.289,1.849) -- (2.350,1.870) -- (2.405,1.905) -- (3.341,2.625) -- (3.387,2.671) -- (3.421,2.726) -- (3.443,2.787) -- (3.450,2.852) -- (3.443,2.916) -- (3.421,2.977) -- (3.387,3.032) -- (3.341,3.078) -- (3.286,3.113) -- (3.225,3.134) -- (3.160,3.142) -- (1.360,3.142) -- (1.295,3.134) -- (1.234,3.113) -- (1.179,3.078) -- (1.133,3.032) -- (1.099,2.977) -- (1.077,2.916) -- (1.070,2.852) -- cycle;
\path[cssVbox, sharp corners, draw=none] (0.270,2.132) -- (0.279,2.049) -- (0.307,1.971) -- (0.351,1.901) -- (0.409,1.842) -- (0.479,1.798) -- (0.558,1.771) -- (0.640,1.762) -- (0.722,1.771) -- (0.801,1.798) -- (0.871,1.842) -- (0.929,1.901) -- (0.973,1.971) -- (1.001,2.049) -- (1.010,2.132) -- (1.010,2.852) -- (1.001,2.934) -- (0.973,3.012) -- (0.929,3.082) -- (0.871,3.141) -- (0.801,3.185) -- (0.722,3.212) -- (0.640,3.222) -- (0.558,3.212) -- (0.479,3.185) -- (0.409,3.141) -- (0.351,3.082) -- (0.307,3.012) -- (0.279,2.934) -- (0.270,2.852) -- cycle;
\path[cssRbox, sharp corners, draw=none] (0.350,2.852) -- (0.357,2.787) -- (0.379,2.726) -- (0.413,2.671) -- (0.459,2.625) -- (0.514,2.590) -- (0.575,2.569) -- (0.640,2.562) -- (1.360,2.562) -- (1.425,2.569) -- (1.486,2.590) -- (1.541,2.625) -- (1.587,2.671) -- (2.199,3.391) -- (2.233,3.446) -- (2.255,3.507) -- (2.262,3.572) -- (2.255,3.636) -- (2.233,3.697) -- (2.199,3.752) -- (2.153,3.798) -- (2.098,3.833) -- (2.037,3.854) -- (1.972,3.862) -- (0.640,3.862) -- (0.575,3.854) -- (0.514,3.833) -- (0.459,3.798) -- (0.413,3.752) -- (0.379,3.697) -- (0.357,3.636) -- (0.350,3.572) -- cycle;
\path[cssRbox, sharp corners, fill=none] (0.350,1.412) -- (0.357,1.347) -- (0.379,1.286) -- (0.413,1.231) -- (0.459,1.185) -- (0.514,1.150) -- (0.575,1.129) -- (0.640,1.122) -- (1.360,1.122) -- (1.425,1.129) -- (1.486,1.150) -- (1.541,1.185) -- (2.405,1.905) -- (2.451,1.951) -- (2.485,2.006) -- (2.507,2.067) -- (2.514,2.132) -- (2.507,2.196) -- (2.485,2.257) -- (2.451,2.312) -- (2.405,2.358) -- (2.350,2.393) -- (2.289,2.414) -- (2.224,2.422) -- (0.640,2.422) -- (0.575,2.414) -- (0.514,2.393) -- (0.459,2.358) -- (0.413,2.312) -- (0.379,2.257) -- (0.357,2.196) -- (0.350,2.132) -- cycle;
\path[cssRbox, sharp corners, fill=none] (1.070,2.132) -- (1.077,2.067) -- (1.099,2.006) -- (1.133,1.951) -- (1.179,1.905) -- (1.234,1.870) -- (1.295,1.849) -- (1.360,1.842) -- (2.224,1.842) -- (2.289,1.849) -- (2.350,1.870) -- (2.405,1.905) -- (3.341,2.625) -- (3.387,2.671) -- (3.421,2.726) -- (3.443,2.787) -- (3.450,2.852) -- (3.443,2.916) -- (3.421,2.977) -- (3.387,3.032) -- (3.341,3.078) -- (3.286,3.113) -- (3.225,3.134) -- (3.160,3.142) -- (1.360,3.142) -- (1.295,3.134) -- (1.234,3.113) -- (1.179,3.078) -- (1.133,3.032) -- (1.099,2.977) -- (1.077,2.916) -- (1.070,2.852) -- cycle;
\path[cssVbox, sharp corners, fill=none] (0.270,2.132) -- (0.279,2.049) -- (0.307,1.971) -- (0.351,1.901) -- (0.409,1.842) -- (0.479,1.798) -- (0.558,1.771) -- (0.640,1.762) -- (0.722,1.771) -- (0.801,1.798) -- (0.871,1.842) -- (0.929,1.901) -- (0.973,1.971) -- (1.001,2.049) -- (1.010,2.132) -- (1.010,2.852) -- (1.001,2.934) -- (0.973,3.012) -- (0.929,3.082) -- (0.871,3.141) -- (0.801,3.185) -- (0.722,3.212) -- (0.640,3.222) -- (0.558,3.212) -- (0.479,3.185) -- (0.409,3.141) -- (0.351,3.082) -- (0.307,3.012) -- (0.279,2.934) -- (0.270,2.852) -- cycle;
\path[cssRbox, sharp corners, fill=none] (0.350,2.852) -- (0.357,2.787) -- (0.379,2.726) -- (0.413,2.671) -- (0.459,2.625) -- (0.514,2.590) -- (0.575,2.569) -- (0.640,2.562) -- (1.360,2.562) -- (1.425,2.569) -- (1.486,2.590) -- (1.541,2.625) -- (1.587,2.671) -- (2.199,3.391) -- (2.233,3.446) -- (2.255,3.507) -- (2.262,3.572) -- (2.255,3.636) -- (2.233,3.697) -- (2.199,3.752) -- (2.153,3.798) -- (2.098,3.833) -- (2.037,3.854) -- (1.972,3.862) -- (0.640,3.862) -- (0.575,3.854) -- (0.514,3.833) -- (0.459,3.798) -- (0.413,3.752) -- (0.379,3.697) -- (0.357,3.636) -- (0.350,3.572) -- cycle;
\node[cssnode, minimum size=4.2mm, font=\scriptsize, cssfill=cssOI1!55!white] (ca0) at (0.640,1.412) {$a$};
\node[cssnode, minimum size=4.2mm, font=\scriptsize, cssfill=cssOI1!55!white] (ca1) at (0.640,2.132) {$a$};
\node[cssnode, minimum size=4.2mm, font=\scriptsize, cssfill=cssOI1!55!white] (ca2) at (2.224,2.132) {$a$};
\node[cssnode, minimum size=4.2mm, font=\scriptsize, cssfill=cssOI1!55!white] (ca3) at (2.224,2.852) {$a$};
\node[cssnode, minimum size=4.2mm, font=\scriptsize, cssfill=cssOI1!55!white] (ca4) at (3.160,2.852) {$a$};
\node[cssnode, minimum size=4.2mm, font=\scriptsize, cssfill=cssOI2!55!white] (cb0) at (0.640,2.852) {$a'$};
\node[cssnode, minimum size=4.2mm, font=\scriptsize, cssfill=cssOI2!55!white] (cb1) at (0.640,3.572) {$a'$};
\node[cssnode, minimum size=4.2mm, font=\scriptsize, cssfill=cssOI2!55!white] (cb2) at (1.972,3.572) {$a'$};
\node[cssnode, minimum size=4.2mm, font=\scriptsize, cssfill=cssGrey!75!white] (cr0) at (1.360,1.412) {$r$};
\node[cssnode, minimum size=4.2mm, font=\scriptsize, cssfill=cssGrey!75!white] (cr1) at (1.360,2.132) {$r$};
\node[cssnode, minimum size=4.2mm, font=\scriptsize, cssfill=cssGrey!75!white] (cr2) at (1.360,2.852) {$r$};
\node[cssnode, minimum size=4.2mm, font=\scriptsize, cssfill=cssGrey!75!white] (cr3) at (1.360,3.572) {$r$};
\draw[csscont] (ca0) -- (ca1);
\draw[cssdep] (ca0) -- (ca2);
\draw[csscont] (ca1) -- (cb0);
\draw[csscont] (ca2) -- (ca3);
\draw[cssdep] (ca2) -- (ca4);
\draw[csscont] (cb0) -- (cb1);
\draw[cssdep] (cb0) -- (cb2);
\draw[cssdep] (cr0) -- (ca1);
\draw[cssdep] (cr0) -- (ca2);
\draw[csscont] (cr0) -- (cr1);
\draw[cssdep] (cr1) -- (ca3);
\draw[cssdep] (cr1) -- (ca4);
\draw[csscont] (cr1) -- (cr2);
\draw[cssdep] (cr2) -- (cb1);
\draw[cssdep] (cr2) -- (cb2);
\draw[csscont] (cr2) -- (cr3);
\node[anchor=north west, inner sep=0pt, align=left, font=\scriptsize, text=black] at (0.350,1.042) {R, V and lineage hold, $L = 2$\\no T: only copies; no comparative rule\\\textbf{no family}};
\draw[cssdep, draw=black!45] (0.130,1.412) -- (0.130,2.162);
\node[text=black!60, font=\scriptsize, anchor=south] at (0.130,2.162) {$t$};
\end{tikzpicture}%
\caption{\new{Conventions as in Figure~\ref{fig:ex-6.1}. Figure~\ref{fig:ex-6.2} without $Q_5$, $Q_6$ and the instances $q_0$--$q_2$, $u$ and $v$, with $c_1$, $c_2$ and $c_3$ renamed $a$, $a'$ and $r$, $I = \{a, a'\}$: R, V and lineage continuation hold, but $a'$ instances produce nothing but further $a'$ instances, whose label is already on the input boundary, and no stipulated rule makes any product depend on the $a$/$a'$ distinction, so T fails and no family is assigned.}}
\label{fig:ex-6.3}
\end{figure}

\subsection{Inert copying}\label{sec:6.4}

A machine repeatedly copies arbitrary data strings \new{(Figure~\ref{fig:ex-6.4})}. Both $i_1$ and $i_2$ are replicated. Within the observed episode, however, the differing contents never produce \del{distinguishable downstream realised structures}\new{a downstream structure of a type absent from the input boundary}. The CSS therefore identifies R but not T.

\begin{figure}[!htbp]
\centering
\begin{tikzpicture}[cssfig, x=1cm, y=1cm]
\path[cssRbox, sharp corners, draw=none] (0.350,2.132) -- (0.357,2.067) -- (0.379,2.006) -- (0.413,1.951) -- (0.459,1.905) -- (1.239,1.185) -- (1.294,1.150) -- (1.355,1.129) -- (1.420,1.122) -- (2.200,1.122) -- (2.265,1.129) -- (2.326,1.150) -- (2.381,1.185) -- (2.427,1.231) -- (2.461,1.286) -- (2.483,1.347) -- (2.490,1.412) -- (2.490,2.132) -- (2.483,2.196) -- (2.461,2.257) -- (2.427,2.312) -- (2.381,2.358) -- (2.326,2.393) -- (2.265,2.414) -- (2.200,2.422) -- (0.640,2.422) -- (0.575,2.414) -- (0.514,2.393) -- (0.459,2.358) -- (0.413,2.312) -- (0.379,2.257) -- (0.357,2.196) -- cycle;
\path[cssRbox, sharp corners, draw=none] (1.910,2.132) -- (1.917,2.067) -- (1.939,2.006) -- (1.973,1.951) -- (2.019,1.905) -- (2.074,1.870) -- (2.135,1.849) -- (2.200,1.842) -- (2.980,1.842) -- (3.045,1.849) -- (3.106,1.870) -- (3.161,1.905) -- (3.941,2.625) -- (3.987,2.671) -- (4.021,2.726) -- (4.043,2.787) -- (4.050,2.852) -- (4.043,2.916) -- (4.021,2.977) -- (3.987,3.032) -- (3.941,3.078) -- (3.886,3.113) -- (3.825,3.134) -- (3.760,3.142) -- (2.200,3.142) -- (2.135,3.134) -- (2.074,3.113) -- (2.019,3.078) -- (1.973,3.032) -- (1.939,2.977) -- (1.917,2.916) -- (1.910,2.852) -- cycle;
\path[cssRbox, sharp corners, fill=none] (0.350,2.132) -- (0.357,2.067) -- (0.379,2.006) -- (0.413,1.951) -- (0.459,1.905) -- (1.239,1.185) -- (1.294,1.150) -- (1.355,1.129) -- (1.420,1.122) -- (2.200,1.122) -- (2.265,1.129) -- (2.326,1.150) -- (2.381,1.185) -- (2.427,1.231) -- (2.461,1.286) -- (2.483,1.347) -- (2.490,1.412) -- (2.490,2.132) -- (2.483,2.196) -- (2.461,2.257) -- (2.427,2.312) -- (2.381,2.358) -- (2.326,2.393) -- (2.265,2.414) -- (2.200,2.422) -- (0.640,2.422) -- (0.575,2.414) -- (0.514,2.393) -- (0.459,2.358) -- (0.413,2.312) -- (0.379,2.257) -- (0.357,2.196) -- cycle;
\path[cssRbox, sharp corners, fill=none] (1.910,2.132) -- (1.917,2.067) -- (1.939,2.006) -- (1.973,1.951) -- (2.019,1.905) -- (2.074,1.870) -- (2.135,1.849) -- (2.200,1.842) -- (2.980,1.842) -- (3.045,1.849) -- (3.106,1.870) -- (3.161,1.905) -- (3.941,2.625) -- (3.987,2.671) -- (4.021,2.726) -- (4.043,2.787) -- (4.050,2.852) -- (4.043,2.916) -- (4.021,2.977) -- (3.987,3.032) -- (3.941,3.078) -- (3.886,3.113) -- (3.825,3.134) -- (3.760,3.142) -- (2.200,3.142) -- (2.135,3.134) -- (2.074,3.113) -- (2.019,3.078) -- (1.973,3.032) -- (1.939,2.977) -- (1.917,2.916) -- (1.910,2.852) -- cycle;
\node[cssnode, minimum size=4.2mm, font=\scriptsize, cssfill=cssGrey!75!white] (dM0) at (2.200,1.412) {$m$};
\node[cssnode, minimum size=4.2mm, font=\scriptsize, cssfill=cssGrey!75!white] (dM1) at (2.200,2.132) {$m$};
\node[cssnode, minimum size=4.2mm, font=\scriptsize, cssfill=cssGrey!75!white] (dM2) at (2.200,2.852) {$m$};
\node[cssnode, minimum size=4.2mm, font=\scriptsize, cssfill=cssOI1!55!white] (dx0) at (1.420,1.412) {$i_1$};
\node[cssnode, minimum size=4.2mm, font=\scriptsize, cssfill=cssOI1!55!white] (dx1) at (1.420,2.132) {$i_1$};
\node[cssnode, minimum size=4.2mm, font=\scriptsize, cssfill=cssOI1!55!white] (dxc) at (0.640,2.132) {$i_1$};
\node[cssnode, minimum size=4.2mm, font=\scriptsize, cssfill=cssOI2!55!white] (dz0) at (2.980,2.132) {$i_2$};
\node[cssnode, minimum size=4.2mm, font=\scriptsize, cssfill=cssOI2!55!white] (dz1) at (2.980,2.852) {$i_2$};
\node[cssnode, minimum size=4.2mm, font=\scriptsize, cssfill=cssOI2!55!white] (dzc) at (3.760,2.852) {$i_2$};
\draw[csscont] (dM0) -- (dM1);
\draw[cssdep] (dM0) -- (dx1);
\draw[cssdep] (dM0) -- (dxc);
\draw[csscont] (dM1) -- (dM2);
\draw[cssdep] (dM1) -- (dz1);
\draw[cssdep] (dM1) -- (dzc);
\draw[csscont] (dx0) -- (dx1);
\draw[cssdep] (dx0) -- (dxc);
\draw[csscont] (dz0) -- (dz1);
\draw[cssdep] (dz0) -- (dzc);
\node[anchor=north west, inner sep=0pt, align=left, font=\scriptsize, text=black] at (0.350,1.042) {R for $i_1$ and $i_2$; no V; $L = 1$\\no T: copy labels on input boundary\\\textbf{no family}};
\draw[cssdep, draw=black!45] (0.130,1.412) -- (0.130,2.162);
\node[text=black!60, font=\scriptsize, anchor=south] at (0.130,2.162) {$t$};
\end{tikzpicture}%
\caption{\new{Conventions as in Figure~\ref{fig:ex-6.1}. $I = \{i_1, i_2\}$: R holds for both strings, with corresponding operator label $m$, but T fails because the labels of the copies are already on the input boundary; no V process connects $i_1$ and $i_2$, $L = 1$, and no family is assigned.}}
\label{fig:ex-6.4}
\end{figure}

The patterns are copied, but this episode does not satisfy the information-family criterion, and so no information-carrier role is assigned. The additional copies do not supply T: their structural labels are already present on the corresponding input boundaries. T requires a variant-dependent product of a structural type absent from the input boundary.

\new{The case predicts R without T for any copied payload whose content is never used.}

\subsection{Translation without hereditary continuation}\label{sec:6.5}

Suppose alternative configurations
\new{$i_1 \to y_1$ and $i_2 \to y_2$}{\delmark}
through the same machinery \new{(Figure~\ref{fig:ex-6.5})}. This is T\new{, provided the machinery satisfies the operator condition in both processes with corresponding organisation and each product type is absent from its input boundary, and replacing $i_1$ by $i_2$, all else corresponding, changes the product from $y_1$ to $y_2$ (Section~\ref{sec:4.4})}.

\begin{figure}[!htbp]
\centering
\begin{tikzpicture}[cssfig, x=1cm, y=1cm]
\path[cssTbox, sharp corners, draw=none] (0.350,1.412) -- (0.357,1.347) -- (0.379,1.286) -- (0.413,1.231) -- (0.459,1.185) -- (0.514,1.150) -- (0.575,1.129) -- (0.640,1.122) -- (1.420,1.122) -- (1.485,1.129) -- (1.546,1.150) -- (1.601,1.185) -- (1.647,1.231) -- (1.681,1.286) -- (1.703,1.347) -- (1.710,1.412) -- (1.710,2.132) -- (1.703,2.196) -- (1.681,2.257) -- (1.647,2.312) -- (1.601,2.358) -- (1.546,2.393) -- (1.485,2.414) -- (1.420,2.422) -- (0.640,2.422) -- (0.575,2.414) -- (0.514,2.393) -- (0.459,2.358) -- (0.413,2.312) -- (0.379,2.257) -- (0.357,2.196) -- (0.350,2.132) -- cycle;
\path[cssTbox, sharp corners, draw=none] (1.130,2.132) -- (1.137,2.067) -- (1.159,2.006) -- (1.193,1.951) -- (1.239,1.905) -- (1.294,1.870) -- (1.355,1.849) -- (1.420,1.842) -- (2.200,1.842) -- (2.265,1.849) -- (2.326,1.870) -- (2.381,1.905) -- (2.427,1.951) -- (2.461,2.006) -- (2.483,2.067) -- (2.490,2.132) -- (2.490,2.852) -- (2.483,2.916) -- (2.461,2.977) -- (2.427,3.032) -- (2.381,3.078) -- (2.326,3.113) -- (2.265,3.134) -- (2.200,3.142) -- (1.420,3.142) -- (1.355,3.134) -- (1.294,3.113) -- (1.239,3.078) -- (1.193,3.032) -- (1.159,2.977) -- (1.137,2.916) -- (1.130,2.852) -- cycle;
\path[cssTbox, sharp corners, fill=none] (0.350,1.412) -- (0.357,1.347) -- (0.379,1.286) -- (0.413,1.231) -- (0.459,1.185) -- (0.514,1.150) -- (0.575,1.129) -- (0.640,1.122) -- (1.420,1.122) -- (1.485,1.129) -- (1.546,1.150) -- (1.601,1.185) -- (1.647,1.231) -- (1.681,1.286) -- (1.703,1.347) -- (1.710,1.412) -- (1.710,2.132) -- (1.703,2.196) -- (1.681,2.257) -- (1.647,2.312) -- (1.601,2.358) -- (1.546,2.393) -- (1.485,2.414) -- (1.420,2.422) -- (0.640,2.422) -- (0.575,2.414) -- (0.514,2.393) -- (0.459,2.358) -- (0.413,2.312) -- (0.379,2.257) -- (0.357,2.196) -- (0.350,2.132) -- cycle;
\path[cssTbox, sharp corners, fill=none] (1.130,2.132) -- (1.137,2.067) -- (1.159,2.006) -- (1.193,1.951) -- (1.239,1.905) -- (1.294,1.870) -- (1.355,1.849) -- (1.420,1.842) -- (2.200,1.842) -- (2.265,1.849) -- (2.326,1.870) -- (2.381,1.905) -- (2.427,1.951) -- (2.461,2.006) -- (2.483,2.067) -- (2.490,2.132) -- (2.490,2.852) -- (2.483,2.916) -- (2.461,2.977) -- (2.427,3.032) -- (2.381,3.078) -- (2.326,3.113) -- (2.265,3.134) -- (2.200,3.142) -- (1.420,3.142) -- (1.355,3.134) -- (1.294,3.113) -- (1.239,3.078) -- (1.193,3.032) -- (1.159,2.977) -- (1.137,2.916) -- (1.130,2.852) -- cycle;
\node[cssnode, minimum size=4.2mm, font=\scriptsize, cssfill=cssGrey!75!white] (eK0) at (1.420,1.412) {$k$};
\node[cssnode, minimum size=4.2mm, font=\scriptsize, cssfill=cssGrey!75!white] (eK1) at (1.420,2.132) {$k$};
\node[cssnode, minimum size=4.2mm, font=\scriptsize, cssfill=cssGrey!75!white] (eK2) at (1.420,2.852) {$k$};
\node[cssnode, minimum size=4.2mm, font=\scriptsize, cssfill=cssOI1!55!white] (eX) at (0.640,1.412) {$i_1$};
\node[cssnode, minimum size=4.2mm, font=\scriptsize, cssfill=cssOI6!60!white] (eY1) at (0.640,2.132) {$y_1$};
\node[cssnode, minimum size=4.2mm, font=\scriptsize, cssfill=cssOI2!55!white] (eZ) at (2.200,2.132) {$i_2$};
\node[cssnode, minimum size=4.2mm, font=\scriptsize, cssfill=cssOI4!55!white] (eY2) at (2.200,2.852) {$y_2$};
\draw[csscont] (eK0) -- (eK1);
\draw[cssdep] (eK0) -- (eY1);
\draw[csscont] (eK1) -- (eK2);
\draw[cssdep] (eK1) -- (eY2);
\draw[cssdep] (eX) -- (eY1);
\draw[cssdep] (eZ) -- (eY2);
\node[anchor=north west, inner sep=0pt, align=left, font=\scriptsize, text=black] at (0.350,1.042) {T under rule $F$; no V\\no R, no lineage: $i_1$, $i_2$ not copied\\\textbf{no family}};
\draw[cssdep, draw=black!45] (0.130,1.412) -- (0.130,2.162);
\node[text=black!60, font=\scriptsize, anchor=south] at (0.130,2.162) {$t$};
\end{tikzpicture}%
\caption{\new{Conventions as in Figure~\ref{fig:ex-6.1}. $I = \{i_1, i_2\}$: T holds under the stipulated rule $F(i_1; k) = y_1$, $F(i_2; k) = y_2$, but neither variant is replicated, so no R process and no lineage continuation exist, no V process connects the variants, and no family is assigned.}}
\label{fig:ex-6.5}
\end{figure}

If the \new{information }carrier configurations are not themselves causally replicated into continuing descendant processes, the observation supplies no \del{deep hereditary lineage}\new{R process and no lineage continuation, and no information family is assigned, however distinct the products are}. Thus variant-dependent causation is not sufficient for sustained heredity.

\new{Sections~\ref{sec:6.1} and~\ref{sec:6.3} to~\ref{sec:6.5} thus show that no single component suffices: the operator condition alone, R and V without T, R without T, and T without R or lineage continuation each fail the joint criterion, which Section~\ref{sec:6.2} satisfies only because one family of \new{information }carrier variants satisfies every component coherently.}

\begin{newtext}
\textbf{What Section~\ref{sec:6} shows.} RVT requires a coherent set of relations rather than independently present motifs. Every case of this section starts from a CSS that we stipulate, i.e., from the output of stage 5 of Figure~\ref{fig:pipeline}, and runs only stage 6. Sections~\ref{sec:cultural} and~\ref{sec:comp} run all six stages, each in one domain and each guided by its own version of Figure~\ref{fig:pipeline}: a cultural episode reconstructed under stated hypotheses, and digital-evolution episodes constructed from a recorded history.
\end{newtext}

\section[A cultural episode: conditional reconstruction of recipe transmission]{\new{A cultural episode: conditional reconstruction of recipe transmission}}\label{sec:cultural}

\textbf{The episode in ordinary terms:} \emph{A cook, A, knows how to prepare a particular dish. Over the course of an afternoon A explains and demonstrates the recipe to B, who did not previously know it: A talks, B listens and watches. When the session ends, A can still prepare the dish and B now can too. Later, B prepares it. Later still, B teaches the same recipe to a third person, D. Separately, another cook, C, who knows a different recipe, prepares a different dish in the same kitchen using the same pots and stove.}

\revnote{[v7 paragraph on the question the episode poses moved unchanged to Appendix~\ref{app:G.1}.]}

{\delmark}
\begin{newtext}
We reconstruct the episode twice: Case A is the episode as stated, and Case B extends it by stipulated events. Both cases share the declared perspective, the physical hypotheses, the structural types and the classes of teaching and preparation processes, and both are evaluated by the same frozen criterion under the conditional procedure of Appendix~\ref{app:F.3}. No recorded episode is analysed. Appendix~\ref{app:G} gives both constructions in full, each value checked by a script.
Unlike the cases of Section~\ref{sec:6}, which start from a stipulated CSS, the reconstruction runs all six stages of Figure~\ref{fig:pipeline}. Figure~\ref{fig:pipeline-cultural} shows how each stage is implemented in both cases, and the subsections follow its stages: stages 1 to 3, which the two cases share (Section~\ref{sec:cultural.1}); stages 4 and 5, in which they differ (Section~\ref{sec:cultural.2}); stage 6 (Section~\ref{sec:cultural.3}), and; the validation that a stipulated history does not permit (Section~\ref{sec:cultural.4}).
\end{newtext}

\begin{figure}[t]
\begin{newtext}
\centering
\begin{tikzpicture}[cssfig,
  pipenum/.style={circle, minimum size=4.6mm, inner sep=0pt, line width=0.5pt,
    draw=cssOI5!70!black, fill=cssOI5!35, font=\footnotesize\bfseries},
  pipeinf/.style={circle, minimum size=4.6mm, inner sep=0pt, line width=0.5pt,
    draw=cssOI2!70!black, fill=cssOI2!35, font=\footnotesize\bfseries}]

\node[anchor=north west, inner sep=0pt] (h) at (0,0) {%
\begin{tabular}{@{\hspace{0.6cm}}>{\raggedright}p{2.85cm}@{\hspace{0.12cm}}>{\raggedright}p{5.75cm}@{\hspace{0.12cm}}>{\raggedright}p{5.75cm}@{}}
\rule{0pt}{0.3cm}\textbf{Stage}
&
\rule{0pt}{0.3cm}\textbf{Case A: the episode as stated}\par
{\scriptsize\color{black!65}stipulated under H1--H4\par}
&
\rule{0pt}{0.3cm}\textbf{Case B: a completed history}\par
{\scriptsize\color{black!65}Case A plus H5--H8, also stipulated\par}
\end{tabular}};
\draw[line width=0.5pt] ([yshift=-0.03cm]h.south west) -- ([yshift=-0.03cm]h.south east);

\node[anchor=north west, inner sep=0pt] (r1) at ([yshift=-0.08cm]h.south west) {%
\begin{tabular}{@{\hspace{0.6cm}}>{\raggedright}p{2.85cm}@{\hspace{0.12cm}}>{\raggedright}p{11.62cm}@{}}
\rule{0pt}{0.45cm}\textbf{Perspective}\par
{\scriptsize\color{black!60}Sect.~\ref{sec:cultural.1}\par}
&
\rule{0pt}{0.45cm}{\scriptsize Measured: positions and configurations of participants, utensils and foodstuffs; the acoustic field; the procedural sequences of food preparation. Resolution of ordinary observation and of one session; two declared recipes. H1: each participant's state factorises into a capability component and a remainder component.\par}
\end{tabular}};

\node[anchor=north west, inner sep=0pt] (r2) at ([yshift=-0.08cm]r1.south west) {%
\begin{tabular}{@{\hspace{0.6cm}}>{\raggedright}p{2.85cm}@{\hspace{0.12cm}}>{\raggedright}p{11.62cm}@{}}
\rule{0pt}{0.45cm}\textbf{Structures}\par
{\scriptsize\color{black!60}Sect.~\ref{sec:cultural.1}\par}
&
\rule{0pt}{0.45cm}{\scriptsize H2: capability and remainder components as persistent, non-overlapping world-tubes, cut into sections; apparatus (pots and stove) as one persistent structure; ingredients, dishes and utterances as single sections.\par}
\end{tabular}};

\node[anchor=north west, inner sep=0pt] (r3) at ([yshift=-0.08cm]r2.south west) {%
\begin{tabular}{@{\hspace{0.6cm}}>{\raggedright}p{2.85cm}@{\hspace{0.12cm}}>{\raggedright}p{11.62cm}@{}}
\rule{0pt}{0.45cm}\textbf{Persistence and types}\par
{\scriptsize\color{black!60}Sect.~\ref{sec:cultural.1}\par}
&
\rule{0pt}{0.45cm}{\scriptsize H3: capability sets $c_0$, $c_1$, $c_2$, $c_{12}$; remainder $m$; apparatus $k$, ingredients $g$, utterance $s$, dishes $y_1$, $y_2$. No AIR partition is established: behavioural competence does not establish one.\par}
\end{tabular}};

\node[anchor=north west, inner sep=0pt] (r4) at ([yshift=-0.08cm]r3.south west) {%
\begin{tabular}{@{\hspace{0.6cm}}>{\raggedright}p{2.85cm}@{\hspace{0.12cm}}>{\raggedright}p{5.75cm}@{\hspace{0.12cm}}>{\raggedright}p{5.75cm}@{}}
\rule{0pt}{0.45cm}\textbf{Nodes and labels}\par
{\scriptsize\color{black!60}Sect.~\ref{sec:cultural.2}\par}
&
\rule{0pt}{0.45cm}{\scriptsize 27 instances: sections of A, B, C and D; apparatus, ingredient and utterance instances; dishes $d_1$, $d_2$.\par}
&
\rule{0pt}{0.45cm}{\scriptsize 47 instances: Case A plus the events of H5--H7, i.e., B modifies the recipe, B teaches E, E teaches F, and B prepares $d_3$.\par}
\end{tabular}};

\node[anchor=north west, inner sep=0pt] (r5) at ([yshift=-0.08cm]r4.south west) {%
\begin{tabular}{@{\hspace{0.6cm}}>{\raggedright}p{2.85cm}@{\hspace{0.12cm}}>{\raggedright}p{5.75cm}@{\hspace{0.12cm}}>{\raggedright}p{5.75cm}@{}}
\rule{0pt}{0.45cm}\textbf{Causal order and CSS}\par
{\scriptsize\color{black!60}Sect.~\ref{sec:cultural.2}\par}
&
\rule{0pt}{0.45cm}{\scriptsize 26 stipulated dependencies, among them H4: the bodies of teacher and learner contribute to the learner's new capability section. $\prec\,=D_A^{+}$.\par}
&
\rule{0pt}{0.45cm}{\scriptsize 53 dependencies: Case A plus those of H5--H7, and, from H8, the cook's capability section to the dish. $\prec\,=D_B^{+}$.\par}
\end{tabular}};

\node[anchor=north west, inner sep=0pt] (r6) at ([yshift=-0.08cm]r5.south west) {%
\begin{tabular}{@{\hspace{0.6cm}}>{\raggedright}p{2.85cm}@{\hspace{0.12cm}}>{\raggedright}p{5.75cm}@{\hspace{0.12cm}}>{\raggedright}p{5.75cm}@{}}
\rule{0pt}{0.45cm}\textbf{Processes, motifs, roles}\par
{\scriptsize\color{black!60}Sect.~\ref{sec:cultural.3}\par}
&
\rule{0pt}{0.45cm}{\scriptsize Four processes: $Q_{AB}$, $Q_{d_1}$, $Q_{d_2}$, $Q_{BD}$. R for $c_1$, lineage $L = 2$. No R for $c_2$, no V, no T: no complete family for $I = \{c_1, c_2\}$, and no roles assigned.\par}
&
\rule{0pt}{0.45cm}{\scriptsize Eight processes. V: $Q_{\mathrm{mod}}$; R for $c_2$: $Q_{BE}$, $Q_{EF}$; T: $Q_{d_1}$, $Q_{d_3}$, by H8. Family $I = \{c_1, c_2\}$ and roles, conditional on H1--H8.\par}
\end{tabular}};

\foreach \rowname in {r1, r2, r4}{
  \draw[black!30, line width=0.3pt] ([yshift=-0.04cm]\rowname.south west) -- ([yshift=-0.04cm]\rowname.south east);
}
\draw[black!55, line width=0.5pt] ([yshift=-0.04cm]r3.south west) -- ([yshift=-0.04cm]r3.south east);
\draw[black!70, line width=0.7pt] ([yshift=-0.04cm]r5.south west) -- ([yshift=-0.04cm]r5.south east);
\draw[black!30, line width=0.3pt] ([xshift=9.38cm, yshift=-0.04cm]r3.south west) -- ([xshift=9.38cm]r6.south west);

\node[pipenum] (c1) at ([xshift=0.3cm, yshift=-0.32cm]r1.north west) {1};
\node[pipenum] (c2) at ([xshift=0.3cm, yshift=-0.32cm]r2.north west) {2};
\node[pipenum] (c3) at ([xshift=0.3cm, yshift=-0.32cm]r3.north west) {3};
\node[pipenum] (c4) at ([xshift=0.3cm, yshift=-0.32cm]r4.north west) {4};
\node[pipenum] (c5) at ([xshift=0.3cm, yshift=-0.32cm]r5.north west) {5};
\node[pipeinf] (c6) at ([xshift=0.3cm, yshift=-0.32cm]r6.north west) {6};
\foreach \cfrom/\cto in {c1/c2, c2/c3, c3/c4, c4/c5, c5/c6}{
  \draw[cssdep, draw=black!55] (\cfrom) -- (\cto);
}

\draw[line width=0.9pt, cssOI5!70!black] ([xshift=-0.06cm, yshift=-0.03cm]r1.north west) -- ([xshift=-0.17cm, yshift=-0.03cm]r1.north west)
  -- ([xshift=-0.17cm, yshift=0.03cm]r5.south west) -- ([xshift=-0.06cm, yshift=0.03cm]r5.south west);
\path (r1.north west) -- (r5.south west) coordinate[pos=0.5] (m1);
\node[rotate=90, anchor=south, text=cssOI5!45!black, inner sep=1pt] at ([xshift=-0.24cm]m1) {construction of the CSS};
\draw[line width=0.9pt, cssOI2!70!black] ([xshift=-0.06cm, yshift=-0.03cm]r6.north west) -- ([xshift=-0.17cm, yshift=-0.03cm]r6.north west)
  -- ([xshift=-0.17cm, yshift=0.03cm]r6.south west) -- ([xshift=-0.06cm, yshift=0.03cm]r6.south west);
\path (r6.north west) -- (r6.south west) coordinate[pos=0.5] (m2);
\node[rotate=90, anchor=south, text=cssOI2!55!black, inner sep=1pt] at ([xshift=-0.24cm]m2) {inference};

\node[anchor=north west, draw=black!60, densely dashed, line width=0.6pt, rounded corners=1mm,
  inner xsep=4pt, inner ysep=7pt, text width=14.86cm, align=left, font=\scriptsize] (ft) at ([yshift=-0.4cm]r6.south west)
  {\textbf{Post-inference validation} (Sect.~\ref{sec:cultural.4}): not available. Both histories are stipulated, and recovering a motif component that a hypothesis entails is not confirmation of it.};
\node[csstag, fill=white, anchor=east, inner sep=2pt] at ([xshift=-0.35cm]ft.north east) {assesses the output of stage 6};

\end{tikzpicture}
\caption{\textbf{The six stages of Figure~\ref{fig:pipeline}, implemented for the cultural episode.} Stages 1 to 3 are shared by Case A and Case B; stages 4 to 6 differ, because Case B adds the stipulated events and dependencies of H5 to H8. Every entry is stipulated or hypothesised, not observed, and below each stage name is the subsection that implements it. Numbers count the instances, dependencies and processes of the stipulated CSS; Appendix~\ref{app:G} gives both constructions in full.}
\label{fig:pipeline-cultural}
\end{newtext}
\end{figure}

\subsection{\new{Stages 1 to 3, shared by both cases: perspective, structures and structural types}}\label{sec:cultural.1}

\revnote{[v7 perspective and physical hypotheses H1--H4, with their ordinary-language explanations, moved unchanged to Appendix~\ref{app:G.1}.]}

\begin{newtext}
\textbf{Physical hypotheses.} Capability configurations are physically ordinary but not observable at the declared resolution. Both cases rest on four hypotheses: H1 (factorisation), each participant's state factorises into a capability component and a remainder component; H2 (individuation), these components admit distinct, persistent, non-overlapping physical instances; H3 (persistence and structural typing), capability instances occupy persistent structural types, and remainder instances keep one structural type $m$ across a process, and; H4 (bodily contribution), in an instruction process the $m$ instances contribute causally to the learner's acquired capability instance. H1--H4 are hypotheses, not established facts: behavioural competence does not establish a physically admissible structural type, such as an AIR, or a separable neural structure.
\end{newtext}

\begin{newtext}
\textbf{Stage 1: perspective.} Figure~\ref{fig:pipeline} asks for the episode $\mathcal{R}$ and the perspective $P = (\pi_P, D_P, J_P, \rho_P, \Sigma_P)$ of Section~\ref{sec:3.1}. The episode $\mathcal{R}$ comprises A teaching B, B preparing the dish, B teaching D and C preparing another dish in the same kitchen. The state map $\pi_P$ declares as measured variables the positions and configurations of participants, utensils and foodstuffs, the acoustic field and the procedural sequences that participants perform when they prepare food. The resolution $\rho_P$ is that of ordinary observation in space and of one session in time, and the declared recipe repertoire contains two recipes. The perspective specifies no dynamics $D_P$ and no interventions $J_P$ beyond the dependencies of stage 5: each stipulated dependency is a hypothesis about what an admissible intervention would change (Section~\ref{sec:3.5}). The individuation rule $\Sigma_P$ is supplied by H1--H3 (Appendix~\ref{app:G.1}).

\textbf{Stage 2: structures.} Figure~\ref{fig:pipeline} asks for the structural domain $Y_P$ and the candidate structures, i.e., world-tube sections in $Y_P$. Under H1 and H2, each participant X has a capability world-tube $X^{\mathrm{cap}}$ and a remainder world-tube $X^{m}$, which do not overlap. The perspective declares four further kinds of structure: the pots and stove, treated as one persistent structure $K$; ingredient instances, each consumed by one preparation; dish instances, and; utterances, whose individuation as persistent structures is itself hypothesised. Each world-tube is cut into sections at the processes in which it takes part: $X_j^{\mathrm{cap}}$ and $X_j^{m}$ are the $j$-th sections of the tubes of X, $K_1, K_2, \ldots$ those of the apparatus, and ingredient, dish and utterance instances are single sections $G_j$, $d_j$ and $u_{XY}$.

\textbf{Stage 3: persistence and structural types.} Figure~\ref{fig:pipeline} asks for the persistent structures and the label set $\mathcal{L}_P$. Under H3, capability and remainder sections are persistent, and remainder sections keep the one type $m$ across a process. The types are hypothesised, not derived from an AIR partition (Section~\ref{sec:3.3}).
\end{newtext}
Capability structural types range over capability sets: $c_0$ for neither recipe, $c_1$ for the first recipe only, $c_2$ for the second only, and $c_{12}$ for both. \new{Further labels are $k$ (apparatus), $g$ (ingredients), $s$ (acoustic pattern of an utterance) and $y_1$, $y_2$ (dish types). Instances are distinct from labels: $B_1^{\mathrm{cap}}$ and $B_2^{\mathrm{cap}}$ are two sections of B's capability world-tube, of types $c_0$ and $c_1$.} The candidate information family assessed below is {\new{$I = \{c_1, c_2\}$}}.

\subsection{\new{Stages 4 and 5, in which the cases differ: nodes and stipulated dependencies}}\label{sec:cultural.2}

\begin{newtext}
\textbf{Stage 4: nodes and labels.} Figure~\ref{fig:pipeline} maps each persistent section to one CSS node, $M_P(S_j) = s_j$, and each node to its structural label $\ell(s_j)$. \emph{Case A}, the episode as stated, has 27 nodes: eight capability sections of A, B, C and D; ten remainder sections; three apparatus sections; two ingredient instances; the dishes $d_1$ and $d_2$, and; the utterances $u_{AB}$ and $u_{BD}$. Five capability sections carry the label $c_1$, two carry $c_0$, and C's carries $c_2$ (Appendix~\ref{app:G.2}).

\emph{Case B}, a completed history, keeps every node, label and dependency of Case A and adds four stipulated hypotheses. Three of them add events, and with them nodes:

\emph{H5 (variation by replacement).} After teaching D, B modifies the recipe, and B's capability section changes type: $B_3^{\mathrm{cap}}\,D\,B_4^{\mathrm{cap}}$ with $\ell(B_4^{\mathrm{cap}}) = c_2$ (process $Q_{\mathrm{mod}}$). We read the modification as replacement, i.e., B now prepares only the modified recipe; acquisition of an additional recipe would instead give the type $c_{12}$. That B's modified configuration has C's type $c_2$ is an idealised convergence, stipulated rather than expected.

\emph{H6 (replication of $c_2$).} B teaches the modified recipe to E ($Q_{BE}$), and E teaches it to F ($Q_{EF}$), each with the dependency pattern of $Q_{AB}$.

\emph{H7 (preparation with $c_2$).} After the modification, B prepares a dish $d_3$ with apparatus and ingredients of the same types as in $Q_{d_1}$ ($Q_{d_3}$).

Case B thus has 47 nodes (Appendix~\ref{app:G.3}).

\textbf{Stage 5: causal order and the CSS.} Figure~\ref{fig:pipeline} asks for the direct dependencies $D$ and the order ${\prec} = D^{+}$. Here, every dependency is stipulated, by one pattern per class of process (Appendix~\ref{app:G.1}). In a teaching process, the sections of teacher and learner continue, the utterance depends on the teacher's capability and body, and the learner's new capability section depends on the utterance, on the learner's body (H4) and on the learner's previous capability section. In a preparation process, body and apparatus continue and contribute to the dish, and the ingredient instance contributes to the dish; no dish depends on the cook's capability section. In Case A, the four processes of the episode give 26 dependencies; $D_A$ is acyclic, every dependency runs forward in time, and ${\prec} = D_A^{+}$ (Appendix~\ref{app:G.2}).

In Case B, the events of H5--H7 add their dependencies by the same patterns, and the fourth hypothesis adds two more:

\emph{H8 (comparative rule).} With the apparatus type $k$, the bodily type $m$ and the ingredient type $g$ fixed, the capability type fixes the dish type, $F(c_1; k, m, g) = y_1$ and $F(c_2; k, m, g) = y_2$: replacing the $c_1$ section in $Q_{d_1}$ by a $c_2$ section, all else corresponding, changes the realised dish type from $y_1$ to $y_2$. By the intervention criterion of Section~\ref{sec:3.5}, H8 entails that the cook's capability section contributes causally to the dish, so Case B also contains $B_2^{\mathrm{cap}}\,D\,d_1$ and $C_1^{\mathrm{cap}}\,D\,d_2$, the only relations that Case B adds between Case A instances.

Case B thus has 53 dependencies; $D_B$ is acyclic and respects temporal precedence, and ${\prec} = D_B^{+}$ (Appendix~\ref{app:G.3}).
\end{newtext}

\subsection{\new{Stage 6: an incomplete and a completed information family}}\label{sec:cultural.3}

\begin{newtext}
\textbf{Case A.} Stage 6 identifies four processes: A teaches B ($Q_{AB}$); B prepares dish $d_1$ ($Q_{d_1}$); C prepares dish $d_2$ with the same apparatus ($Q_{d_2}$), and; B teaches D ($Q_{BD}$). $Q_{AB}$ is an R process for $c_1$: $n^{-}(c_1) = 1$, $n^{+}(c_1) = 2$, and both outputs, $A_2^{\mathrm{cap}}$ and $B_2^{\mathrm{cap}}$, descend from $A_1^{\mathrm{cap}}$; the label $m$ is conserved, $n^{-}(m) = n^{+}(m) = 2$, contributes to $B_2^{\mathrm{cap}}$ (H4) and satisfies the operator condition. $Q_{BD}$ has the corresponding organisation, and its teacher input is $B_2^{\mathrm{cap}}$, produced in $Q_{AB}$, so $A_1^{\mathrm{cap}} \rightsquigarrow B_2^{\mathrm{cap}} \rightsquigarrow D_2^{\mathrm{cap}}$ gives $L = 2$ for $c_1$.

For $I = \{c_1, c_2\}$, however, three conditions remain unestablished. First, no R process for $c_2$ exists: C's capability section is the only $c_2$ instance. Second, no V process connects $c_1$ and $c_2$: no $c_1$ instance precedes a $c_2$ instance, or the reverse. Third, T is not established. $Q_{d_1}$ and $Q_{d_2}$ have corresponding operator labels $k$ and $m$, and their dish types are absent from their input boundaries, but no dish depends on a capability instance, and no comparative rule attributes the dish distinction to the capability distinction rather than to ingredients, quantities or other conditions; different cooks producing different dishes does not by itself establish T. The three gaps hold for every process of the Case A CSS (Appendix~\ref{app:G.2}), so no complete two-variant information family is established. The gaps are absences from the stipulated description, not relations shown to be physically absent: Case A shows that the supplied reconstruction is insufficient, not that recipe teaching lacks hereditary organisation.

\textbf{Case B.} Stage 6 identifies eight processes, the four of Case A and $Q_{\mathrm{mod}}$, $Q_{BE}$, $Q_{EF}$ and $Q_{d_3}$, all causally convex (Appendix~\ref{app:G.3}). The joint criterion of Section~\ref{sec:5} holds for $I = \{c_1, c_2\}$: 1) $Q_{\mathrm{mod}}$ is a V process from $c_1$ to $c_2$; 2) $Q_{AB}$ and $Q_{BD}$ are R processes for $c_1$, and $Q_{BE}$ and $Q_{EF}$ for $c_2$, each with \new{information }carrier multiplicity $1 \to 2$ and with corresponding boundary patterns under operator label $m$; 3) $Q_{d_1}$ and $Q_{d_3}$ take $c_1$ and $c_2$ as inputs and produce $y_1$ and $y_2$, which are absent from their input boundaries, and the apparatus and bodily labels $k$ and $m$ satisfy the operator condition in both, with corresponding patterns; 4) by H8, the product distinction depends on the capability distinction, and; 5) $A_1^{\mathrm{cap}} \rightsquigarrow B_2^{\mathrm{cap}} \rightsquigarrow D_2^{\mathrm{cap}}$ and $B_4^{\mathrm{cap}} \rightsquigarrow E_2^{\mathrm{cap}} \rightsquigarrow F_2^{\mathrm{cap}}$ give $L = 2$ for each variant. $Q_{EF}$ is not required by the criterion, since Section~\ref{sec:5} requires $L \geq 2$ for the family as a whole, not for each variant. H7 is not required under the reading of Section~\ref{sec:4.6} that associates a translation witness with a lineage by its structural label (Appendix~\ref{app:G.4}). $Q_{EF}$ and H7 give $c_2$ its own lineage of depth 2 and a translation witness inside that lineage. Only now are roles assigned: under the Case B description, $c_1$ and $c_2$ occupy information-carrier roles, $m$ replication-operator and translation-operator roles, $k$ a translation-operator role, and $y_1$ and $y_2$ realised-product roles. The assignment is a conditional formal result, not empirical evidence that human skill transmission has this organisation.
\end{newtext}

\begin{figure}[t]
\centering
\begin{tikzpicture}[cssfig, x=0.93cm, y=0.60cm]
\useasboundingbox (-0.45,-1.70) rectangle (7.1,10.05);
\node[font=\footnotesize\bfseries, anchor=west, text=black] at (-0.4,9.55) {(A) Case A: incomplete reconstruction};
\node[text=black] at (0.00,-1.05) {A};
\node[text=black] at (1.70,-1.05) {B};
\node[text=black] at (2.75,-1.05) {D};
\node[text=black] at (6.40,-1.05) {C};
\draw[cssdep, draw=black!45] (-0.35,-0.4) -- (-0.35,1.4);
\node[text=black!60] at (-0.35,1.85) {$t$};
\node[cssnode, cssfill=cssOI1!55!white] (A1c) at (0.00,0.00) {$c_1$};
\node[cssnode, cssfill=cssOI1!55!white] (A2c) at (0.00,1.00) {$c_1$};
\node[cssnode, cssfill=white] (B1c) at (1.70,0.00) {$c_0$};
\node[cssnode, cssfill=cssOI1!55!white] (B2c) at (1.70,1.00) {$c_1$};
\node[cssnode, cssfill=cssOI1!55!white] (B3c) at (1.70,4.00) {$c_1$};
\node[cssnode, cssfill=cssOI2!55!white] (C1c) at (6.40,1.00) {$c_2$};
\node[cssnode, cssfill=white] (D1c) at (2.75,3.00) {$c_0$};
\node[cssnode, cssfill=cssOI1!55!white] (D2c) at (2.75,4.00) {$c_1$};
\node[cssnode, cssfill=cssOI4!55!white] (d1) at (0.85,2.00) {$y_1$};
\node[cssnode, cssfill=cssOI3!50!white] (d2) at (7.20,3.00) {$y_2$};
\draw[csscont] (A1c) -- (A2c);
\draw[cssdep] (A1c) -- (B2c);
\draw[csscont] (B1c) -- (B2c);
\draw[csscont] (B2c) -- (B3c);
\draw[cssdep] (B2c) -- (D2c);
\draw[csscont] (D1c) -- (D2c);
\node[csstag, anchor=west] at (B2c.east) {$Q_{AB}$};
\node[csstag, anchor=west] at (D2c.east) {$Q_{BD}$};
\node[csstag, anchor=south] at (d1.north) {$Q_{d_1}$};
\node[csstag, anchor=south] at (d2.north) {$Q_{d_2}$};
\node[csstag, anchor=north] at (C1c.south) {(i), (ii)};
\node[csstag, anchor=east] at (d1.west) {(iii)};
\node[csstag, anchor=east] at (d2.west) {(iii)};
\node[font=\scriptsize, text=black, align=left, anchor=south west, text width=7.2cm, inner sep=0pt] at (-0.4,4.7) {\textit{Not established in Case A (annotations, not edges):}\\(i) no R process for $c_2$;\\(ii) no V process connecting $c_1$ and $c_2$;\\(iii) no dependence of a dish on a capability\\\hspace*{1.3em}instance and no comparative rule, hence no T.\\[2pt]\textit{Supported conditionally:} R for $c_1$ in $Q_{AB}$ and $Q_{BD}$;\\lineage $A_1^{\mathrm{cap}} \rightsquigarrow B_2^{\mathrm{cap}} \rightsquigarrow D_2^{\mathrm{cap}}$, $L = 2$.};
\end{tikzpicture}%
\hfill\begin{tikzpicture}[cssfig, x=0.93cm, y=0.60cm]
\useasboundingbox (-0.45,-1.70) rectangle (7.1,10.05);
\node[font=\footnotesize\bfseries, anchor=west, text=black] at (-0.4,9.55) {(B) Case B: completed reconstruction};
\node[text=black] at (0.00,-1.05) {A};
\node[text=black] at (1.70,-1.05) {B};
\node[text=black] at (2.75,-1.05) {D};
\node[text=black] at (3.80,-1.05) {E};
\node[text=black] at (5.35,-1.05) {F};
\node[text=black] at (6.40,-1.05) {C};
\draw[cssdep, draw=black!45] (-0.35,-0.4) -- (-0.35,1.4);
\node[text=black!60] at (-0.35,1.85) {$t$};
\node[cssnode, cssfill=cssOI1!55!white] (A1c) at (0.00,0.00) {$c_1$};
\node[cssnode, cssfill=cssOI1!55!white] (A2c) at (0.00,1.00) {$c_1$};
\node[cssnode, cssfill=white] (B1c) at (1.70,0.00) {$c_0$};
\node[cssnode, cssfill=cssOI1!55!white] (B2c) at (1.70,1.00) {$c_1$};
\node[cssnode, cssfill=cssOI1!55!white] (B3c) at (1.70,4.00) {$c_1$};
\node[cssnode, cssfill=cssOI2!55!white, cssadded] (B4c) at (1.70,5.00) {$c_2$};
\node[cssnode, cssfill=cssOI2!55!white, cssadded] (B5c) at (1.70,7.00) {$c_2$};
\node[cssnode, cssfill=cssOI2!55!white] (C1c) at (6.40,1.00) {$c_2$};
\node[cssnode, cssfill=white] (D1c) at (2.75,3.00) {$c_0$};
\node[cssnode, cssfill=cssOI1!55!white] (D2c) at (2.75,4.00) {$c_1$};
\node[cssnode, cssfill=white, cssadded] (E1c) at (3.80,5.00) {$c_0$};
\node[cssnode, cssfill=cssOI2!55!white, cssadded] (E2c) at (3.80,7.00) {$c_2$};
\node[cssnode, cssfill=cssOI2!55!white, cssadded] (E3c) at (3.80,8.00) {$c_2$};
\node[cssnode, cssfill=white, cssadded] (F1c) at (5.35,7.00) {$c_0$};
\node[cssnode, cssfill=cssOI2!55!white, cssadded] (F2c) at (5.35,8.00) {$c_2$};
\node[cssnode, cssfill=cssOI4!55!white] (d1) at (0.85,2.00) {$y_1$};
\node[cssnode, cssfill=cssOI3!50!white] (d2) at (7.20,3.00) {$y_2$};
\node[cssnode, cssfill=cssOI3!50!white, cssadded] (d3) at (0.85,6.00) {$y_2$};
\draw[csscont] (A1c) -- (A2c);
\draw[cssdep] (A1c) -- (B2c);
\draw[csscont] (B1c) -- (B2c);
\draw[csscont] (B2c) -- (B3c);
\draw[cssdep] (B2c) -- (D2c);
\draw[cssdep, densely dashed] (B2c) -- (d1);
\draw[csscont, densely dashed] (B3c) -- (B4c);
\draw[csscont, densely dashed] (B4c) -- (B5c);
\draw[cssdep, densely dashed] (B4c) -- (E2c);
\draw[cssdep, densely dashed] (B4c) -- (d3);
\draw[cssdep, densely dashed] (C1c) -- (d2);
\draw[csscont] (D1c) -- (D2c);
\draw[csscont, densely dashed] (E1c) -- (E2c);
\draw[csscont, densely dashed] (E2c) -- (E3c);
\draw[cssdep, densely dashed] (E2c) -- (F2c);
\draw[csscont, densely dashed] (F1c) -- (F2c);
\node[csstag, anchor=west] at (B2c.east) {$Q_{AB}$};
\node[csstag, anchor=north east] at ([xshift=-1.5mm]B4c.south west) {$Q_{\mathrm{mod}}$};
\node[csstag, anchor=west] at (D2c.east) {$Q_{BD}$};
\node[csstag, anchor=north west] at (E2c.south east) {$Q_{BE}$};
\node[csstag, anchor=west] at (F2c.east) {$Q_{EF}$};
\node[csstag, anchor=south] at (d1.north) {$Q_{d_1}$};
\node[csstag, anchor=south] at (d2.north) {$Q_{d_2}$};
\node[csstag, anchor=south] at (d3.north) {$Q_{d_3}$};
\end{tikzpicture}%
\caption{\new{\textbf{Conditional CSS reconstructions of recipe transmission: (A) incomplete information family; (B) completed information family.} Both are conditional on H1--H4, and Case B also on H5--H8. Colour and symbol give the structural label of an instance only, never a hereditary role. Each panel is the Hasse diagram of $\prec$ restricted to the capability and dish instances, i.e., the transitive reduction of the induced suborder; an edge records ancestry, possibly through omitted bodily, apparatus, ingredient or acoustic instances. Precedence runs upward; grey bars join sections of one capability world-tube; dashed elements are added in Case B; tags name the process that produces an instance. Figure~\ref{fig:recipe-full} gives the complete Hasse diagrams. Case A supplies no R process for $c_2$, no V process between $c_1$ and $c_2$ and no variant-dependent T (annotations, not edges). In Case B, $Q_{\mathrm{mod}}$ supplies V, $Q_{BE}$ and $Q_{EF}$ supply R and lineage continuation for $c_2$, and $Q_{d_1}$ and $Q_{d_3}$ supply T because H8 stipulates that replacing $c_1$ by $c_2$ in $Q_{d_1}$, other input types fixed, changes the dish type from $y_1$ to $y_2$; the diagram does not show this comparative dependence.}}
\label{fig:recipe}
\end{figure}

\begin{table}[t]
\begin{newtext}
\centering
\small
\begin{tabularx}{\textwidth}{>{\raggedright\arraybackslash}p{0.34\textwidth}>{\raggedright\arraybackslash}X>{\raggedright\arraybackslash}X}
\toprule
\textbf{Condition (Sections~\ref{sec:4}, \ref{sec:5})} & \textbf{Case A} & \textbf{Case B} \\
\midrule
Replication (R) of $c_1$ & Supported conditionally ($Q_{AB}$, $Q_{BD}$) & Supported conditionally ($Q_{AB}$, $Q_{BD}$) \\
Replication (R) of $c_2$ & Not established & Supplied by H6 ($Q_{BE}$, $Q_{EF}$) \\
Variation connectivity, $c_1$ to $c_2$ & Not established & Supplied by H5 ($Q_{\mathrm{mod}}$) \\
Shared replication organisation & Operator label $m$ for $c_1$ only; incomplete for the family & Operator label $m$ in all four R processes \\
Shared translation organisation & Candidates exist: $k$, $m$ in $Q_{d_1}$, $Q_{d_2}$ & Supplied: $k$, $m$ in $Q_{d_1}$, $Q_{d_3}$ \\
Variant-dependent downstream realisation & Not established & Supplied by H8 \\
Observed lineage continuation & $c_1$: $L = 2$, conditionally & $c_1$ and $c_2$: $L = 2$ each, conditionally \\
\midrule
Complete two-variant information family & Not established & Established conditionally on H1--H8 \\
\bottomrule
\end{tabularx}
\caption{\textbf{Case A and Case B evaluated by the joint criterion for $I = \{c_1, c_2\}$.} Every entry is conditional on H1--H4 and, for Case B, on H5--H8; ``supplied'' means supplied by stipulation, not observed. Appendix~\ref{app:G.4} gives the condition-by-condition verification.}
\label{tab:recipe}
\end{newtext}
\end{table}

\subsection{\new{Validation: what a stipulated history can and cannot show}}\label{sec:cultural.4}

\begin{newtext}
\textbf{No post-inference validation.} Section~\ref{sec:comp} checks its inferred roles against a withheld record and by interventions. The cultural episode offers neither: every instance and dependency of both cases is stipulated, so stage 6 returns only what the stipulations entail.

\textbf{What the comparison shows.} The two cases use the same physically hypothesised structural types. The difference in inferred hereditary organisation follows from the specified causal history, not from assigning hereditary meaning to the labels (Figure~\ref{fig:recipe}, Table~\ref{tab:recipe}). Case B succeeds, however, partly because H5--H8 were chosen to supply the missing conditions. Each of H5, H6 and H8 is necessary for the assignment and entails a motif component in full, and recovering an entailed component is not confirmation of it (Appendices~\ref{app:F.3} and~\ref{app:G.5}). The pair demonstrates how the criterion operates, not that the criterion is empirically confirmed.
\end{newtext}

\revnote{[v7 Stages 1--5, now Steps 1--5, accounting table and closing discussion (``What the example shows'') moved unchanged to Appendix~\ref{app:G.2}.]}

\begin{newtext}
\textbf{What Section~\ref{sec:cultural} shows.} Two points follow. First, whether an information family is identifiable depends on structural resolution, causal evidence and the realised history: Cases A and B share their structural types, differ only in stipulated events and the dependencies that these events entail, and receive different verdicts. Second, a formally complete hypothetical reconstruction demonstrates logical consistency but does not constitute empirical validation. Section~\ref{sec:comp} therefore turns to episodes whose causal history is recorded rather than stipulated: it constructs a CSS from the complete execution record of a digital-evolution simulator, searches it for R, V and T, and tests the inferred roles by intervention.
\end{newtext}

\FloatBarrier%
\begin{newtext}
\section{Digital-evolution episodes: a Causal Structure Set from executable dynamics}\label{sec:comp}

One run of the simulator studied here holds $3.9 \times 10^{8}$ variable-step pairs, and its record states the causes of every change among them. %
Here, we construct a CSS from that record, search the CSS for R, V and T without any knowledge of the simulator's hereditary entities, and check the result against the simulator's own record, which the search never sees.
The simulator is Avida-like \citep{Ofria2005}: self-copying programs written in the 26-instruction Avida heads instruction set live on a $6 \times 6$ toroidal grid of 36 cells.
\new{An Avida-like simulator suffices, because the test needs a digital world whose design fixes the hereditary roles and whose every step can be replayed, not Avida in particular. The simulator keeps the Avida heads instruction set, the default ancestor and the default mutation rates. It adds what the test needs, i.e., a log of every write with its data parents and random draws keyed to each cell's own step, so that a single step can be replayed with one input changed and every chance event fixed. It simplifies what the test does not need, e.g., an ancestor with 6 instead of about 80 filler sites and constant time slices instead of merit, which keeps the runs short enough to log in full. The inferred roles are checked against this simulator's own record, so its known deviations from Avida (Appendix~\ref{app:H.1}) limit what the results say about Avida, not what they say about the method.} %
They start from a 20-site ancestor, i.e., Avida's default ancestor with 6 instead of about 80 filler sites, and mutate at Avida's default rates (copy errors 0.0075 per site; insertions and deletions 0.05 per division). %
One run lasts 100 updates; no task is rewarded, so selection acts only through replication speed. %
As Avida does \citep{Ofria2005,Lenski2003}, our simulator records organisms, memory contents, births and mutations; we withhold that record during inference and use it only afterwards (Section~\ref{sec:1.2}).
We report two levels of description: organisms, whose births, memory contents at birth and times to first division the search finds and whose T we test by memory-content swaps (Section~\ref{sec:comp.2}), and single executed instructions (Section~\ref{sec:comp.3}).
Each run is one digital-evolution episode, i.e., one bounded episode $\mathcal{R}$ of Figure~\ref{fig:pipeline}, and at each level of description the analysis runs all six stages of that figure. Figure~\ref{fig:pipeline-digital} shows how, one column per level, and the subsections follow its stages: stages 1 to 5 at the level of instructions (Section~\ref{sec:comp.1}); all six stages at the level of organisms (Section~\ref{sec:comp.2}); stage 6 at the level of instructions (Section~\ref{sec:comp.3}); the validation of the stage-6 output (Section~\ref{sec:comp.4}), and; the sensitivity of the results to the choices made at each stage (Section~\ref{sec:comp.5}).
Three limits apply to every result below: 1) the results below come from development runs on seeds 1--5, and the held-out runs on seeds 101--140 test only their reproducibility (\new{Appendix~\ref{app:H.8}}); 2) we reconstructed the simulator's semantics from the Avida heads hardware without checking them against the Avida source, so the results concern this simulator, not Avida, and; 3) each run is one small world of 36 cells. %

\begin{figure}[t]
\begin{newtext}
\centering
\begin{tikzpicture}[cssfig,
  pipenum/.style={circle, minimum size=4.6mm, inner sep=0pt, line width=0.5pt,
    draw=cssOI5!70!black, fill=cssOI5!35, font=\footnotesize\bfseries},
  pipeinf/.style={circle, minimum size=4.6mm, inner sep=0pt, line width=0.5pt,
    draw=cssOI2!70!black, fill=cssOI2!35, font=\footnotesize\bfseries}]

\node[anchor=north west, inner sep=0pt] (h) at (0,0) {%
\begin{tabular}{@{\hspace{0.6cm}}>{\raggedright}p{2.85cm}@{\hspace{0.12cm}}>{\raggedright}p{5.75cm}@{\hspace{0.12cm}}>{\raggedright}p{5.75cm}@{}}
\rule{0pt}{0.3cm}\textbf{Stage}
&
\rule{0pt}{0.3cm}\textbf{Level of instructions}\par
{\scriptsize\color{black!65}stages 1--5: Sect.~\ref{sec:comp.1}; stage 6: Sect.~\ref{sec:comp.3}\par}
&
\rule{0pt}{0.3cm}\textbf{Level of organisms}\par
{\scriptsize\color{black!65}stages 1--6: Sect.~\ref{sec:comp.2}\par}
\end{tabular}};
\draw[line width=0.5pt] ([yshift=-0.03cm]h.south west) -- ([yshift=-0.03cm]h.south east);

\node[anchor=north west, inner sep=0pt] (r1) at ([yshift=-0.08cm]h.south west) {%
\begin{tabular}{@{\hspace{0.6cm}}>{\raggedright}p{2.85cm}@{\hspace{0.12cm}}>{\raggedright}p{5.75cm}@{\hspace{0.12cm}}>{\raggedright}p{5.75cm}@{}}
\rule{0pt}{0.45cm}\textbf{Perspective}
&
\rule{0pt}{0.45cm}{\scriptsize Logged run: 36 cells $\times$ 132 state variables, 81,410 steps; $J_P$: one-step replay with the same random draws. Seeds 1--5 and a no-mutation control; the simulator's own record is withheld.\par}
&
\rule{0pt}{0.45cm}{\scriptsize The level-1 CSS, read through rules we declared: births, organisms, memory content at birth G and time to first division P.\par}
\end{tabular}};

\node[anchor=north west, inner sep=0pt] (r2) at ([yshift=-0.08cm]r1.south west) {%
\begin{tabular}{@{\hspace{0.6cm}}>{\raggedright}p{2.85cm}@{\hspace{0.12cm}}>{\raggedright}p{5.75cm}@{\hspace{0.12cm}}>{\raggedright}p{5.75cm}@{}}
\rule{0pt}{0.45cm}\textbf{Structures}
&
\rule{0pt}{0.45cm}{\scriptsize Value instances of the variables of occupied cells and of memory sites below \texttt{LEN}, cut into sections at every read.\par}
&
\rule{0pt}{0.45cm}{\scriptsize The memory content of one organism between two divisions, and its time to first division.\par}
\end{tabular}};

\node[anchor=north west, inner sep=0pt] (r3) at ([yshift=-0.08cm]r2.south west) {%
\begin{tabular}{@{\hspace{0.6cm}}>{\raggedright}p{2.85cm}@{\hspace{0.12cm}}>{\raggedright}p{5.75cm}@{\hspace{0.12cm}}>{\raggedright}p{5.75cm}@{}}
\rule{0pt}{0.45cm}\textbf{Persistence and types}
&
\rule{0pt}{0.45cm}{\scriptsize $k = 1$: digital memory holds a written value. Scheme B: 74 declared labels, 66 of which occur; no AIR partition.\par}
&
\rule{0pt}{0.45cm}{\scriptsize G-label: the instruction labels of the 40 memory sites at birth, 152 G-labels; P in the organism's own steps.\par}
\end{tabular}};

\node[anchor=north west, inner sep=0pt] (r4) at ([yshift=-0.08cm]r3.south west) {%
\begin{tabular}{@{\hspace{0.6cm}}>{\raggedright}p{2.85cm}@{\hspace{0.12cm}}>{\raggedright}p{5.75cm}@{\hspace{0.12cm}}>{\raggedright}p{5.75cm}@{}}
\rule{0pt}{0.45cm}\textbf{Nodes and labels}
&
\rule{0pt}{0.45cm}{\scriptsize 847,911 nodes from 386,860,320 variable-step pairs, a 456-fold reduction.\par}
&
\rule{0pt}{0.45cm}{\scriptsize 1,710 nodes, i.e., 1,409 G-nodes and 301 P-nodes, of 706 organisms.\par}
\end{tabular}};

\node[anchor=north west, inner sep=0pt] (r5) at ([yshift=-0.08cm]r4.south west) {%
\begin{tabular}{@{\hspace{0.6cm}}>{\raggedright}p{2.85cm}@{\hspace{0.12cm}}>{\raggedright}p{5.75cm}@{\hspace{0.12cm}}>{\raggedright}p{5.75cm}@{}}
\rule{0pt}{0.45cm}\textbf{Causal order and CSS}
&
\rule{0pt}{0.45cm}{\scriptsize $D = D_{\mathrm{base}} \cup D_{\mathrm{maint}} \cup D_{\mathrm{pers}}$ by replay; $\prec\,=D^{+}$, i.e., order ALL.\par}
&
\rule{0pt}{0.45cm}{\scriptsize Reachability in the level-1 CSS: a path from the parent's to the child's memory content in 705 of 705 births.\par}
\end{tabular}};

\node[anchor=north west, inner sep=0pt] (r6) at ([yshift=-0.08cm]r5.south west) {%
\begin{tabular}{@{\hspace{0.6cm}}>{\raggedright}p{2.85cm}@{\hspace{0.12cm}}>{\raggedright}p{5.75cm}@{\hspace{0.12cm}}>{\raggedright}p{5.75cm}@{}}
\rule{0pt}{0.45cm}\textbf{Processes, motifs, roles}
&
\rule{0pt}{0.45cm}{\scriptsize Blind search, passes A--D: R in 17,133 of 81,410 single-step processes; lineage; V; T by \new{information }carrier swap replay. Families of 5, 4, 6 and 4 labels in seeds 1--4.\par}
&
\rule{0pt}{0.45cm}{\scriptsize R in 538 births, V in 167; T by memory-content swaps, of which 808 of 824 follow. Family of 13 G-labels; 13, 21, 14 and 17 in seeds 1--4.\par}
\end{tabular}};

\foreach \rowname in {r1, r2, r3, r4}{
  \draw[black!30, line width=0.3pt] ([yshift=-0.04cm]\rowname.south west) -- ([yshift=-0.04cm]\rowname.south east);
}
\draw[black!70, line width=0.7pt] ([yshift=-0.04cm]r5.south west) -- ([yshift=-0.04cm]r5.south east);
\draw[black!30, line width=0.3pt] ([xshift=9.38cm, yshift=-0.04cm]h.south west) -- ([xshift=9.38cm]r6.south west);

\node[pipenum] (c1) at ([xshift=0.3cm, yshift=-0.32cm]r1.north west) {1};
\node[pipenum] (c2) at ([xshift=0.3cm, yshift=-0.32cm]r2.north west) {2};
\node[pipenum] (c3) at ([xshift=0.3cm, yshift=-0.32cm]r3.north west) {3};
\node[pipenum] (c4) at ([xshift=0.3cm, yshift=-0.32cm]r4.north west) {4};
\node[pipenum] (c5) at ([xshift=0.3cm, yshift=-0.32cm]r5.north west) {5};
\node[pipeinf] (c6) at ([xshift=0.3cm, yshift=-0.32cm]r6.north west) {6};
\foreach \cfrom/\cto in {c1/c2, c2/c3, c3/c4, c4/c5, c5/c6}{
  \draw[cssdep, draw=black!55] (\cfrom) -- (\cto);
}

\draw[line width=0.9pt, cssOI5!70!black] ([xshift=-0.06cm, yshift=-0.03cm]r1.north west) -- ([xshift=-0.17cm, yshift=-0.03cm]r1.north west)
  -- ([xshift=-0.17cm, yshift=0.03cm]r5.south west) -- ([xshift=-0.06cm, yshift=0.03cm]r5.south west);
\path (r1.north west) -- (r5.south west) coordinate[pos=0.5] (m1);
\node[rotate=90, anchor=south, text=cssOI5!45!black, inner sep=1pt] at ([xshift=-0.24cm]m1) {construction of the CSS};
\draw[line width=0.9pt, cssOI2!70!black] ([xshift=-0.06cm, yshift=-0.03cm]r6.north west) -- ([xshift=-0.17cm, yshift=-0.03cm]r6.north west)
  -- ([xshift=-0.17cm, yshift=0.03cm]r6.south west) -- ([xshift=-0.06cm, yshift=0.03cm]r6.south west);
\path (r6.north west) -- (r6.south west) coordinate[pos=0.5] (m2);
\node[rotate=90, anchor=south, text=cssOI2!55!black, inner sep=1pt] at ([xshift=-0.24cm]m2) {inference};

\node[anchor=north west, draw=black!60, densely dashed, line width=0.6pt, rounded corners=1mm,
  inner xsep=4pt, inner ysep=7pt, text width=14.86cm, align=left, font=\scriptsize] (ft) at ([yshift=-0.4cm]r6.south west)
  {\textbf{Post-inference validation} (Sect.~\ref{sec:comp.4}): the withheld record (706 of 706 organisms match; 17,131 of the 17,133 steps with R are copy steps), founding mutations, label permutations and a no-mutation control. Sect.~\ref{sec:comp.5} varies the choices made at each stage.};
\node[csstag, fill=white, anchor=east, inner sep=2pt] at ([xshift=-0.35cm]ft.north east) {assesses the output of stage 6};

\end{tikzpicture}
\caption{\textbf{The six stages of Figure~\ref{fig:pipeline}, implemented for the digital-evolution episodes.} The pipeline runs twice, once per level of description: at the level of instructions on the logged run, and at the level of organisms on the level-1 CSS, read through rules we declared. Every entry is computed from the record; values are for seed 1 unless stated, and the column heads name the subsections that implement each stage.}
\label{fig:pipeline-digital}
\end{newtext}
\end{figure}

\subsection{Stages 1 to 5 at the level of instructions: a CSS from executable dynamics}\label{sec:comp.1}

\textbf{Stage 1: perspective.} Figure~\ref{fig:pipeline} asks for the episode $\mathcal{R}$ and the perspective $P = (\pi_P, D_P, J_P, \rho_P, \Sigma_P)$ of Section~\ref{sec:3.1}. The episode $\mathcal{R}$ is one run of the simulator, and the state map $\pi_P$ is its record.
The simulator logs every executed instruction, i.e., one step, with its cell, its control reads, every write with its value and data parents, cell resets and random draws. %
Each cell holds 132 state variables, among them 80 memory sites, four heads, three registers and two stacks. %
The dynamics $D_P$ is the simulator's transition rule, which executes one instruction of one cell per step.
Three tests make this record a complete causal record: 1) replaying the log reproduces the final state exactly; 2) re-executing each of the 81,410 steps of seed 1 with every non-parent variable of the cell and its neighbours randomised leaves all writes unchanged (0 failures), and; 3) random draws are keyed to the seed, the cell and the cell's own step counter, so a replay reuses the same chance events. %
The admissible interventions $J_P$ are one-step replays, in which a variable read at a step takes an alternative value and the step is re-executed with the same random draws (stage 5). The resolution $\rho_P$ is one value of one variable of one cell and one executed step, and the individuation rule $\Sigma_P$ is that of stages 2 and 3.
The platform's descriptions of organisms, memory contents, lineages and mutations sit in a separate file that the CSS code may not open, and the perspective $P$ admits every recorded state variable (Definition~\ref{def:1}). %

\textbf{Stage 2: structures.} Figure~\ref{fig:pipeline} asks for the structural domain $Y_P$ and the candidate structures, i.e., world-tube sections in $Y_P$. A value instance is one value of one variable of one cell, from its creation by a write, a reset or the initial state to the next write or reset of that variable.
The admitted structural domain $Y_P^{\mathrm{acc}}$ (Section~\ref{sec:3.3}) contains the state variables of occupied cells and the memory sites below the current memory size \texttt{LEN}. An instance outside this domain when its creating step ends, i.e., in an empty cell (\texttt{ALIVE} = 0) or at a memory index at or beyond \texttt{LEN}, is not a node. The rule depends only on the state at the instance's creation, not on whether the value is later read, so the node set is complete relative to the declared domain; such values have no counterpart in Avida, whose memory ends at \texttt{LEN} (physical grounding, Definition~\ref{def:1}). %
Each instance is split into sections at every step that reads the instance; these sections are the candidate structures.

\textbf{Stage 3: persistence and structural types.} Figure~\ref{fig:pipeline} asks for the persistent structures and the label set $\mathcal{L}_P$. Every instance in the admitted domain is persistent. A value instance is held in the memory of the computer that runs the simulator, and digital memory is built to hold a written value: the physical state of a memory cell settles into the basin of its logical value within a few clock cycles, whereas a change without a write occurs only as a rare hardware error. \new{We therefore adopt a declared persistence rule: a value persists until its next write, and every section of at least one executed step is admitted, $k = 1$. The stability of digital storage motivates this rule, since it leads us to expect $\tau_{\mathrm{mix}} \ll \tau_{\mathrm{esc}}$ (Section~\ref{sec:3.3}) for every value between two writes, however few simulator steps the value lasts; the rule does not establish an AIR partition, and we did not measure these times.} Section~\ref{sec:comp.5} compares $k = 10$. %
Under the labelling we use, scheme B, a memory site is labelled by its instruction, one of 26. %
Every other variable is labelled by its name and whether its value is zero or non-zero, e.g., (\texttt{BX}, non-zero), with the 20 stack slots and the 10 read-label slots as one kind each. %
Scheme B allows $26 + 24 \times 2 = 74$ labels, of which 66 occur in seed 1. %
These labels are declared value classes, not derived from an AIR partition (Section~\ref{sec:3.3}), which we did not implement. %

\textbf{Stage 4: nodes and labels.} Figure~\ref{fig:pipeline} maps each persistent section to one CSS node and each node to its structural label. One section is one node, so a site that a process reads appears on both its input and its output boundary (Section~\ref{sec:3.9}).
The run of seed 1 holds $36 \times 132 = 4{,}752$ state variables at each of 81,410 steps, i.e., $4{,}752 \times 81{,}410 = 386{,}860{,}320$ variable-step pairs. %
The CSS condenses them into 847,911 nodes, a $386{,}860{,}320 / 847{,}911 = 456$-fold reduction (Table~\ref{tab:comp-css}, Figure~\ref{fig:pipeline-digital}). %
Seeds 2--5 give 879,538, 883,939, 865,934 and 3,498 nodes. %

\textbf{Stage 5: causal order and the CSS.} Figure~\ref{fig:pipeline} asks for the direct dependencies $D$ and the order ${\prec} = D^{+}$. We evaluate the dependence $D$ of Section~\ref{sec:3.5} by replay, as the union of three kinds of edge. %
For every node $x$ that ends in a read at step $s$, and for every alternative value of $x$'s variable, we re-execute step $s$ alone, with every other variable at its factual value and the same random draws. %
A written node whose label changes and a write that no longer happens (prevention) each give an edge from $x$ to the node that the factual write created; these label-change and prevention edges form $D_{\mathrm{base}}$. %
A write that happens only in the replay, with a label other than that of the variable's current section, gives a maintenance edge from $x$ to that section; these edges form $D_{\mathrm{maint}}$. %
One alternative suffices for an edge; the alternatives are a declared set, e.g., the 25 other instructions for a memory site, and random draws are not intervened on. %
Persistence edges join successive sections of one instance and form $D_{\mathrm{pers}}$. %
The complete accepted direct-edge relation is thus $D = D_{\mathrm{base}} \cup D_{\mathrm{maint}} \cup D_{\mathrm{pers}}$, and the order of the CSS is its transitive closure, $\prec = D^{+}$ (Section~\ref{sec:3.6}), which we call order ALL; the transitive closure of $D_{\mathrm{base}} \cup D_{\mathrm{pers}}$, i.e., without maintenance edges, we call order DP. %
Unless stated otherwise, every result below uses order ALL; Section~\ref{sec:comp.5} compares order DP. %
An edge records whether one node causes another, nothing more; separating \new{information }carriers from operators is left to the RVT criterion. %

In seed 1, the builder stores 2,363,872 write edges, i.e., edges from changed or prevented writes, 9,260,100 maintenance edges and the 453,875 persistence edges of $D_{\mathrm{pers}}$. %
The first two counts also include edges whose write changes an exact value but not the label; such edges are not part of $D$ (Appendix~\ref{app:H.2}). %
Every stored edge points forward in node numbering (0 violations), so $\prec$ is a strict partial order. %
Every child site is reached from its source site in the parent, 13,915 of 13,915 in seed 1, and an unmutated copy depends on its template site while a mutated copy does not. %
Building this CSS took 48 s with 4 processes on an Apple M2 (132--144 s with 1 process for seeds 2--4); the run with its checks and a comparison against the $k = 10$ build took 195 s. %

\textbf{The order is dense.} In seed 1, 62.6\% of node pairs are ordered through all stored edges, against 13.0\% through the stored write edges alone, and on average 94.5\% of the sections of a parent's memory sites precede any one site of its child (specificity 0.945). %
Inside a replication, nearly every parent site therefore precedes every child site, so ancestry clauses such as condition 2 of R (Section~\ref{sec:4.1}) or $x \prec y$ in V (Section~\ref{sec:4.2}) restrict only weakly. %

\begin{table}[htbp]
\begin{newtext}
\centering
\small
\caption{CSS construction, development seeds 1--5 ($k = 1$, scheme B). Births and mutation events come from the withheld simulator record and are not used to build the CSS. Stored write and maintenance edges include edges that change an exact value but not a label, which are not part of $D$ (Appendix~\ref{app:H.2}).}
\label{tab:comp-css}
\begin{tabularx}{\textwidth}{@{}>{\raggedright\arraybackslash}Xrrrrr@{}}%
\toprule
Quantity & Seed 1 & Seed 2 & Seed 3 & Seed 4 & Seed 5 \\
\midrule
Executed steps & 81,410 & 84,089 & 83,914 & 83,000 & 401 \\ %
Births (record) & 705 & 724 & 750 & 726 & 0 \\ %
Mutation events (record) & 228 & 241 & 248 & 256 & 3 \\ %
CSS nodes & 847,911 & 879,538 & 883,939 & 865,934 & 3,498 \\ %
\midrule
\multicolumn{6}{@{}l}{\emph{Seed 1 only}} \\
Variable-step pairs & \multicolumn{5}{l}{$4{,}752 \times 81{,}410 = 386{,}860{,}320$} \\ %
Labels occurring (of 74) & \multicolumn{5}{l}{66} \\ %
Stored edges: write / maintenance / persistence & \multicolumn{5}{l}{2,363,872 / 9,260,100 / 453,875} \\ %
Node pairs ordered, stored edges: write / + persistence / all & \multicolumn{5}{l}{13.0\% / 39.5\% / 62.6\%} \\ %
Child sites reached from their source site & \multicolumn{5}{l}{13,915 of 13,915} \\ %
Specificity, all stored edges & \multicolumn{5}{l}{0.945} \\ %
Build time, 4 processes & \multicolumn{5}{l}{48 s} \\
Run with checks and the $k = 10$ comparison & \multicolumn{5}{l}{195 s} \\ %
\bottomrule
\end{tabularx}
\end{newtext}
\end{table}

\begin{figure}[p]
\begin{newtext}
\centering
\input{figures/css-L1-seed1}
\caption{\textbf{All nodes of the level-1 CSS of seed 1, without edges.} Each column is one state variable of one cell (36 cells $\times$ 132 variables), time runs upward over 81,410 steps, and colour gives the scheme-B label; a node is a vertical run of one colour in one column, and the 847,911 nodes fill the image. a) The whole run, downsampled by a majority rule. b) Cells 21 and 22, steps 0--3,000, one pixel per step: the copy loop fills the memory sites of the parent as a staircase, and each division appears as a cut, marked at the right as parent$\rightarrow$child cell.} %
\label{fig:css-L1}
\end{newtext}
\end{figure}

\subsection{Stages 1 to 6 at the level of organisms: an information family tested by memory-content swaps}\label{sec:comp.2}

At the level of organisms, the six stages run on the level-1 CSS rather than on the record (Figure~\ref{fig:pipeline-digital}, right column).

\textbf{Stage 1: perspective.} The level-2 perspective reads the level-1 CSS of Section~\ref{sec:comp.1} through four rules that we declare on the CSS alone: 1) a birth as a step executed by one cell that creates new instances in another cell; 2) an organism as lasting as long as the \texttt{ALIVE} instance created at its birth; 3) its memory content at birth, G, i.e., the instruction labels of its 40 memory sites at birth, whose label we call its G-label, and; 4) its time to first division, P, i.e., its age in own steps at its first successful division. %
Avida calls the memory content a genome and its label a genotype, and the time to first division a gestation time; biology would speak of genotype and phenotype; the names used here assign no hereditary role, because whether G occupies \del{a}\new{an} \new{information }carrier role and P a product role is what the criterion tests.
We declared G and P ourselves rather than deriving them from the CSS, so results at this level are conditional on that description (Appendix~\ref{app:F.3}). %
The admissible interventions $J_P$ include the memory-content swaps of stage 6.

\textbf{Stage 2: structures.} The structures are the memory content of one organism between two divisions and its time to first division.

\textbf{Stage 3: persistence and structural types.} Both types are declared rather than derived from an AIR partition: a G-node carries the G-label of its organism's memory content at birth, and a P-node its value of P. Seeds 1--4 hold 152, 176, 162 and 164 distinct G-labels, and P of the unmutated ancestor is 69 steps. %

\textbf{Stage 4: nodes and labels.} Seed 1 has 1,710 level-2 nodes, i.e., 1,409 G-nodes and 301 P-nodes (Figure~\ref{fig:css-L2}). The organisms, births, G-labels, values of P and life endings found in this way match the simulator's record in every run, e.g., 706 of 706 organisms in seed 1. %

\begin{figure}[p]
\begin{newtext}
\centering
\input{figures/css-L2-seed1}
\caption{\textbf{All nodes of the level-2 CSS of seed 1, without edges.} Two level-2 variables per cell: G, the memory content of one organism between two divisions, coloured by G-label in the colours of Figure~\ref{fig:family}, and P, the time to first division, drawn after the first division in grey; time runs upward. All 1,710 nodes, i.e., 1,409 G-nodes and 301 P-nodes, are drawn. a) The whole run, in which the ancestor's G-label gives way to those of its descendant clades. b) Cells 14--27, steps 0--3,000, the time span of Figure~\ref{fig:css-L1}b.} %
\label{fig:css-L2}
\end{newtext}
\end{figure}

\textbf{Stage 5: causal order.} At this level, the order is read from reachability in the level-1 CSS. The relation that R and V need, a causal path from the parent's memory content to the child's, exists in every birth of every run, e.g., in 705 of 705 births of seed 1. %

\textbf{Stage 6: R, V and lineage.} R is a birth whose child has the parent's G-label, and V a birth whose child has another; both require the causal path of stage 5. %
Seed 1 has 538 R births and 167 V births; seeds 2--4 have 545, 574 and 548 R births and 179, 176 and 178 V births (Figure~\ref{fig:family}). %
Lineage is the longest line of successive R births within one G-label; in seeds 1--4, 13, 21, 14 and 17 G-labels reach $L \geq 2$, with longest lineages of 17, 18, 20 and 20 generations. %

\textbf{Stage 6: T by memory-content swaps.} The swap test declares G as candidate \new{information }carrier, the processor, reset at every birth and identical in every cell, as candidate operator, and P as candidate product; the swap tests whether P depends on G, and these roles follow only if the criterion holds. %
A memory-content swap tests each candidate pair, i.e., two G-labels with $L \geq 2$ and different values of P. %
At an organism's birth, we write the memory content of the other G-label into its memory and re-run the simulator with the logged schedule and the same random draws until the organism's first division, age-limit reset or overwrite. %
Before its first division an organism cannot change another cell, so every other cell runs as recorded. %
Replays without a swap reproduce every tested value of P, 25 of 25 per run. %
In seed 1, P follows the inserted memory content in 808 of 824 swaps and none gives another value of P; the other 16 end censored, because the organism is overwritten before the inserted memory content completes a division. %
Over seeds 1--4, 4,135 of 4,314 swaps follow ($4{,}135 / 4{,}314 = 95.9\%$), 166 end censored and 13 end sterile (Figure~\ref{fig:validation}b), and all 13 sterile swaps fall on 4 organisms whose own random draws include copy errors during gestation. %

\textbf{Stage 6: the organism-level information family.} A G-label is a member when it has $L \geq 2$, a swap-confirmed T partner and a V birth that connects it to another member. %
The families hold 13, 21, 14 and 17 G-labels in seeds 1--4 ($n = 4$ runs), with P from 61 to 93 steps; each family is the set of all G-labels with $L \geq 2$, and the ancestor is a member in every run. %
These results use gestation processes and reachability in the CSS, not single-step processes, and are the same in order ALL and in order DP. %
The no-mutation control has one G-label, and seed 5 one organism without a birth, so neither has V, a T pair or a family at this level (Section~\ref{sec:comp.4}). %

\begin{figure}[htbp]
\begin{newtext}
\centering
\input{figures/avida-genealogy-seed1}
\caption{\textbf{Genealogy of seed 1 at the level of organisms.} All 706 organisms found blind in the CSS of seed 1, as the Hasse diagram of the organism-level order, i.e., parent below child; time runs upward. Node colour gives the G-label only: each of the 152 distinct G-labels has its own colour, related G-labels share a hue family, and a mutant is lighter or darker than its parent; shape gives the fate (circle: divided, square: censored, diamond: sterile), all at one size; no two nodes overlap, and horizontal position only keeps relatives together. Thin grey edges are R births, in which the child has its parent's G-label (538); bold vermilion edges are V births, in which it has another (167), and a vermilion boundary marks the parent and the child of each V birth. The organism-level information family consists of the 13 G-labels with $L \geq 2$; it is the result of the criterion and is not drawn.}
\label{fig:family}
\end{newtext}
\end{figure}

\begin{figure}[htbp]
\begin{newtext}
\centering
\begin{minipage}{11.25cm}{\footnotesize a)}\\[2pt]\scalebox{1.5}{\begin{tikzpicture}[cssfig, x=1cm, y=1cm]
\useasboundingbox (0,0) rectangle (7.5,4.816);
\draw[cssRbox] (0.85,0.962) rectangle (7.25,4.462);
\node[cssnode, minimum size=5mm, font=\fontsize{6}{7}\selectfont, cssfill=cssOI5!55!white] (zb0) at (3.05,1.562) {hc};
\node[cssnode, minimum size=5mm, font=\fontsize{6}{7}\selectfont, cssfill=cssOI6!60!white] (zb1) at (6.75,1.562) {nA};
\node[cssnode, rectangle, rounded corners=1.2mm, minimum width=8mm, minimum height=4.4mm, font=\fontsize{6}{7}\selectfont, cssfill=cssGrey!70!white] (zb4) at (5.3,1.562) {RH+};
\node[cssnode, rectangle, rounded corners=1.2mm, minimum width=8mm, minimum height=4.4mm, font=\fontsize{6}{7}\selectfont, cssfill=cssGrey!70!white] (zb6) at (2.1,1.562) {COP+};
\node[cssnode, rectangle, rounded corners=1.2mm, minimum width=8mm, minimum height=4.4mm, font=\fontsize{6}{7}\selectfont, cssfill=cssGrey!70!white] (zb7) at (4.3,1.562) {RLN+};
\node[cssnode, minimum size=5mm, font=\fontsize{6}{7}\selectfont, cssfill=cssOI1!45!white] (zb10) at (1.3,1.962) {if};
\node[cssnode, minimum size=5mm, font=\fontsize{6}{7}\selectfont, cssfill=cssOI6!60!white] (zb11) at (6.75,4.112) {nA};
\node[cssnode, rectangle, rounded corners=1.2mm, minimum width=8mm, minimum height=4.4mm, font=\fontsize{6}{7}\selectfont, cssfill=white] (zb15) at (3.85,4.112) {RL0};
\node[cssnode, rectangle, rounded corners=1.2mm, minimum width=8mm, minimum height=4.4mm, font=\fontsize{6}{7}\selectfont, cssfill=cssGrey!70!white] (zb16) at (4.75,4.112) {RLN+};
\node[cssnode, minimum size=5mm, font=\fontsize{6}{7}\selectfont, cssfill=cssOI3!50!white] (zb18) at (1.3,2.712) {mh};
\node[cssnode, minimum size=5mm, font=\fontsize{6}{7}\selectfont, cssfill=cssOI5!55!white] (zb22) at (3.05,4.112) {hc};
\node[cssnode, rectangle, rounded corners=1.2mm, minimum width=8mm, minimum height=4.4mm, font=\fontsize{6}{7}\selectfont, cssfill=cssGrey!70!white] (zb23) at (1.3,4.112) {RH+};
\node[cssnode, rectangle, rounded corners=1.2mm, minimum width=8mm, minimum height=4.4mm, font=\fontsize{6}{7}\selectfont, cssfill=cssGrey!70!white] (zb25) at (2.2,4.112) {COP+};
\node[cssnode, minimum size=5mm, font=\fontsize{6}{7}\selectfont, cssfill=cssOI6!60!white] (zb26) at (6.05,4.112) {nA};
\draw[cssdep] (zb0) -- (zb15);
\draw[cssdep] (zb0) -- (zb16);
\draw[cssdep] (zb0) -- (zb18);
\draw[csscont] (zb0) -- (zb22);
\draw[csscont] (zb1) -- (zb11);
\draw[cssdep] (zb1) -- (zb15);
\draw[cssdep] (zb1) -- (zb16);
\draw[cssdep] (zb1) -- (zb26);
\draw[cssdep] (zb4) -- (zb15);
\draw[cssdep] (zb4) -- (zb16);
\draw[cssdep] (zb4) -- (zb26);
\draw[cssdep] (zb7) -- (zb15);
\draw[cssdep] (zb7) -- (zb16);
\draw[cssdep] (zb10) -- (zb18);
\draw[cssdep] (zb18) -- (zb23);
\draw[cssdep] (zb18) -- (zb25);
\draw[cssdep, densely dotted] (zb18) -- (zb26);
\node[anchor=east, inner sep=0pt, align=right, font=\tiny\itshape, text=black!75] at (0.78,1.562) {in};
\node[anchor=east, inner sep=0pt, align=right, font=\tiny\itshape, text=black!75] at (0.78,4.112) {out};
\node[anchor=north, inner sep=0pt, align=center, font=\tiny\itshape, text=black!75] at (2.1,1.292) {in \& out};
\node[anchor=center, inner sep=0pt, align=center, font=\tiny\itshape, text=black!75] at (7.07,2.837) {R};
\node[anchor=north west, inner sep=0pt, align=left, font=\tiny, text=black] at (0.15,0.812) {R for nA: $n^{-} = 1 \to n^{+} = 2$; $\mathrm{Op}(Q) = \{$hc, RH+, RLN+$\}$\\($n^{-} = n^{+} = 1$). Step 6,535, cell 21; closure in ALL: 27 nodes\\(median 25).};
\end{tikzpicture}%
}\end{minipage}\\[8pt]
\begin{minipage}{11.25cm}{\footnotesize b)}\\[2pt]\scalebox{1.5}{\begin{tikzpicture}[cssfig, x=1cm, y=1cm]
\useasboundingbox (0,0) rectangle (7.5,4.816);
\draw[cssVbox] (0.3,1.566) rectangle (2.8,3.716);
\node[cssnode, minimum size=5mm, font=\fontsize{6}{7}\selectfont, cssfill=cssOI1!45!white] (zcx) at (0.9,2.116) {if};
\node[cssnode, minimum size=5mm, font=\fontsize{6}{7}\selectfont, cssfill=cssOI2!55!white] (zcy) at (0.9,3.166) {nB};
\node[cssnode, minimum size=5mm, font=\fontsize{6}{7}\selectfont, cssfill=cssOI2!55!white] (zcyc) at (0.9,4.366) {nB};
\draw[cssdep] (zcx) -- (zcy);
\draw[cssdep] (zcy) -- (zcyc);
\node[anchor=west, inner sep=0pt, align=left, font=\tiny\itshape, text=black!75] at (1.25,2.116) {template\\step 35{,}482};
\node[anchor=west, inner sep=0pt, align=left, font=\tiny\itshape, text=black!75] at (1.25,3.166) {mutated copy};
\node[anchor=west, inner sep=0pt, align=left, font=\tiny\itshape, text=black!75] at (1.25,4.366) {child at birth};
\node[anchor=center, inner sep=0pt, align=center, font=\tiny\itshape, text=black!75] at (2.62,3.536) {V};
\node[anchor=center, inner sep=0pt, align=center, font=\tiny\itshape, text=black!75] at (5.3,4.566) {T};
\draw[cssTbox] (3.2,1.566) rectangle (5.15,4.716);
\draw[cssTbox] (5.45,1.566) rectangle (7.4,4.716);
\draw[cssproc] (3.1,1.266) rectangle (7.45,1.886);
\node[cssnode, rectangle, rounded corners=1.2mm, minimum width=8mm, minimum height=4.4mm, font=\fontsize{6}{7}\selectfont, cssfill=cssGrey!70!white] (zcopRH) at (4.85,1.576) {RH+};
\node[cssnode, rectangle, rounded corners=1.2mm, minimum width=8mm, minimum height=4.4mm, font=\fontsize{6}{7}\selectfont, cssfill=cssGrey!70!white] (zcopRLN) at (5.8,1.576) {RLN+};
\node[anchor=west, inner sep=0pt, align=left, font=\tiny\itshape, text=black!75] at (3.2,1.576) {operators};
\node[cssnode, minimum size=5mm, font=\fontsize{6}{7}\selectfont, cssfill=cssOI6!60!white] (zcca) at (3.75,2.716) {nA};
\node[cssnode, minimum size=5mm, font=\fontsize{6}{7}\selectfont, cssfill=cssOI4!55!white] (zccb) at (6.85,2.716) {nC};
\node[cssnode, rectangle, rounded corners=1.2mm, minimum width=8mm, minimum height=4.4mm, font=\fontsize{6}{7}\selectfont, cssfill=white] (zcpa) at (4.2,3.966) {RL0};
\node[cssnode, rectangle, rounded corners=1.2mm, minimum width=8mm, minimum height=4.4mm, font=\fontsize{6}{7}\selectfont, cssfill=cssGrey!70!white] (zcpb) at (6.4,3.966) {RL+};
\draw[cssdep] (zcca) -- (zcpa);
\draw[cssdep] (zccb) -- (zcpb);
\draw[cssdep] (zcopRH) -- (zcpa);
\draw[cssdep] (zcopRH) -- (zcpb);
\draw[cssdep] (zcopRLN) -- (zcpa);
\draw[cssdep] (zcopRLN) -- (zcpb);
\draw[cssadded, <->, draw=cssT, >={Stealth[length=1.7mm,width=1.3mm]}] (zcca) -- (zccb);
\node[anchor=center, inner sep=1pt, fill=white, align=center, font=\tiny\itshape, text=black!75] at (5.31,2.716) {swap};
\node[anchor=north west, inner sep=0pt, align=left, font=\tiny\itshape, text=black!75] at (3.3,4.636) {$\times$1{,}614};
\node[anchor=north east, inner sep=0pt, align=right, font=\tiny\itshape, text=black!75] at (7.3,4.636) {$\times$1{,}364};
\node[anchor=north west, inner sep=0pt, align=left, font=\tiny, text=black] at (0.15,1.116) {V: copy error at step 35,482 (withheld record): template if $\to$\\copy nB; template precedes copy in ALL only (one edge).};
\node[anchor=north west, inner sep=0pt, align=left, font=\tiny, text=black] at (0.15,0.558) {T: data reads under $\{$RH+, RLN+$\}$; swap nA $\to$ nC turns RL0 into\\RL+ (20 of 20 replays), nC $\to$ nA the reverse (20 of 20).};
\end{tikzpicture}%
}\end{minipage}
\caption{\textbf{Zooms from the run of seed 1 into single CSS processes, in order ALL.} a) One copy step of seed 1 (step 6,535): its process in order ALL, i.e., the convex closure of the step, 27 nodes, of which 14 are shown; the fork of \texttt{nop-A} into two \texttt{nop-A} instances is R, $n^{-} = 1 \rightarrow n^{+} = 2$. b) Left: a copy error at step 35,482, in which an \texttt{if-label} template was copied as \texttt{nop-B}, i.e., V between two family members, read post hoc from the withheld record. Right: one T pair, \texttt{nop-A} versus \texttt{nop-C} read as copied data under the shared operators (\texttt{RH}, non-zero) and (\texttt{RLN}, non-zero); the product, the read label, is zero or non-zero, and swapping the \new{information }carriers swaps the product in 20 of 20 replays in each direction. In both panels, time runs upward; a solid edge is a direct dependence, a dotted edge records ancestry through omitted nodes, and a grey bar joins two sections of a value that persists. In b), a colour change along an edge is V; two processes on a grey operator rail, with their products above, are T; the dashed arrow marks the \new{information }carrier swap. Both panels show declared subsets of nodes. a) 14 of the 27 nodes are shown, i.e., memory sites, operators and outputs; 5 inputs (\texttt{AGE}, \texttt{GL}, \texttt{IP}, \texttt{LEN} and \texttt{WH}, each non-zero) and 8 inner CPU nodes are omitted; 17 edges are drawn, 16 direct and 1 dotted, against 91 edges in the process. b) The V box is the convex closure of its 2 nodes in order ALL, 308 nodes; each T box draws one of the records counted in its corner ($\times$1,614 and $\times$1,364), at steps 8,993 and 53,289, with 4 of 27 and 4 of 50 nodes shown.} %
\label{fig:zooms}
\end{newtext}
\end{figure}

\subsection{Stage 6 at the level of instructions: blind search for R, V, T and an information family}\label{sec:comp.3}

\textbf{Procedure.} The blind search uses the CSS and the record format only, without instruction names or \del{simulator annotations}\new{designated roles (Section~\ref{sec:comp.4})}; Avida knowledge enters afterwards, in comparisons we mark as guided. %
This search is stage~6 of Figures~\ref{fig:pipeline} and~\ref{fig:pipeline-digital}, and it runs hierarchically in four search passes (Appendices~\ref{app:D} and~\ref{app:H.4}): pass~A finds R in every single-step process; pass~B finds lineage, i.e., a copy that moves into another cell and is read there as a template, chained over generations; pass~C tests V among the labels that reach $L \geq 2$ in pass~B, i.e., the candidates, and; pass~D tests T by \del{a}\new{an} \new{information }carrier swap replay. %
In order ALL, the nodes read at one step, their continuations and the nodes the step creates do not form a causally convex set, so the process of a step is their convex closure. %
We analysed this closure for every step of every run, with none skipped; its median is 25--28 nodes per run, but 40--46 steps per run have closures above 50,000 nodes, up to 67,856. %
From pass~B on, \new{information }carrier variants are restricted to memory labels by construction; other kinds enter only as operators and products. %

\textbf{Replication and lineage.} Pass~A finds R in 17,133 of 81,410 single-step processes of seed 1 (21.0\%), against at most 8.9\% and 3.9\% in 20 permutations of labels over all nodes or within each kind, with the order kept. %
\new{Figure~\ref{fig:zooms}a shows one of them, a copy step in which \texttt{nop-A} forks into two instances.}
Over seeds 1--4 the share is 20.7--21.3\%, against permutation maxima of 8.7--9.0\% and 3.8--4.0\% ($n = 4$ runs). %
\new{R records whether the instances of a label multiply with descent, not how; copying and re-computation of a label both satisfy it.} %
Pass~B finds 5,987 moves of a finished copy into another cell in seed 1, and 14 memory labels re-enter replication in a later generation, i.e., reach an observed lineage depth $L \geq 2$ (Section~\ref{sec:4.7}); seeds 2--4 give 13, 15 and 12 such labels. %
At each child's first division, a guided check verified node by node that all 6,070 child sites that are outputs of the parent's process are inputs of the child's own (Section~\ref{sec:4.6}). %
Lineages verified in this way reach 8 generations in seed 1 and 8--12 in seeds 1--4; later divisions link through persistence chains. %

\textbf{A single step is not a closed process in order ALL.} A maintenance edge runs from a node read at a later step to a section that is still current at that step. %
Edges of $D_{\mathrm{pers}}$ and $D_{\mathrm{base}}$ end at nodes that start where their sources end, so only edges of $D_{\mathrm{maint}}$ lead from a later step back to the nodes of step $s$, and through them the convex closure of the step takes in later steps. %
A guided diagnosis of seed 1, on 40 sampled copy steps per template label and 30 sampled divisions, shows two consequences. %
First, when the copied site holds an instruction of the copy loop, the site is executed again at the next pass of the loop, and the read of the template's continuation there leads back to the copy step through maintenance edges. %
The continuation is then not an output of the closure, $n^{+} = n^{-}$, and R fails: R passes in order DP and fails in order ALL at 40 of 40 sampled \texttt{h-copy} templates, 39 of 40 \texttt{if-label} templates and 16 of 40 \texttt{mov-head} templates, whereas the sampled \texttt{nop-A}, \texttt{nop-C}, \texttt{h-alloc}, \texttt{h-search} and \texttt{h-divide} templates give the same result in both orders. %
Over the whole run of seed 1, the R events of \texttt{h-copy} fall from 906 to 24, and those of \texttt{if-label} from 897 to 15. %
Second, at a division, a section of a child's memory site that the child reads before the other sections created at the division is not an output of the division's closure, which holds for 256 of 423 child sections in the 21 sampled divisions with a child. %
Moves therefore fall from 14,834 to 5,987 in seed 1, i.e., by $(14{,}834 - 5{,}987) / 14{,}834 = 60\%$, and lineage chains thin: R events at $L \geq 2$ fall from 7,743 to 2,136 for \texttt{nop-C}, from 2,125 to 71 for \texttt{h-search} and from 1,228 to 48 for \texttt{h-alloc}, and \texttt{h-copy} loses $L \geq 2$ altogether in seeds 1, 2 and 4 and in the no-mutation control. %
Pass~D records only candidates, so translation follows: records with the executed site as \new{information }carrier and a product fall from 23,832 to 7,174 in seed 1, and in seed 3, where \texttt{h-copy} stays a candidate, it forms no confirmed pair. %
This is a finding about the process class: in the order as defined, a single executed instruction is not causally closed, because maintenance edges pull the next pass of the copy loop and the child's first steps into its closure, and single-step processes therefore rarely show R for the instructions that run the copy loop. %
The organism-level results of Section~\ref{sec:comp.2} use gestation processes, which contain the whole copy loop, and are the same in both orders; at the level of instructions, we did not search processes that span a pass of the copy loop. %

\textbf{Variation.} Pass~C reads V as Section~\ref{sec:4.2} states it, at the level of instances. %
V holds from \new{information }carrier $x$ to \new{information }carrier $y$ when a path of CSS edges runs from a particular instance labelled $x$ to a particular instance labelled $y$, through any intermediate labels and without need of a mutation. %
We fixed this instance-path reading during development, before any held-out run.
In the dense order, V links 173 of the 182 ordered pairs of candidate labels in seed 1 (95\%) and 92--97\% in seeds 1--4, and connects all candidates in every run. %
V thus removes no candidate, and each family is the set of \new{information }carriers in swap-confirmed T pairs that also share a replication operator set. %
\new{Figure~\ref{fig:zooms}b, left, shows one V pair between two family members, located post hoc in the withheld record.}

\textbf{Translation.} Pass~D groups processes with a memory \new{information }carrier input by their operator labels and takes two \new{information }carriers with different products under the same operators as a candidate pair. %
In order DP, the executed site is an operator of its own step in 99.9\% of records, so under the full operator set two \new{information }carriers would never share operators; we therefore use the non-memory operators as the non-empty corresponding subset that Appendix~\ref{app:C.1} admits. %
\del{A}\new{An} \new{information }carrier swap replay tests each candidate for variant-dependent downstream realisation (Section~\ref{sec:5}): with the first \new{information }carrier's value replaced by the second's, the product must change to the second's type, and conversely. %
\new{Figure~\ref{fig:zooms}b, right, shows one confirmed pair.}
\del{Carriers}\new{Information carriers} are split by read type, i.e., read as the executed instruction (control read) or as copied data (data read); we made this split post hoc, after a first version without it lost \texttt{h-search} and \texttt{nand}. %
In seed 1, the replay confirms 27 of 94 candidate pairs (7 control, 20 data); seeds 2--4 give 23, 24 and 23 confirmed pairs. %
Translation thus rests mainly on data reads: the confirmed data-read pairs number 20, 21, 22 and 23 in seeds 1--4, the control-read pairs 7, 2, 2 and 0. %
In seed 1, 6 of the 7 control-read pairs have an empty operator signature, i.e., in their closures no CPU variable meets the operator condition; these 6 pairs pass the swap test but lack the non-empty corresponding operator organisation that Section~\ref{sec:4.4} requires. We call them T-like: they behave like T in the swap test but are not T as defined, and the family rule does not count them. No other run has such pairs. %

\textbf{The information family of seed 1.} A memory label is a member when it 1) re-enters replication with $L \geq 2$; 2) lies in the V-connected group of the members; 3) is \del{a}\new{an} \new{information }carrier in a swap-confirmed T pair, each \new{information }carrier with at least 2 instances, i.e., records of the \new{information }carrier under the pair's operator signature, and; 4) shares a replication operator set with another member, each at least twice. %
In seed 1 the family has 5 members (Table~\ref{tab:comp-family}): the ancestral instructions \texttt{nop-A}, \texttt{nop-B}, \texttt{nop-C}, \texttt{h-alloc} and \texttt{if-label}. %
Counting the 6 T-like pairs would add \texttt{h-search} and the mutants \texttt{get-head}, \texttt{nand}, \texttt{inc} and \texttt{swap-stk}, which are \new{information }carriers only in those pairs, and give 10 members. %
With at least 10 instances per member, \texttt{if-label} drops out and 4 members remain (Section~\ref{sec:comp.5}). %
We use a minimum of 2 instances, so that young variants, which have had few chances to act, are not excluded by construction. %
The ancestral \texttt{h-copy}, the instruction that copies one site, is not a member, for the reason given above, and neither are \texttt{h-divide} and \texttt{mov-head}. %
\del{No family-level null exists yet: the label permutation tests R only.}\new{The label permutation tests R only; Section~\ref{sec:comp.4} adds an exploratory family-level null.} %

\textbf{Roles.} The replication operator set that the members of seed 1 share is the read head being non-zero, (\texttt{RH}, non-zero). %
Translation comes in two kinds. %
Through control reads, executing an instruction changes the CPU state: execution is translation, as in the printer example of Section~\ref{sec:4.4}, with the instructions as files and the CPU state as printer. %
In order ALL, control-read T rests on one pair in seed 1, \texttt{nop-A} and \texttt{if-label} with 3 instances each, and on 2, 2 and 0 pairs in seeds 2--4. %
Through data reads, copying a nop writes its value into the read label that marks the end of the memory content; \texttt{nop-A} writes zero, \texttt{nop-B} and \texttt{nop-C} non-zero, so data-read T partly reflects that \texttt{nop-A} has the value 0. %
The products are CPU values that become non-zero or zero and blank memory after allocation; they are short-lived and matter only through how fast an organism copies itself, so the family is a formal RVT detection, not verified heredity. %

\textbf{Seeds 1--4.} In the four runs with births, the families hold 5, 4, 6 and 4 labels ($n = 4$ runs; 4, 4, 5 and 4 with at least 10 instances per member; Table~\ref{tab:comp-family}). %
The labels common to all four families are \texttt{nop-A}, \texttt{nop-B}, \texttt{nop-C} and \texttt{h-alloc}. %
The other members differ per run: \texttt{if-label} in seed 1, and \texttt{h-search} and the mutant \texttt{get-head} in seed 3, the only mutant member in any run. %
Of the 9 ancestral instructions, \texttt{h-copy}, \texttt{h-divide} and \texttt{mov-head} are members in no run. %
\new{In the held-out runs, \texttt{h-copy} is a member in 17 of 39, \texttt{mov-head} in 3 and \texttt{h-divide} in 1 (Appendix~\ref{app:H.8}).} %

\textbf{Guided comparison.} A guided check, which locates processes from Avida knowledge but evaluates the conditions on the CSS alone, finds R at 83.0\% of the 20,649 copy steps of seed 1 (81.0--83.0\% in seeds 1--4) and in 704 of 705 births (99.9\%), whose gestation process includes the parent's memory content after division. %

\subsection{Post-inference validation against \del{withheld annotations}\new{designated roles}, interventions and controls}\label{sec:comp.4}

\new{We call the roles that the simulator's design fixes and its records document, i.e., which structures are organisms, hereditary programs, copies and mutations, \emph{designated roles}, and the roles that the blind search assigns from the CSS \emph{inferred roles}.}
Three checks compare the blind results with what is known independently: 1) the simulator's withheld record, read post hoc; 2) interventions by replay, i.e., the memory-content swaps of Section~\ref{sec:comp.2} and the \new{information }carrier swaps of Section~\ref{sec:comp.3}, and; 3) controls, i.e., label permutations and a run without mutations.
\new{The first compares inferred with designated roles; the other two test the inferred roles by intervention and against controls.}

\textbf{Withheld record.} At the level of organisms, the organisms, births, G-labels, values of P and life endings found blind match the record in every run (Section~\ref{sec:comp.2}). %
\new{At the level of organisms, the inferred family also equals the family that the criterion assigns when its conditions are read from the withheld record: 13, 21, 14 and 17 G-labels in seeds 1--4, with none missed and none added. Because the label of an organism, its memory content at birth, is the genotype that the design defines, this agreement is expected once births and memory contents are read correctly. Its content is existential, in the sense of Appendix~\ref{app:B.11}: there is an admissible labelling at which the criterion assigns the information carrier role to exactly the G-labels that the record shows passed on unchanged over two or more births, and choosing that labelling with the design in mind adds no hereditary content (Section~\ref{sec:3.3}).} %
At the level of instructions, 17,131 of the 17,133 single-step processes of seed 1 that show R are copy steps, and the moves found in pass~B fall in 704 of the 705 divisions and in no other step. %
A circular alignment of the memory contents of parent and child at every birth found in the CSS, a diagnostic that the family rule does not use, recovers 422 of 423 logged copy errors in seeds 1--4, with one false pairing. %

\textbf{Founding mutations.} A guided trace follows every site of every organism of seed 1 through the withheld record back to the mutation event that wrote its label. %
For children copied from the middle of the parent's memory content, it infers 92 site sources from the offset of the organism's mapped sites; 20 sites stay unresolved, none of which holds or was copied from \texttt{get-head}, \texttt{nand} or \texttt{inc}. %
Seed 1 has no mutant member under the family rule used here; if the T-like pairs are counted, it has four, \texttt{get-head}, \texttt{nand}, \texttt{inc} and \texttt{swap-stk}, and we traced the first three: \texttt{get-head} entered through 9 events, of which 7 reached a child and 1 founded a lineage, i.e., reached a grandchild; \texttt{nand} through 13, 9 and 3, and; \texttt{inc} through 8, 5 and 2. %
\del{The founding events with most carriers are insertions at a division: \texttt{get-head} with 144 carrier organisms over 11 generations, \texttt{nand} with 59 over 12, and \texttt{inc} with 7 over 4.}\new{The founding events whose mutant label is carried by the most organisms are insertions at a division: \texttt{get-head}, carried by 144 organisms over 11 generations, \texttt{nand} by 59 over 12, and \texttt{inc} by 7 over 4.} %
An inserted value comes from a random draw, so these founding events have no original label and no template node. %
\texttt{nand} also has two founding copy errors, from \texttt{nop-C} and from \texttt{if-label}; at the \texttt{nop-C} event, the template node precedes the \texttt{nand} node in neither order, i.e., the mutated copy does not depend on its template. %
Under the instance-path reading, the first node of the mutated site is reached at all six founding events from instances of 13--19 instruction labels, so V into a mutant runs through other instances, not from a template to the mutant. %

\textbf{Controls.} The label-permutation null keeps the order and permutes labels over nodes; the null tests R only. %
\new{A no-mutation control, i.e., seed 1 with all mutation rates set to 0, removes one source of variation, mutation; what it tests therefore depends on the level of description.} %
\new{At the level of organisms, mutation is the only source of new variants: memory sites still change label in every gestation, from \texttt{nop-A}, which \texttt{h-alloc} writes into each new site, to the instruction that the copy loop writes there, but every child's memory content at birth equals its parent's, so the control has one G-label, hence no V, no T pair and no family, as expected.} %
\new{At the level of instructions, the variants are instruction types, and the ancestral program contains several of them from the start; mutation only adds further types.}
\new{At this level the control} shows R in 20,153 of 83,236 steps (24.2\%), $L \geq 2$ for 4 memory labels, \texttt{h-divide}, \texttt{nop-A}, \texttt{nop-B} and \texttt{nop-C}, 7 swap-confirmed T pairs, all through data reads, and no heritable variation: no child site differs from its parent's. %
\new{V, which Section~\ref{sec:4.2} defines structurally, needs no mutation: causal paths run between instances of different ancestral labels within one unchanging program, so the control has a family of 3 ancestral labels, \texttt{nop-A}, \texttt{nop-B} and \texttt{nop-C}.} %
\new{This family is expected, and it says nothing about the specificity of the criterion at the level of instructions, because removing mutations does not remove variation there; conclusions from the control are drawn at the level of organisms only.}
\new{Readings of V that add a correspondence between an original and its variant (Section~\ref{sec:comp.5}) give the control no family at the level of instructions either, because no child site differs from its parent's: they turn V at this level into a test for mutation.} %
\new{Specificity at the level of instructions must therefore come from other controls. The label permutations above test R only; an exploratory family-level null on the development runs extends them to the structural part of the family.}
\new{It reassigns the labels at random among the persistent instances of each kind, keeping the frequency of each label and the causal order, and runs passes A--C and pass D up to the candidate T pairs, without swap replays, because a replay creates nodes that carry no reassigned label; the real CSS is scored the same way.}
\new{The null is a graph-randomisation diagnostic, not a physical negative control: the edges of the CSS record whether a target's type is realised, so a relabelled graph is not the CSS of any history.}
\new{The pass criterion was fixed before any relabelled full CSS was scored, but the definition of the null was revised during development, so the result is exploratory.}
\new{The real CSS gives a structural family of 13, 13, 14 and 10 memory labels in seeds 1--4; 20 reassignments per seed give at most 4, 5, 5 and 4 (one-sided $p = 1/21$ per seed).}
\new{A family of 2--5 labels appears in every one of the 80 reassignments, always with \texttt{nop-A} and \texttt{nop-C}, the most frequent memory labels, so the null distinguishes the size of the family, not its existence.} %
\new{In every reassignment, the structural family consists of exactly the labels that reach $L \geq 2$: in this dense order, V and the candidate T pairs filter nothing, so the null tests whether the labels align with replication and continuing lineages, not the specificity of the full criterion.}
\new{The swap test for T does filter: of the 13 labels of the structural family of seed~1, it confirms 5.}

\begin{figure}[htbp]
\begin{newtext}
\centering
\begin{minipage}[t]{0.48\textwidth}{\footnotesize a)}\\[2pt]\centering\definecolor{avAxis}{HTML}{444444}%
\definecolor{avGhost}{HTML}{DCDCDC}%
\definecolor{avGhostEdge}{HTML}{8C8C8C}%
\definecolor{avMute}{HTML}{6E6E6E}%
\definecolor{cssOI1}{HTML}{0072B2}%
\begin{tikzpicture}[x=1cm, y=1cm, line cap=butt, line join=round]
\useasboundingbox (0,0) rectangle (8,7.59);
\node[anchor=base west, inner sep=0pt, font=\scriptsize, text=avMute] at (0.15,7.36) {seed 1 (bars)};
\node[anchor=base east, inner sep=0pt, font=\scriptsize, text=avMute] at (4.15,7.36) {seed 2};
\node[anchor=base east, inner sep=0pt, font=\scriptsize, text=avMute] at (5.1,7.36) {seed 3};
\node[anchor=base east, inner sep=0pt, font=\scriptsize, text=avMute] at (6.05,7.36) {seed 4};
\node[anchor=base east, inner sep=0pt, font=\scriptsize, text=avMute] at (7,7.36) {seed 5};
\node[anchor=base east, inner sep=0pt, font=\scriptsize, text=avMute] at (7.95,7.36) {control};
\draw[draw=avAxis, line width=0.3pt] (0.1,7.32) -- (7.95,7.32);
\node[anchor=base west, inner sep=0pt, font=\scriptsize, text=black] at (0.15,6.91) {variable-step pairs};
\node[anchor=base west, inner sep=0pt, font=\scriptsize, text=avMute] at (2.616,6.91) {(4,752 $\times$ 81,410)};
\fill[fill=cssOI1] (0.15,6.83) rectangle (4.959,6.63);
\node[anchor=base west, inner sep=0pt, font=\scriptsize\bfseries, text=black] at (5.039,6.64) {386,860,320};
\node[anchor=base west, inner sep=0pt, font=\scriptsize, text=black] at (0.15,6.35) {CSS nodes};
\fill[fill=cssOI1] (0.15,6.27) rectangle (3.47,6.07);
\node[anchor=base west, inner sep=0pt, font=\scriptsize\bfseries, text=black] at (3.55,6.08) {847,911};
\node[anchor=base west, inner sep=0pt, font=\scriptsize, text=black] at (0.15,5.79) {single-step processes};
\node[anchor=base east, inner sep=0pt, font=\scriptsize, text=black] at (4.15,5.79) {84,089};
\node[anchor=base east, inner sep=0pt, font=\scriptsize, text=black] at (5.1,5.79) {83,914};
\node[anchor=base east, inner sep=0pt, font=\scriptsize, text=black] at (6.05,5.79) {83,000};
\node[anchor=base east, inner sep=0pt, font=\scriptsize, text=black] at (7,5.79) {401};
\node[anchor=base east, inner sep=0pt, font=\scriptsize, text=black] at (7.95,5.79) {83,236};
\fill[fill=cssOI1] (0.15,5.71) rectangle (2.9,5.51);
\node[anchor=base west, inner sep=0pt, font=\scriptsize\bfseries, text=black] at (2.98,5.52) {81,410};
\node[anchor=base west, inner sep=0pt, font=\scriptsize, text=black] at (0.15,5.23) {processes with R};
\node[anchor=base east, inner sep=0pt, font=\scriptsize, text=black] at (4.15,5.23) {17,390};
\node[anchor=base east, inner sep=0pt, font=\scriptsize, text=black] at (5.1,5.23) {17,889};
\node[anchor=base east, inner sep=0pt, font=\scriptsize, text=black] at (6.05,5.23) {17,289};
\node[anchor=base east, inner sep=0pt, font=\scriptsize, text=black] at (7,5.23) {66};
\node[anchor=base east, inner sep=0pt, font=\scriptsize, text=black] at (7.95,5.23) {20,153};
\fill[fill=cssOI1] (0.15,5.15) rectangle (2.521,4.95);
\node[anchor=base west, inner sep=0pt, font=\scriptsize\bfseries, text=black] at (2.601,4.96) {17,133};
\node[anchor=base west, inner sep=0pt, font=\scriptsize, text=avMute] at (3.616,4.96) {(21.0\% of steps)};
\filldraw[fill=avGhost, draw=avGhostEdge, line width=0.4pt, dash pattern=on 2pt off 1.2pt] (0.15,4.84) rectangle (2.312,4.74);
\node[anchor=base west, inner sep=0pt, font=\scriptsize, text=avMute] at (2.392,4.7) {null max 8.9\% $\approx$ 7,250; runs 8.7--11.2\%};
\node[anchor=base west, inner sep=0pt, font=\scriptsize, text=black] at (0.15,4.41) {memory labels, $L \geq 2$};
\node[anchor=base east, inner sep=0pt, font=\scriptsize, text=black] at (4.15,4.41) {13};
\node[anchor=base east, inner sep=0pt, font=\scriptsize, text=black] at (5.1,4.41) {15};
\node[anchor=base east, inner sep=0pt, font=\scriptsize, text=black] at (6.05,4.41) {12};
\node[anchor=base east, inner sep=0pt, font=\scriptsize, text=black] at (7,4.41) {8};
\node[anchor=base east, inner sep=0pt, font=\scriptsize, text=black] at (7.95,4.41) {4};
\fill[fill=cssOI1] (0.15,4.33) rectangle (0.792,4.13);
\node[anchor=base west, inner sep=0pt, font=\scriptsize\bfseries, text=black] at (0.872,4.14) {14};
\node[anchor=base west, inner sep=0pt, font=\scriptsize, text=black] at (0.15,3.85) {T candidate pairs};
\node[anchor=base east, inner sep=0pt, font=\scriptsize, text=black] at (4.15,3.85) {97};
\node[anchor=base east, inner sep=0pt, font=\scriptsize, text=black] at (5.1,3.85) {114};
\node[anchor=base east, inner sep=0pt, font=\scriptsize, text=black] at (6.05,3.85) {73};
\node[anchor=base east, inner sep=0pt, font=\scriptsize, text=black] at (7,3.85) {8};
\node[anchor=base east, inner sep=0pt, font=\scriptsize, text=black] at (7.95,3.85) {14};
\fill[fill=cssOI1] (0.15,3.77) rectangle (1.255,3.57);
\node[anchor=base west, inner sep=0pt, font=\scriptsize\bfseries, text=black] at (1.335,3.58) {94};
\node[anchor=base west, inner sep=0pt, font=\scriptsize, text=black] at (0.15,3.29) {swap-confirmed pairs};
\node[anchor=base east, inner sep=0pt, font=\scriptsize, text=black] at (4.15,3.29) {23};
\node[anchor=base east, inner sep=0pt, font=\scriptsize, text=black] at (5.1,3.29) {24};
\node[anchor=base east, inner sep=0pt, font=\scriptsize, text=black] at (6.05,3.29) {23};
\node[anchor=base east, inner sep=0pt, font=\scriptsize, text=black] at (7,3.29) {2};
\node[anchor=base east, inner sep=0pt, font=\scriptsize, text=black] at (7.95,3.29) {7};
\fill[fill=cssOI1] (0.15,3.21) rectangle (0.952,3.01);
\node[anchor=base west, inner sep=0pt, font=\scriptsize\bfseries, text=black] at (1.032,3.02) {27};
\fill[fill=cssOI1] (0.15,2.91) rectangle (0.89,2.81);
\node[anchor=base west, inner sep=0pt, font=\scriptsize, text=avMute] at (0.97,2.77) {21 with non-empty operator signature};
\filldraw[fill=white, draw=cssOI1, line width=0.5pt] (0.15,2.65) rectangle (0.586,2.55);
\draw[draw=cssOI1, line width=0.35pt] (0.15,2.59) -- (0.21,2.65) (0.17,2.55) -- (0.27,2.65) (0.23,2.55) -- (0.33,2.65) (0.29,2.55) -- (0.39,2.65) (0.35,2.55) -- (0.45,2.65) (0.41,2.55) -- (0.51,2.65) (0.47,2.55) -- (0.57,2.65) (0.53,2.55) -- (0.586,2.606);
\node[anchor=base west, inner sep=0pt, font=\scriptsize, text=avMute] at (0.666,2.51) {6 T-like (empty signature); 0 in other runs};
\node[anchor=base west, inner sep=0pt, font=\scriptsize, text=black] at (0.15,2.22) {family members};
\node[anchor=base east, inner sep=0pt, font=\scriptsize, text=black] at (4.15,2.22) {4};
\node[anchor=base east, inner sep=0pt, font=\scriptsize, text=black] at (5.1,2.22) {6};
\node[anchor=base east, inner sep=0pt, font=\scriptsize, text=black] at (6.05,2.22) {4};
\node[anchor=base east, inner sep=0pt, font=\scriptsize, text=black] at (7,2.22) {2};
\node[anchor=base east, inner sep=0pt, font=\scriptsize, text=black] at (7.95,2.22) {3};
\fill[fill=cssOI1] (0.15,2.14) rectangle (0.541,1.94);
\node[anchor=base west, inner sep=0pt, font=\scriptsize\bfseries, text=black] at (0.621,1.95) {5};
\filldraw[fill=white, draw=cssOI1, line width=0.5pt] (0.15,1.84) rectangle (0.71,1.74);
\draw[draw=cssOI1, line width=0.35pt] (0.15,1.78) -- (0.21,1.84) (0.17,1.74) -- (0.27,1.84) (0.23,1.74) -- (0.33,1.84) (0.29,1.74) -- (0.39,1.84) (0.35,1.74) -- (0.45,1.84) (0.41,1.74) -- (0.51,1.84) (0.47,1.74) -- (0.57,1.84) (0.53,1.74) -- (0.63,1.84) (0.59,1.74) -- (0.69,1.84) (0.65,1.74) -- (0.71,1.8);
\node[anchor=base west, inner sep=0pt, font=\scriptsize, text=avMute] at (0.79,1.7) {10 if the T-like pairs are counted};
\draw[draw=avAxis, line width=0.5pt] (0.15,1.58) -- (4.78,1.58);
\draw[draw=avAxis, line width=0.5pt] (0.15,1.58) -- (0.15,1.52);
\node[anchor=base, inner sep=0pt, font=\scriptsize, text=black] at (0.15,1.29) {1};
\draw[draw=avAxis, line width=0.5pt] (1.27,1.58) -- (1.27,1.52);
\node[anchor=base, inner sep=0pt, font=\scriptsize, text=black] at (1.27,1.29) {10$^{2}$};
\draw[draw=avAxis, line width=0.5pt] (2.39,1.58) -- (2.39,1.52);
\node[anchor=base, inner sep=0pt, font=\scriptsize, text=black] at (2.39,1.29) {10$^{4}$};
\draw[draw=avAxis, line width=0.5pt] (3.51,1.58) -- (3.51,1.52);
\node[anchor=base, inner sep=0pt, font=\scriptsize, text=black] at (3.51,1.29) {10$^{6}$};
\draw[draw=avAxis, line width=0.5pt] (4.63,1.58) -- (4.63,1.52);
\node[anchor=base, inner sep=0pt, font=\scriptsize, text=black] at (4.63,1.29) {10$^{8}$};
\node[anchor=base west, inner sep=0pt, font=\scriptsize, text=avMute] at (5.13,1.29) {count (log scale)};
\node[anchor=base west, inner sep=0pt, font=\scriptsize, text=avAxis] at (0.15,0.97) {How to read: bars = seed 1, log scale; numbers by a};
\node[anchor=base west, inner sep=0pt, font=\scriptsize, text=avAxis] at (0.15,0.67) {stage name = other runs; hatched = T-like pairs.};
\node[anchor=base west, inner sep=0pt, font=\scriptsize, text=avAxis] at (0.15,0.37) {Limits: order ALL, threshold 2, T-like pairs not counted};
\node[anchor=base west, inner sep=0pt, font=\scriptsize, text=avAxis] at (0.15,0.07) {as T; development runs only; null = 20 permutations.};
\end{tikzpicture}\end{minipage}\hfill
\begin{minipage}[t]{0.5\textwidth}{\footnotesize b)}\\[2pt]\centering\resizebox{\linewidth}{!}{\definecolor{avAxis}{HTML}{444444}%
\definecolor{avGrid}{HTML}{E4E4E4}%
\definecolor{avMute}{HTML}{6E6E6E}%
\definecolor{cssOI1}{HTML}{0072B2}%
\definecolor{cssOI2}{HTML}{E69F00}%
\definecolor{cssOI3}{HTML}{009E73}%
\definecolor{cssOI4}{HTML}{CC79A7}%
\begin{tikzpicture}[x=1cm, y=1cm, line cap=butt, line join=round]
\useasboundingbox (0,0) rectangle (8,7);
\draw[draw=avGrid, line width=0.3pt] (1.652,0.95) -- (1.652,6.85);
\draw[draw=avGrid, line width=0.3pt] (1.15,1.392) -- (7.85,1.392);
\draw[draw=avGrid, line width=0.3pt] (3.327,0.95) -- (3.327,6.85);
\draw[draw=avGrid, line width=0.3pt] (1.15,2.867) -- (7.85,2.867);
\draw[draw=avGrid, line width=0.3pt] (5.002,0.95) -- (5.002,6.85);
\draw[draw=avGrid, line width=0.3pt] (1.15,4.342) -- (7.85,4.342);
\draw[draw=avGrid, line width=0.3pt] (6.677,0.95) -- (6.677,6.85);
\draw[draw=avGrid, line width=0.3pt] (1.15,5.817) -- (7.85,5.817);
\draw[draw=avAxis, line width=0.5pt] (1.15,0.95) rectangle (7.85,6.85);
\draw[draw=avAxis, line width=0.5pt] (1.652,0.95) -- (1.652,0.88);
\draw[draw=avAxis, line width=0.5pt] (1.15,1.392) -- (1.08,1.392);
\node[anchor=base, inner sep=0pt, font=\scriptsize, text=black] at (1.652,0.64) {60};
\node[anchor=base east, inner sep=0pt, font=\scriptsize, text=black] at (1.02,1.302) {60};
\draw[draw=avAxis, line width=0.5pt] (3.327,0.95) -- (3.327,0.88);
\draw[draw=avAxis, line width=0.5pt] (1.15,2.867) -- (1.08,2.867);
\node[anchor=base, inner sep=0pt, font=\scriptsize, text=black] at (3.327,0.64) {70};
\node[anchor=base east, inner sep=0pt, font=\scriptsize, text=black] at (1.02,2.777) {70};
\draw[draw=avAxis, line width=0.5pt] (5.002,0.95) -- (5.002,0.88);
\draw[draw=avAxis, line width=0.5pt] (1.15,4.342) -- (1.08,4.342);
\node[anchor=base, inner sep=0pt, font=\scriptsize, text=black] at (5.002,0.64) {80};
\node[anchor=base east, inner sep=0pt, font=\scriptsize, text=black] at (1.02,4.252) {80};
\draw[draw=avAxis, line width=0.5pt] (6.677,0.95) -- (6.677,0.88);
\draw[draw=avAxis, line width=0.5pt] (1.15,5.817) -- (1.08,5.817);
\node[anchor=base, inner sep=0pt, font=\scriptsize, text=black] at (6.677,0.64) {90};
\node[anchor=base east, inner sep=0pt, font=\scriptsize, text=black] at (1.02,5.727) {90};
\node[anchor=base, inner sep=0pt, font=\scriptsize, text=black] at (4.5,0.11) {P of the inserted memory content (steps)};
\node[anchor=center, inner sep=0pt, font=\scriptsize, text=black, rotate=90] at (0.2,3.9) {P after the swap (steps)};
\draw[draw=avAxis, line width=0.5pt] (1.15,0.95) -- (7.85,6.85);
\fill[fill=cssOI1] (3.16,2.72) -- (3.305,2.759) arc (15.161:90:0.15cm) -- cycle;
\fill[fill=cssOI2] (3.16,2.72) -- (3.105,2.58) arc (-111.613:15.161:0.15cm) -- cycle;
\fill[fill=cssOI3] (3.16,2.72) -- (3.018,2.669) arc (-160.323:-111.613:0.15cm) -- cycle;
\fill[fill=cssOI4] (3.16,2.72) -- (3.16,2.87) arc (-270:-160.323:0.15cm) -- cycle;
\draw[draw=avAxis, line width=0.3pt] (3.16,2.72) circle (0.15cm);
\fill[fill=cssOI1] (2.992,2.572) -- (3.132,2.583) arc (4.444:90:0.14cm) -- cycle;
\fill[fill=cssOI2] (2.992,2.572) -- (2.951,2.439) arc (-107.407:4.444:0.14cm) -- cycle;
\fill[fill=cssOI3] (2.992,2.572) -- (2.876,2.495) arc (-146.296:-107.407:0.14cm) -- cycle;
\fill[fill=cssOI4] (2.992,2.572) -- (2.992,2.713) arc (-270:-146.296:0.14cm) -- cycle;
\draw[draw=avAxis, line width=0.3pt] (2.992,2.572) circle (0.14cm);
\fill[fill=cssOI1] (3.662,3.162) -- (3.788,3.164) arc (0.691:90:0.126cm) -- cycle;
\fill[fill=cssOI2] (3.662,3.162) -- (3.556,3.095) arc (-147.545:0.691:0.126cm) -- cycle;
\fill[fill=cssOI3] (3.662,3.162) -- (3.557,3.232) arc (-213.376:-147.545:0.126cm) -- cycle;
\fill[fill=cssOI4] (3.662,3.162) -- (3.662,3.288) arc (-270:-213.376:0.126cm) -- cycle;
\draw[draw=avAxis, line width=0.3pt] (3.662,3.162) circle (0.126cm);
\fill[fill=cssOI1] (2.49,2.13) -- (2.594,2.191) arc (30.251:90:0.12cm) -- cycle;
\fill[fill=cssOI2] (2.49,2.13) -- (2.438,2.021) arc (-115.356:30.251:0.12cm) -- cycle;
\fill[fill=cssOI3] (2.49,2.13) -- (2.371,2.11) arc (-170.586:-115.356:0.12cm) -- cycle;
\fill[fill=cssOI4] (2.49,2.13) -- (2.49,2.25) arc (-270:-170.586:0.12cm) -- cycle;
\draw[draw=avAxis, line width=0.3pt] (2.49,2.13) circle (0.12cm);
\fill[fill=cssOI1] (3.83,3.31) -- (3.884,3.329) arc (18.889:90:0.057cm) -- cycle;
\fill[fill=cssOI3] (3.83,3.31) -- (3.83,3.367) arc (-270:18.889:0.057cm) -- cycle;
\draw[draw=avAxis, line width=0.3pt] (3.83,3.31) circle (0.057cm);
\filldraw[fill=cssOI3, draw=avAxis, line width=0.3pt] (2.322,1.982) circle (0.045cm);
\fill[fill=cssOI2] (1.82,1.54) -- (1.782,1.524) arc (-157.229:90:0.041cm) -- cycle;
\fill[fill=cssOI3] (1.82,1.54) -- (1.82,1.581) arc (-270:-157.229:0.041cm) -- cycle;
\draw[draw=avAxis, line width=0.3pt] (1.82,1.54) circle (0.041cm);
\filldraw[fill=cssOI2, draw=avAxis, line width=0.3pt] (2.657,2.277) circle (0.038cm);
\filldraw[fill=cssOI3, draw=avAxis, line width=0.3pt] (6.342,5.522) circle (0.034cm);
\filldraw[fill=cssOI4, draw=avAxis, line width=0.3pt] (3.495,3.015) circle (0.025cm);
\filldraw[fill=cssOI3, draw=avAxis, line width=0.3pt] (4.332,3.752) circle (0.022cm);
\filldraw[fill=cssOI3, draw=avAxis, line width=0.3pt] (7.18,6.26) circle (0.021cm);
\node[anchor=base west, inner sep=0pt, font=\scriptsize, text=black] at (1.27,6.56) {Seeds 1--4: 4,314 swaps};
\node[anchor=base west, inner sep=0pt, font=\scriptsize, text=black] at (1.27,6.24) {4,135 follow (on $y=x$)};
\node[anchor=base west, inner sep=0pt, font=\scriptsize, text=avMute] at (1.27,5.92) {166 censored, 13 sterile};
\node[anchor=base west, inner sep=0pt, font=\scriptsize, text=black] at (1.27,5.68) {Circle area = swaps};
\draw[draw=avAxis, line width=0.3pt] (1.52,5.45) circle (0.014cm);
\node[anchor=base, inner sep=0pt, font=\scriptsize, text=black] at (1.52,5.06) {10};
\draw[draw=avAxis, line width=0.3pt] (2.22,5.45) circle (0.045cm);
\node[anchor=base, inner sep=0pt, font=\scriptsize, text=black] at (2.22,5.06) {100};
\draw[draw=avAxis, line width=0.3pt] (3.12,5.45) circle (0.142cm);
\node[anchor=base, inner sep=0pt, font=\scriptsize, text=black] at (3.12,5.06) {1000};
\node[anchor=base west, inner sep=0pt, font=\scriptsize, text=black] at (4.15,2.88) {seed};
\node[anchor=base east, inner sep=0pt, font=\scriptsize, text=black] at (5.78,2.88) {follow};
\node[anchor=base east, inner sep=0pt, font=\scriptsize, text=avMute] at (6.68,2.88) {cens.};
\node[anchor=base east, inner sep=0pt, font=\scriptsize, text=avMute] at (7.78,2.88) {sterile};
\fill[fill=cssOI1] (4.22,2.65) circle (0.06cm);
\node[anchor=base west, inner sep=0pt, font=\scriptsize, text=black] at (4.33,2.56) {1};
\node[anchor=base east, inner sep=0pt, font=\scriptsize, text=black] at (5.78,2.56) {808};
\node[anchor=base east, inner sep=0pt, font=\scriptsize, text=avMute] at (6.68,2.56) {16};
\node[anchor=base east, inner sep=0pt, font=\scriptsize, text=avMute] at (7.78,2.56) {0};
\fill[fill=cssOI2] (4.22,2.35) circle (0.06cm);
\node[anchor=base west, inner sep=0pt, font=\scriptsize, text=black] at (4.33,2.26) {2};
\node[anchor=base east, inner sep=0pt, font=\scriptsize, text=black] at (5.78,2.26) {1,435};
\node[anchor=base east, inner sep=0pt, font=\scriptsize, text=avMute] at (6.68,2.26) {74};
\node[anchor=base east, inner sep=0pt, font=\scriptsize, text=avMute] at (7.78,2.26) {9};
\fill[fill=cssOI3] (4.22,2.05) circle (0.06cm);
\node[anchor=base west, inner sep=0pt, font=\scriptsize, text=black] at (4.33,1.96) {3};
\node[anchor=base east, inner sep=0pt, font=\scriptsize, text=black] at (5.78,1.96) {867};
\node[anchor=base east, inner sep=0pt, font=\scriptsize, text=avMute] at (6.68,1.96) {29};
\node[anchor=base east, inner sep=0pt, font=\scriptsize, text=avMute] at (7.78,1.96) {2};
\fill[fill=cssOI4] (4.22,1.75) circle (0.06cm);
\node[anchor=base west, inner sep=0pt, font=\scriptsize, text=black] at (4.33,1.66) {4};
\node[anchor=base east, inner sep=0pt, font=\scriptsize, text=black] at (5.78,1.66) {1,025};
\node[anchor=base east, inner sep=0pt, font=\scriptsize, text=avMute] at (6.68,1.66) {47};
\node[anchor=base east, inner sep=0pt, font=\scriptsize, text=avMute] at (7.78,1.66) {2};
\draw[draw=avAxis, line width=0.4pt] (4.15,1.58) -- (7.8,1.58);
\node[anchor=base west, inner sep=0pt, font=\scriptsize, text=black] at (4.15,1.26) {all};
\node[anchor=base east, inner sep=0pt, font=\scriptsize, text=black] at (5.78,1.26) {4,135};
\node[anchor=base east, inner sep=0pt, font=\scriptsize, text=avMute] at (6.68,1.26) {166};
\node[anchor=base east, inner sep=0pt, font=\scriptsize, text=avMute] at (7.78,1.26) {13};
\end{tikzpicture}}\end{minipage}\\[4pt]
\begin{minipage}[t]{0.95\textwidth}{\footnotesize c)}\\[2pt]\centering\input{figures/avida-membership}\end{minipage}
\caption{\textbf{Validation at a glance, development runs.} b) Memory-content swaps at the level of organisms: each circle is one combination of P of the inserted memory content and P after the swap, with P the time to first division in the organism's own steps; circle area gives the number of swaps, and the wedges give the seeds. All 4,135 swaps with a value of P lie on the diagonal: P follows the inserted memory content. 166 swaps end censored and 13 sterile and have no value of P; the table gives them per seed, of 4,314 swaps in total. a) From the record of seed 1 to its information family on a log scale, with the counts of the other runs beside each stage; the ghost bar is the maximum of 20 label permutations, and the hatched bars are the 6 T-like pairs. c) Family membership per run in order ALL: filled squares are members at both instance thresholds, half-filled squares at threshold 2 only, hatched squares only if the T-like pairs are counted, and DP marks members in order DP only; the tinted block is the core of seeds 1--4.} %
\label{fig:validation}
\end{newtext}
\end{figure}

\subsection{Sensitivity to the choices made at each stage}\label{sec:comp.5}

The results of Sections~\ref{sec:comp.1} to~\ref{sec:comp.4} follow from declared choices of the perspective $P$, i.e., the variables, persistence rule, labels and alternative values that build the CSS, of the process class searched, and of the reading of the criterion (Table~\ref{tab:comp-sens}). Each paragraph below names the stage of Figure~\ref{fig:pipeline-digital} at which its choice is made.

\textbf{Persistence (stage 3).} A CSS with a minimum dwell time of $k = 10$ steps has 534,870 nodes instead of 847,911. %
On nodes present in both, 92.6\% of the $k = 1$ order relations hold at $k = 10$, and 97.3\% of the $k = 10$ relations at $k = 1$; we have not rerun the motif search at $k = 10$. %

\textbf{Order and process class (stages 5 and 6).} The process class of the instruction-level search depends on the order. %
In order DP, i.e., without the maintenance edges $D_{\mathrm{maint}}$, a single step is causally convex and its process is the step's own nodes; in order ALL, it is the step's convex closure (Section~\ref{sec:comp.3}). %
On the same CSSs, order DP gives R in 19,756 of 81,410 steps of seed 1 (24.3\%; 23.8--24.6\% in seeds 1--4, against permutation maxima of 5.9--6.1\% and 3.4--3.6\%), 14,834 moves, 19 labels with $L \geq 2$, and 21 of 146 candidate T pairs confirmed in seed 1 (13 control, 8 data), against 19, 23 and 14 confirmed pairs in seeds 2--4. %
The families then hold 11, 8, 10 and 8 labels (8, 7, 8 and 6 with at least 10 instances per member), with 6 ancestral instructions in every family, \texttt{h-search} and \texttt{h-copy} in addition to the 4 of order ALL; the mutant members are \texttt{get-head} (seeds 1--4), \texttt{swap} (seeds 2--4), \texttt{nand} and \texttt{inc} (seed 1), and \texttt{swap-stk} and \texttt{push} (seed 3) (Table~\ref{tab:comp-family}). %
In order DP, seed 1 has two doubtful members, each with T only as the site after another instruction, in fewer than 10 steps: after \texttt{h-search}, \texttt{h-divide} ends the search label, so \texttt{h-search} sets \texttt{BX} and \texttt{CX} to zero, unlike \texttt{nop-C} in the same position, and after \texttt{pop}, \texttt{if-label} selects no register, so \texttt{pop} writes \texttt{BX} where \texttt{nop-A} in the same position selects \texttt{AX}. %
One replication operator set covers all 11 members of seed 1 in order DP: \texttt{LEN}, \texttt{GL}, \texttt{AGE}, \texttt{IP}, \texttt{WH} and \texttt{RH}, all non-zero, i.e., memory size, length of the memory content at birth, age, instruction pointer and the write and read heads; in order ALL the shared set shrinks to \texttt{RH}. %
\texttt{AGE} and \texttt{GL} act as operators in order DP only because every step checks the logged age limit, $\texttt{AGE} > 20\,\texttt{GL}$; in Avida the age limit is a rule of the world, not copying machinery. %
The guided check finds R at 95.4\% of the 20,649 copy steps of seed 1 in order DP, against 83.0\% in order ALL, and at 703 instead of 704 of 705 births. %
Order ALL costs more computation: with five runs in parallel on an Apple M2 with 8 cores, passes~A and~B took 2,120--2,419 s per run and pass~D took 2,034--2,278 s, against 51--54 s and 20--22 s in order DP. %
Node-verified lineage, the V connectivity of all candidates and every organism-level result are the same in both orders. %

\textbf{Labels (stage 3).} Scheme B and the restriction of \new{information }carrier variants to memory labels from pass~B on (Section~\ref{sec:comp.3}) are declared choices informed by the platform: a memory site has 26 labels, every other variable 2. The search is blind to instruction names and \del{simulator annotations}\new{designated roles}, not to these choices. %
Scheme A, which labels every non-memory variable by its exact value, gives 2,997 labels in seed 1, 2,118 of them from the three random inputs; we have not analysed it. %
Under scheme A, \texttt{IP}, \texttt{AGE} and the heads would change label at almost every step, and translation through execution would largely disappear. %
Scheme B, by contrast, hides one translation: in order DP, \texttt{mov-head} is copied 1,927 times along label chains of 21 generations, but the heads it moves stay non-zero, so its distinct products share one label. %
The merger of these products is the outcome collapse of Corollary~\ref{cor:2}; whether a finer labelling recovers T for \texttt{mov-head} has not been tested. %

\textbf{Operator correspondence (stage 6).} With the full operator set, the guided check in order DP finds 1 candidate pair type in 225 operator groups, i.e., no T. %
With the non-memory subset of Appendix~\ref{app:C.1}, the swap replay confirms 32 of 65 candidate pair types in both directions, i.e., $32/65 = 49\%$, so every family here rests on the subset reading. %
In order ALL, that subset is empty for 6 of the 7 control-read pairs of seed 1 (Section~\ref{sec:comp.3}). %

\textbf{Reading of V (stage 6).} \new{At the level of instructions, the} reading of V decides whether the no-mutation control has a family and changes the size of the families in runs with mutations (Table~\ref{tab:comp-family}). %
The instance-path reading is the literal reading of Section~\ref{sec:4.2}; in a gestation nearly every input precedes nearly every output, so V holds in every birth, including the 549 of 705 births in seed 1 without any mutation. %
Four alternatives add a correspondence between an original and its variant; we evaluated them with pass~D in order DP and use them as diagnostics only. %
1) Positional comparison of parent and child sites, plus $x \prec y$, finds 878 mismatched sites in seed 1, of which at most 154 can be copy errors. %
At least 878~$-$~154~$=$~724 (82\%) are therefore sites shifted by insertions and deletions, and the families hold 11, 8, 10 and 7 labels. %
2) Logged mutations with $x \prec y$, the guided reading, hold at only 22--41\% of copy errors and give families of 8, 4, 5 and 0. %
3) Correspondence within the copy step finds no V, because a mutated copy here, unlike in Avida, does not read its template; the founding copy error from \texttt{nop-C} to \texttt{nand} in seed 1 shows this in both orders (Section~\ref{sec:comp.4}). %
4) Circular alignment at birth recovers 422 of 423 copy errors and gives the families of reading 1). %
None of the four alternatives gives the no-mutation control a family, because no child site differs from its parent's\new{: they make V at this level a test for mutation, whereas the instance-path reading also counts the variation already present in the ancestral program (Section~\ref{sec:comp.4})}. %

\textbf{Threshold (stage 6).} We count instances, i.e., records of \del{a}\new{an} \new{information }carrier under the pair's operator signature, control and data reads alike. %
With at least 10 instances per member instead of 2, applied after the swap test, the families of seeds 1--4 shrink from 5, 4, 6 and 4 to 4, 4, 5 and 4 labels: \texttt{if-label} leaves the family of seed 1, and \texttt{get-head} that of seed 3. %
Requiring 10 instances of both \new{information }carriers of each pair also removes \texttt{h-search} in seed 3, which leaves 4 members in every run with births. %
In order DP, the same change gives 8, 7, 8 and 6 labels instead of 11, 8, 10 and 8. %

\textbf{Lineage measure (stage 6).} Lineage depth depends on its measure: 8--12 generations node-verified in seeds 1--4, 21--24 along the label chains of pass~B, and 17--20 along G-label lineages. %
Section~\ref{sec:comp.3} quotes the first, the smallest of the three, and Section~\ref{sec:comp.2} the third. %

\textbf{Level of description (stage 1).} The no-mutation control has a family of 3 labels at the level of instructions and none at the level of organisms, where the instruction labels of a memory content collapse into one G-label. %
The control is thus a computed case of Corollary~\ref{cor:1}: variant collapse destroys detectable V. %
\new{The level decides what the control tests: mutation is the only source of new variants among organisms, but not among instructions, so the control tests the criterion at the level of organisms only.}
At a third level, colonies, i.e., connected patches of at least 3 organisms of one G-label, exist in 94--96 of 100 updates and hold 60--67\% of organisms in seeds 1--4. %
Colony R occurs 1--8 times per run, all by fission except once, but colonies that arise by R stay small (3--7 members) and short-lived (1--5 updates), so no colony lineage extends beyond one step. %
T is not specified, because a colony's doubling time depends on its neighbourhood and is not a stable property of its G-label, and the 36-cell world may be too small for colony lineages; Appendix~\ref{app:H} gives the details. %

\textbf{Seed 5 (stage 6).} Seed 5 shows what a lineage rule without a division requirement admits. %
The founder of seed 5 miscopied its end label, copied past the end of its memory content, read its own copies as templates and died at the age limit after 401 steps without offspring. %
Pass~B counts each re-copy of a copy as a generation, and 8 memory labels reach $L \geq 2$. %
Both swap-confirmed T pairs are data reads at copy steps, in which \texttt{h-copy} writes a zero read label for \texttt{nop-A} and a non-zero one for \texttt{nop-C}. %
None of the copies was executed; all 870 control reads of memory fall on the founder's own 20 sites. %
Under the instance-path reading, the run thus has a family of 2 labels, \texttt{nop-A} and \texttt{nop-C}, within one organism that never divided; with at least 10 instances per member it has none. %
We keep the rule: the criterion decides whether R, V and T hold, and lineage depth is the evidence for reading a family as heredity or as an accidental motif (Section~\ref{sec:4.8}). %

\begin{table}[htbp]
\begin{newtext}
\centering
\small
\caption{Information-family members per development run under the instance-path reading of V. $\bullet$: member in order ALL with at least 2 and at least 10 instances per member; $\circ$: member in order ALL with at least 2 instances only; t: member in order ALL only if the 6 T-like pairs of seed 1, i.e., pairs with an empty operator signature, are counted (each t entry except \texttt{swap-stk} is also a member in order DP); DP: member in order DP only (at least 2 instances). Below, family sizes under both process classes, both thresholds and alternative readings of V; --: not run.}
\label{tab:comp-family}
\begin{tabular}{@{}llcccccc@{}}
\toprule
Label & Origin & Seed 1 & Seed 2 & Seed 3 & Seed 4 & No mutation & Seed 5 \\
\midrule
\texttt{nop-A} & ancestral & $\bullet$ & $\bullet$ & $\bullet$ & $\bullet$ & $\bullet$ & $\circ$ \\
\texttt{nop-B} & ancestral & $\bullet$ & $\bullet$ & $\bullet$ & $\bullet$ & $\bullet$ & \\
\texttt{nop-C} & ancestral & $\bullet$ & $\bullet$ & $\bullet$ & $\bullet$ & $\bullet$ & $\circ$ \\
\texttt{h-alloc} & ancestral & $\bullet$ & $\bullet$ & $\bullet$ & $\bullet$ & & \\
\texttt{h-search} & ancestral & t & DP & $\bullet$ & DP & & \\
\texttt{h-copy} & ancestral & DP & DP & DP & DP & DP & \\
\texttt{h-divide} & ancestral & DP & & & & & \\
\texttt{if-label} & ancestral & $\circ$ & & & & & \\
\texttt{get-head} & mutant & t & DP & $\circ$ & DP & & \\
\texttt{nand} & mutant & t & & & & & \\
\texttt{inc} & mutant & t & & & & & \\
\texttt{swap} & mutant & & DP & DP & DP & & \\
\texttt{swap-stk} & mutant & t & & DP & & & \\
\texttt{push} & mutant & & & DP & & & \\ %
\midrule
\multicolumn{2}{@{}l}{Order ALL, $\geq 2$ instances (used here)} & 5 & 4 & 6 & 4 & 3 & 2 \\ %
\multicolumn{2}{@{}l}{Order ALL, $\geq 10$ instances} & 4 & 4 & 5 & 4 & 3 & 0 \\ %
\multicolumn{2}{@{}l}{Order ALL, $\geq 2$, T-like pairs counted} & 10 & 4 & 6 & 4 & 3 & 2 \\ %
\multicolumn{2}{@{}l}{Order DP, $\geq 2$ instances} & 11 & 8 & 10 & 8 & 4 & 2 \\ %
\multicolumn{2}{@{}l}{Order DP, $\geq 10$ instances} & 8 & 7 & 8 & 6 & 4 & 1 \\ %
\multicolumn{2}{@{}l}{Order DP, positional reading} & 11 & 8 & 10 & 7 & 0 & 0 \\ %
\multicolumn{2}{@{}l}{Order DP, alignment at birth} & 11 & 8 & 10 & 7 & 0 & 0 \\ %
\multicolumn{2}{@{}l}{Order DP, guided, logged mutations} & 8 & 4 & 5 & 0 & -- & -- \\ %
\bottomrule
\end{tabular}
\end{newtext}
\end{table}

\begin{table}[htbp]
\begin{newtext}
\centering
\small
\caption{Declared choices and their effect on R, V, T and the information family.}
\label{tab:comp-sens}
\begin{tabularx}{\textwidth}{@{}>{\raggedright\arraybackslash}p{0.17\textwidth}>{\raggedright\arraybackslash}p{0.19\textwidth}>{\raggedright\arraybackslash}p{0.19\textwidth}>{\raggedright\arraybackslash}X@{}}
\toprule
Choice & Used here & Alternative & Effect of the alternative \\
\midrule
Persistence & $k = 1$ step & $k = 10$ steps & 534,870 nodes; 92.6\% and 97.3\% of order relations shared; motifs not rerun \\ %
\midrule
Order and process class at the level of instructions & order ALL, $\prec = D^{+}$, as defined: convex closure of a step, median 25--28 nodes, up to 67,856 & order DP, $(D_{\mathrm{base}} \cup D_{\mathrm{pers}})^{+}$: the step's own nodes & seed 1: R in 24.3\% instead of 21.0\% of steps; 21 instead of 27 confirmed T pairs, 13 instead of 7 through control reads; families of 11, 8, 10, 8 instead of 5, 4, 6, 4 labels; 6 instead of 4 ancestral labels in every family, \texttt{h-copy} among them; organism level unchanged \\ %
\midrule
Labels & scheme B & scheme A & 2,997 labels; not analysed; under scheme B, T of \texttt{mov-head} is hidden \\ %
\midrule
Operator correspondence & non-memory subset (Appendix~\ref{app:C.1}) & full operator set & order DP: 1 candidate pair type in 225 groups, i.e., no T \\ %
\midrule
Reading of V & instance path & positional; logged mutations; copy step; alignment & order DP: families of 7--11; 0--8; no V; as positional. The no-mutation control has no family \\ %
\midrule
T threshold & 2 instances & 10 instances & order ALL: 5, 4, 6, 4 $\rightarrow$ 4, 4, 5, 4 members in seeds 1--4 \\ %
\midrule
Operator signature of a T pair & non-empty, as Section~\ref{sec:4.4} requires & empty admitted (T-like pairs) & seed 1: 5 $\rightarrow$ 10 members; no other run has such pairs \\ %
\midrule
Lineage measure & node-verified, 8--12 & label chains; G-label lineage & 21--24; 17--20 \\ %
\midrule
Level & instructions & organisms; colonies & control: family of 3 $\rightarrow$ none; colonies: lineage of one step, T not specified \\ %
\midrule
Division requirement & none & required & seed 5 has no lineage, hence no family of 2 \\ %
\bottomrule
\end{tabularx}
\end{newtext}
\end{table}

On development seeds 1--5, the organisms, births, G-labels and values of P found blind in a CSS built by one-step replay match the withheld record, P follows the inserted memory content in 95.9\% of 4,314 memory-content swaps in seeds 1--4, and the organism-level families hold 13, 21, 14 and 17 G-labels. %
At the level of instructions, a role-blind search in the order of the CSS as defined finds one family in each of the four runs with births, of 5, 4, 6 and 4 labels with \texttt{nop-A}, \texttt{nop-B}, \texttt{nop-C} and \texttt{h-alloc} in every family; \texttt{h-copy} is a member of none, because a single executed instruction is not a causally closed process in that order. %
\new{In an exploratory family-level null, the structural part of the search finds 2--5 labels after random reassignment, against 10--14 in the real CSS of seeds 1--4; the null tests whether the labels align with replication and lineage, and the swap test for T then narrows the 13 to 5 in seed~1 (Section~\ref{sec:comp.4}).}
\new{On the 39 held-out runs with births, the frozen pipeline finds a swap-confirmed family of 3--9 labels in every run, with \texttt{h-copy} a member in 17, and a structural family of 8--16 labels against at most 4--6 after reassignment; 99.76\% of 29,321 decided memory-content swaps follow the inserted memory content. Under the frozen rule the held-out test fails, on its check of the organism-level reading in 2 runs, because the reference definition gives one heritable genotype no phenotype, whereas the reading agrees with the record (Appendix~\ref{app:H.8}).} %
The significance of the test lies in that the hereditary roles, i.e., which instructions act as \new{information }carriers, which CPU state copies them and which products their execution yields, are outputs of the search rather than inputs to it; the organisms are reconstructed blind from the CSS, whereas G and P follow rules we declared (Section~\ref{sec:comp.2}).
The composition of each instruction-level family rests on declared choices: the single-step process class, the instance-path reading of V, the lineage rule without division requirement, scheme B and the subset reading of operator correspondence\new{. The no-mutation control has no family at the level of organisms, as expected, and a family of 3 ancestral labels at the level of instructions, where removing mutations leaves the variation among the ancestral instruction types in place}.
Instruction-level processes that span a pass of the copy loop and real Avida remain to be done, and testing the specificity of the criterion needs an experiment with independently characterised positive and negative cases (Section~\ref{sec:8.world}).
\end{newtext}

\section{What the framework does and does not claim}\label{sec:7}

Because the framework deliberately addresses a broad problem, its boundaries should be explicit. Two components must be distinguished\del{.}:
\begin{enumerate}[label=\arabic*)]
  \item The CSS representation is a general construction: any bounded episode observed under an admissible perspective can be mapped to a labelled strict partial order of persistent structure instances.
  \item RVT plus lineage continuation is a specific hypothesis about which organisation within such a representation constitutes hereditary organisation.
\end{enumerate}
Limitations of the second are not limitations of the first.

\noindent\textbf{1. The CSS representation:}

\textbf{is not ontology-free.} Observation requires a perspective, state variables, primitive structures and resolution. The claim is narrower: hereditary roles are not supplied in that ontology{\delmark}.{\delmark}

\textbf{is not unique.} An episode may admit several physically legitimate spatial and temporal coarse-grainings. These belong to a constrained admissible family.

\textbf{represents only what was observed.} Only realised causal relations within the available observations are represented. No unobserved capacity, and no continuation beyond the observation window, is inferred.

\textbf{carries no semantic content.} Structural labels encode structural type and nothing further. Whether a configuration ``means'' anything to an external interpreter is not represented and not required.

\noindent\textbf{2. The RVT criterion, and what follows from \new{it}:}

\textbf{does not equate copying with heredity.} R alone is insufficient. R and V alone are also insufficient under the initial criterion.

\textbf{does not require identical mechanisms across substrates.} Equivalence is evaluated after inference, on the causal boundary organisation of the inferred processes, not on material identity (Section~\ref{sec:8.4}, Appendix~\ref{app:C}).

\textbf{does not attribute semantics to inferred \new{information }carriers.} A structure may become a candidate information carrier in virtue of its realised causal role, without the framework saying what its configuration signifies.

\textbf{does not separately identify selection.} Differential propagation of inferred variants can be analysed subsequently. It is not required in order to identify the underlying hereditary architecture.

\textbf{does not operationalise cumulative evolutionary accumulation.} Lineage-closed RVT identifies organisation capable of propagating consequential variation. It does not by itself establish the historical accumulation and retention of increasingly organised modification across many generations. The criterion is explicitly limited in this way.

\textbf{does not predict hereditary stability.} The stability, fragility or collapse of an inferred hereditary organisation is a downstream question. The present paper stops at identification and representation.

\subsection{Hereditary systems that the RVT criterion may not capture}\label{sec:7.1}

Several physically documented hereditary systems press on the distinction drawn above. In each case the CSS representation appears adequate, while the RVT criterion as formulated here either fails or is undetermined. The representation is intended to be general; the criterion is explicitly provisional, and the cases below are the evidence for treating it that way.

\textbf{Compositional and collective inheritance.} Autocatalytic assemblies can propagate their statistical composition through growth and division without any discrete structure whose variants are copied. In compositional-inheritance models, what is transmitted is the makeup of the collective rather than the configuration of an individual \new{information }carrier \citep{Segre2000}, and related autocatalytic formalisms describe closure and self-maintenance without a distinguished informational component \citep{Hordijk2018,Huson2024}. Such systems provide no \new{information }carrier distinct from the assembly, no replication operator distinct from the assembly's own chemistry, and no realised structure distinguishable from the \new{information }carrier itself. R and V may be identifiable at the level of the collective; T, as defined here, has no evident referent.

\textbf{Structural and cortical inheritance.} Cortical patterning in ciliates is inherited by direct structural templating: experimentally inverted cortical rows propagate faithfully across many generations without corresponding change in the genome \citep{Beisson1965}. Here the inherited structure is the realised structure. There is no configuration that is interpreted to produce something else, and consequently no separation between \new{information }carrier and product for T to relate. The organisation is nonetheless unambiguously hereditary, and the more general structural type of non-genetic inheritance systems raises the same issue \citep{Jablonka2006}.

\textbf{Selection on intrinsic replicator properties.} In quasispecies and \emph{in vitro} replicator systems, variants differ in replication rate because of their own physical properties rather than through any translated product \new{\citep{Eigen1971,Mills1967,Biebricher2005}}. Variation is inherited and causally consequential, but its consequence is the replication of the \new{information }carrier itself. Such systems satisfy the standard conditions for evolution by natural selection \citep{Lewontin1970} while failing T as stated. Section~\ref{sec:4.3} motivates T by excluding copying of \emph{causally inert} patterns; that motivation does not cover this case, since the variation here is not inert. Such systems satisfy R, V and lineage continuation while failing T as stated.

\textbf{Ambiguous cases.} Viroids replicate as short non-coding RNA using host polymerase, and sequence variants produce different host effects through host machinery \citep{Diener2003}. Prion propagation proceeds by conformational templating without a conserved replication operator distinct from the propagating structure \citep{Prusiner1998}. Whether these satisfy T depends on whether host or environmental machinery may serve as the shared operator through which variant-dependent downstream structures are realised. The present formulation does not settle this, and the ambiguity is itself informative: it indicates that the boundary between replication machinery and translation machinery requires sharper specification than ``distinguishable downstream realised structures'' currently provides.

These cases share a common feature. None of them challenges R, V or lineage continuation. All of them press on T, and specifically on its requirement that variant differences act \emph{through} shared operator organisation to produce structures distinct from the \new{information }carrier. T is therefore both the most distinctive component of the criterion and the component most exposed to counterexample.

We regard this as the central open question rather than as a defect to be repaired by weakening the criterion. The question is not merely whether RVT can be extended to admit these systems, but whether they should be admitted at all. Each of the cases above is hereditary in the ordinary sense of transmitting organisation across generations. It is much less clear that each is \textbf{information-bearing} in the sense the present framework attempts to make precise: that some physically instantiated configuration is interpreted by organisation other than itself, and thereby produces consequences that are not simply its own perpetuation.

If that separation between \new{information }carrier and interpreter is the physically significant property, then the exclusion of compositional, structural and intrinsic-selection inheritance is principled rather than accidental, and RVT delimits a narrower structural type than heredity in general: the structural type in which hereditary variation is interpreted rather than merely propagated. This narrower structural type is also the one associated with the emergence of new organisational levels, where innovations in storage, transmission and interpretation have repeatedly been implicated \new{\citep{Szathmary1995,Jablonka2006}}.

If, instead, these systems are judged to carry information in a defensible physical sense, then T requires either weakening or replacement by a more general condition relating inherited variation to differentiated causal consequence. Determining which of these holds requires adversarial testing against precisely the cases listed above, and is deliberately left open.

\subsection{What can legitimately be called a result}\label{sec:7.2}

\new{The paper reports results of three kinds, which differ in evidential status: mathematical results, formal consequences of stipulated histories, and computational results. A fourth list gives the claims that remain unestablished.}

\new{\textbf{Mathematical results.}} We \textbf{can} presently give:
\begin{itemize}
  \item the formal definition of admissibility (Appendix~\ref{app:A});
  \item the refinement relation (Appendix~\ref{app:B});
  \item the sufficient motif-faithful invariance proposition (Appendix~\ref{app:B});
  \item the simple corollaries about information loss by variant, outcome, replication-boundary or lineage collapse (Appendix~\ref{app:B});
  \item the physical-witness proposition (Appendix~\ref{app:B}).
\end{itemize}
Those follow from the definitions.

\begin{newtext}
\textbf{Formal consequences of stipulated histories.} Sections~\ref{sec:6} and~\ref{sec:cultural} apply the criterion to synthetic configurations and to conditional reconstructions of recipe transmission. Each result has the form of Appendix~\ref{app:F.3}: under the stated description, these structures occupy these roles. The synthetic configurations show which organisations the criterion accepts and which it rejects. The paired recipe reconstructions show that its verdict follows the declared causal history: Case A establishes no complete information family, and Case B establishes one conditionally, in part because its added relations are stipulated to satisfy the conditions. Neither is evidence that a physical system, human skill transmission included, has the stipulated organisation.

\textbf{Computational results.} Section~\ref{sec:comp} reports results obtained from executable dynamics: a CSS constructed by intervention replay from the complete causal record of an Avida-like simulator; the agreement of the organisms found in the CSS with the withheld simulator record; memory-content swaps in which the time to first division follows the inserted memory content, and; a role-blind search of single executed instructions that finds one information family in each development run with births. These results hold for that simulator, on development runs, under the declared choices listed in Section~\ref{sec:comp.5}, including the process class searched. They are formal RVT detections, i.e., the conditions of the criterion hold in the reconstructed CSS; they are not verified heredity.
\end{newtext}

We \textbf{cannot yet claim}:
\begin{itemize}
  \item a complete characterisation of admissible perspectives;
  \item invariance under arbitrary refinement;
  \item uniqueness of a \del{minimal}\new{minimally} sufficient CSS;
  \item a proof that the existential criterion over all admissible CSSs is equivalent to an independently defined representation-free hereditary invariant\del{.}\new{;}
  \item \new{that the RVT criterion identifies hereditary organisation in general: the computational test covers one simulator and one world of 36 cells;} %
  \item \new{that the computational search is specific at the level of instructions: the exploratory family-level null of Section~\ref{sec:comp.4} tests only whether the labels align with replication and lineage, on development runs, and the control run without mutations is not a specificity test at that level, because removing mutations leaves the variation among the ancestral instruction types in place;}
  \item \new{structural labels derived from an AIR partition: the computational CSS uses declared labels;}
  \item \new{correspondence across substrates on the basis of the recipe reconstructions, whose structures and causal relations are stipulated rather than reconstructed independently from observation.}
\end{itemize}

\section{Discussion and open problems}\label{sec:8}

\subsection{Hereditary roles are relational physical properties}\label{sec:8.1}

The central conceptual inversion of CSS is that structures do not become hereditary because an observer names them \emph{genes}, \emph{information}, \emph{programs} or \emph{memes}. They acquire these roles because of the causal organisation in which realised instances participate.

A candidate \new{information }carrier is recognisable because:
\begin{itemize}
  \item variants of its structural type are causally related;
  \item they are replicated;
  \item their differing configurations cause differing realised downstream structures;
  \item descendants continue to participate in corresponding replication organisation.
\end{itemize}

Similarly, replication and translation operators are identified from what their instances do in processes, rather than from their biochemical, mechanical or computational names.

Information carriers and hereditary organisation are therefore identified \textbf{jointly}. This mutual dependence reflects an underlying physical interdependence: stable heredity requires sufficiently persistent information carriers, while hereditary organisation in turn promotes both their continued propagation and persistence.

\subsection{Why instance identity matters}\label{sec:8.2}

A type-level graph saying

\noindent\textbf{DNA $\rightarrow$ DNA}

\noindent records that structures of one type give rise to structures of the same type. It cannot distinguish a single continuing lineage from the repeated independent creation of similar DNA structures, because both produce the same type-level statement.

CSS retains individual physical instances, and the two cases are then distinct. Repetition of a structural type is a set of instances sharing a label,
\[
  \ell(x) = i \ \text{for many}\ x \in W,
\]
\noindent which by itself establishes only that structures of that type occurred. A lineage is a sequence of distinct instances related by realised causal descent,
\[
  x_0 \rightsquigarrow x_1 \rightsquigarrow x_2 \rightsquigarrow \cdots,\ \ell(x_g) = i,
\]
\noindent in which each instance is produced from its predecessor through a replication process, as defined in Section~\ref{sec:4.6}.

The first case is recoverable from a type-level representation; the second is not. This is why the framework cannot collapse all instances of a structural type into a single node.

\subsection{Origin-of-life and unfamiliar systems}\label{sec:8.3}

The inverse formulation is particularly relevant to origins-of-life research. A theory that begins with ``the genome'' or ``the organism'' cannot use those same categories to identify the physical transition at which genomes or organisms first emerge without introducing assumptions about the answer.

In principle, a sufficiently resolved observation can instead be transformed into an admissible CSS and searched for the first occurrence of coherent consequential replication and increasing lineage depth.

The question becomes: \emph{When does realised causal organisation first support persistent \new{information }carrier variants whose differences are propagated and made causally consequential?}

Threshold questions of this kind are central to origin-of-life scenarios \citep{Jeancolas2020}.

Digital-evolution systems such as Avida offer a particularly clean \del{future} test because complete causal histories, replication, mutation and lineage relations can be recorded while native labels such as \emph{genome}, \emph{organism}, \emph{program} and \emph{lineage} are withheld during inference \new{\citep{Ofria2005,Lenski2003,Batut2013,Wilke2001}}. \del{Empirical validation is outside the scope of the present definitional paper.} \new{Section~\ref{sec:comp} runs a first such test on an Avida-like simulator: a role-blind search of the reconstructed CSS finds an information family in each development run with births, and the organisms found in the CSS match the withheld record. The test establishes computational feasibility on one simulator; it does not answer the origin-of-life question above. The next steps are experiments with independently characterised positive and negative cases (Section~\ref{sec:8.world}), real Avida and larger worlds. Chemical and prebiotic systems additionally require observations whose causal record, unlike a simulator's, is incomplete.}

\subsection{Different substrates can instantiate the same organisation}\label{sec:8.4}

Nothing in the CSS formalism requires an information carrier to be nucleic acid.

After inference, familiar interpretations might include:

\begin{table}[htbp]
  \centering
  \small
  \begin{tabularx}{\textwidth}{@{}l>{\raggedright\arraybackslash}X>{\raggedright\arraybackslash}X>{\raggedright\arraybackslash}X>{\raggedright\arraybackslash}X@{}}
    \toprule
    \textbf{Domain} & \textbf{Possible \new{information }carrier interpretation} & \textbf{Replication operator} & \textbf{Translation operator} & \textbf{Possible realised structure} \\
    \midrule
    Biological & nucleic-acid variant & replication machinery & expression machinery & proteins, cellular structures \\
    Cultural & persistent learned pattern & teaching\slash transmission organisation & action\slash production machinery & \del{Cultural}\new{cultural} expression \\
    Civilisational & durable record & copying\slash archive machinery & institutions\slash tools\slash manufacturing equipment & implemented artefacts \\
    Cybernetic & machine-readable state & copying\slash network machinery & processor\slash controller\slash robot & machine state \\
    \bottomrule
  \end{tabularx}
  \label{tab:c4-1}
\end{table}

These interpretations are illustrative only.

They are \textbf{not CSS inputs}.

A particular structure occupies a listed role only if the realised causal motif supports the assignment.

Correspondence of this kind requires a criterion, and the criterion applies to inferred processes rather than to whole systems. Complete CSSs of a cell, an institution and a software deployment are not expected to resemble one another. What can be compared, after hereditary roles have been inferred, is the boundary organisation of the processes bearing those roles: the input and output label multiplicities, the correspondence between \new{information }carrier variants, the operator organisation, the relevant causal incidences and, where the role requires it, the lineage relations. Two processes in physically distinct CSSs are hereditarily equivalent with respect to a motif when a role-preserving relabelling maps one onto the other while preserving these features. Substrate-specific internal detail is allowed to differ. Appendix~\ref{app:C} develops the criterion; the point required here is that the claim is not that two systems are the same, but that they instantiate the same specified hereditary causal organisation, and that this is decidable rather than asserted.

\new{The conditional recipe reconstructions of Section~\ref{sec:cultural} apply the same criterion to a cultural case; in Case B, capability configurations, cooking apparatus and dishes take the \new{information }carrier, translation-operator and product roles. They do not establish correspondence across substrates: their structures and causal relations are stipulated under hypotheses H1--H4 rather than reconstructed independently from observation, and Case B satisfies the criterion in part because its added relations were chosen to satisfy the criterion. The computational test of Section~\ref{sec:comp} concerns a single substrate, a simulated processor, and does not bear on correspondence either. Establishing correspondence requires CSSs reconstructed independently in each substrate and compared under Appendix~\ref{app:C}.}

\subsection{Relation between inferred roles and domain concepts}\label{sec:8.5}

Domains have independently developed concepts of information, replication, lineage, parasitism and maintenance. These concepts are not interchangeable across fields, and many arose in response to particular disciplinary problems. Where a CSS-inferred role and an established domain concept diverge, the divergence requires interpretation rather than automatic resolution in either direction.

Interpreting such disagreement requires a standard stated independently of the formal criterion. We take the scientific target of the present framework to be the identification of physical organisation in which variation is retained, transmitted across generations and made causally consequential, this being the organisation associated with the emergence of new organisational levels \new{\citep{Szathmary1995,Jablonka2006}}. RVT plus lineage continuation is a hypothesis about how to detect that organisation. Two questions must therefore be kept apart: whether the criterion correctly detects the causal pattern it specifies, and whether that pattern is adequate to the target. A criterion may succeed entirely at the first and fail at the second, and only the second is settled by disagreement with domain concepts.

Four cases should be distinguished, and the first should be assessed before the others.

\textbf{Observational insufficiency.} A domain may classify a system as hereditary on the basis of extensive independent evidence while a bounded CSS episode contains no observed replication, no resolved variant distinction or no recoverable causal relation. Section~\ref{sec:4.9} restricts claims to what is realised within the available observation, and Appendix~\ref{app:F.2} treats the resulting limits on identifiability. A disagreement of this kind reflects the observation rather than the concept, and observational sufficiency should be established before any conceptual interpretation is attempted. Failures of detection are otherwise readily mistaken for principled exclusions.

\new{The paired recipe reconstructions of Section~\ref{sec:cultural} illustrate the difference between formal motif assignment and ordinary-language interpretation. In ordinary language, teaching a recipe is cultural inheritance in both cases. In Case A the criterion assigns no information family, because the reconstruction supplies no R process for $c_2$, no V process connecting $c_1$ and $c_2$, and no comparison showing that the dish depends on the capability distinction. In Case B, which adds these relations by stipulation, the criterion assigns the family $\{c_1, c_2\}$. The ordinary-language description is the same in both cases; the formal assignment differs only through the supplied causal history, and Case A is therefore an instance of observational insufficiency, not evidence that recipe teaching lacks hereditary organisation.}

\textbf{The domain concept is heterogeneous.} A single term may bundle causally distinct phenomena that a formal criterion separates. Parasitism may be an example \new{of} such \new{a} case. Within CSS one might ask whether \del{a}\new{an} \new{information }carrier family propagates through replication or translation operators associated with another family, as distinct from whether it draws on resources while replicating through operators of its own. This distinguishes dependence on machinery from dependence on resources, and the two are combined under the ordinary term.

The formulation is not yet a criterion for parasitism, and several questions remain open: what makes an operator belong to one family rather than another, whether relevant contribution is direct, indirect or net, over what interval and process boundary contribution is assessed, and why any positive contribution should disqualify an otherwise exploitative relation. Two unrelated \new{information }carrier families propagating through shared operators may exhibit ordinary dependence on common infrastructure rather than exploitation. Whether either form of dependence warrants the parasitism label is a further question, and one that cannot be posed without a criterion sharp enough to separate the cases. Consistent with the treatment of roles throughout, any such determination would apply to observed episodes rather than to object types.

\textbf{The divergence is deliberate.} Some domain concepts contain content the present framework does not require for role assignment. Two examples: i) Information-theoretic treatments of adaptation quantify the information a genome carries about its environment \citep{Adami2012}; CSS \new{information }carrier identification neither requires nor quantifies that relation, which may be studied after \new{information }carriers have been identified. ii) Some accounts of reproduction require material overlap between parent and offspring, excluding manufacture \citep{Griesemer2000}; R imposes no such condition, though manufacture must still satisfy its multiplicity and causal-contribution requirements to qualify.

A deliberate exclusion is not justified by the reality of the causal distinction alone. Almost any two physical cases differ causally at some resolution. Justification requires showing that the distinction bears on the declared target: through explanatory consequences, through cross-domain correspondence, or through another stated objective. Properties the framework does not require for role assignment should also be distinguished from properties it cannot represent or analyse, which is a different and stronger claim.

\textbf{The criterion is inadequate to the target.} A divergence indicates a defect when the criterion's extension fails to track the organisation described above. Both directions count. False exclusion occurs where a system retains, transmits and renders consequential its variation, yet fails the criterion. False inclusion occurs where the criterion is satisfied by organisation that does not do so. Section~\ref{sec:7.1} examines several systems that appear to challenge the criterion in the first direction, and those cases are best understood as tests of whether translation belongs in a general hereditary criterion rather than as demonstrations that the criterion classifies equivalent organisation inconsistently.

The inferred roles are therefore offered as a distinct set of concepts, not as corrections to established usage. Their value is to be judged by whether they are recovered consistently across substrates, whether the distinctions they draw correspond to differences in mechanism rather than to representational artefact, whether they are robust under admissible changes of representation, and whether they support subsequent explanatory analysis. These are matters for application rather than assertion.

\subsection{Translation and environmental coupling}\label{sec:8.6}

Section~\ref{sec:4.5} established T as the outward causal branch that makes \new{information }carrier variation environmentally consequential. A natural later extension could identify closed loops of the form
\[
  i_k \rightarrow y_k \rightarrow E_k \rightarrow R_k^{\mathrm{future}},
\]
\noindent where $E_k$ denotes changed environmental structures that modify subsequent \new{information }carrier propagation.

The current theory stops before requiring this return pathway, because the two branches answer different questions. T establishes that a variant's configuration is read out into differentiated consequence, which is what makes the structure \del{a}\new{an} \new{information }carrier at all. The return pathway establishes that those consequences act back on the variant's own propagation, which is what makes the system selective. Requiring both would identify a narrower class: hereditary organisation under selection, rather than information-bearing hereditary organisation as such.

\subsection{Selection is downstream of hereditary identification}\label{sec:8.7}

The standard formulation for evolution by natural selection requires variation, heredity and differential fitness \citep{Lewontin1970}. The third condition presupposes that one already knows which entities to count descendants of, and therefore cannot form part of a criterion whose purpose is to identify those entities. CSS separates the two questions accordingly.

Only once the \new{information }carrier family has been inferred can one ask whether variants differ in:
\begin{itemize}
  \item replication rate;
  \item persistence;
  \item survival;
  \item lineage continuation;
  \item downstream environmental effect.
\end{itemize}

Those differences would provide the starting point for a substrate-independent treatment of selection. Copy-count fitness measures are intentionally not imported into the present paper.

\subsection{Cumulative evolution is a further problem}\label{sec:8.8}

The present criterion does not determine whether:
\begin{itemize}
  \item a new modification persists for many subsequent generations;
  \item multiple modifications accumulate in the same lineage;
  \item historically acquired organisation increases;
  \item apparent long-term change reflects genuine cumulative preservation rather than repeated replacement.
\end{itemize}

These questions should not be smuggled into the definition of RVT.

Instead, once CSS has inferred \new{information }carriers and lineages, future work can ask whether later states retain identifiable causal organisation originating from progressively earlier modifications.

This separates two difficult problems:

\emph{What is the hereditary organisation?}

from

\emph{What does that organisation cumulatively retain?}

The second is explicitly left open.

\begin{newtext}%
\subsection{A first digital world tests feasibility and reproducibility, not specificity}\label{sec:8.world}

The computational test of Section~\ref{sec:comp} asks three questions, which the world of that section answers to different degrees. The first is feasibility, i.e., whether a CSS can be built from a complete causal record, searched without designated roles and tested by intervention. The second is reproducibility, i.e., whether the frozen method gives the same results on runs it has not seen. The third is discriminative validity, i.e., whether the criterion separates information carriers from other causal organisation. The development runs answer the first: the CSS of seed 1 holds 847,911 nodes, the organisms found blind match the withheld record, 706 of 706 in seed 1, and in 95.9\% of 4,314 memory-content swaps the time to first division follows the inserted memory content. The held-out runs on seeds 101--140 answer the second\new{: the swap-confirmed information family and its excess over the family-level null recur in each of the 39 runs with births, and 99.76\% of 29,321 decided memory-content swaps follow the inserted memory content. Under the frozen rule the held-out test nevertheless fails, on its check of the organism-level reading: in 2 runs, the reference definition gives one heritable genotype no phenotype, whereas the reading agrees with the record (Appendix~\ref{app:H.8})}. Neither\new{ set of runs} answers the third.

At the level of organisms, the perspective selects the candidates. We declared the memory content at birth as the candidate information carrier and the time to first division as its candidate product (Section~\ref{sec:comp.2}). Within that choice, the results are not given in advance: the swaps show by intervention that the memory content fixes the time to first division, and the criterion admits 13 of the 152 G-labels of seed 1. The agreement with the record is therefore a conditional validation with informed candidate selection, and an existence result in the sense of Appendix~\ref{app:B.11} (Section~\ref{sec:comp.4}). It does not show that a CSS can select these structures from a wider set of candidates, or reject competing structures that carry no information.

At the level of instructions, the search is blind to instruction names and designated roles, but no complete reference exists. The record allows single functions to be checked: 17,131 of the 17,133 steps with inferred R are copy steps. The design, however, fixes no information family at this level, so a detected family cannot be scored against a classification into members and non-members. The family-level null shows that the labels align with replication and lineage beyond chance, with 10--14 labels in the real CSS against at most 5 after reassignment, and the swap test rejects 8 of the 13 structural candidates of seed 1. The null does not test the full criterion, because V links 92--97\% of the candidate pairs and so removes almost no candidate.

Two limits come from the world. First, with 36 cells and 100 updates, the population is too small for lineages above the organism level: colonies form, but no colony lineage extends beyond one replication transition (Section~\ref{sec:comp.5}). Second, nothing is rewarded, so the downstream consequences that we characterised concern mainly execution, copying and division. Five limits come from our own choices and would remain in any other world: 1) the instance-path reading of V connects nearly all candidates; 2) the maintenance edges of $D$ enlarge the closure of a single step, to a median of 25--28 nodes and up to 67,856; 3) the single-step process class misses roles in copying, e.g., \texttt{h-copy} shows R 24 times in seed 1 against 906 times in order DP\new{, and is a family member in 17 of the 39 held-out runs in order ALL against 39 of 39 in order DP}; 4) scheme~B hides the translation of \texttt{mov-head}, and; 5) every family rests on the subset reading of operator correspondence (Section~\ref{sec:comp.5}).

More runs of the same world test reproducibility but cannot supply the reference cases that specificity needs. A discriminating experiment therefore needs, at the resolution searched, independently characterised positive and negative cases, withheld from the search and classified by properties of the dynamics, not by whether they meet RVT. The negative cases must be ones that the criterion could wrongly accept, i.e., organisations that meet parts of R, V and T without carrying hereditary information. Copies whose differences have no effect beyond copying fail T by construction, and so test the implementation rather than the criterion. Such cases could be built into the present simulator or into another digital or chemical world. Rewarded tasks, copying and interpreting processes that can be identified separately, longer histories and structural types derived from the dynamics, e.g., from almost-invariant regions, each address a further question, best taken one at a time once the criterion has been shown to separate known positive and negative cases. The first world thus tests whether the method can be executed and repeated, and its limits specify what a discriminating second experiment must contain.
\end{newtext}

\subsection{Open theoretical and computational problems}\label{sec:8.9}

Two kinds of question remain open, for different reasons. Some can be settled only empirically or by counterexample: whether RVT is sufficient, and what any extension would require. Resolving these by stipulation would make the criterion true by construction rather than testable, which is the failure the inverse formulation exists to avoid. Others are unsolved mathematical and algorithmic problems. These are not left open by choice, but specifying the representation is what makes them stateable.

\textbf{Sufficiency of RVT.} RVT is an initial candidate criterion. Adversarial systems may show that additional causal constraints are needed. Such a result would refine the hereditary criterion rather than invalidate the CSS representation.

\textbf{Mathematical theory of admissible coarse-graining.} The spatial and temporal refinement relation introduced here should be developed formally. Questions include the existence of maximal and minimal admissible representations; equivalence classes under neutral refinements; invariants under world-tube subdivision; and conditions under which aggregation necessarily destroys a hereditary motif. Appendix~\ref{app:B} states these as Open Problems~\ref{op:1} to~\ref{op:5}.

\textbf{AIR partition construction.} Practical algorithms must produce complete mutually exclusive structural partitions at the chosen resolution and quantify uncertainty in their boundaries. \new{The computational test of Section~\ref{sec:comp} used declared value classes as labels, not an AIR partition, and Section~\ref{sec:comp.5} shows that the labelling decides which translation relations are visible.}

\textbf{Efficient motif search.} The worst-case combinatorial complexity of process and motif search remains to be characterised (Appendix~\ref{app:D}). \new{Section~\ref{sec:comp} shows that a bounded, role-blind search is practical for one simulator run: the CSS of seed 1, with 847,911 nodes, was built in 48 s, and the search covered every single-step process in the order as defined, i.e., the convex closure of each step, up to 67,856 nodes, in 2,120--2,419 s per run for R and lineage. %
The search is not complete: it examined declared process classes only, so its detections hold within those classes and it says nothing about organisation outside them. The single-step class is one example: in the order as defined, single-step processes show R for the copy instruction \texttt{h-copy} only 24 times in seed 1, against 906 times in order DP, because the closure of a single step takes in the next iteration of the copy loop; gestation processes, which contain the whole loop, were searched only at the level of organisms (Section~\ref{sec:comp.3}). %
Scaling to larger worlds, longer runs and processes spanning many steps remains open.}

\new{\textbf{Robust motif identification in the computational test.} \new{The held-out runs on seeds 101--140 show that the information family recurs on unseen runs but that its membership varies: a family exists in 39 of 39 runs with births, \texttt{nop-A}, \texttt{nop-B} and \texttt{nop-C} are members in every run and \texttt{h-alloc} in 36, whereas \texttt{h-copy}, a member in no development run, is a member in 17 (Appendix~\ref{app:H.8}).}
\new{The reference pairs, i.e., copy errors in which the original and the mutant label are each inherited over at least 2 successive births and translated differently, are recovered at a threshold of 2 instances in 19 of the 39 held-out runs in order ALL and in 34 of 39 in order DP; the difference comes from the maintenance edges, which only order ALL contains (Section~\ref{sec:8.world}).} %
The exploratory family-level null of Section~\ref{sec:comp.4} shows that the labels align with replication and lineage more than by chance; the specificity of the full criterion remains open, and Section~\ref{sec:8.world} states what an experiment that tests it must contain. %
Real Avida and larger worlds are the next test environments.}

\textbf{Partial observability.} Identifiability limits should be quantified for cases where nodes, causal relations and structural distinctions are lost (Appendix~\ref{app:F}).

\textbf{Multiple-observation composition.} Formal consistency conditions are needed for combining independently reconstructed CSSs using observational bridge relations (Appendix~\ref{app:F}).

\textbf{Stronger hereditary criteria.} RVT may eventually require extension to account explicitly for maintenance, environmental feedback, selection or other causal structure. Any extension should preserve the separation between role-neutral CSS construction and post-CSS role inference.

\textbf{Roles other than hereditary ones.} The separation between role-neutral CSS construction and post-CSS motif inference is not specific to heredity: other relational roles, such as tool, scaffold or boundary, are candidates for the same treatment, and whether the inference strategy generalises beyond RVT is left to future work.

\section{Conclusion}\label{sec:9}

The identification of hereditary organisation is usually performed with an ontology already containing the relevant hereditary entities.

Genes, organisms, cultural traits, records and programs are enormously useful scientific categories. But if they are supplied before analysis, they cannot simultaneously constitute the result of an inverse search for previously unknown hereditary organisation.

The Causal Structure Set reverses this logic.

A bounded physical observation is first represented through persistent realised world-tubes. Spatial and temporal divisions provide constrained alternative representations. Structural labels are assigned independently of hereditary function through complete exclusive typing of the accepted persistent instances. AIR methods provide one possible implementation where the required partition can be established. Realised structure-creating dependencies define a strict partial causal order.

Only afterwards are hereditary roles sought.

Causally convex processes expose input/output organisation. Structural labels conserved across a process while contributing causally to outputs of other labels satisfy the operator condition. Replication establishes causal copying; variation establishes alternative configurations of the same structural family; translation establishes variant-dependent downstream realisation. The same \new{information }carrier family must participate coherently in these motifs, and the strength of observed hereditary continuation increases with the number of causally connected generations through which descendants re-enter replication organisation.

The resulting information-carrier, operator and realised-structure roles are therefore \textbf{relational properties of observed causal organisation rather than intrinsic labels attached to material objects}.

\new{The examples of Sections~\ref{sec:6} and~\ref{sec:cultural} show how the criterion operates and where its verdict is conditional. Synthetic configurations separate complete from incomplete organisation, and the paired recipe reconstructions show that the verdict follows the declared causal history rather than the labels; neither is empirical evidence.}

\new{Section~\ref{sec:comp} shows that the inversion can be carried out on executable dynamics. A CSS built by intervention replay from the complete record of an Avida-like simulator holds $8.5$--$8.8 \times 10^{5}$ nodes per run. %
The organisms found blind in the CSS match the withheld simulator record, and in memory-content swaps the time to first division follows the inserted memory content in 95.9\% of 4,314 cases. %
At the level of instructions, a role-blind search finds an information family in each of the four development runs with births, with the same 4 ancestral instructions at its core; the copy instruction itself is a member of none, because in the order as defined a single executed instruction is not a causally closed process. %
\new{The frozen analysis finds a family again in all 39 held-out runs with births, with \texttt{h-copy} a member in 17 of them; the preregistered held-out test nevertheless fails in 2 runs, because of a choice we made in defining the designated family, not because of an error in the inference.} %
The families are formal RVT detections on development \new{and held-out }runs, under declared choices of process class, labelling and the reading of variation, not verified heredity.}

CSS makes no claim about unobserved capacities. It does not yet identify selection, cumulative evolutionary accumulation or stability. Instead, it provides a representation in which those subsequent questions can be asked without having first supplied the hereditary ontology they concern. \new{Whether the criterion identifies hereditary organisation in general, whether it is specific, i.e., rejects copying without consequence and runs without heritable variation, and whether it applies across substrates also remain open.}

If this inverse programme succeeds, hereditary organisation can be treated as a discoverable class of realised physical causal organisation across substrates, scales and dynamics.

\section*{\new{Code and data availability}}
\addcontentsline{toc}{section}{Code and data availability}
\begin{newtext}
The Avida-like simulator, the CSS builder, the search passes A--D, the organism-level analysis\new{, the preregistration with its freeze archive, the held-out results} and the scripts that produce every number and figure of Section~\ref{sec:comp} are archived at Zenodo, \new{\href{https://doi.org/10.5281/zenodo.23798795}{doi:10.5281/zenodo.23798795}}.
The scripts use Python with the standard library only.
The runs are deterministic, so the code regenerates every log and every CSS from its seed; Appendix~\ref{app:H.9} gives the commands for seed 1.
The script that checks the recipe reconstructions of Section~\ref{sec:cultural} and Appendix~\ref{app:G}, \texttt{recipe\_css.py}, is in the same archive.
\end{newtext}

\clearpage
\begin{appendices}
\section{Admissible observational perspectives}\label{app:A}

\delmark\new{This appendix specifies the admissibility requirements for the observational perspective $P = (\pi_P, D_P, J_P, \rho_P, \Sigma_P)$ introduced in Section~\ref{sec:3.1}. Figure~\ref{fig:pipeline} summarises how that perspective enters the six-stage construction and inference procedure, and Sections~\ref{sec:3} to~\ref{sec:5} define the resulting mathematical operations. A representation is not admissible merely because it produces a well-formed labelled partial order: the observational variables, represented structures, structural types and causal relations must also satisfy the following physical and epistemic conditions.}

\begin{definition}[Admissible perspective]\label{def:1}
A perspective $P$ is \textbf{admissible for the observed episode} $\mathcal{R}$ if the following conditions hold.

\textbf{Physical grounding.} Every represented variable and structure instance is grounded in physical or recorded degrees of freedom belonging to $\mathcal{R}$. The perspective may coarse-grain these degrees of freedom, but it may not introduce unobserved physical entities or causal events.

\textbf{Declared resolution.} The spatial, temporal and structural distinctions resolvable under $P$ are specified independently of the later hereditary-role inference. Unresolved distinctions are treated as unresolved rather than reconstructed from domain knowledge.

\textbf{Persistent structural individuation.} The structural rule $\Sigma_P$ generates persistent world-tubes, or finite sections thereof, satisfying the declared persistence criterion at that resolution.

\textbf{Completeness and non-overlap.} The resulting structure set is complete relative to the declared structural rule: every qualifying realised structure is represented. Incompatible structure instances may not assign the same physical degrees of freedom simultaneously to more than one CSS node. Completeness is relative to the declared individuation rule, which specifies which wholes or constituents are represented; it does not require simultaneous representation of every persistent structure at every organisational level.

\textbf{Exclusive structural typing.} Every accepted structure instance receives exactly one hereditary-role-neutral structural type. Where AIRs supply the structural types, they must form a qualifying mutually exclusive partition covering the admitted structural-state domain $Y_P^{\mathrm{acc}}$, as specified in Section~\ref{sec:3.3}; coverage of every dynamical state is not required. Exclusive structural typing concerns membership of the represented instances in structural types, not their material composition. For example, ``woman'' and ``human being,'' interpreted as their ordinary membership structural types, overlap and cannot serve as mutually exclusive types. By contrast, ``human'' and ``water parcel'' can coexist as types: a human body contains water but is not itself a water parcel. A CSS describing a person swimming may therefore represent the person and surrounding water parcels as separate nodes. Water within the person is included in the person node and is not simultaneously represented by additional constituent nodes, as required by the separate physical non-overlap condition.

\textbf{Causal consistency.} The causal relations available under $P$ are grounded in $D_P$ and $J_P$. Interventions on a represented source may alter the physical degrees of freedom constituting that source but may not directly impose the later target state whose causal dependence is being evaluated. Causal order must be compatible with temporal precedence and with observed physical identity.

\textbf{Hereditary-role blindness.} The construction of $P$, its structural rules, labels and causal criteria may not depend on the hereditary-role assignment subsequently inferred from the CSS. In particular, structures may not be distinguished because they are known to be \new{information }carriers, variants, replication machinery, translated products, members of a lineage, or because such a distinction causes an RVT motif to appear.
\end{definition}

The last requirement concerns the information used to construct the representation, not the vocabulary in which it is expressed. A hereditary distinction does not become neutral merely by translating it into physical language. Suppose two structural types were defined by which downstream structures their members later give rise to. The definition would be stated in physical terms, but the distinction it draws is the T relation itself. Such a labelling is inadmissible.

Where AIRs supply the structural types, this particular construction is excluded twice over. An AIR is defined by intrinsic properties of a region, fast internal relaxation and slow escape, which make no reference to what its occupants later produce; structural types defined by downstream consequence are therefore generally not almost-invariant, and splitting an existing AIR by the outputs its occupants generate yields subregions that the criterion rejects (Section~\ref{sec:3.3}). Blindness must be established separately for the alternative classification procedures that Section~\ref{sec:3.3} admits, and for the construction of $Y_P$ itself, since a coordinate recording a configuration's later products would fail physical grounding rather than blindness.

These conditions do not make $P$ observer-independent. They instead constrain the observer's freedom so that hereditary organisation is not supplied as part of the representation it is subsequently meant to infer.

\subsection{A family rather than a privileged perspective}\label{app:A.1}

An observed physical episode may satisfy these conditions under more than one perspective. A cell may be represented as one persistent world-tube at one level and as a collection of molecular world-tubes at another. A persisting structure may likewise be represented by one long temporal section or by several successive sections when its interactions require finer causal ordering.

We therefore associate the episode not with one privileged CSS, $C_P$, but with the family
\[
  C_{\mathrm{adm}}(\mathcal{R}) = \{C_P : P \text{ is admissible for } \mathcal{R}\}.
\]

Perspective dependence is consequently moved from an implicit modelling choice into an explicit object of the theory. The physical episode is fixed. Its admissible representations need not be unique.

\subsection{Perspective dependence is not a nuisance parameter}\label{app:A.2}

Perspective dependence is not treated as a nuisance parameter to be chosen away. A perspective determines what part of the realised physical organisation is resolved, while admissibility constrains which descriptions count as legitimate representations of the same episode. Hereditary-role inference is invariant under coarsenings that preserve the motif-defining distinctions and causal boundary organisation, and can be lost when those distinctions are erased. The deeper claim that hereditary existence itself is representation-independent remains a precise mathematical research problem: it requires a characterisation of the admissible representation family and proof that the existential CSS criterion cannot be satisfied by representational artefact alone.

\section{Coarse-graining, refinement and the physical-witness property}\label{app:B}

A physical episode does not in general determine one unique scientifically useful representation. The same physical degrees of freedom may be analysed at molecular, cellular or organismal scales; the same persistent object may be represented over one long temporal interval or through successive causally ordered sections.

This appendix attempts to formalise the relation between such representations: when two of them describe the same episode at different resolutions, which transformations relate them, and what survives those transformations. We call $\mathcal{C}(\mathcal{R}, P)$ the family of CSSs admissible for $\mathcal{R}$ under perspective $P$, generated from the declared primitive structures by permissible spatial and temporal transformations.

The treatment is preliminary. The definitions below are proposals rather than settled formulations, and the results that follow from them are correspondingly provisional: they establish what holds \emph{given} these definitions, not that the definitions are the right ones. Whether the admissible family is well characterised, whether the refinement relation has the structure assumed, and whether the preservation conditions are necessary as well as sufficient are all open, and are stated as such in Appendix~\del{B.13.}\new{\ref{app:B.12}.}
What the appendix does establish is that coarse-graining is not unrestricted redescription, and that the negative results, on what representational change destroys, follow directly from the definitions whatever their eventual form.

\subsection{Temporal division}\label{app:B.1}

Consider a persistent world-tube $S \subset \mathcal{R}$. An admissible temporal refinement divides it into consecutive non-overlapping interiors:
\[
S = S_1 \cup S_2 \cup \cdots \cup S_k,
\]
ordered in time. This operation is required whenever treating the whole persistent structure as one CSS node would obscure causal ordering among its successive effects.

Suppose, for example, structure A affects B, B later affects the persisting A, and A subsequently affects C. At the level of physical objects this appears to involve feedback. In the CSS representation we instead use successive sections
\[
A_1 \prec B \prec A_2 \prec C.
\]

No graph cycle is required. Persistence can therefore be unfolded into explicit time-indexed worldline sections without altering the relevant causal conclusions, and the strict partial-order formulation adopts precisely this representation for feedback and reciprocal interaction.

Temporal aggregation is the inverse operation where causally relevant ordering is not lost.

\subsection{Spatial division and aggregation}\label{app:B.2}

A persistent macrostructure may itself consist of several spatially distinguishable persistent constituents. A spatial refinement replaces a world-tube $S$ by constituent world-tubes{\delmark} ${\delmark}S_1, S_2, \cdots, S_k$ such that, over the represented interval,
\[
S = \bigcup_{j=1}^{k} S_j
\]
at the chosen physical resolution, with mutually disjoint represented degrees of freedom except at permitted boundaries.

Conversely, an aggregate of primitive structures may be represented as one macrostructure if:
\begin{itemize}
  \item the aggregate is itself persistent under the declared perspective;
  \item its constituent organisation defines an admissible structural state;
  \item the aggregation does not contradict the recorded identity or causal ordering required at that level.
\end{itemize}

Thus a cell might be one structure at one CSS resolution and a collection of molecular structures at a finer resolution.

It cannot simultaneously be both at the same spacetime location within the same CSS if doing so assigns the same physical degrees of freedom to overlapping nodes.

Completeness and non-overlap, as required by Definition~\ref{def:1}, prevent selective construction of a CSS around a desired hereditary motif. A representation cannot omit qualifying persistent instances, or add an extra structural distinction, because doing so helps produce R, V or T.

\subsection{Formal refinement and coarsening maps}\label{app:B.3}

Let $C_1 = (W_1, \prec_1, \ell_1)$, $C_2 = (W_2, \prec_2, \ell_2)$ and $C_1,C_2 \in \mathcal{C}(\mathcal{R}, P)$.

We write
\[
C_1 \preceq_{\mathrm{ref}} C_2
\]
when $C_1$ is a refinement of $C_2$. Thus $C_2$ is obtained from $C_1$ by admissible loss of spatial, temporal, or structural resolution.

For $\mathcal{L}_k = \ell_k(W_k)$, a refinement/coarsening relation requires maps
\[
g \colon W_1 \to W_2
\]
and
\[
q \colon {\new{\mathcal{L}}}_1 \to {\new{\mathcal{L}}}_2
\]
from fine-grained structure instances and labels to their coarse-grained counterparts.

The definition below covers coarsenings for which the coarse structural type is determined by the fine structural label through $q$. It does not cover every spatial aggregation: assigning a type to an aggregate may require its constituent organisation rather than the label of an individual constituent.

Structural coarsening requires distinguishing two operations. One may merge labels in an existing CSS while retaining its causal order, thereby preserving ancestry established under the finer classification. Alternatively, one may reconstruct the causal dependencies using the coarser structural types as the target outcomes of the causal tests. These operations need not yield the same order, even when the realised node set is unchanged, because a retained causal relation may have been established through a structural distinction that the coarser classification no longer resolves.

These operations must be distinguished when assessing admissibility. A graph obtained by retaining the finer order after merging labels is not automatically a CSS reconstructed under the coarser classification. To qualify as such, its order must satisfy the causal construction specified for that classification. Alternatively, retained relations may be carried as explicitly identified evidence from the finer representation, in which case the distinction between the displayed classification and the classification supporting those relations must remain explicit.

\begin{definition}[Admissible coarsening]\label{def:2}
The definition below states the conditions on a coarsening under either construction. Which construction is performed must be declared, since the conditions are checked against different objects: on the re-derived order the causal clause constrains dependencies established under the coarse classification, while on the retained order it constrains dependencies inherited from the fine one.

The pair $(g, q)$ is an \textbf{admissible coarsening} when:
\begin{itemize}
  \item \textbf{Spacetime coverage is preserved.} The physical world-tube represented by each coarse node $w \in W_2$ is the admissible union of the fine world-tube sections in its fibre $g^{-1}(w) \subseteq W_1$.
  \item \textbf{No physical history is added.} Coarse nodes may represent persistent aggregates of fine instances, but every such aggregate must be physically grounded in the same episode. Coarsening may not invent physical occurrences or causal events.
  \item \textbf{Non-overlap is preserved.} Distinct coarse nodes do not acquire incompatible claims on the same physical degrees of freedom.
  \item \textbf{Structural typing commutes with coarsening:} $\ell_2 \circ g = q \circ \ell_1$.
  \item \textbf{Causal order is not reversed.} If $x \prec_1 y$ and $g(x) \neq g(y)$, the coarsened representation may retain or lose the distinction required to expose this causal path, but it may not imply $g(y) \prec_2 g(x)$.
  \item \textbf{The dependency relation is not enlarged.} If the coarse representation carries an intervention-supported dependency between coarse nodes, it must correspond to a dependency holding between their fine constituents. Preserving $\prec$ does not by itself secure this, since $\prec$ is the transitive closure of $D$.
  \item \textbf{Identity and ancestry are respected.} Temporal aggregation may join successive sections of the same persisting physical structure, and spatial aggregation may join physical constituents into a persistent macrostructure, but neither operation may create an ancestry relation absent from $C_1$.
\end{itemize}
\end{definition}

These requirements define which representations describe the same realised physical episode at different levels of resolution. They do \textbf{not} imply that every property inferred at one level must remain detectable at every other level.

Indeed, arbitrary coarsening can erase hereditary organisation.

Admissible representations need not be comparable. A perspective may individuate persons by the skills they can perform or by some other persistent property, yielding partitions that cross-cut: some instances share a class under one and differ under the other. Neither is a coarsening of the other, so no admissible $(g, q)$ relates them in either direction, and a hereditary motif visible in one may be invisible in the other without either being defective. The two do share a common coarsening, namely the partition that merges all such classes, but that is where the motif is least likely to survive, since merging \new{information }carrier variants destroys detectable variation (Corollary~\ref{cor:1}). This is why the criterion quantifies existentially over the admissible family rather than requiring invariance across it: there is no privileged member and no route along which every motif can be tracked.

These spatial and temporal operations naturally induce a refinement relation among admissible representations.

These conditions specify a constrained refinement relation. Establishing its partial-order structure, including the identification of equivalent representations, requires further analysis.

A complete mathematical theory of this refinement family, including canonical extrema, equivalence classes of representations and invariants, is beyond the immediate requirements of the present paper. The point needed here is narrower: \textbf{coarse-graining is not unrestricted redescription}. It arises from constrained physical division and aggregation of observed world-tubes.

\subsection{Role invariance under coarsening}\label{app:B.4}

The appropriate invariance question can only be posed after hereditary motifs have been defined.

Suppose CSS $C$ contains an information family $F$ with candidate \new{information }carrier labels
\[
I = \{i_1, \ldots, i_m\},
\]
together with its associated R, V and T processes, operator organisation and observed lineage relations.

Let
\[
h = (g, q) \colon C \to C'
\]
be an admissible coarsening in the sense of Definition~\ref{def:2}.

\begin{definition}[Motif-faithful coarsening]\label{def:3}
The coarsening $h$ is \textbf{faithful to information family} $F$ when it preserves every distinction and boundary relation required to witness $F$. For every witnessing process $Q$, its image $Q' = g(Q)$ must be a causally convex process in $C'$; all conditions below are evaluated on its actual input and output boundaries. Specifically:
\begin{enumerate}[label=\arabic*.]
  \item distinct \new{information }carrier variants required by V remain distinguishable: $i_j \neq i_k \Longrightarrow q(i_j) \neq q(i_k)$ for every variant pair used by the motif;
  \item translated outcome labels remain distinguishable from one another, and each remains absent from the input boundary of its corresponding translation process;
  \item the R multiplicity increases remain satisfied, the additional \new{information }carrier occurrences remain physically distinct, and each designated T product remains present on its corresponding output boundary;
  \item the selected non-empty operator organisations remain qualifying and corresponding under Appendix~\ref{app:C.1};
  \item the causal requirements of the R, V and T witnesses remain satisfied in the target CSS, evaluated on $\prec$ as specified in Section~\ref{sec:3.6}, including the comparative causal assessment establishing variant-dependent output differentiation\del{.}\new{;}
  \item \new{witnessed lineages remain lineages of the same depth: for every replication transition $x_g \rightsquigarrow x_{g+1}$ of a witnessed lineage, the image $h(x_{g+1})$ is an output of $g(R_g)$ and an input of $g(R_{g+1})$, so that $h(x_g) \rightsquigarrow h(x_{g+1})$, and distinct transitions map to distinct transitions.} %
\end{enumerate}
\end{definition}

\begin{proposition}[Sufficient condition for hereditary-role invariance]\label{prop:1}
Let $F$ be a coherent information family witnessed in CSS $C$, with \new{information }carrier labels $I$. If
\[
h \colon C \to C'
\]
is an admissible coarsening faithful to $F$, then the mapped witnesses satisfy the joint criterion for $q(I)$ in $C'$, preserving their hereditary-role assignments under $q$.

A preserved lineage witness of depth $d$ establishes
\[
L\big(C', q(I)\big) \geq d.
\]
\end{proposition}

Equality of the maximal depths in the two representations requires an additional argument excluding longer qualifying lineages in $C'$. Preservation of the required causal relations is an assumption of the proposition, not a consequence of label merging.

\begin{proof}[Proof sketch]
R depends only on the relevant \new{information }carrier boundary multiplicities and causal contribution of \new{information }carrier inputs to \new{information }carrier outputs. These are preserved by conditions 3 and 5.

V depends on the causal transformation of one distinguishable \new{information }carrier label into another. Conditions 1 and 5 preserve both the distinction and its causal relation.

T depends on distinguishable input variants, corresponding operator organisation and distinguishable variant-dependent outputs. These are preserved by conditions 1\new{--}5.

Lineage continuation depends on causal descent followed by re-entry into corresponding replication organisation; \del{condition}\new{conditions} 4 and 6 \del{preserves}\new{preserve} that sequence.

The mapped processes therefore satisfy the joint criterion, and each preserved lineage retains its witnessed number of replication transitions.
\end{proof}

This proposition gives a \textbf{sufficient}, not necessary, characterisation of role-preserving coarsening.

\subsection{Information loss under non-faithful coarsening and refinement}\label{app:B.5}

Representational change in either direction can destroy the information required to witness a hereditary motif, and the two directions threaten different parts of the criterion. A coarsening in the sense of Definition~\ref{def:2} comprises a node map $g$ and a label map $q$, and may aggregate spatio-temporal instances, coarsen the structural classification, or both. Corollaries~\ref{cor:1} and~\ref{cor:2} below concern the label map, Corollary~\ref{cor:4} the node map, and Corollary~\ref{cor:3} either. Corollaries~\ref{cor:5} to~\ref{cor:7} concern refinement and are stated after Definition~\ref{def:4}. The statements below concern designated witnesses and assume admissible source and target CSSs. A graph obtained by retaining the order after label merging is not automatically admissible under the coarser classification (Appendix~\ref{app:B.3}). Loss of a designated witness does not establish that no alternative witness exists.

\begin{corollary}[Variant collapse destroys detectable V]\label{cor:1}
If a coarsening identifies two \new{information }carrier variants required by a V relation, $q(i_1) = q(i_2)$, then that V relation is not detectable in the coarsened CSS.
\end{corollary}

\begin{corollary}[Outcome collapse destroys detectable T]\label{cor:2}
If $q(i_1) = q(i_2)$ for the distinct downstream structures through which two variants satisfied T, then that T relation is not detectable at the coarsened resolution. The designated T witness is also lost if a translated product label becomes present on its corresponding input boundary.
\end{corollary}

\begin{corollary}[Replication can be lost under temporal or structural aggregation]\label{cor:3}
If coarsening collapses the process boundaries required to observe $n^{+}_{Q}(i) > n^{-}_{Q}(i)$, then the corresponding R relation may no longer be identifiable.
\end{corollary}

\begin{corollary}[Temporal coarsening can shorten a witnessed lineage]\label{cor:4}
If temporal aggregation removes boundaries required to distinguish descendant re-entry events, the corresponding lineage witness is no longer represented at its original depth. This does not by itself establish an ordering between the maximal lineage depths inferred independently in the two CSSs.
\end{corollary}

Thus hereditary-role inference is \textbf{not invariant under arbitrary coarsening}. This is not paradoxical: coarse-graining can destroy the information required to witness the physical organisation.

The corollaries above concern a family inferred in the finer representation and then coarsened. The converse case, a family inferred in a given representation and then refined, is not covered by Definition~\ref{def:3}, since there the family belongs to the source of the coarsening rather than its target. It requires its own condition.

\begin{definition}[Preservation of label coincidences]\label{def:4}
For the following results, consider admissible CSSs
\[
C = (W, \prec, \ell),\; C' = (W, \prec, \ell')
\]
related by a label coarsening map $q$, with
\[
\ell = q \circ \ell'.
\]
Let $W_F$ contain the instances used by the designated witnesses of $F$. Their label coincidences are preserved when
\[
\ell(x) = \ell(y) \;\Longrightarrow\; \ell'(x) = \ell'(y),\; x, y \in W_F.
\]
Distinct coarse labels cannot merge under this refinement. The condition concerns label coincidences only; preservation of the complete hereditary witness additionally requires its causal conditions to remain valid at the refined classification.
\end{definition}

\begin{corollary}[Separating input and output \new{information }carrier classes destroys an R witness]\label{cor:5}
If the fine labels assigned to the input occurrences of a replicated coarse type are disjoint from those assigned to its output occurrences, no fine label supplies the required positive input count and output increase in that witness. Partial splitting need not destroy R.
\end{corollary}

\begin{corollary}[Separating input and output operator classes destroys an operator witness]\label{cor:6}
If the fine labels assigned to the input occurrences of a selected coarse operator type are disjoint from those assigned to its output occurrences, none satisfies
\[
n^{-}(o) = n^{+}(o) > 0.
\]
Splitting a coarse operator class into subtypes conserved across the boundary need not destroy the operator condition.
\end{corollary}

\begin{corollary}[Label refinement can invalidate a replication-lineage witness]\label{cor:7}
Relabelling on a fixed node set and order preserves the instance-level re-entry relation
\[
x \in O(Q) \cap I(Q').
\]
It can nevertheless invalidate the lineage witness by destroying a required R condition or the correspondence of its operator organisation.
\end{corollary}

These results are complementary to Corollaries~\ref{cor:1} to~\ref{cor:4}. The motif conditions fall into those requiring structural classes to coincide --- the replication multiplicity, operator conservation and the R and operator conditions supporting lineage continuation --- and those requiring them to differ, namely variation and the distinguishability of translated outcomes. Coarsening principally threatens the second group and refinement the first, but the division is a tendency rather than a partition: Corollary~\ref{cor:3} shows that replication, a condition of the first kind, can also be lost under coarsening when the process boundaries at which multiplicities are evaluated are aggregated. What follows in general is that neither direction of representational change is safe, and that no maximally refined representation is guaranteed to expose a hereditary organisation that a coarser one resolves.

\subsection{Why existential search over admissible representations does not manufacture heredity}\label{app:B.6}

Defining hereditary organisation through the existence of an admissible representation,
\[
H(\mathcal{R}) = 1 \;\Longleftrightarrow\; \exists\, C \in \mathcal{C}(\mathcal{R}, P):\; H_{\mathrm{RVT}}(C) = 1,
\]
raises an immediate concern. If an episode admits many spatial resolutions, temporal segmentations and structural partitions, could a sufficiently large representation family make hereditary organisation almost inevitable? In other words, could one search enough representations until some representation happens to resemble replication, variation and translation?

This objection would be decisive if an admissible CSS were an arbitrary graph constructed from the data. It is not.

The distinction is between \textbf{representational multiplicity} and \textbf{physical multiplicity}. Alternative CSSs may expose different aspects of the same physical history, but every element used to establish a hereditary motif must have a corresponding physical witness in $\mathcal{R}$.

Let
\[
C = (W, \prec, \ell) \in \mathcal{C}(\mathcal{R}, P)
\]
be an admissible AIR-derived CSS. Associated with $C$ is a physical realisation map
\[
\pi_C \colon W \to \mathcal{R}
\]
taking every CSS node to the finite realised world-tube section in $\mathcal{R}$ that it represents. $\pi_C$ is the inverse of the mapping $M_P$ of Section~\ref{sec:3.4}, composed with the inclusion of accepted sections into $\mathcal{R}$; it is well defined because $M_P$ is bijective.

Admissibility constrains this map in several ways. Nodes may represent only realised physical structures in $\mathcal{R}$; AIR labels must arise from the declared dynamics rather than from the hereditary role subsequently inferred; causal relations must correspond to realised structure-creating causal dependencies realised in $\mathcal{R}$; and spatial or temporal aggregation may remove distinctions but may not introduce physical events or causal ancestry that were absent from the history.

Consequently, an RVT motif in an admissible CSS is not merely a pattern in an abstract graph. It is shorthand for a collection of realised physical relations in $\mathcal{R}$.

\subsection{Physical witnesses of hereditary motifs}\label{app:B.7}

For a candidate \new{information }carrier family
\[
I = {\new{\{}}i_1{\new{,}} \ldots{\new{,}} i_m{\new{\}}},
\]
an admissible CSS can establish replication only if the physical history contains distinct realised \new{information }carrier occurrences such that an input occurrence causally contributes to additional output occurrences of the corresponding structural type. Temporal subdivision of a single persisting occurrence does not satisfy this requirement.

Likewise, variation requires physically distinguishable persistent structural types,
\[
i_j \neq i_k,
\]
identified independently through the structural partition and connected by realised causal history.

Translation requires more than a choice of labels. Alternative \new{information }carrier states must participate through corresponding physical operator organisation and causally contribute to physically distinguishable downstream realised structures,
\[
i_j \longrightarrow y_j,\; i_k \longrightarrow y_k,\; y_j \neq y_k.
\]

Each designated translated product has a structural type absent from the corresponding process's input boundary. Absence concerns structural type rather than physical instance identity: an additional copy of an input type does not satisfy this requirement.

Finally, lineage continuation requires actual causal descent: output \new{information }carrier occurrences must subsequently enter further replication processes. Repeated observation of structurally similar objects without causal parent--descendant connection does not establish a lineage.

Thus each component of the inferred motif has a corresponding witness in the underlying history:

R $\longleftrightarrow$ realised causal production of additional \new{information }carrier occurrences, V $\longleftrightarrow$ realised persistent structural differentiation, T $\longleftrightarrow$ realised variant-dependent downstream differentiation, L $\longleftrightarrow$ realised causal descent across successive generations.

An admissible representation determines \textbf{how these relations are resolved}, not whether the underlying physical events occurred.

\begin{proposition}[Physical-witness property]\label{prop:2}
Let $C \in \mathcal{C}(\mathcal{R}, P)$ be an admissible CSS satisfying the following conditions:
\begin{itemize}
  \item every node represents a realised physical world-tube occurrence or section thereof;
  \item structural types are obtained through a hereditary-role-independent, complete and mutually exclusive structural partition;
  \item replication multiplicities count distinct physical occurrences rather than temporal subdivisions of one persisting occurrence;
  \item every causal relation used in R, V, T or lineage inference is supported by a realised structure-creating causal dependence;
  \item process boundaries and representation rules are fixed independently of whether the resulting process satisfies a hereditary motif; and
  \item refinement and aggregation do not introduce physical occurrences or causal relations absent from ${\new{\mathcal{R}}}$.
\end{itemize}
Then
\[
H_{\mathrm{RVT}}(C) = 1
\]
implies that $\mathcal{R}$ contains a physical witness consisting of the corresponding replication, variation, variant-dependent realisation, and genealogical relations.
\end{proposition}
\del{The physical-witness property is relative to a declared admissible structural classification. Every relation required by the inferred motif must have the physical support specified by its predicate: an intervention-supported dependency where required, or a path of such dependencies where ancestry suffices. Changing the classification may change which motifs are detectable, but labels alone do not supply the required physical support. The proposition does not establish invariance under arbitrary repartitioning, nor the sufficiency of RVT as a criterion for heredity.}
The physical-witness property is relative to a declared admissible structural classification. Every relation required by the inferred motif must have the physical support specified by its predicate: an intervention-supported dependency where required, or a path of such dependencies where ancestry suffices.

Two limits follow. Changing the classification may change which motifs are detectable, so the proposition establishes no invariance across admissible partitions; Appendix~\ref{app:B.5} gives the corresponding negative results in both directions. But at any given admissible classification, no motif can be produced by the choice of labels or boundaries alone, since each of its constituent relations must be independently witnessed in $\mathcal{R}$. What the proposition secures is therefore physical grounding within a representation, not independence from it.

The proposition also establishes nothing about whether RVT is the correct criterion for heredity. That question is empirical and is treated in Section~\ref{sec:7.1}.

\subsection{More representations do not substitute for missing physical organisation}\label{app:B.8}

The importance of this distinction can be illustrated by a limiting example. Suppose an arbitrarily detailed measurement campaign records the physical evolution of a lifeless region of the Moon for $10^{6}$ hours. Increasing the duration and resolution of the observation may identify enormous numbers of persistent structures and causal processes: rocks fracture, dust moves, temperature cycles, impacts create craters, chemical structures form and decay, and many metastable configurations appear.

The resulting AIR-CSS family may therefore be extremely large.

That fact alone does not generate heredity.

If the physical history contains no structures that causally generate additional occurrences of variant structural states, no admissible decomposition can supply R. If structurally different descendants are absent, no decomposition can supply V. If alternative replicated states do not generate distinguishable downstream structures through corresponding operator organisation, no decomposition can supply T. If descendants do not themselves re-enter corresponding replication processes, no increase in observation duration can supply the missing causal genealogy.

A million descriptions of a history lacking these relations do not create them.

Conversely, if an admissible representation of that lunar history did reveal a sustained structure of
\[
\text{R+V+T+causal lineage,}
\]
the appropriate conclusion would not be that a convenient representation had manufactured heredity. It would be that the observation contained previously unrecognised physical hereditary organisation. Whether one additionally chose to call that organisation \emph{life} would require whatever further criteria one's definition of life imposes, but the hereditary causal organisation itself would be a physical discovery.

The same logic applies to less dramatic cases. Searching multiple admissible resolutions is not equivalent to searching arbitrary graph partitions until a desired pattern occurs. Each positive inference carries a physical witness.

\subsection{What AIR contributes}\label{app:B.9}

The independence requirement of Definition~\ref{def:1} excludes structural classifications chosen to reproduce the hereditary relations subsequently sought. For example, one could define $i_1$ as structures that later produce $y_1$, $i_2$ as structures that later produce $y_2$, and subsequently ``discover''
\[
i_1 \to y_1,\; i_2 \to y_2.
\]
Such a construction would build T into the structural labels.

Where a qualifying AIR partition is available, structural distinctions can instead be established from the dynamics independently of hereditary-role inference. At the declared resolution $\rho_P$, the partition of Section~\ref{sec:3.3} is
\[
\mathcal{A}(P) = \{A_c : c \in {\new{\mathcal{L}}}_P\},
\]
with
\[
\bigcup_{c \in {\new{\mathcal{L}}}_P} A_c = Y^{\mathrm{acc}}_P,\; A_c \cap A_{c'} = \emptyset \text{ for } c \neq c'.
\]

Each partition cell must satisfy the declared persistence requirements.

AIR methods can supply structural distinctions without hereditary labels, but this independence must also hold for the choice of variables, resolution, algorithm and parameters. It does not follow from automation alone.

Hereditary organisation is subsequently assessed through the causal and genealogical criterion.

\subsection{Representation as exposure rather than construction}\label{app:B.10}

The resulting interpretation of the existential criterion is therefore:
\[
\exists P : \exists C \in \mathcal{C}(\mathcal{R}, P) : H_{\mathrm{RVT}}(C) = 1
\]
does not mean

\emph{there exists some clever way of describing the history that makes it look hereditary}.

It means

\emph{there exists a physically admissible resolution of the episode at which a set of realised physical structures and causal relations satisfying the hereditary criterion can be resolved}.

Different representations can hide or expose the same physical organisation in the same way that changing observational scale can make a physical pattern easier or harder to identify. Coarse-graining may merge variants, translated outcomes or successive generations and thereby destroy identifiability. Fine-graining may reveal the physical relations on which the higher-level organisation depends. Neither operation is permitted to create relations not instantiated by the physical history.

This is why a large family of admissible representations is not itself problematic. Indeed, if hereditary organisation is emergent, multiple physically meaningful scales should be expected. The burden on the theory is not to select one privileged scale, but to ensure that every positive inference remains tethered to a physical causal witness.

The residual scientific question is therefore not whether representational freedom can manufacture heredity, but whether the proposed physical witness pattern,
\[
\text{coherent RVT organisation + causal lineage,}
\]
is itself a sufficiently general and discriminating criterion for hereditary organisation.

That question is deliberately left open to theoretical counterexample and empirical test.

\subsection{An existential rather than universal criterion}\label{app:B.11}

Because several admissible CSS representations may correspond to the same physical episode, an important question is how hereditary inference depends on representation.

It would be inappropriate to demand that an RVT organisation appear under every admissible coarsening. Corollaries~\ref{cor:1}, \ref{cor:2} and~\ref{cor:4} show why: variant collapse erases V, outcome collapse erases T, and temporal aggregation of successive replication processes reduces observed lineage depth. Coarse descriptions can remove precisely the distinctions required to witness a hereditary motif, so universal invariance is too strong a requirement.

The criterion is therefore existential: the physically relevant question is whether some admissible, sufficiently informative representation supports the organisation.

This is not optimisation over arbitrary descriptions. Admissibility is constrained independently of any motif, by primitive observations, spatial and temporal resolution, persistence, complete mutually exclusive hereditary-role-neutral structural labelling, non-overlap, completeness, physical identity and realised causal order. Selection among admissible representations is accordingly unrestricted, since admissibility is established for each member independently of which is examined. A motif cannot be manufactured by inventing labels or omitting structures.

\subsection{What is not yet proved}\label{app:B.12}

\begin{openproblem}[Characterisation of the admissible perspective family]\label{op:1}
The conditions above constrain admissible perspectives but do not yet constitute a complete mathematical characterisation. A complete theory would need to establish when two observational descriptions are admissible representations of the same realised physical history and when a spatial, temporal or structural transformation between them constitutes a valid refinement rather than a change of physical ontology. In particular, it remains to be determined whether a canonical maximal refinement or canonical equivalence class exists for general dynamical systems.
\end{openproblem}

\begin{openproblem}[Characterisation of all role-preserving transformations]\label{op:2}
Proposition~\ref{prop:1} gives sufficient conditions for hereditary-role invariance under coarsening, and Definition~\ref{def:4} the corresponding conditions for refinement, but neither is shown to be necessary. The stronger problem is to characterise the complete invariance class
\[
\mathrm{Inv}(F) = \{\, (h, C') \mid C' \in \mathcal{C}(\mathcal{R}, P),\ h \colon C \to C' \text{ admissible, and } F \text{ survives } h \,\}.
\]
This requires first a general notion of survival covering all admissible transformations, of which the coarsening and refinement conditions above are two special cases. Such a result would establish which spatial, temporal and structural distinctions are irrelevant to a particular hereditary organisation, and which constitute its minimal observational requirements.
\end{openproblem}

\begin{openproblem}[Minimal sufficient representations]\label{op:3}
For an inferred hereditary organisation $F$, define a CSS $C_{\mathrm{min}}$ to be \textbf{minimally sufficient} if $F$ remains inferable in $C_{\mathrm{min}}$, but is no longer inferable under any strict admissible coarsening of $C_{\mathrm{min}}$. It remains open whether \del{minimal}\new{minimally} sufficient CSS representations always exist, whether they are unique up to motif-preserving equivalence, and how they can be constructed efficiently. This question is important because it asks precisely how much physical detail is necessary to expose a hereditary organisation.
\end{openproblem}

\begin{openproblem}[Monotonicity under refinement]\label{op:4}
It is tempting to assume that if a hereditary motif is visible in a representation $C$, it must remain visible under every finer admissible representation. That has not been established. Fine-graining may split a structural label into several dynamically distinguishable types, divide an apparent common operator class, or alter the process boundaries at which multiplicities are evaluated. Consequently, arbitrary refinement need not preserve the exact symbolic motif. A central theoretical problem is therefore to identify a subclass of \textbf{heredity-faithful refinements} for which
\[
H_{\mathrm{RVT}}(C) = 1 \;\Longrightarrow\; H_{\mathrm{RVT}}(C') = 1
\]
whenever
\[
C' \preceq_{\mathrm{ref}} C.
\]
Such a result would establish a genuine refinement-invariance theorem for hereditary organisation.
\end{openproblem}

\begin{openproblem}[Representation-independent hereditary existence]\label{op:5}
The strongest ontological claim of the framework is that heredity is a property of realised physical causal organisation rather than of a chosen representation. Operationally, the present framework expresses this claim existentially:
\[
\mathrm{Her}(\mathcal{R}) = 1 \;\Longleftrightarrow\; \exists\, P \text{ admissible for } \mathcal{R},\ \exists\, C \in \mathcal{C}(\mathcal{R}, P) : H_{\mathrm{RVT}}(C) = 1.
\]
For this expression to become a true representation-invariance theorem rather than an operational definition, one would need to prove that:
\begin{itemize}
  \item $\mathcal{C}(\mathcal{R}, P)$, and the class of admissible perspectives, are characterised independently of hereditary-role inference;
  \item no sequence of admissible transformations within or between representations can produce an RVT organisation absent from $\mathcal{R}$. Proposition~\ref{prop:2} establishes this for a single admissible representation under a declared structural typing; the general claim across the family is not established;
  \item representations exposing the same physical hereditary organisation form a well-defined equivalence class, which requires the general notion of survival sought in Open Problem~\ref{op:2};
  \item the existential criterion corresponds to a property of $\mathcal{R}$ stated without reference to any representation. No such general theorem is claimed here.
\end{itemize}
Establishing it would provide the rigorous analogue of coordinate invariance suggested by the geometric intuition motivating the framework.
\end{openproblem}

\section{Cross-substrate process equivalence}\label{app:C}

The purpose of substrate independence is not to claim that complete physical systems are graph-isomorphic. A cell, an institution and a software deployment process may contain entirely different structures and internal dynamics. The appropriate object of comparison is the \textbf{inferred hereditary process}.

Crucially, this comparison occurs \textbf{after hereditary role inference}: structural types and causal order are defined role-neutrally first, and hereditary correspondence is evaluated afterwards. This appendix contains two distinct constructions. Section~\ref{app:C.1} defines operator correspondence between processes, which is tested during motif detection and before any role is assigned; it is used by the criterion itself. Section~\ref{app:C.2} defines equivalence between hereditary organisations that have already been inferred, and is therefore a post-inference comparison. Both rest on the same boundary descriptions.

\subsection{Boundary descriptions and operator correspondence}\label{app:C.1}

For a process $Q$, define the \textbf{boundary description}
\[
B(Q) = ( L^{-}(Q), L^{+}(Q), \Gamma(Q), \mathrm{Op}(Q) ),
\]
where $L^{-}(Q)$ and $L^{+}(Q)$ are the multisets of structural labels carried by the instances in $I(Q)$ and $O(Q)$, with \del{respective equal} multiplicities $n^{-}_{Q}(c)$ and $n^{+}_{Q}(c)$ \new{respectively} as defined in Section~\ref{sec:3.8};
\[
\Gamma(Q) = \{ (x, y) \in I(Q) \times O(Q) : x \prec y \}
\]
records instance-level input-output ancestry, the relation used being $\prec$, so that $\Gamma(Q)$ records ancestry rather than intervention-supported dependency (Section~\ref{sec:3.6}).

The boundary description therefore records labelled instances. The instances appearing in $\Gamma(Q)$ retain their structural labels, so the label multisets and the ancestry relation are two views of the same labelled boundary rather than independent components. Both are listed because the correspondence conditions below refer to each.

\textbf{Testing operator correspondence.} In each process, select a non-empty subset of $\mathrm{Op}(Q)$ and designate the candidate \new{information }carrier and product witnesses. The compared patterns contain all boundary instances of the selected operator types and the designated witnesses, together with the restriction of $\Gamma(Q)$ to those instances.

Two such patterns \textbf{correspond} when there exist bijections between their instances and between their structural types that preserve input and output membership, structural labelling, membership of the selected operator set, and causal incidence under $\Gamma(Q)$. Within one CSS, operator types are preserved. Candidate \new{information }carrier and product types may be matched to different types, which is what permits comparison across physically unrelated processes. Unselected boundary detail may differ.

The correspondence is structural and can be tested before hereditary roles are assigned. It is used in this way by the criterion itself: Section~\ref{sec:4.1} requires corresponding operator organisation across the R processes of a candidate family, Section~\ref{sec:4.6} requires a non-empty corresponding operator set for lineage continuation, and Definition~\ref{def:3} requires operator correspondence to survive coarsening.

Where lineage correspondence is required, the boundary descriptions are supplemented by the descendant re-entry relations between processes, which the matching must also preserve.

\subsection{Boundary-pattern equivalence}\label{app:C.2}

Let $Q$ and $Q'$ be processes in physically distinct CSSs. Let $S_{Q}$ and $S_{Q'}$ denote the boundary-instance sets selected for comparison under Section~\ref{app:C.1}, and $C_{Q}$ and $C_{Q'}$ their structural-label sets.

After hereditary inference, their boundary patterns are hereditarily equivalent with respect to a motif when there are bijections
\[
f\colon C_{Q} \to C_{Q'},\ \phi\colon S_{Q} \to S_{Q'}
\]
preserving input/output membership, selected operator membership, and inferred roles, with
\[
\ell'\bigl(\phi(x)\bigr) = f\bigl(\ell(x)\bigr)
\]
for every selected instance $x$, and
\[
(x, y) \in \Gamma_{Q} \ \Longleftrightarrow\ (\phi(x), \phi(y)) \in \Gamma_{Q'}
\]
for selected input-output pairs.

These conditions preserve the multiplicities and causal incidences of the compared patterns. For lineage-related comparisons, the mappings must additionally extend consistently across the participating processes and preserve descendant re-entry.

For example,
\[
r_{i} \to r_{ii}
\]
and
\[
r'_{j} \to r'_{jj}
\]
instantiate equivalent replication boundary organisation under
\[
f(i) = j,\ f(r) = r'.
\]
This is the minimal notion of equivalence. Exact boundary isomorphism is the minimal case; internal topology is retained only where necessary to preserve the inferred role.

The scientific claim is therefore not: \emph{these systems are the same}; it is: \emph{these physically different systems instantiate the same specified hereditary causal organisation}.

\section{Computational search}\label{app:D}

The formal existence of a motif does not imply that finding it is computationally trivial.

A CSS with {\new{$N$}} nodes may contain a very large number of causally convex subsets. Candidate structural families, process boundaries, operator sets and lineage continuations create a combinatorial search problem.

The present paper does not claim to solve the optimal search problem. It specifies what a valid solution must satisfy.

A comparison clarifies the distinction between formal specification and computational tractability. Exact treatment of realistic quantum many-body systems is generally impractical, and applications therefore rely on approximations whose accuracy is assessed through theoretical analysis, benchmarking and experiment. Computational difficulty alone does not invalidate the underlying theory.

CSS faces a comparable distinction. A representation and its motif predicates may be precisely specified even when exhaustive search over a realistically sized episode is infeasible. Such infeasibility does not itself show that the specification is inconsistent or that the criterion is scientifically inadequate. It does, however, constrain the framework's practical usefulness, which must be demonstrated rather than inferred from the analogy.

For CSS search, soundness and completeness must be distinguished. A method that verifies every returned motif against the full criterion may be sound while remaining incomplete because pruning omits other valid motifs. Positive detections then remain valid within the supplied CSS, whereas failure to detect a motif does not establish its absence. Pruning rules proved to discard only candidates that cannot satisfy the criterion can preserve completeness; other strategies require an explicit account of what their outputs establish.

CSS does not yet possess a record of successful approximation comparable to established physical theories. Developing tractable search methods, proving guarantees where possible, and characterising missed detections through suitable benchmarks are therefore substantive parts of the research programme.

A natural computational strategy\new{ for the motif search, i.e., stage~6 of Figure~\ref{fig:pipeline},} is hierarchical:

R-compatible boundaries $\rightarrow$ shared operator organisation $\rightarrow$ V relations $\rightarrow$ variant-dependent T $\rightarrow$ lineage extension.

The order is important.

Searching first for candidate replication boundaries eliminates large parts of the process space before more expensive family matching is attempted. Shared operator structure further groups candidates. Only those families then need to be examined for V, T and continuing genealogical organisation.

\new{Section~\ref{sec:comp} implements a hierarchical search procedure of this kind, in four search passes, on the CSS of a digital-evolution simulator; Appendix~\ref{app:H} gives the details.
The order of its passes differs from the one above: R-compatible boundaries (pass~A) $\rightarrow$ lineage continuation (pass~B) $\rightarrow$ V relations (pass~C) $\rightarrow$ variant-dependent T (pass~D), with shared operator organisation tested within pass~D and in the family rule, rather than R $\rightarrow$ operators $\rightarrow$ V $\rightarrow$ T $\rightarrow$ lineage.
Lineage was searched second, in pass~B, so that V and T are examined only for labels whose copies re-enter replication.
The search covered declared process classes only, i.e., single steps and chains of copy, move and re-read at the level of instructions, and births at the level of organisms. %
In the sense of the soundness paragraph above, the search's positive detections therefore hold within the searched process classes only, and a label the search does not detect is not thereby shown to lack a hereditary role, e.g., the copy instruction \texttt{h-copy}, which no single-step process in the order as defined places in a family \new{in the development runs }(Section~\ref{sec:comp.3}).} %

\begin{newtext}
The hierarchical search procedure of Section~\ref{sec:comp} is thus an initial, practical search, not a general exact or complete search algorithm.
Together with the construction of the CSS on which it runs, the procedure partially addresses three problems.
First, the construction gives any search its input: one-step intervention replay gives the 847,911 nodes of seed 1 and their direct-edge relation $D$ in 48 s (Section~\ref{sec:comp.1}). %
Second, the procedure reduces the process space hierarchically, as proposed above, by restricting candidate processes to declared classes and by testing V and T only for labels whose copies re-enter replication.
Third, the procedure characterises what it misses against independent knowledge of the simulator, i.e., the withheld record and the instruction semantics; e.g., the labelling used hides the translation of \texttt{mov-head}, an ancestral instruction that is copied but never a family member (Section~\ref{sec:comp.5}). %
Five problems remain open: 1) search over arbitrary causally convex subsets rather than declared process classes, which the dense order of this CSS makes large: in the order as defined, the convex closure of a single step has a median of only 25--28 nodes per run, but 40--46 steps per run have closures above 50,000 nodes, up to 67,856 (Section~\ref{sec:comp.3}); 2) pruning rules proved to preserve completeness; 3) the complexity of CSS search in the variables listed below, and its scaling beyond one world of 36 cells; 4) structural labels derived from the dynamics, e.g., from an AIR partition (Section~\ref{sec:3.3}), rather than declared value classes, and; 5) a null model for the full criterion, since the \new{exploratory family-level null of Section~\ref{sec:comp.4} tests only whether the labels align with replication and lineage}\del{label-permutation null of Section~\ref{sec:comp.4} tests R only}. %
\end{newtext}

Possible computational tools include:
\begin{itemize}
  \item persistence-based pruning;
  \item causal-window restriction;
  \item partial-order reachability indices;
  \item minimal causal neighbourhoods;
  \item bounds on motif depth;
  \item parallel search over independent regions;
  \item branch-and-bound or constraint-satisfaction methods.
\end{itemize}
The relevant complexity variables are likely to include:
\begin{itemize}
  \item node number;
  \item structural-label diversity;
  \item causal width;
  \item number of candidate process boundaries;
  \item lineage depth.
\end{itemize}

A further reduction is available at the level of representation rather than of search within one. Because admissibility is established for each representation independently (Section~\ref{sec:3.3}), the search may be guided by expectations about where hereditary organisation is likely, and admissible representations examined in whatever order those expectations suggest. This does not affect the status of any motif found, since each representation is admissible on its own terms.

Efficient implementation will therefore require careful use of computational resources. That requirement is not a defect in the definition: many precisely defined combinatorial objects are expensive to find.

The algorithmic and complexity theory of CSS search remains a distinct research problem.

\section{Extended comparison with the closest precedents}\label{app:E}

Section~\ref{sec:2.6} states the essential boundary between CSS and the frameworks that make the most closely related inverse moves. This appendix develops each comparison.

Several existing frameworks move substantially beyond theories that simply assume genes, organisms or other evolutionary units. In particular, the information theory of individuality, algorithmic accounts of the origin of life, Assembly Theory and Constructor Theory each remove an important piece of domain-specific ontology. They therefore provide especially relevant precedents for the inverse problem considered here. The distinction from CSS is not that these approaches are domain specific while CSS is generic. Rather, each solves a different part of the inverse problem and stops before the joint identification of hereditary \new{information }carriers and hereditary process roles from a realised causal history.

\subsection{Inferring the individual: Krakauer et al.}\label{app:E.1}

Krakauer, Bertschinger, Olbrich, Flack and Ay \citep{Krakauer2020} make perhaps the closest methodological move to the one pursued here. Rather than assuming that the biological individual is known, they explicitly seek to \textbf{infer or detect evolved individuals}. Given measurements over time, candidate aggregations or coarse-grainings are evaluated according to how much information their present state carries about their own future. System--environment boundaries can therefore emerge from an algorithmic decomposition rather than being stipulated beforehand, and individuality can be graded, nested and instantiated at different organisational levels.

This is important for CSS because it demonstrates that a seemingly primitive biological category can legitimately be made the \textbf{output of an inverse inference problem}. The information-theoretic individual is not defined by cell walls, reproduction or a conventional organismal ontology; it is an aggregation with sufficient temporal informational integrity. Krakauer et al.\ even propose adjusting the system--environment boundary until adding further processes no longer improves prediction of the candidate system's future.

The inference target, however, is different from that of CSS. Krakauer et al.\ ask approximately:

\emph{Which aggregation of variables behaves as an individual?\del{.}}

Their quantities partition predictive information between a candidate system $S$ and environment $E$,
\[
I(S_{n}, E_{n}, S_{n+1}),
\]
and distinguish organismal, colonial, and environmentally driven forms of individuality according to the relative contributions of the system and environment.

CSS does \textbf{not} seek primarily to recover a system--environment boundary. Its inverse problem is instead:

\emph{Which realised structures occupy the mutually dependent hereditary roles?\del{.}}

A CSS may therefore contain no uniquely privileged ``individual'' at all. What must be recovered are structural variants participating jointly in causal replication, variation and variant-dependent realisation, together with the operator structures implementing those processes and the causal ancestry connecting successive generations.

This creates a precise boundary between the approaches. The information theory of individuality can infer \textbf{which collection of variables has temporal integrity as an individual} without requiring replication. CSS asks whether, within or across whatever individuals may exist, one can infer \textbf{which structures are hereditary \new{information }carriers, which are replication and translation operators, and which downstream structures are variant-dependent realisations}. The former discovers a boundary; the latter attempts to discover a \textbf{relational division of hereditary labour}.

A second distinction is methodological. Krakauer et al.'s primary criterion is information-theoretic predictability across time. CSS retains individual realised structure instances and their causal ancestry because repeated instances of a type are not sufficient to establish genealogy. Statistical temporal dependence can therefore be important evidence in constructing a physical description, but the CSS hereditary criterion additionally asks whether particular realised instances causally produced descendants and participated in the required process organisation.

Thus CSS extends the inverse logic exemplified by Krakauer et al.\ rather than claiming that inference without prespecified biological individuals is itself new:
\begin{itemize}
  \item Krakauer et al.: \emph{infer the individual from temporal informational organisation};
  \item CSS: \emph{infer hereditary roles from realised causal organisation}.
\end{itemize}

\textbf{Formal relation.} The two frameworks describe different aspects of a system. The individuality measures require joint probability distributions over time-indexed system and environment variables. A CSS records a realised causal history and does not, by itself, specify those distributions. Their estimation may use repeated observations or a sufficiently informative time series under appropriate sampling assumptions.

The probabilistic intervention criterion of Section~\ref{sec:3.5} does not automatically provide the required stochastic process. Causal evaluations may use deterministic dynamics, uncertainty over initial conditions, or particular intervention experiments without specifying a joint distribution over system and environment histories.

Joint application therefore requires an additional observational or dynamical model specifying the relevant variables, their probability distribution and their correspondence to CSS instances. Individuality measures could then identify candidate system\new{--}environment partitions, while CSS identifies hereditary organisation in the corresponding realised histories. One could ask whether an inferred information family lies within, spans or cuts across an informational individual. This is a proposed comparison between the frameworks, not a correspondence determined by their definitions alone.

\subsection{Inferring informational control: Walker and Davies}\label{app:E.2}

Walker and Davies \citep{Walker2013} move closer to the specific issue of hereditary causal organisation. Their \emph{Algorithmic Origins of Life} explicitly argues that the origin of life should not be identified merely with self-replication or Darwinian evolution. Instead, they propose a transition in \textbf{causal architecture} in which information acquires context-dependent causal efficacy over the matter in which it is instantiated. Their central problem is how informational and mechanical aspects of living organisation separate into interacting causal roles, and they explicitly describe the challenge as explaining how information-control systems emerge from ordinary molecular dynamics.

This is much closer to the present motivation than a generic statement that Walker and Davies study information. They distinguish trivial replication from \textbf{non-trivial replication}, emphasise the separation of stored instructions from the machinery implementing them, and identify translation between informational and material organisation as a key transition. In their account, modern life is characterised by distributed information control, dual hardware/software roles and causal influence running both upward and downward through organisational hierarchies.

The similarity to the role assigned to T in CSS is therefore real rather than superficial. Both approaches reject mere copying as sufficient. Both require information-bearing differences to become causally active in the construction or control of other physical organisation.

The edge lies in \textbf{what is operationally inferred}.

Walker and Davies provide a proposed criterion for the transition to living informational organisation:

ordinary dynamics $\rightarrow$ context-dependent information control.

They suggest that changes in causal architecture, including a shift toward top-down information flow, could provide measurable signatures of this transition. But the framework does not provide an instance-level inverse procedure that begins with hereditary-role-neutral persistent structures and returns:

\new{information }carrier variants, replication operators, translation operators, realised products, genealogical relations.

Indeed, the conceptual analysis can legitimately talk about algorithms, genetic information, hardware, software and translation because its objective is to characterise \textbf{what type of causal organisation distinguishes life}. CSS poses an additional inverse problem: if none of those roles has yet been identified in the observation, can their physical bearers be recovered jointly from the realised causal organisation itself?

The distinction can therefore be expressed as:
\begin{itemize}
  \item Walker--Davies: \emph{What causal transition makes information an active controller of matter?}
  \item CSS: \emph{Which observed structures instantiate the \new{information }carrier/control roles in such an organisation?}
\end{itemize}

Walker and Davies consequently provide an important \textbf{conceptual antecedent to} T and to the idea that life requires more than copying. CSS attempts to turn part of that causal insight into a role-discovery criterion.

\textbf{Formal relation.} The algorithmic-origins account does not specify a unique formal object directly corresponding to a CSS. It nevertheless discusses quantitative approaches to changes in causal architecture, including transfer entropy in coupled dynamical systems and possible applications of integrated information. The relationship to CSS is therefore conceptual but can support quantitative comparison once a particular operational measure is chosen.

Joint application would require an explicit hypothesis connecting that measure to CSS role inference. For example, one could ask whether a measured transition in information flow coincides with the appearance of a coherent \new{information }carrier-operator organisation in a reconstruction of the same system. Such a comparison requires sufficiently informative data for both analyses and a declared correspondence between their descriptions. Agreement would support the proposed connection; disagreement would require examining the measurement, reconstruction and hypothesised relationship. Neither outcome establishes an equivalence between the frameworks without further argument.

\subsection{Inferring historical selection: Assembly Theory}\label{app:E.3}

Assembly Theory \citep{Sharma2023} constitutes another important inverse approach and must be treated particularly carefully because it explicitly claims not to require prespecification of the units of selection. It characterises an object through its possible construction histories and assigns an \textbf{assembly index} according to the shortest recursive pathway capable of constructing it. At the ensemble level, assembly combines object copy number with assembly index to quantify evidence that selection and historical contingency were required to produce the observed population of complex objects.

The underlying insight is highly relevant to CSS: present physical organisation may retain objective evidence of \textbf{history}, even when the historical process itself is not directly observed. Assembly Theory deliberately changes the ontology of an object from a point-like entity to something characterised by its construction possibilities. It thereby seeks to infer selection without first identifying conventional biological individuals or units of selection.

The difference from CSS is therefore \textbf{not} that Assembly Theory presupposes a biological individual. Its inference target is different. For an observed object $x$, Assembly Theory asks approximately:

\emph{How much historically contingent construction and selection is evidenced by $x$ and its copy number?}

The framework starts from distinguishable objects and a declared construction alphabet or set of elementary joining operations; for molecular implementations, for example, molecular bonds provide the construction operations. It then evaluates assembly paths and copy numbers.

What it does not attempt to recover is the \textbf{hereditary causal role decomposition responsible for those copies}. High assembly and high copy number can provide evidence that selection and memory have shaped an ensemble without identifying which physical structures stored transmissible variants, which structures copied them, which machinery interpreted them, or which realised outputs differed because of those variants.

This difference is particularly visible in Assembly Theory's own treatment of function: the formulation explicitly notes that its selection model does not model the functional differences on which selection might act; it captures the specificity of selection through historical reuse and object abundance.

CSS instead asks for a realised organisation of the form $i_{1},i_{2},\ldots$ participating in corresponding R processes and producing $y_{1},y_{2},\ldots$ through corresponding T operators, with realised genealogical continuation.

Thus:
\begin{itemize}
  \item Assembly Theory: \emph{infer historical selection from the construction complexity and abundance of objects.}
  \item CSS: \emph{infer the causal hereditary roles through which variant organisation is propagated and realised.}
\end{itemize}

\textbf{Formal relation.} Both frameworks can represent dependencies within construction processes, but a CSS ancestry order is not automatically an assembly pathway. CSS nodes represent persistent structure instances, and their ancestry may include maintenance, catalytic and environmental dependencies as well as construction. The observation window may also begin after some constituent structures were produced. Assembly calculations additionally require specified building blocks, joining operations and rules for reusing previously constructed components. Relating the two would require a mapping from a CSS history to a construction within a declared assembly scheme, which the present framework does not supply; in particular, the assembly index does not bound CSS node count, causal depth or ancestry length.

Joint application would therefore combine complementary descriptions. Assembly measures could characterise construction complexity and, together with copy number, the proposed selection signatures of objects whose hereditary roles CSS has inferred. CSS would retain realised instance-level ancestry that those measures do not summarise. Neither description generally determines the other: an assembly calculation may require compositional information absent from the CSS, while the assembly index does not specify realised genealogy.

\subsection{Defining physical information tasks: Constructor Theory}\label{app:E.4}

Constructor Theory approaches the problem from yet another edge. Rather than treating information as an abstract mathematical quantity, Constructor Theory of Information \citep{Deutsch2015} defines informational properties through \textbf{which physical transformations are possible and impossible}. A variable is an information variable when its attributes support the relevant computation and cloning tasks, and an information medium is a physical substrate possessing such a variable. The cloning task for a candidate variable $X$ has the form
\[
\mathcal{T}_{X} = \{(x, x_{0}) \to (x, x) \mid x \in X\},
\]
where $x_{0}$ is a receptive attribute of a second instance of the substrate, preparable from generic resources. The task specifies the required transformation for every attribute $x$ in $X$.

This is a genuinely physical, substrate-independent definition. It is therefore inaccurate to describe Constructor Theory as merely taking information as a semantic primitive. Its purpose is almost the opposite: to characterise information through objective physical regularities concerning possible transformations.

Constructor Theory of Life \citep{Marletto2015} goes further. Under no-design laws, Marletto analyses accurate replication and self-reproduction and argues that an accurate replicator requires a \textbf{vehicle} that, together with the replicator, constitutes the self-reproducer. This provides a substrate-independent physical separation between recipe-like information and the machinery capable of constructing and copying it, an important formal antecedent to the \new{information }carrier/operator distinction used here.

The key difference is the contrast between \textbf{counterfactual capability} and \textbf{realised role discovery}.

Constructor Theory asks questions of the form:

\emph{For this substrate and variable, which transformation tasks are possible or impossible?}

Its fundamental explanatory objects are tasks, substrates and the possible/impossible distinction.

CSS asks:

\emph{In this realised physical causal history, which structures actually occupied the hereditary roles?}

To establish that a substrate is an information medium in Constructor Theory, one specifies candidate attributes/variables and asks whether the appropriate cloning and computational tasks are possible. To establish \del{a}\new{an} \new{information }carrier in CSS, one does not begin by designating an information variable; one searches a role-neutral realised causal representation for structural variants that \textbf{jointly} satisfy replication, variation, variant-dependent translation and genealogical continuation.

Likewise, Constructor Theory of Life derives constraints on the logical architecture required for self-reproduction once replication/self-reproduction tasks are considered. CSS instead attempts to discover the corresponding physical participants from an episode in which no replicator, vehicle, recipe or constructor has been labelled in advance.

The distinction is therefore:
\begin{itemize}
  \item Constructor Theory: \emph{characterise information and replication by counterfactual physical task capability.}
  \item CSS: \emph{infer realised hereditary roles from instance-level causal history.}
\end{itemize}

The approaches intersect strongly, but their directions of inference differ.

\textbf{Formal relation.} A constructor-theoretic task specifies a set of input-output attribute pairs for declared substrates. With an explicit mapping from CSS boundary structures and labels to substrates and attributes, a realised CSS process may correspond to an observed transition matching one such pair. This correspondence does not establish the task's possibility in the constructor-theoretic sense. That requires that physical laws permit arbitrarily accurate performance across the task's legitimate inputs and arbitrarily good retention of the ability to perform it again.

The replication comparison is similarly conditional. Some CSS R processes may be mapped to observed instances of a cloning specification, but R's general multiplicity and causal conditions do not themselves specify the receptive substrate, the complete attribute variable or the accuracy requirements of constructor-theoretic cloning. A finite collection of successful copying events therefore does not establish the corresponding constructor capability. Conversely, impossibility of a task does not generally exclude every individual transition matching one of its input-output pairs.

Joint application would connect constructor-theoretic constraints on repeatable transformations with CSS descriptions of particular realised histories. Such a comparison must specify the substrates, attributes, resources and process boundaries involved. The replicator-vehicle organisation analysed in Constructor Theory of Life provides a candidate architectural correspondence, but the vehicle is not automatically a CSS operator. Its copying or construction components may satisfy CSS operator conditions in particular processes; across a complete self-reproduction process, the vehicle's multiplicity may instead increase. Whether the derived architecture and inferred roles correspond must therefore be established at declared process boundaries.

\section{Composition, identifiability and quantitative extensions}\label{app:F}

\subsection{Multiple observations and composition of CSS histories}\label{app:F.1}

A strict observational framework raises a practical problem: many real datasets do not provide one uninterrupted causal record.

Instead, we may have several bounded observations
\[
\mathcal{R}_{1},\ldots, \mathcal{R}_{n},
\]
yielding
\[
\mathcal{C}(\mathcal{R}_{1}, P_{1}), \ldots, \mathcal{C}(\mathcal{R}_{n}, P_{n}).
\]

Two cases must be distinguished.

\subsubsection{Independent observations}\label{app:F.1.1}

If no observed evidence links the physical instances in the different episodes, the CSSs remain separate.

The fact that they contain structures with the same AIR label does \textbf{not} establish a shared lineage.

They can nevertheless be compared through the post-inference process-equivalence relation of Appendix~\ref{app:C}.

\subsubsection{Observationally connected episodes}\label{app:F.1.2}

If additional observation establishes continuity between episodes, for example:
\begin{itemize}
  \item overlapping temporal windows;
  \item tracked persistence of the same physical object;
  \item a documented transfer between endpoint and starting point;
  \item an independently observed parent--descendant connection;
  \item another recorded causal bridge,
\end{itemize}
then the episode representations may be combined.

Schematically,
\[
C^{\star} = C_{1} \sqcup C_{2} \sqcup \cdots \sqcup C_{n} + E_{\mathrm{bridge}},
\]
where
\[
E_{\mathrm{bridge}}
\]
contains only observationally supported cross-episode causal relations.

Shared physical instances observed in overlapping windows can be identified consistently, after which the transitive causal order is completed.

This permits long hereditary histories to be reconstructed from distributed observations without inventing causal links through unobserved periods.

The principle remains strict:

structural similarity $\nRightarrow$ genealogical continuity.

This composition problem is likely to be important for biological, historical and institutional data and deserves a fuller mathematical treatment in future work.

\subsection{Incomplete observation and identifiability}\label{app:F.2}

CSS provides an ideal representation relative to the information actually available.

Real observations will often fail to resolve:
\begin{itemize}
  \item relevant persistent structures;
  \item causal intermediates;
  \item variant differences;
  \item process boundaries;
  \item parts of a lineage.
\end{itemize}

These limitations should not be corrected by external semantic knowledge.

If a distinction is not observable at the chosen perspective, it cannot appear in {\new{$\ell$}}.

If a causal bridge is not supported by the observation, it cannot be inserted into $\prec$.

If a lineage leaves the observation window, its continuation is unknown.

The appropriate scientific consequence is therefore reduced \textbf{identifiability}, not semantic completion.

This gives the framework a natural way of expressing observational limits. A coarse representation may support only R; a richer observation may reveal V and T; a longer observation may increase lineage depth.

Variable-level causal-discovery methods face a related but distinct problem: graphs inferred from statistical relations among observed variables may be underdetermined or distorted by incomplete observation and temporal aggregation \new{\citep{Andersson1997,Squires2023,Fan2024}}. CSS instead uses realised structure instances and their causal organisation, allowing additional constraints from temporal precedence, persistence, ancestry and physical structure, together with interventions where available. Evidence that structural prior information can restore causal recoverability from temporally aggregated data \citep{Fan2024} suggests that a CSS reconstructed from incomplete observations may still retain enough causal organisation to support role inference.

Future work can systematically degrade complete observations and quantify at which point each inferred role is lost.

\subsection{Conditional role inference and sensitivity to hypothetical reconstructions}\label{app:F.3}

The construction of Section~\ref{sec:3} maps a description of an episode to a CSS, and the criterion of Sections~\ref{sec:4} and~\ref{sec:5} maps that CSS to a role assignment. Neither operation inspects where the description came from. A description supported by observation and one containing stipulated elements are processed identically; what differs is what may be concluded from the result.

This section sets out how to use that fact. \textbf{Where the required observations are unavailable, a reconstruction may be stated explicitly as a hypothesis, put through the ordinary construction, and its role assignment reported as conditional on the hypothesis rather than as a finding about the episode.} Repeating this across alternative reconstructions then shows which hereditary conclusions depend on which assumptions, and which would be changed by evidence that is not currently available. An example where this is especially valuable is in the analysis of cultural episodes involving the (hidden) skill sets of actors -- discussed in more detail below.

The section is placed here because the difficulty it addresses is the one treated throughout Appendix~\ref{app:F}: what can be established as observation becomes incomplete. Section~\ref{app:F.2} states the resulting limits on identifiability. Conditional analysis does not remove those limits. It provides a disciplined way of working within them, at the cost of a weaker class of conclusion, and its principal requirement is that the weaker status remain attached to the result wherever it is subsequently used.

\textbf{Two levels of requirement.} The construction imposes requirements of two kinds.
\begin{enumerate}[label=\arabic*)]
  \item \textbf{Formal \emph{well-formedness}} requires that a description satisfy the applicable representational conditions, including persistence, completeness and non-overlap relative to the declared individuation rule, exclusive structural typing, and acyclic dependencies consistent with temporal precedence. Dependencies may be stipulated as hypotheses or derived from a hypothetical dynamical model; their status must be explicit, and they must be consistent with any dynamics and intervention rules supplied.
  \item \textbf{Empirical admissibility}, as set out in Definition~\ref{def:1}, additionally requires independent evidential support for the represented structures and relations in the episode described, including support for any model-based causal evaluations.
\end{enumerate}
A description may satisfy the first without satisfying the second. Applying the representation and motif-evaluation operations to such a description yields a well-defined result whose evidential status differs from that of an observationally grounded analysis: if the description held, these structures would occupy these hereditary roles. The mathematical operations are unchanged; the evidential requirements are not, and the results of the two modes are not interchangeable.

\textbf{What the hypothetical reconstruction mode adds.} Observational analysis under incomplete evidence already supports partial conclusions: particular motifs may be supported while others are not, a lineage may have an established minimum depth, and specific causal relations may remain unresolved. Conditional analysis complements this rather than replacing it. Observational analysis establishes what the evidence supports; conditional analysis examines which additional assumptions would complete or alter that interpretation, and which hereditary conclusions are sensitive to them.

\textbf{Preserving role neutrality.} A hypothetical reconstruction may stipulate structures and causal dependencies, but hereditary roles must remain outputs of the criterion. A proposed role assignment may motivate the analysis or serve as a hypothesis to be tested; it must not determine the structural classification or causal relations supplied to the inference. Hypothesising that an instructional interaction alters another participant's persistent configuration is admissible as a hypothesis; declaring that the interaction transmits hereditary information is not, since the latter assigns the role that the criterion is meant to determine. Hypothesised elements should therefore be described as structures and relations subsequently assigned roles, not as hypothesised \new{information }carriers or operators.

A hypothesis may directly entail a motif component where it specifies all of that component's defining conditions, including the process boundary multiplicities. Any such entailment should be identified when the description is stated. The derivation remains logically legitimate; what it cannot provide is independent empirical support for the hypothesised organisation, and recovery of an entailed component must not be reported as confirmation of it.

\textbf{Procedure.}
\begin{enumerate}[label=\arabic*.]
  \item State the description $\Delta$ in full, marking each element as observationally supported or hypothesised, and identify any motif component whose defining conditions are entailed in full by hypothesised elements.
  \item Verify formal well-formedness. Explicitness does not guarantee it: a stated description may contain overlapping instances, incompatible structural assignments or cyclic dependencies.
  \item Construct the CSS and apply the causal motif criterion. Record the role assignment as conditional on $\Delta$.
  \item Repeat across a declared set of alternative descriptions $\Delta_{1}, \ldots, \Delta_{n}$, stating the basis on which the set was selected.
  \item Establish correspondence between reconstructions. Where these contain different structures or classes, comparison of role assignments requires an explicit mapping between them, for which the boundary equivalence of Appendix~\ref{app:C} provides one criterion.
  \item Report which role assignments, or equivalent organisations under that mapping, persist across the tested set; which individual assumptions or combinations of assumptions alter them; and which candidate observations or interventions, where any exist, would discriminate between reconstructions that differ.
\end{enumerate}

Step 6 is where the analysis becomes operational. Agreement across the tested set establishes robustness \emph{within that set}, not across every reconstruction compatible with the evidence, and a set of alternatives sharing one unsupported assumption may exhibit apparent robustness while leaving the decisive uncertainty untested. A result proved across a formally specified class of evidence-compatible reconstructions would be a stronger identifiability claim, and is not what this procedure provides. The analysis may also establish that two reconstructions are indistinguishable on the available evidence. Whether they are observationally equivalent under the declared observational interface, meaning that no observation the interface permits could distinguish them, is a stronger claim requiring more than a comparison of their fit to the existing record.

\textbf{Cultural episodes as the principal case.} Transmission through externalised media --- documents, artefacts, recordings, institutional archives --- leaves persistent and observable candidate structures, which makes several requirements easier to satisfy, though the relevant processes and causal dependencies must still be reconstructed rather than read off the record. Transmission of internalised capability is harder in three specific respects. Structural individuation is difficult, since the almost-invariant-region procedure of Section~\ref{sec:3.3} requires a dynamical description that behavioural repertoire does not readily supply, and any alternative classification must still deliver persistence, completeness, mutual exclusivity and role independence. Persistence between demonstrations may lack independent evidential support, even where similar performances are observed repeatedly. And ethnographic records rarely supply the counterfactual evidence on which Section~\ref{sec:3.5} depends. These are difficulties in obtaining sufficient evidence in particular episodes, not properties of cultural systems as such.

Where the evidence is insufficient, the framework already permits withholding judgement, reporting the motifs that are supported, or seeking further evidence. Two further courses are tempting and both conceal the evidential gap rather than closing it: encoding the missing dependence into the observational perspective, and positing an unobserved internal structure as a node. Both are inadmissible under Definition~\ref{def:1}, since each presents a hypothesis as an empirically established input. Assumptions embedded in a perspective can of course be varied; the difficulty is that their assumed status is not visible, so that variation is neither prompted nor reported. Conditional analysis does not uniquely make assumptions variable. It makes their status explicit and subjects them to the sensitivity analysis of step 6.

\textbf{Reporting.} Every reported result must remain explicitly linked to the assumptions on which it depends and to its evidential status, including in tables, figures and summary statistics. Combined presentation of conditional and observationally grounded results is acceptable where that distinction is unmistakable, for instance in separated columns or panels. Different conclusions within a single analysis may depend on different subsets of the stated assumptions, and step 6 is intended to establish which. Where a result is conditional, the form \emph{under description $\Delta$, structures $X$ occupy \new{information }carrier roles} is preferred to \emph{$X$ is \del{a}\new{an} \new{information }carrier, assuming $\Delta$}, since the latter more readily loses its qualification in subsequent use.

\new{\textbf{A worked application.} Section~\ref{sec:cultural} applies this procedure to recipe transmission, and Appendix~\ref{app:G} gives the details. Both descriptions declare the same physical hypotheses H1--H4, the same structural types and the same classes of teaching and cooking processes (step 1). Case A, description $\Delta_{A}$, contains the stipulated episode only. Case B, description $\Delta_{B}$, adds identified hypothetical structures and causal relations, H5--H8: a modification of a $c_1$ configuration into a $c_2$ configuration (H5), two further teaching processes for $c_2$ (H6), one further cooking process (H7), and a comparative rule under which replacing $c_1$ by $c_2$, other conditions corresponding, changes the dish (H8). Both descriptions are checked for well-formedness and evaluated by the same frozen criterion (steps 2 and 3). Because $\Delta_{B}$ extends $\Delta_{A}$, the mapping of step 5 is the identity on their shared structures. The comparison of step 6 gives: under $\Delta_{A}$, no complete information family is assigned; under $\Delta_{B}$, $c_1$ and $c_2$ occupy \new{information }carrier roles. The difference depends on the added relations, which are stipulated to supply the missing conditions; recovering those conditions in Case B is therefore not confirmation of them, and satisfying the criterion under $\Delta_{B}$ gives H1--H4 and the added relations no empirical support.}

\textbf{What is not claimed.} A conditional analysis establishes no hereditary organisation beyond what the supporting evidence for each result allows. Where the physical witnesses required by a motif depend on hypothesised structures or hypothesised causal relations, that motif has no established physical witness and the property of Appendix~\ref{app:B.7} does not apply to it. A motif whose required witnesses are all observationally supported retains its observational status, even where the same analysis contains hypothesised elements elsewhere; a single analysis may therefore recover one information family observationally while exploring another conditionally. In every case, a conditional result establishes the roles assigned by the frozen criterion under the stated description, and as Section~\ref{sec:7.1} sets out, does not establish that the criterion captures every intended sense of hereditary organisation.

\textbf{Whether the mode remains necessary.} Some of the difficulties above reflect present measurement capability. Structural observation of the relevant internal states, independent evidence bearing on their persistence between demonstrations, and settings supplying counterfactual variation in transmission would each reduce the reliance on hypothesised elements, and where the full admissibility conditions of Definition~\ref{def:1} are then satisfied, the analysis becomes an ordinary application of the framework. Other difficulties may not be removable in this way: historical episodes may have lost the evidence that would have settled them, and some reconstructions may remain indistinguishable under any observation the available interface permits. Persistence is in either case established by inference rather than direct tracking, since tracking is itself inferential; what differs between the cases is whether that inference has independent support. The boundary between episodes admitting direct analysis and those requiring hypothetical descriptions is therefore likely to move unevenly and not to disappear, and the reporting requirements are intended to keep the two modes distinguishable as it does.

\subsection{From CSS to quantitative spaces of hereditary organisation}\label{app:F.4}

CSS provides more information than the categorical existence of R, V and T.

Once hereditary organisation has been inferred, its causal architecture can be measured through quantities such as:
\begin{itemize}
  \item pathway redundancy;
  \item dependence on shared operators;
  \item causal bottlenecks;
  \item modularity or causal insulation;
  \item lineage branching;
  \item ancestry depth;
  \item alternative causal support.
\end{itemize}

Related quantities have proved informative in the study of robustness and structured fragility in other classes of complex system \new{\citep{Carlson2000,Albert2000,Ay2007,PateCornell2004}}, and low-dimensional resilience relations can emerge across physically different systems \citep{Gao2016}.

A collection of such CSS-derived quantities could, in principle, define coordinates in a \textbf{Heredity Order Parameter Space (HOPS)}.

Physically distinct hereditary systems could then be located according to causal organisation rather than substrate-specific variables, potentially allowing persistence, fragility or transitions to be compared across systems. Stability regimes of this kind have been identified separately across biology \citep{Biebricher2005}, culture, institutions, software \citep{Schulte2014} and digital life.

This possibility is a downstream motivation for CSS, not part of the present theory.

We do not define HOPS coordinates here, do not assume that a low-dimensional representation exists, and do not develop a theory of hereditary stability.

\section{\new{Detailed conditional reconstructions of recipe transmission}}\label{app:G}

\begin{newtext}
This appendix gives the full reconstructions behind Section~\ref{sec:cultural}. The appendix is a worked application of the conditional-inference procedure of Appendix~\ref{app:F.3} and does not restate that procedure. Appendix~\ref{app:G.1} states the description common to both cases (step 1 of the procedure). Appendices~\ref{app:G.2} and~\ref{app:G.3} construct the two CSSs, verify their well-formedness and apply the criterion (steps 2 and 3). Appendix~\ref{app:G.4} compares the two role assignments, identifies the hypotheses on which the difference rests (steps 5 and 6) and checks the synthetic example of Section~\ref{sec:6.2}, and Appendix~\ref{app:G.5} reports the remaining sensitivities and evidential limits. Every instance, dependency, boundary, multiplicity and verdict stated here was computed from the explicit instance and dependency lists by a script that implements the definitions of Sections~\ref{sec:3.5}--\ref{sec:3.9}, \ref{sec:4.1}--\ref{sec:4.7} and~\ref{sec:5} and Appendix~\ref{app:C.1} directly, including an exhaustive search for the bijections of Appendix~\ref{app:C.1}.
\end{newtext}

\subsection{\new{Observational perspective and physical hypotheses}}\label{app:G.1}

\revnote{[Moved unchanged from v7 Section 6.6.]} \new{The episode itself is stated at the start of Section~\ref{sec:cultural}.}

Nothing in this is unusual, and most people would say without hesitation that something was passed from A to B and that it was then passed on again. The question is what, if anything, a framework that is not told about teaching, recipes or knowledge would find in such an episode, and what it would need in order to find it.

\textbf{Perspective.} $P$ declares as measured variables the positions and configurations of participants, utensils and foodstuffs; the acoustic field; and the procedural sequences participants perform when preparing food. Spatial resolution is that of ordinary observation, temporal resolution that of a session. The declared recipe repertoire contains two recipes.

\textbf{Physical hypotheses.} During the lesson, A speaks and demonstrates, while B listens and watches. Afterwards, B can prepare a dish that B could not prepare before. We want to describe two features of this episode physically: the lasting change associated with B's acquired capability, and the continued presence of the bodily structures through which instruction took place.

A sufficiently detailed physical description could describe the relevant brain states, bodily structures and their interactions without calling anything an information carrier or a hereditary operator. We do not have access to that description. Rather than reconstructing individual cells or molecules, we therefore hypothesise a coarser physical description: persistent configurations associated with the participants' capabilities, and the remaining bodily structures involved in speaking, demonstrating, listening and watching.

The question is whether these proposed components can be treated as distinct persistent physical structures, classified independently of hereditary role, and connected through specified causal dependencies. The hypotheses below state what this reconstruction assumes. Their ordinary-language explanations motivate the proposed description; they do not establish that its physical assumptions hold.

\textbf{H1 (factorisation).} At the declared resolution, each participant's state space factorises into a capability component and a remainder component. In ordinary terms: whatever it is about a person that constitutes knowing how to prepare a dish, presumably some configuration of neural connectivity, can be described separately from everything else about that person's body.

\textbf{H2 (individuation).} These capability and remainder components admit physically grounded, persistent, mutually non-overlapping instances in the sense of Section~\ref{sec:3.2} and Definition~\ref{def:1}. In ordinary terms: we assume that the physical structure whose configuration distinguishes what a person knows how to cook can be separated from the remaining bodily structures. Whatever is included in the capability structure must be excluded from the bodily remainder. This is a physical separation we are hypothesising; simply distinguishing ``what someone knows'' from ``the body through which they act'' does not establish it. Describing different properties of the same constituents would not by itself establish two non-overlapping instances. H2 is a substantive individuation hypothesis about the physics of the participants, not a consequence of distinguishing knowledge from bodily machinery.

\textbf{H3 (persistence and structural typing).} The capability instances occupy persistent structural types, individuated by the procedural sequences a participant is capable of performing, which change during instruction; the remainder instances occupy a persistent structural type $m$, unchanged across such a process. In ordinary terms: knowing the recipe is assumed to correspond to a physical configuration that lasts between occasions of cooking. Meanwhile, the bodily remainder continues to belong to the same structural type throughout the lesson. Speaking, listening and moving change its physical state without necessarily changing its type.

\textbf{H4 (bodily contribution).} In an instruction process, the input $m$ instances contribute causally to the output capability instance of the learner. In ordinary terms: the teacher's voice and hands and the learner's ears and eyes are part of what brings the learner's new knowledge about, not bystanders to it.

H4 is required because H3 supplies only conservation. A label conserved across a process satisfies the operator condition only if its instances also contribute causally to an output of another structural type, and that contribution is a separate claim about the episode.

If H1 or H2 fails, the proposed decomposition is inadmissible. This does not establish whether another admissible representation of the episode contains qualifying operators.

\begin{newtext}
\textbf{Further structures and notation.} Both cases type four further kinds of instance that the perspective above declares: apparatus sections, with label $k$, the pots and stove being treated as one persistent structure $K$; ingredient instances, with label $g$ for one declared set of foodstuffs, each consumed by one preparation; dish instances, with labels $y_1$ and $y_2$ for the two dish types, and; utterances, with label $s$ for the acoustic pattern, whose individuation as persistent structures is itself hypothesised (Step 1 below). Each participant X has a capability world-tube $X^{\mathrm{cap}}$ and a remainder world-tube $X^{m}$, and $X_j^{\mathrm{cap}}$ and $X_j^{m}$ denote their $j$-th sections. The apparatus sections are $K_1, K_2, \ldots$; ingredient, dish and utterance instances are single sections $G_j$, $d_j$ and $u_{XY}$. Each instance carries a time index $t$, the start of its section, shown as the vertical coordinate of Figure~\ref{fig:recipe-full}. Two macros give the dependency patterns of the two process classes. A teaching process with teacher sections $x, \mu$, learner sections $z, \nu$ and utterance $u$ contributes
\begin{multline*}
\mathcal{D}_{\mathrm{T}}(x, \mu, z, \nu, u;\, x', \mu', z', \nu') \\
= \{(x, x'), (\mu, \mu'), (z, z'), (\nu, \nu'), (x, u), (\mu, u), (u, z'), (\nu, z')\},
\end{multline*}
i.e., the four sections continue, the utterance depends on the teacher's capability and body, and the learner's new capability section depends on the utterance, on the learner's body (H4) and on the learner's previous capability section. A preparation process with bodily section $\mu$, apparatus section $\kappa$ and ingredient instance $\gamma$ contributes
\[
\mathcal{D}_{\mathrm{P}}(\mu, \kappa, \gamma;\, \mu', \kappa', \delta) = \{(\mu, \mu'), (\kappa, \kappa'), (\mu, \delta), (\kappa, \delta), (\gamma, \delta)\},
\]
i.e., body and apparatus continue and contribute to the dish $\delta$, and the ingredient instance contributes to the dish. Neither macro contains a dependence of the dish on the cook's capability section. Case A withholds the capability-to-dish dependence; including it on suppression grounds (Section~\ref{sec:3.5}) changes no Case A verdict, since the comparative rule, R for $c_2$ and V remain missing.
\end{newtext}

\subsection{\new{Case A: incomplete reconstruction}}\label{app:G.2}

\revnote{[Moved unchanged from v7 Section 6.6. The v7 Stages 1--5 are renamed Steps 1--5, to keep them apart from the six stages of Figure~\ref{fig:pipeline}.]} \new{Notes in square brackets relate the steps below to the formal construction that follows them.}

\paragraph{\del{Stage}\new{Step} 1. A instructs B; B subsequently prepares the dish.}

Under H1 and H2 each participant-section is represented by two nodes. Instruction: $A_1^{\mathrm{cap}}$, $A_1^{m}$, $B_1^{\mathrm{cap}}$, $B_1^{m}$ on the input side; $A_2^{\mathrm{cap}}$, $A_2^{m}$, $B_2^{\mathrm{cap}}$, $B_2^{m}$ on the output side. Labels $\ell(A_1^{\mathrm{cap}}) = \ell(A_2^{\mathrm{cap}}) = c_1$, $\ell(B_1^{\mathrm{cap}}) = c_0$, $\ell(B_2^{\mathrm{cap}}) = c_1$, and $\ell = m$ for all four bodily nodes. Also present are the apparatus, ingredients, the prepared dish, and the acoustic pattern during instruction.

The acoustic node requires a coherence criterion declared independently of the transmission relation it is used to support, since occupying a world-tube for a finite interval does not itself satisfy persistence. Absent such a criterion it is a further hypothesised element.

\emph{Stipulated relations.} The dependency relation $D$ for the instruction process comprises: $A_1^{\mathrm{cap}}\,D\,A_2^{\mathrm{cap}}$ and $B_1^{m}\,D\,B_2^{m}$ and $A_1^{m}\,D\,A_2^{m}$, representing continuation of persisting occurrences; $B_1^{{\new{\mathrm{cap}}}}\,D\,B_2^{{\new{\mathrm{cap}}}}$ representing reconfiguration of $B$'s pre-existing capability component; $A_1^{\mathrm{cap}}\,D\,u$, the utterance produced from $A$'s capability configuration acting through $A$'s bodily apparatus; $A_1^{m}\,D\,u$, the bodily contribution to the utterance; $u\,D\,B_2^{\mathrm{cap}}$ and $B_1^{m}\,D\,B_2^{\mathrm{cap}}$, the acquisition of the capability by $B$ through $B$'s own receiving apparatus. The relation
\[ A_1^{\mathrm{cap}} \prec B_2^{\mathrm{cap}} \]
holds only by transitive closure and is not itself stipulated. H4 is instantiated by the path $A_1^{m}\,D\,u\,D\,B_2^{{\new{\mathrm{cap}}}}$ and the dependence $B_1^{m}\,D\,B_2^{{\new{\mathrm{cap}}}}$.

The additional dependence $B_1^{{\new{\mathrm{cap}}}}\,D\,B_2^{{\new{\mathrm{cap}}}}$ is a physical hypothesis: removing or disrupting B's pre-existing capability component would reduce the probability of the later configuration being realised. It does not assert that lacking the recipe is necessary for learning it. The source is a physical instance typed as $c_0$, not the absence of a structure. Physical continuity alone does not establish this dependence.

The instruction process is
\[ Q = \{A_1^{{\new{\mathrm{cap}}}}, A_1^{m}, B_1^{{\new{\mathrm{cap}}}}, B_1^{m}, u, A_2^{{\new{\mathrm{cap}}}}, A_2^{m}, B_2^{{\new{\mathrm{cap}}}}, B_2^{m}\}. \]
\new{[In the formal construction below, $Q$ is $Q_{AB}$ and $u$ is $u_{AB}$.]}
For the stipulated graph, its boundaries are
\begin{align*}
I(Q) &= \{A_1^{{\new{\mathrm{cap}}}}, A_1^{m}, B_1^{{\new{\mathrm{cap}}}}, B_1^{m}\},\\
O(Q) &= \{A_2^{{\new{\mathrm{cap}}}}, A_2^{m}, B_2^{{\new{\mathrm{cap}}}}, B_2^{m}\},
\end{align*}
with $u$ interior. In particular, $B_1^{{\new{\mathrm{cap}}}}$ is no longer maximal because it precedes $B_2^{{\new{\mathrm{cap}}}}$. Suppression of instruction as a whole would not separately establish every listed dependency; each requires its own causal justification and is stipulated here.

\emph{Replication.} Taking $Q$ as the instruction process,
\[ n^{-}(c_1) = 1 \]
and
\[ n^{+}(c_1) = 2. \]

The output contains two distinct occurrences: $A_2^{\mathrm{cap}}$, the continuation, and $B_2^{\mathrm{cap}}$, which newly acquires the structural type. Temporal subdivision alone would have given no increase, so the increase is genuine. $Q$ satisfies the multiplicity condition for R; its causal condition rests on the stipulated dependence.

\emph{Operator.} Under H3,
\[ n^{-}(m) = n^{+}(m) = 2. \]

Under H4, $m$-instances contribute causally to $B_2^{\mathrm{cap}}$, an output of a different structural type. Structural type $m$ therefore satisfies the operator condition\del{, and $Q$ has a candidate replication operator}\new{ in $Q$}.

\del{The assignment}\new{That $m$ satisfies the operator condition} is derived rather than declared, but \del{its}\new{the} defining conditions are substantially hypothesised, and as Appendix~\ref{app:F.3} notes, deriving a role from hypothesised conditions provides no independent support for the organisation assumed. Note also that $m$ is not privileged. The apparatus, and the physical fabric of the room, would equally satisfy the condition if individuated as persistent non-overlapping structures and shown to contribute causally. A room represented as an encompassing region would overlap its contents and fail non-overlap; a representation of its fabric need not. Background conditions therefore require the same individuation discipline as the participants, and where they satisfy it the operator set is large.

\emph{Not supported by \del{Stage}\new{Step} 1.} $B$ preparing a dish is not $B$ re-entering replication, so no lineage continuation. Only one candidate recipe variant has been transmitted, so no variation between recipe variants or translation comparison is established.

\paragraph{\del{Stage}\new{Step} 2. C, whose capability component has structural type $c_2$, prepares a different dish using the same apparatus.}

The apparatus appears on both boundaries of two preparation processes with unchanged multiplicity, contributing causally to outputs of a different label, and \del{is a candidate translation operator}\new{satisfies the operator condition in both}.

For T, each designated dish type must be absent from its preparation process's input boundary and present on its output boundary. The two processes must also satisfy the specified correspondence of operator organisation; using the same apparatus does not alone establish that correspondence.

This does not establish T. T requires that the variant distinction produce the distinguishable outcomes, and two participants of different structural types producing different dishes is consistent with the difference arising from ingredients, quantities or other conditions. Establishing T requires evidence that the outcome difference depends on the capability variant.

T is moreover evaluated over $c_1$ and $c_2$, whose individuation and classification depend on H1--H3. Observing participants, apparatus and dishes does not remove that dependence. Components of the motif, such as the apparatus contribution and the physical distinguishability of the dishes, may be observationally supported while the complete motif remains conditional on the decomposition.

\paragraph{\del{Stage}\new{Step} 3. B, now in structural type $c_1$, instructs D.}

The reconstruction extends the \del{Stage}\new{Step} 1 hypotheses to this event: D's capability component changes from $c_0$ to $c_1$, with the corresponding continuation, utterance and bodily-contribution relations stipulated as before. The instructor's capability instance is stipulated to continue the particular occurrence acquired by B in the first event. Any intervening sections are linked by continuation dependencies; matching structural labels alone would not establish this connection.

\new{[In the formal construction below, $B_2^{\mathrm{cap}}$ is one section from the end of $Q_{AB}$ to the start of $Q_{BD}$, so that $B_2^{\mathrm{cap}} \in O(Q_{AB}) \cap I(Q_{BD})$, as the re-entry condition of Section~\ref{sec:4.6} requires. With intervening sections, a different instance would lie in $I(Q_{BD})$.]}

Section~\ref{sec:4.6} requires re-entry through corresponding replication-operator organisation with a non-empty corresponding operator set. For the two instruction processes, map the instructor's and learner's capability and bodily boundary instances to their respective counterparts. The extended hypotheses stipulate preservation of structural labels, boundary multiplicities and the required input--output causal incidences, with $m$ as the corresponding operator label. Conditional on these hypotheses and acoustic individuation, the two events satisfy the lineage-continuation requirement.

Background structures may also supply corresponding operators where their boundary multiplicities and causal contributions satisfy the same requirements in both events. Their common presence alone is insufficient.

\paragraph{\del{Stage}\new{Step} 4. A written recipe is present, used by A and by B.}

Two situations must be distinguished. If the document was present throughout and is recognised only now, this is a revised reconstruction of the same episode and Appendix~\ref{app:B.3} applies. If it is introduced into the situation, the physical episode differs.

\new{[The formal construction below does not include Step 4.]}

Persisting does not make the document an operator: its causal contribution and boundary multiplicities must be established in each process, and its representation must satisfy non-overlap.

Adding the document requires re-evaluating the intervention-supported dependencies in $D$ under the revised description and then reconstructing the ancestry order,
\[ \prec = D^{+}. \]
In particular, whether
\[ A_1^{\mathrm{cap}} \prec B_2^{\mathrm{cap}} \]
still holds depends on whether a path of supported dependencies remains. Suppressing $A$'s contribution while leaving the document available may leave $B$'s acquisition unchanged, but that endpoint test alone does not determine whether the ancestry path survives. Each dependency on the proposed path must be assessed under the declared intervention rules, including how alternative routes respond.

\paragraph{\del{Stage}\new{Step} 5. Accounting.}

\begin{table}[!htbp]
\centering
\begin{tabularx}{\textwidth}{>{\raggedright\arraybackslash}p{0.38\textwidth}X}
\toprule
\multicolumn{1}{c}{\textbf{Condition}} & \multicolumn{1}{c}{\textbf{Status in the reconstruction}} \\
\midrule
Capability individuation and classification & Conditional on H1--H3 \\
\midrule
R multiplicity increase for $c_1$ in \del{Stage}\new{Step} 1 & Satisfied under H1--H3 and the stated boundaries \\
\midrule
R causal requirement in \del{Stage}\new{Step} 1 & Conditional on the stipulated dependencies and acoustic individuation \\
\midrule
Replication operator $m$ & Conditional on H1--H4 and the stated boundaries \\
\midrule
Replication of $c_2$ & No witness supplied \\
\midrule
Variation connecting $c_1$ and $c_2$ & No witness supplied \\
\midrule
T product boundaries and operator correspondence & Requirements stated; explicit witnesses remain to be specified \\
\midrule
T dependence of outcomes on the capability variant & Not established \\
\midrule
Lineage continuation & Conditional on H1--H4, acoustic individuation, cross-event continuity and the stipulated relations in both events \\
\bottomrule
\end{tabularx}
\caption{\new{\textbf{Accounting of the v7 reconstruction, i.e., Case A.}}}
\label{tab:c3-sec6-7-1}
\end{table}

No coherent information family is established for $I = \{c_1, c_2\}$. No replication process is supplied for $c_2$, no variation witness connects $c_1$ and $c_2$, and the required T comparison remains unestablished. The learner's reconfiguration from $c_0$ to $c_1$ supplies a generic V witness, but $c_0$ is outside the candidate family and this transition does not connect its two members.

\textbf{What the example shows.} The principal lesson concerns neither observation nor motif detection but individuation. Whether an operator is recoverable at all turns on H2, the hypothesis that capability and bodily configuration admit distinct non-overlapping persistent instances. That is a claim about the physics of participants, and it is prior to every question about evidence for particular relations. Under it, and under the further hypotheses, the reconstruction yields \del{a candidate replication operator}\new{a structural type, $m$, that satisfies the operator condition,} and lineage continuation; without it, this representation yields neither.

The three missing family conditions are of different kinds. The absence of a variant-transition event is a property of the episode as stipulated here, and no further description of \emph{this} episode would supply it. In an actual incompletely observed episode the corresponding absence would establish nothing, since additional observation could reveal a transition not previously resolved. The unevidenced variant-dependence of T, by contrast, is a gap that better evidence about this same episode could close. Distinguishing the two is what an incremental reconstruction makes possible.

\begin{newtext}
\paragraph{Formal construction of Case A.} The CSS of Case A covers Steps 1--3 and C's preparation; Step 4 describes a revised or different episode and is not included. Its instances are
\begin{align*}
W_A = {}& \{A_1^{\mathrm{cap}}, A_2^{\mathrm{cap}}, B_1^{\mathrm{cap}}, B_2^{\mathrm{cap}}, B_3^{\mathrm{cap}}, C_1^{\mathrm{cap}}, D_1^{\mathrm{cap}}, D_2^{\mathrm{cap}}\} \\
& \cup \{A_1^{m}, A_2^{m}, B_1^{m}, \ldots, B_4^{m}, C_1^{m}, C_2^{m}, D_1^{m}, D_2^{m}\} \\
& \cup \{K_1, K_2, K_3, G_1, G_2, d_1, d_2, u_{AB}, u_{BD}\},
\end{align*}
27 instances, with $\ell(A_1^{\mathrm{cap}}) = \ell(A_2^{\mathrm{cap}}) = \ell(B_2^{\mathrm{cap}}) = \ell(B_3^{\mathrm{cap}}) = \ell(D_2^{\mathrm{cap}}) = c_1$, $\ell(B_1^{\mathrm{cap}}) = \ell(D_1^{\mathrm{cap}}) = c_0$, $\ell(C_1^{\mathrm{cap}}) = c_2$, $\ell(X_j^{m}) = m$, $\ell(K_j) = k$, $\ell(G_j) = g$, $\ell(d_1) = y_1$, $\ell(d_2) = y_2$ and $\ell(u_{AB}) = \ell(u_{BD}) = s$. The stipulated dependency relation is
\begin{multline*}
D_A = \mathcal{D}_{\mathrm{T}}(A_1^{\mathrm{cap}}, A_1^{m}, B_1^{\mathrm{cap}}, B_1^{m}, u_{AB};\, A_2^{\mathrm{cap}}, A_2^{m}, B_2^{\mathrm{cap}}, B_2^{m}) \cup \mathcal{D}_{\mathrm{P}}(B_2^{m}, K_1, G_1;\, B_3^{m}, K_2, d_1) \\
\cup \mathcal{D}_{\mathrm{P}}(C_1^{m}, K_2, G_2;\, C_2^{m}, K_3, d_2) \cup \mathcal{D}_{\mathrm{T}}(B_2^{\mathrm{cap}}, B_3^{m}, D_1^{\mathrm{cap}}, D_1^{m}, u_{BD};\, B_3^{\mathrm{cap}}, B_4^{m}, D_2^{\mathrm{cap}}, D_2^{m}),
\end{multline*}
26 pairs. $D_A$ is acyclic, every pair runs from a smaller to a larger time index, and no pair is implied by the others, so the Hasse diagram of ${\prec} = D_A^{+}$ has the same 26 edges (Figure~\ref{fig:recipe-full}, solid). The four processes are the instance sets of the four terms, with $B_2^{\mathrm{cap}}$ added to $Q_{d_1}$ and $C_1^{\mathrm{cap}}$ to $Q_{d_2}$: $Q_{AB}$, $Q_{d_1}$, $Q_{d_2}$ and $Q_{BD}$. Each is causally convex. Their boundaries are those of Table~\ref{tab:recipe-boundaries}, except that in Case A $B_2^{\mathrm{cap}}$ and $C_1^{\mathrm{cap}}$ lie on both boundaries of their preparation processes, since no dependency leaves them within the process.

\paragraph{Verdicts for Case A.} First, $Q_{AB}$ and $Q_{BD}$ are R processes for $c_1$: in each, $n^{-}(c_1) = 1$, $n^{+}(c_1) = 2$, and both $c_1$ outputs descend from the $c_1$ input, e.g., $A_1^{\mathrm{cap}} \prec u_{AB} \prec B_2^{\mathrm{cap}}$. In both, $\mathrm{Op}(Q) = \{m\}$, and the patterns of Appendix~\ref{app:C.1} formed by the $m$ and $c_1$ boundary instances correspond under $A_1^{\mathrm{cap}} \mapsto B_2^{\mathrm{cap}}$, $A_2^{\mathrm{cap}} \mapsto B_3^{\mathrm{cap}}$, $B_2^{\mathrm{cap}} \mapsto D_2^{\mathrm{cap}}$, $A_1^{m} \mapsto B_3^{m}$, $A_2^{m} \mapsto B_4^{m}$, $B_1^{m} \mapsto D_1^{m}$, $B_2^{m} \mapsto D_2^{m}$. Second, $A_1^{\mathrm{cap}} \rightsquigarrow B_2^{\mathrm{cap}}$ in $Q_{AB}$ and $B_2^{\mathrm{cap}} \rightsquigarrow D_2^{\mathrm{cap}}$ in $Q_{BD}$ are replication transitions, $B_2^{\mathrm{cap}} \in O(Q_{AB}) \cap I(Q_{BD})$, and the lineage depth for $c_1$ is $L = 2$; the continuations $A_1^{\mathrm{cap}} \to A_2^{\mathrm{cap}}$ and $B_2^{\mathrm{cap}} \to B_3^{\mathrm{cap}}$ are not transitions (Section~\ref{sec:4.6}). Third, the candidate T pair $(Q_{d_1}, Q_{d_2})$ meets bullets 1--3, 5 and 6 of Section~\ref{sec:4.4}: $c_1$ and $c_2$ are inputs; $\mathrm{Op}(Q_{d_1}) = \mathrm{Op}(Q_{d_2}) = \{k, m\}$ with corresponding patterns; $n^{-}(y_k) = 0$ and $n^{+}(y_k) = 1$, and; $y_1 \neq y_2$. The pair fails bullets 4 and 7: no capability input precedes the dish, and no comparative rule is stipulated. Fourth, three whole-CSS facts extend the negative verdicts from the declared processes to every process of the Case A CSS. No pair $x \prec y$ has the labels $c_1$ and $c_2$, in either order, so no process is a V process between $c_1$ and $c_2$. No pair $x \prec y$ has both labels $c_2$, so no process is an R process for $c_2$. $C_1^{\mathrm{cap}}$ has no causal descendant, so in no process does a $c_2$ instance contribute to a product. No information family is assigned for $I = \{c_1, c_2\}$. The three facts concern the stipulated description: an actual episode observed more completely could contain relations that this description lacks.
\end{newtext}

\subsection{\new{Case B: extended reconstruction}}\label{app:G.3}

\begin{newtext}
\textbf{Added hypotheses.} Case B keeps $W_A$, $\ell$ on $W_A$ and $D_A$, and adds four stipulated hypotheses. H1--H4 apply unchanged to the added participants E and F.

\emph{H5 (variation by replacement).} After teaching D, B modifies the recipe. B's capability section $B_3^{\mathrm{cap}}$ is followed by a section $B_4^{\mathrm{cap}}$ of type $c_2$, which depends on $B_3^{\mathrm{cap}}$ and on B's bodily activity $B_4^{m}$. The modification is read as replacement: B's capability set after the modification contains the modified recipe only, so $\ell(B_4^{\mathrm{cap}}) = c_2$. Acquisition of an additional recipe would give $\ell(B_4^{\mathrm{cap}}) = c_{12}$, a different structural type with different consequences (Appendix~\ref{app:G.5}); the reconstruction uses the replacement reading throughout. That $B_4^{\mathrm{cap}}$ has the type of C's capability, $c_2$, is an idealised convergence and is stipulated.

\emph{H6 (replication of $c_2$).} B teaches the modified recipe to E, and E then teaches it to F. Both teaching processes have the dependency pattern $\mathcal{D}_{\mathrm{T}}$ of Step 1. E's acquired section $E_2^{\mathrm{cap}}$ is a single section from the end of the first process to the start of the second, so that $E_2^{\mathrm{cap}}$ is the teacher's capability input of the second.

\emph{H7 (preparation with $c_2$).} After the modification and before teaching E, B prepares a dish $d_3$ with the apparatus $K$ and an ingredient instance $G_3$ of the declared type $g$, i.e., with the same apparatus, bodily and ingredient types as in $Q_{d_1}$.

\emph{H8 (comparative rule).} With the apparatus type $k$, the bodily type $m$ and the ingredient type $g$ fixed, the structural type of the cook's capability section fixes the dish type: $F(c_1; k, m, g) = y_1$ and $F(c_2; k, m, g) = y_2$. Equivalently, an intervention that replaces the $c_1$ section in $Q_{d_1}$ by a $c_2$ section, all else corresponding, changes the realised dish type from $y_1$ to $y_2$. Because such a replacement changes whether the dish realises its type, the intervention criterion of Section~\ref{sec:3.5} makes each cook's capability section a cause of the dish in every preparation process to which the rule applies. Case B therefore contains the dependencies $(B_2^{\mathrm{cap}}, d_1)$, $(C_1^{\mathrm{cap}}, d_2)$ and $(B_4^{\mathrm{cap}}, d_3)$; the first two are the only pairs that Case B adds between instances of Case A. The rule is consistent with every realised preparation, since $F$ gives $y_1$ for $Q_{d_1}$ and $y_2$ for $Q_{d_2}$ and $Q_{d_3}$.

\textbf{Instances and dependencies.} Case B adds 20 instances: $B_4^{\mathrm{cap}}, B_5^{\mathrm{cap}}$ (label $c_2$); $E_1^{\mathrm{cap}}, F_1^{\mathrm{cap}}$ ($c_0$); $E_2^{\mathrm{cap}}, E_3^{\mathrm{cap}}, F_2^{\mathrm{cap}}$ ($c_2$); $B_5^{m}, B_6^{m}, B_7^{m}, E_1^{m}, E_2^{m}, E_3^{m}, F_1^{m}, F_2^{m}$ ($m$); $K_4$ ($k$); $G_3$ ($g$); $d_3$ ($y_2$), and; $u_{BE}, u_{EF}$ ($s$). The dependency relation is
\begin{multline*}
D_B = D_A \cup \{(B_2^{\mathrm{cap}}, d_1), (C_1^{\mathrm{cap}}, d_2)\} \cup \{(B_3^{\mathrm{cap}}, B_4^{\mathrm{cap}}), (B_4^{m}, B_5^{m}), (B_4^{m}, B_4^{\mathrm{cap}})\} \\
\cup \mathcal{D}_{\mathrm{P}}(B_5^{m}, K_3, G_3;\, B_6^{m}, K_4, d_3) \cup \{(B_4^{\mathrm{cap}}, d_3)\} \cup \mathcal{D}_{\mathrm{T}}(B_4^{\mathrm{cap}}, B_6^{m}, E_1^{\mathrm{cap}}, E_1^{m}, u_{BE};\, B_5^{\mathrm{cap}}, B_7^{m}, E_2^{\mathrm{cap}}, E_2^{m}) \\
\cup \mathcal{D}_{\mathrm{T}}(E_2^{\mathrm{cap}}, E_2^{m}, F_1^{\mathrm{cap}}, F_1^{m}, u_{EF};\, E_3^{\mathrm{cap}}, E_3^{m}, F_2^{\mathrm{cap}}, F_2^{m}),
\end{multline*}
53 pairs on 47 instances. $D_B$ is acyclic and respects temporal precedence, and again no pair is implied by the others, so the Hasse diagram of $D_B^{+}$ has 53 edges. Every Hasse edge of Case A is a Hasse edge of Case B, so Figure~\ref{fig:recipe-full} draws both diagrams at once.

\textbf{Processes and boundaries.} Case B adds four processes: $Q_{\mathrm{mod}} = \{B_3^{\mathrm{cap}}, B_4^{m}, B_4^{\mathrm{cap}}, B_5^{m}\}$; $Q_{d_3}$, the instances of its $\mathcal{D}_{\mathrm{P}}$ term together with $B_4^{\mathrm{cap}}$, and; $Q_{BE}$ and $Q_{EF}$, the instances of the two $\mathcal{D}_{\mathrm{T}}$ terms. All eight processes are causally convex. The added dependencies do not break the convexity of the Case A processes: $B_2^{\mathrm{cap}}$ now precedes $d_1$, but $d_1$ precedes no instance of $Q_{BD}$. Table~\ref{tab:recipe-boundaries} lists the boundaries, the boundary multiplicities and $\mathrm{Op}(Q)$ for every process. A capability section may serve as input to two processes, e.g., $B_4^{\mathrm{cap}} \in I(Q_{d_3}) \cap I(Q_{BE})$, because a preparation process leaves the capability section that it uses unsplit.
\end{newtext}

\begin{table}[tbp]
\begin{newtext}
\centering
\footnotesize
\setlength{\tabcolsep}{3pt}
\begin{tabularx}{\textwidth}{l>{\raggedright\arraybackslash}X>{\raggedright\arraybackslash}X>{\raggedright\arraybackslash}p{2.35cm}>{\raggedright\arraybackslash}p{2.0cm}l}
\toprule
$Q$ & $I(Q)$ & $O(Q)$ & $n^{-}_{Q}$ by label & $n^{+}_{Q}$ by label & $\mathrm{Op}(Q)$ \\
\midrule
$Q_{AB}$ & $\{A_{1}^{\mathrm{cap}}, B_{1}^{\mathrm{cap}}, A_{1}^{m}, B_{1}^{m}\}$ & $\{A_{2}^{\mathrm{cap}}, B_{2}^{\mathrm{cap}}, A_{2}^{m}, B_{2}^{m}\}$ & $c_0{:}\,1, c_1{:}\,1, m{:}\,2$ & $c_1{:}\,2, m{:}\,2$ & $\{m\}$ \\
$Q_{d_1}$ & $\{B_{2}^{\mathrm{cap}}, B_{2}^{m}, K_{1}, G_{1}\}$ & $\{B_{3}^{m}, K_{2}, d_{1}\}$ & $c_1{:}\,1, g{:}\,1, k{:}\,1, m{:}\,1$ & $k{:}\,1, m{:}\,1, y_1{:}\,1$ & $\{k, m\}$ \\
$Q_{d_2}$ & $\{C_{1}^{\mathrm{cap}}, C_{1}^{m}, K_{2}, G_{2}\}$ & $\{C_{2}^{m}, K_{3}, d_{2}\}$ & $c_2{:}\,1, g{:}\,1, k{:}\,1, m{:}\,1$ & $k{:}\,1, m{:}\,1, y_2{:}\,1$ & $\{k, m\}$ \\
$Q_{BD}$ & $\{B_{2}^{\mathrm{cap}}, D_{1}^{\mathrm{cap}}, B_{3}^{m}, D_{1}^{m}\}$ & $\{B_{3}^{\mathrm{cap}}, D_{2}^{\mathrm{cap}}, B_{4}^{m}, D_{2}^{m}\}$ & $c_0{:}\,1, c_1{:}\,1, m{:}\,2$ & $c_1{:}\,2, m{:}\,2$ & $\{m\}$ \\
$Q_{\mathrm{mod}}$ & $\{B_{3}^{\mathrm{cap}}, B_{4}^{m}\}$ & $\{B_{4}^{\mathrm{cap}}, B_{5}^{m}\}$ & $c_1{:}\,1, m{:}\,1$ & $c_2{:}\,1, m{:}\,1$ & $\{m\}$ \\
$Q_{d_3}$ & $\{B_{4}^{\mathrm{cap}}, B_{5}^{m}, K_{3}, G_{3}\}$ & $\{B_{6}^{m}, K_{4}, d_{3}\}$ & $c_2{:}\,1, g{:}\,1, k{:}\,1, m{:}\,1$ & $k{:}\,1, m{:}\,1, y_2{:}\,1$ & $\{k, m\}$ \\
$Q_{BE}$ & $\{B_{4}^{\mathrm{cap}}, E_{1}^{\mathrm{cap}}, B_{6}^{m}, E_{1}^{m}\}$ & $\{B_{5}^{\mathrm{cap}}, E_{2}^{\mathrm{cap}}, B_{7}^{m}, E_{2}^{m}\}$ & $c_0{:}\,1, c_2{:}\,1, m{:}\,2$ & $c_2{:}\,2, m{:}\,2$ & $\{m\}$ \\
$Q_{EF}$ & $\{E_{2}^{\mathrm{cap}}, F_{1}^{\mathrm{cap}}, E_{2}^{m}, F_{1}^{m}\}$ & $\{E_{3}^{\mathrm{cap}}, F_{2}^{\mathrm{cap}}, E_{3}^{m}, F_{2}^{m}\}$ & $c_0{:}\,1, c_2{:}\,1, m{:}\,2$ & $c_2{:}\,2, m{:}\,2$ & $\{m\}$ \\
\bottomrule
\end{tabularx}
\caption{\textbf{Processes of Case B: input and output boundaries, boundary multiplicities and operator labels.} Labels with zero multiplicity are omitted. In Case A, $Q_{AB}$ and $Q_{BD}$ are identical; in $Q_{d_1}$ and $Q_{d_2}$, $B_2^{\mathrm{cap}}$ and $C_1^{\mathrm{cap}}$ also lie in $O(Q)$, which adds $c_1{:}\,1$ and $c_2{:}\,1$ to $n^{+}$, and $\mathrm{Op}(Q)$ is unchanged.}
\label{tab:recipe-boundaries}
\end{newtext}
\end{table}

\begin{figure}[tbp]
\centering
\begin{tikzpicture}[cssfig, x=1.05cm, y=0.72cm]
\node[text=black] at (0.00,-1.0) {$C^{\mathrm{cap}}$};
\node[text=black] at (0.85,-1.0) {$C^{m}$};
\node[text=black] at (2.55,-1.0) {$K$};
\node[text=black] at (4.25,-1.0) {$B^{m}$};
\node[text=black] at (5.10,-1.0) {$B^{\mathrm{cap}}$};
\node[text=black] at (6.80,-1.0) {$A^{\mathrm{cap}}$};
\node[text=black] at (7.65,-1.0) {$A^{m}$};
\node[text=black] at (8.50,-1.0) {$D^{\mathrm{cap}}$};
\node[text=black] at (9.35,-1.0) {$D^{m}$};
\node[text=black] at (10.20,-1.0) {$E^{\mathrm{cap}}$};
\node[text=black] at (11.05,-1.0) {$E^{m}$};
\node[text=black] at (12.75,-1.0) {$F^{\mathrm{cap}}$};
\node[text=black] at (13.60,-1.0) {$F^{m}$};
\node[text=black] at (-0.85,13.0) {$t$};
\node[text=black!70] at (-0.85,0) {0};
\node[text=black!70] at (-0.85,1) {1};
\node[text=black!70] at (-0.85,2) {2};
\node[text=black!70] at (-0.85,3) {3};
\node[text=black!70] at (-0.85,4) {4};
\node[text=black!70] at (-0.85,5) {5};
\node[text=black!70] at (-0.85,6) {6};
\node[text=black!70] at (-0.85,7) {7};
\node[text=black!70] at (-0.85,8) {8};
\node[text=black!70] at (-0.85,9) {9};
\node[text=black!70] at (-0.85,10) {10};
\node[text=black!70] at (-0.85,11) {11};
\node[text=black!70] at (-0.85,12) {12};
\node[cssnode, cssfill=cssOI1!55!white] (A1c) at (6.80,0.00) {$c_1$};
\node[cssnode, cssfill=cssOI1!55!white] (A2c) at (6.80,2.00) {$c_1$};
\node[cssnode, cssfill=white] (B1c) at (5.10,0.00) {$c_0$};
\node[cssnode, cssfill=cssOI1!55!white] (B2c) at (5.10,2.00) {$c_1$};
\node[cssnode, cssfill=cssOI1!55!white] (B3c) at (5.10,6.00) {$c_1$};
\node[cssnode, cssfill=cssOI2!55!white] (C1c) at (0.00,2.00) {$c_2$};
\node[cssnode, cssfill=white] (D1c) at (8.50,4.00) {$c_0$};
\node[cssnode, cssfill=cssOI1!55!white] (D2c) at (8.50,6.00) {$c_1$};
\node[cssnode, cssfill=cssGrey!45!white] (A1m) at (7.65,0.00) {$m$};
\node[cssnode, cssfill=cssGrey!45!white] (A2m) at (7.65,2.00) {$m$};
\node[cssnode, cssfill=cssGrey!45!white] (B1m) at (4.25,0.00) {$m$};
\node[cssnode, cssfill=cssGrey!45!white] (B2m) at (4.25,2.00) {$m$};
\node[cssnode, cssfill=cssGrey!45!white] (B3m) at (4.25,3.00) {$m$};
\node[cssnode, cssfill=cssGrey!45!white] (B4m) at (4.25,6.00) {$m$};
\node[cssnode, cssfill=cssGrey!45!white] (C1m) at (0.85,2.00) {$m$};
\node[cssnode, cssfill=cssGrey!45!white] (C2m) at (0.85,4.00) {$m$};
\node[cssnode, cssfill=cssGrey!45!white] (D1m) at (9.35,4.00) {$m$};
\node[cssnode, cssfill=cssGrey!45!white] (D2m) at (9.35,6.00) {$m$};
\node[cssnode, cssfill=cssOI6!60!white] (K1) at (2.55,2.00) {$k$};
\node[cssnode, cssfill=cssOI6!60!white] (K2) at (2.55,3.00) {$k$};
\node[cssnode, cssfill=cssOI6!60!white] (K3) at (2.55,4.00) {$k$};
\node[cssnode, cssfill=cssOI7!35!white] (G1) at (3.40,1.00) {$g$};
\node[csstag, anchor=south west] at (G1.north east) {$G_{1}$};
\node[cssnode, cssfill=cssOI7!35!white] (G2) at (1.70,2.00) {$g$};
\node[csstag, anchor=south west] at (G2.north east) {$G_{2}$};
\node[cssnode, cssfill=cssOI4!55!white] (d1) at (3.40,3.00) {$y_1$};
\node[csstag, anchor=south west] at (d1.north east) {$d_{1}$};
\node[cssnode, cssfill=cssOI3!50!white] (d2) at (1.70,4.00) {$y_2$};
\node[csstag, anchor=south west] at (d2.north east) {$d_{2}$};
\node[cssnode, cssfill=cssOI5!25!white] (uAB) at (5.95,1.00) {$s$};
\node[csstag, anchor=south west] at (uAB.north east) {$u_{AB}$};
\node[cssnode, cssfill=cssOI5!25!white] (uBD) at (5.95,5.00) {$s$};
\node[csstag, anchor=south east] at (uBD.north west) {$u_{BD}$};
\node[cssnode, cssfill=cssOI2!55!white, cssadded] (B4c) at (5.10,7.00) {$c_2$};
\node[cssnode, cssfill=cssOI2!55!white, cssadded] (B5c) at (5.10,10.00) {$c_2$};
\node[cssnode, cssfill=white, cssadded] (E1c) at (10.20,7.00) {$c_0$};
\node[cssnode, cssfill=cssOI2!55!white, cssadded] (E2c) at (10.20,10.00) {$c_2$};
\node[cssnode, cssfill=cssOI2!55!white, cssadded] (E3c) at (10.20,12.00) {$c_2$};
\node[cssnode, cssfill=white, cssadded] (F1c) at (12.75,10.00) {$c_0$};
\node[cssnode, cssfill=cssOI2!55!white, cssadded] (F2c) at (12.75,12.00) {$c_2$};
\node[cssnode, cssfill=cssGrey!45!white, cssadded] (B5m) at (4.25,7.00) {$m$};
\node[cssnode, cssfill=cssGrey!45!white, cssadded] (B6m) at (4.25,8.00) {$m$};
\node[cssnode, cssfill=cssGrey!45!white, cssadded] (B7m) at (4.25,10.00) {$m$};
\node[cssnode, cssfill=cssGrey!45!white, cssadded] (E1m) at (11.05,7.00) {$m$};
\node[cssnode, cssfill=cssGrey!45!white, cssadded] (E2m) at (11.05,10.00) {$m$};
\node[cssnode, cssfill=cssGrey!45!white, cssadded] (E3m) at (11.05,12.00) {$m$};
\node[cssnode, cssfill=cssGrey!45!white, cssadded] (F1m) at (13.60,10.00) {$m$};
\node[cssnode, cssfill=cssGrey!45!white, cssadded] (F2m) at (13.60,12.00) {$m$};
\node[cssnode, cssfill=cssOI6!60!white, cssadded] (K4) at (2.55,8.00) {$k$};
\node[cssnode, cssfill=cssOI7!35!white, cssadded] (G3) at (3.40,6.00) {$g$};
\node[csstag, anchor=south west] at (G3.north east) {$G_{3}$};
\node[cssnode, cssfill=cssOI3!50!white, cssadded] (d3) at (3.40,8.00) {$y_2$};
\node[csstag, anchor=south west] at (d3.north east) {$d_{3}$};
\node[cssnode, cssfill=cssOI5!25!white, cssadded] (uBE) at (5.95,9.00) {$s$};
\node[csstag, anchor=south west] at (uBE.north east) {$u_{BE}$};
\node[cssnode, cssfill=cssOI5!25!white, cssadded] (uEF) at (11.90,11.00) {$s$};
\node[csstag, anchor=south east] at (uEF.north west) {$u_{EF}$};
\draw[csscont] (A1c) -- (A2c);
\draw[cssdep] (A1c) -- (uAB);
\draw[csscont] (A1m) -- (A2m);
\draw[cssdep] (A1m) -- (uAB);
\draw[csscont] (B1c) -- (B2c);
\draw[cssdep] (B1m) -- (B2c);
\draw[csscont] (B1m) -- (B2m);
\draw[csscont] (B2c) -- (B3c);
\draw[cssdep, densely dashed] (B2c) -- (d1);
\draw[cssdep] (B2c) -- (uBD);
\draw[csscont] (B2m) -- (B3m);
\draw[cssdep] (B2m) -- (d1);
\draw[csscont, densely dashed] (B3c) -- (B4c);
\draw[csscont] (B3m) -- (B4m);
\draw[cssdep] (B3m) -- (uBD);
\draw[csscont, densely dashed] (B4c) -- (B5c);
\draw[cssdep, densely dashed] (B4c) -- (d3);
\draw[cssdep, densely dashed] (B4c) -- (uBE);
\draw[cssdep, densely dashed] (B4m) -- (B4c);
\draw[csscont, densely dashed] (B4m) -- (B5m);
\draw[csscont, densely dashed] (B5m) -- (B6m);
\draw[cssdep, densely dashed] (B5m) -- (d3);
\draw[csscont, densely dashed] (B6m) -- (B7m);
\draw[cssdep, densely dashed] (B6m) -- (uBE);
\draw[cssdep, densely dashed] (C1c) -- (d2);
\draw[csscont] (C1m) -- (C2m);
\draw[cssdep] (C1m) -- (d2);
\draw[csscont] (D1c) -- (D2c);
\draw[cssdep] (D1m) -- (D2c);
\draw[csscont] (D1m) -- (D2m);
\draw[csscont, densely dashed] (E1c) -- (E2c);
\draw[cssdep, densely dashed] (E1m) -- (E2c);
\draw[csscont, densely dashed] (E1m) -- (E2m);
\draw[csscont, densely dashed] (E2c) -- (E3c);
\draw[cssdep, densely dashed] (E2c) -- (uEF);
\draw[csscont, densely dashed] (E2m) -- (E3m);
\draw[cssdep, densely dashed] (E2m) -- (uEF);
\draw[csscont, densely dashed] (F1c) -- (F2c);
\draw[cssdep, densely dashed] (F1m) -- (F2c);
\draw[csscont, densely dashed] (F1m) -- (F2m);
\draw[cssdep] (G1) -- (d1);
\draw[cssdep] (G2) -- (d2);
\draw[cssdep, densely dashed] (G3) -- (d3);
\draw[csscont] (K1) -- (K2);
\draw[cssdep] (K1) -- (d1);
\draw[csscont] (K2) -- (K3);
\draw[cssdep] (K2) -- (d2);
\draw[csscont, densely dashed] (K3) -- (K4);
\draw[cssdep, densely dashed] (K3) -- (d3);
\draw[cssdep] (uAB) -- (B2c);
\draw[cssdep] (uBD) -- (D2c);
\draw[cssdep, densely dashed] (uBE) -- (E2c);
\draw[cssdep, densely dashed] (uEF) -- (F2c);
\end{tikzpicture}%
\caption{\new{\textbf{Complete Hasse diagrams of the CSSs of Case A and Case B.} Case A consists of the solid nodes and edges (27 instances, 26 edges); Case B consists of all nodes and edges (47 instances, 53 edges), the dashed ones being added by H5--H8. Every Hasse edge of Case A is a Hasse edge of Case B, and in both cases every Hasse edge is a stipulated direct dependency. Labelled columns are world-tubes, with sections numbered $1, 2, \ldots$ from the bottom; unlabelled columns collect ingredient, dish and utterance instances; grey bars join successive sections of one world-tube, and arrows join instances of different world-tubes. The vertical coordinate is the time index $t$ of a section's start, and time increases upward; every edge points to a larger $t$. Colour and printed symbol give the structural label only: $c_0$, $c_1$, $c_2$ capability types; $m$ bodily remainder; $k$ apparatus; $g$ ingredients; $s$ acoustic pattern; $y_1$, $y_2$ dish types. Both diagrams are conditional on H1--H4, and Case B also on H5--H8.}}
\label{fig:recipe-full}
\end{figure}

\begin{newtext}
\textbf{Verdicts for Case B.} The joint criterion of Section~\ref{sec:5} holds for $I = \{c_1, c_2\}$, condition by condition.
\begin{enumerate}[label=\arabic*), leftmargin=*]
  \item Variation connectivity. $Q_{\mathrm{mod}}$ is a V process from $c_1$ to $c_2$: $B_3^{\mathrm{cap}} \in I(Q_{\mathrm{mod}})$, $B_4^{\mathrm{cap}} \in O(Q_{\mathrm{mod}})$ and $B_3^{\mathrm{cap}} \prec B_4^{\mathrm{cap}}$.
  \item Replication. $Q_{AB}$ and $Q_{BD}$ are R processes for $c_1$, as in Case A; $Q_{BE}$ and $Q_{EF}$ are R processes for $c_2$, with $n^{-}(c_2) = 1$, $n^{+}(c_2) = 2$ and both $c_2$ outputs descending from the $c_2$ input.
  \item Shared replication organisation. All four processes have $\mathrm{Op}(Q) = \{m\}$, and their patterns correspond pairwise with operator label $m$; e.g., $Q_{AB}$ maps to $Q_{BE}$ under $A_1^{\mathrm{cap}} \mapsto B_4^{\mathrm{cap}}$, $A_2^{\mathrm{cap}} \mapsto B_5^{\mathrm{cap}}$, $B_2^{\mathrm{cap}} \mapsto E_2^{\mathrm{cap}}$, $A_1^{m} \mapsto B_6^{m}$, $A_2^{m} \mapsto B_7^{m}$, $B_1^{m} \mapsto E_1^{m}$, $B_2^{m} \mapsto E_2^{m}$ and the type map $c_1 \mapsto c_2$.
  \item Translation. In $Q_{d_1}$ and $Q_{d_3}$, $c_1$ and $c_2$ are inputs, $B_2^{\mathrm{cap}} \prec d_1$ and $B_4^{\mathrm{cap}} \prec d_3$, $n^{-}(y_k) = 0$ and $n^{+}(y_k) = 1$, and $y_1 \neq y_2$.
  \item Shared translation organisation. $\mathrm{Op}(Q_{d_1}) = \mathrm{Op}(Q_{d_3}) = \{k, m\}$, and the patterns correspond under $B_2^{\mathrm{cap}} \mapsto B_4^{\mathrm{cap}}$, $d_1 \mapsto d_3$, $B_2^{m} \mapsto B_5^{m}$, $B_3^{m} \mapsto B_6^{m}$, $K_1 \mapsto K_3$, $K_2 \mapsto K_4$ and the type map $c_1 \mapsto c_2$, $y_1 \mapsto y_2$.
  \item Variant-dependent downstream realisation. The two processes have the same \del{non-carrier input types}\new{input types apart from the information carriers}, $\{g, k, m\}$, and H8 gives $F(c_1; k, m, g) = y_1$ and $F(c_2; k, m, g) = y_2$, the realised product types.
  \item Observed lineage continuation. $B_4^{\mathrm{cap}} \rightsquigarrow E_2^{\mathrm{cap}}$ in $Q_{BE}$ and $E_2^{\mathrm{cap}} \rightsquigarrow F_2^{\mathrm{cap}}$ in $Q_{EF}$, with $E_2^{\mathrm{cap}} \in O(Q_{BE}) \cap I(Q_{EF})$, give $L = 2$ for $c_2$; together with $L = 2$ for $c_1$, $L(C_B, I) = 2$.
\end{enumerate}
The pair $(Q_{d_1}, Q_{d_2})$ also satisfies every T condition under H8. We designate $(Q_{d_1}, Q_{d_3})$ because its $c_2$ \new{information }carrier instance, $B_4^{\mathrm{cap}}$, belongs to the $c_2$ lineage and enters replication in $Q_{BE}$, which keeps the variant identity of that lineage associated with its T relation (Section~\ref{sec:4.6}). Under the Case B description, $c_1$ and $c_2$ occupy information-carrier roles, $m$ replication-operator and translation-operator roles, $k$ a translation-operator role, and $y_1$ and $y_2$ realised-product roles.
\end{newtext}

\subsection{\new{Comparative formal verification}}\label{app:G.4}

\begin{newtext}
Table~\ref{tab:recipe-verification} sets the verdicts side by side, with the witnesses and the added hypothesis on which each change rests. To identify exactly which additional hypotheses change the result, we withdrew each of H5--H8 from Case B in turn and evaluated the remaining description with the same criterion. Withdrawing H5, i.e., removing the dependence of $B_4^{\mathrm{cap}}$ on $B_3^{\mathrm{cap}}$, leaves no pair $x \prec y$ with labels $c_1$ and $c_2$ in the whole CSS, so variation connectivity fails (withdrawing H5 entirely removes every $c_2$ lineage and gives the same verdict). Withdrawing H6 leaves no R process for $c_2$. Withdrawing H8 removes the comparative rule together with the capability--dish dependencies it entails, so T fails. In each of the three cases no information family is assigned. Withdrawing H7 leaves the family assigned, because under H8 the preparation $Q_{d_2}$ by C pairs with $Q_{d_1}$ as a T witness. The $c_2$ \new{information }carrier of $Q_{d_2}$, $C_1^{\mathrm{cap}}$, belongs to no lineage, so this verdict holds under the reading of Section~\ref{sec:4.6} that associates a translation witness with a lineage by its structural label. H5, H6 and H8 are therefore each necessary for the assignment in Case B, and, under that reading, H7 is not; H7 makes the $c_2$ translation witness a member of the $c_2$ lineage.
\end{newtext}

\paragraph{\new{Synthetic example of Section~\ref{sec:6.2}.}} \new{The same script checks the positive control of Section~\ref{sec:6.2}. It reproduces the boundaries printed below, R for $c_1$ in $Q_1$ and $Q_2$ and for $c_2$ in $Q_4$ with operator label $c_3$, V in $Q_3$, T in $Q_5$ and $Q_6$ with operator label $c_4$, and $L = 2$, so that the family $\{c_1, c_2\}$ is assigned. The specification follows.}

\revnote{[Moved unchanged from v7 Section 6.2.]}

\textbf{Instances and labels.} Let
\[ W = \{a_0, \ldots, a_4, b_0, b_1, b_2, p_0, \ldots, p_3, q_0, q_1, q_2, u, v\}, \]
with
\[ \ell(a_j) = c_1, \ \ell(b_j) = c_2, \ \ell(p_j) = c_3, \ \ell(q_j) = c_4, \ \ell(u) = c_5, \ \ell(v) = c_6. \]
For physical identity, $a_1$ continues $a_0$, $a_3$ continues $a_2$, and $b_1$ continues $b_0$. The occurrences $a_2$, $a_4$ and $b_2$ are newly produced, distinct physical occurrences. The transition from $a_1$ to $b_0$ changes structural type. Successive $p_j$ and $q_j$ represent temporal sections of their respective persisting structures.

\textbf{Dependencies.} For compactness, define
\[ \mathcal{D}(g, x; g', y, z) = \{(g, g'), (g, y), (g, z), (x, y), (x, z)\}. \]
The stipulated dependency relation is
\begin{multline*}
D = \mathcal{D}(p_0, a_0; p_1, a_1, a_2) \cup \mathcal{D}(p_1, a_2; p_2, a_3, a_4) \cup \mathcal{D}(p_2, b_0; p_3, b_1, b_2) \\
\cup \{(a_1, b_0), (q_0, q_1), (q_0, u), (a_4, u), (q_1, q_2), (q_1, v), (b_2, v)\}.
\end{multline*}

Set $\prec$ the transitive closure of $D$: ${\prec} = D^{+}$.

The relation is acyclic.

\textbf{Processes and boundaries.} Define each $Q_j$ as the union of the input and output sets listed below. Each subset is causally convex, and its minimal and maximal elements give the stated boundaries.
\begin{align*}
I(Q_1) &= \{p_0, a_0\}, & O(Q_1) &= \{p_1, a_1, a_2\}\\
I(Q_2) &= \{p_1, a_2\}, & O(Q_2) &= \{p_2, a_3, a_4\}\\
I(Q_3) &= \{a_1\}, & O(Q_3) &= \{b_0\}\\
I(Q_4) &= \{p_2, b_0\}, & O(Q_4) &= \{p_3, b_1, b_2\}\\
I(Q_5) &= \{q_0, a_4\}, & O(Q_5) &= \{q_1, u\}\\
I(Q_6) &= \{q_1, b_2\}, & O(Q_6) &= \{q_2, v\}
\end{align*}

\textbf{Comparative causal specification.} The hypothetical production dynamics, with other conditions fixed, obey the product-type rule
\[ F(c_1, c_4) = c_5, \ F(c_2, c_4) = c_6. \]
Both cases are realised, in $Q_5$ and $Q_6$ respectively. This stipulated rule supplies the variant-dependent comparison required by T; the ancestry relations alone would not establish it.

\textbf{Joint evaluation.} Processes $Q_1$ and $Q_2$ satisfy R for $c_1$, and $Q_4$ satisfies R for $c_2$: each has \new{information }carrier multiplicity
\[ 1 \to 2, \]
with causal production of an additional physical occurrence. The label $c_3$ has multiplicity
\[ 1 \to 1 \]
and contributes to the outputs carrying $c_1$ or $c_2$. Its boundary incidences correspond across all three processes, \del{so}\new{and} it satisfies the operator condition in each.

Process $Q_3$ supplies
\[ V : c_1 \to c_2, \]
making the candidate family
\[ I = \{c_1, c_2\} \]
weakly connected.

Processes $Q_5$ and $Q_6$ satisfy T. Their shared label $c_4$ is conserved and contributes to outputs of other labels, the product labels differ, and
\[ n^{-}_{Q_5}(c_5) = n^{-}_{Q_6}(c_6) = 0, \ n^{+}_{Q_5}(c_5) = n^{+}_{Q_6}(c_6) = 1. \]
\del{Carrier}\new{Information carrier} output multiplicity is zero in these processes, which T permits.

Finally,
\[ a_0 \rightsquigarrow a_2 \rightsquigarrow a_4 \]
is a lineage of two replication transitions: the additional occurrence $a_2$ produced by $Q_1$ re-enters $Q_2$. Thus
\[ L = 2. \]

\begin{table}[tbp]
\begin{newtext}
\centering
\footnotesize
\begin{tabularx}{\textwidth}{>{\raggedright\arraybackslash}p{0.2\textwidth}>{\raggedright\arraybackslash}X>{\raggedright\arraybackslash}X>{\raggedright\arraybackslash}p{0.12\textwidth}}
\toprule
\textbf{Condition} & \textbf{Case A} & \textbf{Case B} & \textbf{Changed by} \\
\midrule
Well-formedness & 27 instances, 26 dependencies; acyclic, time-respecting; 4 processes convex & 47 instances, 53 dependencies; acyclic, time-respecting; 8 processes convex & -- \\
R for $c_1$ & $Q_{AB}$, $Q_{BD}$: $1 \to 2$ & $Q_{AB}$, $Q_{BD}$: $1 \to 2$ & -- \\
R for $c_2$ & None in any process & $Q_{BE}$, $Q_{EF}$: $1 \to 2$ & H6 \\
Variation connectivity & No $c_1$--$c_2$ pair in $\prec$ & $Q_{\mathrm{mod}}$: $B_3^{\mathrm{cap}} \prec B_4^{\mathrm{cap}}$ & H5 \\
Shared replication organisation & $\{m\}$, for $c_1$ only & $\{m\}$, all four R processes & H6 \\
Translation & $(Q_{d_1}, Q_{d_2})$: no capability input precedes a dish & $(Q_{d_1}, Q_{d_3})$, also $(Q_{d_1}, Q_{d_2})$ & H8; H7 adds $Q_{d_3}$ \\
Shared translation organisation & $\{k, m\}$, patterns correspond; no T process & $\{k, m\}$, patterns correspond & H8 \\
Variant-dependent downstream realisation & No rule & $F$ of H8 & H8 \\
Observed lineage continuation & $c_1$: $L = 2$ & $c_1$: $L = 2$; $c_2$: $L = 2$ & H6 (for $c_2$) \\
\midrule
Information family $\{c_1, c_2\}$ & Not assigned & Assigned & H5, H6, H8 \\
\bottomrule
\end{tabularx}
\caption{\textbf{Condition-by-condition verification of Case A and Case B.} All entries are conditional on H1--H4, and those of Case B also on H5--H8. ``Changed by'' names the added hypotheses without which the Case B entry reverts to that of Case A; for the family, each of H5, H6 and H8 is necessary, and H7 is not under the reading of Section~\ref{sec:4.6} that associates a translation witness with a lineage by its structural label.}
\label{tab:recipe-verification}
\end{newtext}
\end{table}

\subsection{\new{Sensitivity and evidential limitations}}\label{app:G.5}

\begin{newtext}
\textbf{Dependence on H1--H4.} Both verdicts rest on H1--H4 in the same way, so the difference between them does not depend on these hypotheses, but each verdict does. If H1 or H2 fails, the decomposition into capability and remainder instances is inadmissible, neither CSS exists in the form given here, and whether another admissible representation of the episode contains qualifying operators remains open (Appendix~\ref{app:G.1}). As Appendix~\ref{app:G.2} states, the individuation hypothesis is prior to every question about particular relations.

\textbf{Persistent skills.} H3 treats a capability set as a persistent structural type, i.e., an almost-invariant region of the capability component. Recipe knowledge in practice varies by degree, decays and is recombined, and the partition into $c_0$, $c_1$, $c_2$ and $c_{12}$ is the coarsest one that distinguishes the two recipes. Under a finer partition every process would have to be re-evaluated, since R depends on label multiplicities. H5 adds an idealised convergence: B's modified configuration falls into exactly C's type.

\textbf{Reading of the modification.} Under the acquisition reading, $B_4^{\mathrm{cap}}$ and $B_5^{\mathrm{cap}}$ carry $c_{12}$. The script then gives three changes. $Q_{BE}$ has $n^{-}(c_2) = 0$ and is no R process for $c_2$; instead $c_{12}$ satisfies the operator condition in $Q_{BE}$. $Q_{\mathrm{mod}}$ is a V process from $c_1$ to $c_{12}$ only, so no declared process is a V process between $c_1$ and $c_2$. Under the acquisition reading the family is still assigned, because a process need not be declared (Section~\ref{sec:3.7}): the convex process $\{B_3^{\mathrm{cap}}, B_4^{\mathrm{cap}}, u_{BE}, E_2^{\mathrm{cap}}\}$, with input boundary $\{B_3^{\mathrm{cap}}\}$ and output boundary $\{E_2^{\mathrm{cap}}\}$, is a V process from $c_1$ to $c_2$ through $c_{12}$ and $s$, which Section~\ref{sec:5} admits; $c_2$ has one R process, $Q_{EF}$, so $L = 1$ for $c_2$, and the family has $L = 2$ through $c_1$. T rests on $(Q_{d_1}, Q_{d_2})$, as when H7 is withdrawn, so this assignment holds under the same reading of Section~\ref{sec:4.6} as in Appendix~\ref{app:G.4}. Section~\ref{sec:comp.5} discusses the same reading of V for the simulator.

\textbf{Stipulated causal relations.} The T verdict of Case B rests on H8, a rule about a counterfactual dish type, not on a performed intervention. Actual preparations differ in quantities, timing and execution below the declared resolution, and H8 states that none of these differences changes the dish type. Following Appendix~\ref{app:F.3}, the motif components entailed in full by hypothesised elements are: V, by H5; R for $c_2$ and its lineage continuation, by H6 together with H3 and H4, and; variant-dependent realisation, and with it T, by H8 together with the preparation processes of Case A. H7 adds a second $c_2$ translation witness, which lies inside the $c_2$ lineage. Recovering these components in Case B is not confirmation of them, and the assignment under the Case B description gives H1--H8 no empirical support.

\textbf{Observations that would discriminate.} Step 6 of Appendix~\ref{app:F.3} asks which observations or interventions would discriminate between the descriptions. A controlled comparison in which one cook prepares from one ingredient batch with the same apparatus before and after a modification bears on H8; whether the dishes prepared by E and F follow the modified or the original recipe bears on H6; structural observation of the capability configuration itself bears on H2 and H3.
\end{newtext}

\begin{newtext}
\section{Computational methods}\label{app:H}

Section~\ref{sec:comp} rests on six development runs, i.e., seeds 1--5 and a no-mutation control, each processed by the same chain of scripts. %
This appendix gives the methods behind that chain; the results are stated in Section~\ref{sec:comp} and are not repeated.

\subsection{Simulator and record}\label{app:H.1}

\textbf{Simulator.} We wrote an Avida-like simulator \citep{Ofria2005} in Python, using the standard library only. %
Besides the variables named in Section~\ref{sec:comp.1}, each cell holds \texttt{ALIVE}, the memory size \texttt{LEN}, the length of the memory content at birth \texttt{GL}, the age \texttt{AGE}, the stack pointers and the active-stack flag, three inputs with their pointer, the last output, a copy counter, an allocation flag, and a read label of 10 slots with its length. %
The memory holds at most 80 sites and allocation doubles the memory size, so the memory content at birth is capped at 40 sites. %
The 20-site ancestor is placed in the centre cell, and its time to first division, without mutation, is 69 own steps. %
A copy error replaces the copied instruction by one drawn uniformly from the 26, and a division inserts or deletes one site; the rates of Section~\ref{sec:comp} are Avida's defaults, set in our specification rather than read from an Avida configuration. %
In every update, the occupied cells are put in a keyed random order and execute one instruction each, round-robin, 30 times, without merit; a newborn waits for the next update. %
A division places the child in one of the 8 neighbouring cells, an empty one if any exists, and otherwise overwrites the occupant; a cell is reset, i.e., its organism dies, when \texttt{AGE} exceeds $20\,\texttt{GL}$. %
The documentation of the code lists 16 known or unchecked deviations from Avida, e.g., allocated memory filled with \texttt{nop-A} and no minimum share of executed instructions at a division. %

\textbf{Record.} The simulator passes seven tests, three of which make the record complete (Section~\ref{sec:comp.1}). %
In the parent-sufficiency test, the state before each step is rebuilt, every variable of the executing cell and of its 8 neighbours outside the step's parent set is randomised, and the re-executed step must give identical writes, resets and non-writes. %
Each random draw is a hash (BLAKE2b) of the seed, the cell, the cell's own step counter, the kind of draw and its index within the step, so a replay or an intervention receives the same chance events at the same own step of a cell. %
The withheld file holds organism identity per cell over time, parent identities, the memory content at every birth, the source site of every child site, the mutation events and the executed sites. %
The CSS and search code open the log through a loader and every other file through a guard, both of which raise an error on the withheld file; only scoring code, marked guided, reads the withheld file. %

\subsection{CSS construction}\label{app:H.2}

\textbf{Nodes and labels.} A rewrite with an unchanged value starts a new instance, because the value then has new causes, and a read followed by an overwrite in the same step leaves no continuation, i.e., no zero-length section becomes a node; in seed 1, 478,904 instances give 847,911 nodes. %
The kind of a label is \texttt{memory} for the 80 memory sites and otherwise the variable's name, with the 20 stack slots as one kind and the 10 read-label slots as another, which gives 24 non-memory kinds. %

\textbf{Dependence.} For a node $x$ that ends in a read at step $s$ and an alternative value $a$, step $s$ is replayed alone, from the factual state before $s$, with $x = a$ and the step's own draws: the $n$-th draw of a kind is the factual $n$-th, and a draw beyond the factual count is a fresh one. %
Every other variable thus keeps its factual value, so each edge of $D_{\mathrm{base}}$ and $D_{\mathrm{maint}}$ is a one-step controlled direct effect in the sense of Section~\ref{sec:3.5}. %
The builder compares the final value of every variable written in the step: 1) a factual write whose value changes and 2) a factual write that no longer happens, i.e., a prevention, each give a write edge from $x$ to the instance that the factual write created, and; 3) a write that happens only in the replay, with a value other than the current one, gives a maintenance edge from $x$ to the first section of that variable's current instance that ends after $s$. %
The declared alternatives, each without the factual value $v$, are: 1) the 25 other instructions for a memory site; 2) every value in range for a bounded variable with at most 26 values, and $\{0, 1, v-1, v+1, \mathrm{maximum}\}$ for a larger range, and; 3) $\{0, 1, -1, v+1, v-1, 2^{31}-1\}$ for registers, stack slots, inputs and the output. %
Seed 1 requires 8,230,818 one-step replays. %

\textbf{Direct-edge relation and order.} Every stored write edge and maintenance edge records whether it is a prevention and whether it changes the scheme-B label. %
$D_{\mathrm{base}}$ consists of the write edges that carry either flag, $D_{\mathrm{maint}}$ of the maintenance edges that change the label, and $D_{\mathrm{pers}}$ of every persistence edge; an edge that changes only an exact value, i.e., only a scheme-A label, is stored but is not part of $D$ (Section~\ref{sec:comp.1}). %
Neither order ALL, $D^{+}$, nor order DP, $(D_{\mathrm{base}} \cup D_{\mathrm{pers}})^{+}$, is stored; both are answered by reachability on the edges of $D$. %

\textbf{Checks and build time.} The builder passes six tests, among them that an alternative equal to the factual value gives no edge, that an unmutated copy depends on its template site and a mutated copy does not, and that the source of every write edge lies in the step's logged read set; on the full database of seed 1, the same holds for every maintenance edge. %
A diagnostic then checks that every stored edge runs forward in time and in node numbering, which makes the edge graph acyclic. %
The share of node pairs ordered through the stored edges of each kind is estimated on a sample of 10,000 node pairs, and specificity is the mean share of the sections of a parent's memory sites that precede one child site through all stored edges. %
Both measures use the stored edges without their flags, i.e., a relation that contains order ALL. %
Of the 48 s that the CSS of seed 1 took with 4 worker processes and Python 3.9.6, 43 s went to the one-step replays, the edges and the database; of the 195 s of the whole invocation, the comparison with the $k = 10$ build took 124 s. %

\subsection{Process classes}\label{app:H.3}

In order DP, the process of step $s$ is $Q_s$, i.e., the nodes that end in a read at $s$, whose variables are in the step's logged read set, and all nodes that start at $s$, i.e., continuation sections and created nodes. %
$Q_s$ is causally convex in order DP, because every edge of $D_{\mathrm{pers}}$ and $D_{\mathrm{base}}$ ends at a node that starts where its source ends; the code asserts this convexity at every step. %
In order ALL, the process is the convex closure of $Q_s$, i.e., the nodes that are both descendants and ancestors of members of $Q_s$, computed by a backward search from $Q_s$, pruned at nodes that end before the earliest end of a member, and a forward search inside the backward set. %
We raised the cap on the closure from 50,000 to 2,000,000 nodes and capped the backward search at 5,000,000 nodes; no step reached either cap (Section~\ref{sec:comp.3}). %
Boundaries are taken in the subgraph that the process induces on the edges defining the order: a member is an input when no such edge from another member enters it, and an output when none leaves it. %
The guided checks of births use a gestation process, i.e., the convex closure of 1) the earliest section of each parent memory site that ends after the parent's previous division or birth; 2) the first sections of the child's memory sites, and; 3) the sections of the parent's memory sites after the division, without which $n^{+} = n^{-}$ for every memory label of an error-free birth. %
Beyond single steps and their chains in pass~B (Appendix~\ref{app:H.4}), the blind search used no process class at the level of instructions; in particular, no blind search covered processes that span a pass of the copy loop. %

\subsection{Blind search}\label{app:H.4}

The blind search is stage~6 of Figure~\ref{fig:pipeline}, run in the four search passes of Section~\ref{sec:comp.3}.
Its scripts read the CSS and the record format of the log, i.e., variable identities, cells and read sets; instruction names are attached to labels in the output only, and every comparison that uses them is marked guided. %

\textbf{Pass~A: replication.} In every step's process, each label of every kind is tested for R (Section~\ref{sec:4.1}), and the step shows R when at least one label passes. %
For a memory label, each output instance that is not an input instance is a copy, and the inputs of the same label among its ancestors are its templates. %
The null keeps every process, its boundaries and its ancestry relations, permutes the labels over all nodes or within each kind, and recomputes R; we report the largest share of steps with R in 20 permutations of each type, with fixed seeds. %

\textbf{Pass~B: lineage.} A move is an output memory node of a new instance that has an input of the same label in another cell among its ancestors. %
R events are processed in step order, and the depth of an event is 1 plus the largest depth of an earlier event whose copy is its template, directly or through a move; the lineage depth $L$ of a label is the largest depth of its events, and the labels with $L \geq 2$ are the candidates. %
Because a template can be an earlier copy in the same cell, a re-copy of a copy counts as a generation. %

\textbf{Pass~C: variation.} Every stored edge runs forward in node numbering, so one sweep over the edges of $D$ in source order gives each node the set of candidate labels that have a node upstream of it, itself included. %
V from $x$ to $y$ holds when a node labelled $y$ carries $x$ in its set; reachability is thus computed between particular nodes, never between labels. %
On seed 5, a forward search from every candidate node gave the same sets in both orders. %
Candidates linked by V in either direction form components, and the family rule uses the largest component. %

\textbf{Pass~D: translation.} Every memory input of a process whose label is a candidate is \del{a}\new{an} \new{information }carrier and gives one record, i.e., one instance, with three entries: 1) its read type, control when its variable is among the step's logged control reads, i.e., the executed instruction, a modifier or a label, and data otherwise; 2) the operator signature, i.e., the non-memory labels that meet the operator condition of Section~\ref{sec:3.9}, and; 3) its products, i.e., the outputs it precedes whose labels are absent from the input boundary. %
Records are grouped by read type and signature, and a candidate pair is two \new{information }carriers of one group whose modal, i.e., most frequent, product sets are non-empty and differ, each \new{information }carrier with at least 2 records in the group. %
For each direction, up to 20 records of the first \new{information }carrier with its modal product set are drawn by a seeded random sample, and each of their steps is replayed alone with the \new{information }carrier's site set to the second \new{information }carrier's instruction. %
The products of a replay are the labels of the variables the replayed step writes that are absent from the labels of the variables it reads; a replay passes when these products include a product of the second \new{information }carrier that the first lacks, or any product of the second when its set lies within the first's, and none that only the first \new{information }carrier makes. %
A direction passes when at least half of its replays pass, and a pair is confirmed when both directions pass; the one-step replay is the same in both orders. %
\texttt{blind\_t2.py} does not test the non-empty operator signature that separates T from T-like pairs (Section~\ref{sec:comp.3}), so \texttt{w1\_followup.py} recomputes the families from the stored records, without new replays, and checks that the six largest replication operator sets, the only ones the record files keep, cover every member. %

\textbf{Family.} The replication operator sets of a memory label are the signatures of the processes in which the label passes R. %
In the family rule of Section~\ref{sec:comp.3}, the V-connected group is the largest V component of pass~C, and a shared replication operator set requires at least 2 R steps of each label under that set. %
The threshold of 2 instances applies to both \new{information }carriers of a candidate pair; the sensitivity threshold of 10 applies after the swap test, to each \new{information }carrier separately and, as a stricter variant, to both \new{information }carriers of a pair. %

\subsection{Guided checks}\label{app:H.5}

A guided check uses knowledge of Avida, i.e., the executed instruction of each step and the births, lineages and mutation events of the withheld record, to locate processes, and evaluates R, V and T on the CSS alone; the result measures the sensitivity of the criterion, does not enter the family rule and is never a test. %
Section~\ref{sec:comp} uses five guided checks: 1) R at copy steps, located by instruction name and counted as the copy steps at which any memory label has an R event, a count that differs from that of steps at which the template's own label passes R by 5 of 20,649 steps in seed 1 in order DP; 2) R at births, on the gestation processes of Appendix~\ref{app:H.3}; 3) node-verified lineage, in which a child site must be an output of the gestation process of its birth and an input of that of the child's first division, and the parent site that the withheld source map gives as its source must precede the child site there with the same label, each such link adding 1 to the depth; 4) a diagnosis of the change from order DP to order ALL on 40 sampled copy steps per template label and 30 sampled divisions of seed 1, and; 5) a trace of every site of every organism of seed 1 back, through the withheld source map, to the mutation event that wrote its label. %
Check 1 runs in the guided output fields of \texttt{blind\_rv.py}, checks 2 and 3 in \texttt{gmc.py}, check 4 in \texttt{w1\_diag\_copyloop.py} and check 5 in \texttt{w9\_trace.py}. %
A circular alignment of the memory labels of parent and child at every birth found in the CSS, with unit costs and the cheapest rotation of the child, is blind but outside the criterion; only its comparison with the logged copy errors is guided. %

\subsection{Organism level}\label{app:H.6}

\textbf{Organisms.} \texttt{l2\_genotype.py} builds the organism level from the CSS and the log; the withheld record is opened afterwards, to check every birth, organism, G-label, value of P and end of life. %
A site of the G-label without a node, i.e., at or beyond \texttt{LEN}, is written as a gap, so the G-label records the length of the memory content. %
P uses the actual value of \texttt{AGE}, i.e., scheme A for this one variable; P is recorded as sterile when the age-limit reset comes before a division, and as censored when an overwrite comes first or the organism is alive at the end of the run. %
The G-node of an organism consists of the level-1 nodes of its sites $M_0$ to $M_{39}$ created at its birth, with their later sections, cut at every division; its P-node consists of the level-1 nodes that its first division writes in its own cell, because the memory content causes that step, whereas \texttt{AGE} counts steps whatever the memory content does. %
A level-2 node precedes another when some level-1 node of the first precedes some level-1 node of the second in order ALL; the code asserts acyclicity. %
The causal path of an R or V birth is found by a backward search from the nodes of the child's memory content to a node of the parent's memory content before the birth, in order DP first and in order ALL when DP finds none; for Section~\ref{sec:comp}, the path check was also run in each order alone. %

\textbf{Memory-content swap.} A candidate pair consists of two G-labels with $L \geq 2$ whose modal, i.e., most frequent non-censored, values of P differ. %
For each pair $(g_1, g_2)$, \texttt{l2\_swap.py} takes up to 20 pairs of organisms, $a$ of $g_1$ and $b$ of $g_2$, both with a non-censored value of P, in order of birth. %
At $a$'s birth step, the memory of $a$'s cell is set to the sites of $g_2$ followed by zeros, and \texttt{LEN} and \texttt{GL} to the length of $g_2$; the simulator then runs on with the logged schedule and the keyed draws, so $a$ receives at each own step the draws it would have received. %
The replay stops at $a$'s first division, which gives P, at a reset at the age limit (sterile), at an overwrite (censored) or at the end of the run (censored); $b$ receives $g_1$ in the same way. %
A swap follows when its P equals the modal P of the inserted G-label; a pair is confirmed under rule `all' when every non-censored swap follows, in both directions, with at least one such swap per direction, and under rule `majority' when more than half follow. %
Before the swaps, 25 organisms per run are replayed without a swap, to check the replay. %
The family consists of the G-labels with $L \geq 2$ that have a confirmed partner and a V connection to another member, pruned to a fixed point. %
A swap ignores what a changed timing would do to the neighbours after the first division. %

\subsection{Controls, family-level null and colonies}\label{app:H.7}

\textbf{Controls.} The no-mutation control is seed 1 with the rates of copy errors, insertions and deletions set to 0, and seed 5 is the development seed whose founder never divided; both went through the same chain of scripts as seeds 1--4. %

\textbf{Family-level null.} \texttt{family\_null.py} extends the label permutations of pass~A (Appendix~\ref{app:H.4}) to the structural part of the family rule. %
Because the causal order is kept, the convex closure of every single-step process is the same for every labelling; it is computed once per seed, in order ALL with the caps of Appendix~\ref{app:H.3}, and its boundary is stored with the ancestor relations between outputs and inputs and the read type of each input. %
A labelling is scored on the stored boundaries by passes~A--C, by pass~D up to the candidate T pairs, i.e., without swap replays, and by the family rule, and the real labelling is scored in the same way; the statistic is the number of memory labels in this structural family. %
A reassignment gives every instance of a kind the label of another instance of the same kind, drawn at random, with all sections of an instance sharing that label, so the frequency of each label among the instances of its kind is kept; seeds 2000--2019 give 20 reassignments per run. %
A run passes when its real structural family is larger than in every reassignment, i.e., at a one-sided $p = 1/21$. %
The pass criterion was written before any relabelled full CSS was scored, but the null itself was revised during development, from per-node permutations to per-instance reassignments, so the result is exploratory (Section~\ref{sec:comp.4}). %
On the real labelling, the code reproduces the outputs of passes~A--D of seeds 1--4 exactly, and with the swap-confirmed pairs it gives the families of 5, 4, 6 and 4 labels. %
One closure pass takes 707--807~s per seed and one reassignment 3--6~s; the 80 reassignments of seeds 1--4, with their closure passes, took 57~min. %

\textbf{Colonies.} \texttt{l3\_colony.py} reads the level-2 organisms, the log and the CSS, intervenes nowhere and does not open the withheld record. %
1) A patch is a maximal set of occupied cells at the end of an update whose organisms share one G-label, connected through the 8 neighbours of a cell on the torus, and a colony is a patch of at least 3 organisms, labelled by their G-label. %
2) A colony continues a colony of the same G-label in the previous update when their cell sets overlap, matched one to one by largest overlap; a colony that continues none is new, either founded or split off, i.e., a fission product. %
3) A new colony is downstream of an earlier colony when a level-1 node of the memory contents of the earlier colony's members, during membership, strictly precedes a node of the memory contents of the new colony's members in its first update, in order ALL; one sweep over the edges, with one bit per colony, computes this relation for every node. %
4) R is a new colony downstream of a still present colony of the same G-label, and V a new colony downstream of a colony of another G-label; the level-3 lineage depth is 1 plus the largest depth of the R sources, or 0 without one. %

\subsection{Development and held-out seeds}\label{app:H.8}

\new{The held-out seeds 101--140 were reserved, seeds 101--120 on 2026-10-20 and seeds 121--140 on 2026-10-29, and were run only after the freeze, in the no-task condition. Every result of Section~\ref{sec:comp} comes from development seeds 1--5 unless it is named as held-out.} %
During development, and before any held-out run, we fixed eight choices: 1) $L \geq 2$ (decision 4, 2026-10-20); 2) $k = 1$ and 3) scheme B (decisions 15 and 16, 2026-10-26); 4) every recorded state variable in the perspective; 5) the instance-path reading of V; 6) no false family and no division requirement (decisions 17--19, 2026-10-27); 7) order ALL for every search pass, and; 8) a minimum of 2 instances, with 10 as sensitivity analysis. %
Choices 7 and 8 date from 2026-10-28\new{; the frozen preregistration contains all eight}. %
Two further choices followed first results: the split by read type, and the enforcement of the non-empty operator signature that the preregistration already requires. %
\new{We froze the preregistration on 2026-10-29, before any held-out run, by committing} one authoritative version, with the scoring code, the randomisation seeds and the rules for eligible and failed runs, to the project's git repository\new{ as commit \texttt{4dfe999}, and by archiving a time-stamped copy; the SHA-256 hashes of the preregistration, of the archive and of each of the 27 frozen files are recorded beside the preregistration}; after the freeze, any revision needs new untouched seeds. %
The analysis that the preregistration \new{freezes} consists of the choices above, $D$ as in Appendix~\ref{app:H.2}, the search passes A to D, the non-empty operator subset of Appendix~\ref{app:C.1}, the family rule and the family-level null of Appendix~\ref{app:H.7}; the positional and alignment readings of V stay diagnostics. %
The held-out test scores five predictions on the $m$ runs with births: 1) a swap-confirmed information family of at least 2 instruction types exists; 2) the real structural family is larger than in all 20 reassignments of the family-level null; 3) the organisms, births, genotypes and lineages inferred under the declared rules, and with them the organism-level family, match the record, a check of the reading rather than a test of the criterion, since the label of an organism, its memory content at birth, is the genotype that the design defines; 4) of the decided memory-content swaps, i.e., those that end in a division or in sterility, at least 99\% follow the inserted memory content, pooled over runs, and; 5) at least 99\% of the steps with inferred R are designated copy steps, and R is found at 75\% or more of the designated copy steps, in every run. %
Predictions 1 and 2 pass when they hold in at least $c(m)$ runs, the count that separates a per-run success rate of 0.95 from one of 0.70 with the smallest maximum error under a binomial model, e.g., $c(32) = 28$; both errors stay below 5\% for every $m$ from 24 to 40, and with $m < 24$ the test is inconclusive. Prediction 3 must hold in every run. %
Every observed proportion of runs is reported with its exact (Clopper--Pearson) 95\% confidence interval, and every rate of events within runs with an interval that resamples seeds. %
Reference pairs, i.e., copy errors from an original to a mutant label in which each label is inherited unmutated over at least 2 successive births and each is translated, differently, by the shared processor, are reported per run as a description, not scored: the design fixes copying, mutation and translation for each instruction, but no family at the level of instructions. %
The preregistration fixes order ALL for passes~A, B and~D, a threshold of 2 instances, counted as instances rather than executions, with 10 as sensitivity analysis, the recovery threshold of 75\% for prediction 5, which lies about 3 standard deviations of the spread between development runs below their mean of 81.8\%, and the no-task condition analysed here. %

\new{\textbf{Held-out results.} All 40 runs completed, with no crash and no failed stage, and the 27 frozen files matched their hashes after the batch. Seed 132 has no births, because its founder dies at the age limit before dividing, so $m = 39$ and $c(39) = 34$; at this count, the probability of a false fail is 1.2\% at a per-run success rate of 0.95, and that of a false pass 1.1\% at 0.70. Table~\ref{tab:heldout} gives the results. Predictions 1, 2, 4 and 5 pass, and prediction 1 also passes at the sensitivity threshold of 10 instances, in 39 of 39 runs.} %
\new{Under the frozen rule, the held-out test fails, on prediction 3, which holds in 37 of 39 runs: in seeds 118 and 139, the organism-level family read from the CSS holds one genotype more than the reference family. The cause lies in the reference definition, not in the reading: the reference takes the phenotype of a genotype from its clean organisms only, i.e., organisms without a copy error in their first gestation, and each extra genotype is heritable, with $L = 2$, but has no clean organism, so the reference family omits it, whereas the reading gives the same genotype, the same $L$ and the same phenotype, 72 steps, as the record.} %
\new{Every heritable genotype in development had clean organisms; in the held-out runs, 2 of 533 heritable genotypes have none. The preregistration states that the two families can differ only by a misreading; the held-out runs show a second route, which a working document had named before the freeze but the preregistration did not carry over.} %
\new{Post hoc, and therefore not part of the verdict: if the reference definition fell back to all organisms for a genotype without clean organisms, both genotypes would join the reference family, and prediction 3 would hold in 39 of 39 runs.} %
\new{Two descriptive results differ from development. First, the membership of the family varies: \texttt{nop-A}, \texttt{nop-B} and \texttt{nop-C} are members in 39 of 39 runs and \texttt{h-alloc} in 36, but \texttt{h-copy}, a member in no development run, is a member in 17 and \texttt{h-search} in 19. Second, the reference pairs are recovered at a threshold of 2 instances in 19 of 39 runs in order ALL and in 34 of 39 in order DP, against 3 of 4 and 4 of 4 in development.} %

\begin{table}[htbp]
\begin{newtext}
\centering
\small
\caption{Held-out test on seeds 101--140 with the frozen pipeline, no-task condition: $m = 39$ runs with births and pass count $c(39) = 34$; prediction 5 is scored on all 40 runs. Intervals are exact (Clopper--Pearson) for proportions of runs and resample seeds for rates of events.}
\label{tab:heldout}
\begin{tabularx}{\textwidth}{@{}>{\raggedright\arraybackslash}p{0.12\textwidth}>{\raggedright\arraybackslash}X>{\raggedright\arraybackslash}p{0.15\textwidth}>{\raggedright\arraybackslash}p{0.14\textwidth}l@{}}
\toprule
Prediction & Result & Pass rule & 95\% interval & Verdict \\
\midrule
1 & swap-confirmed family of 3--9 instruction types in 39 of 39 runs & at least 34 of 39 runs & 91.0--100\% & pass \\
2 & real structural family of 8--16 types, against at most 4--6 in the 20 reassignments, in 39 of 39 runs & at least 34 of 39 runs & 91.0--100\% & pass \\
3 & reading matches the record in 37 of 39 runs; seeds 118 and 139 fail & every run & 82.7--99.4\% & fail \\
4 & 29,250 of 29,321 decided swaps follow the inserted memory content (99.76\%) & at least 99\%, pooled & 99.66--99.85\% & pass \\
5, precision & 695,012 of 695,022 steps with R are copy steps (99.9986\%); every run at least 99.994\% & at least 99\% in every run & 99.9978--99.9993\% & pass \\
5, recovery & R at 695,012 of 842,281 copy steps (82.5\%); every run at least 80.4\% & at least 75\% in every run & 82.1--82.9\% & pass \\
\bottomrule
\end{tabularx}
\end{newtext}
\end{table}

\subsection{Reproducibility}\label{app:H.9}

All analysis code is Python with the standard library only, run from the folder of the simulator and chained per run by the shell drivers \texttt{run\_dev\_seeds.sh}, \texttt{run\_order\_all.sh}, \texttt{run\_l2.sh} and \texttt{run\_family\_null.sh}; every script except the CSS builder runs as a single process (Table~\ref{tab:appH-scripts}). %
For seed 1, the blind search in both orders is \texttt{zsh run\_order\_all.sh seed1}, followed by \texttt{python3 w1\_followup.py}. %
The CSS is an SQLite database of nodes and edges, opened read-only by every script that reads the CSS. %
Each log stores a SHA-256 hash of the simulator code, the Python version and the platform. %
Three properties make the chain deterministic: 1) the draws are keyed, and the permutations and swap samples use fixed seeds; 2) the CSS builder gives identical tables with 1 and 4 processes, and; 3) a fresh rerun in order DP reproduced the earlier result files exactly in all six runs. %
The runs and analyses were made on an Apple M2 laptop with 8 cores and 16 GB of memory, identified from the paths in the CSS build log, which does not name the machine. %
Besides the run times of Section~\ref{sec:comp.5}, pass~C takes about 13 s per run, the organism level of seed 1 about 4 s plus 32 s for the swaps, the colony level of seed 1 11 s, and the family-level null of seeds 1--4 57 min (Appendix~\ref{app:H.7}). %

\textbf{Figures.} Figures~\ref{fig:css-L1} and~\ref{fig:css-L2} come from \texttt{label\_images.py}, Figures~\ref{fig:family} and~\ref{fig:validation}b from \texttt{avida\_figures.py}, Figure~\ref{fig:zooms}a and~b from \texttt{avida\_zooms.py}, and Figure~\ref{fig:validation}a and~c from \texttt{avida\_panels.py}. %
\texttt{avida\_figures.py} and \texttt{avida\_panels.py} recompute every drawn number from the result files and compare that number with the source reports and, for \texttt{avida\_panels.py}, with the text and tables of Section~\ref{sec:comp}; a mismatch stops the run before a figure is written. %
\texttt{label\_images.py} checks that the level-2 occupancy of every cell equals the cell's \texttt{ALIVE} row at level 1. %
\texttt{avida\_zooms.py} rebuilds the candidate pairs and the seeded swap sample of \texttt{blind\_t2.py} for seed 1, replays every sampled swap, checks the counts against the stored result and, in a guided step, reads the withheld mutation record to locate the copy error of Figure~\ref{fig:zooms}b. %

The simulator, all scripts, the result files and the figure generators are archived at Zenodo, \new{\href{https://doi.org/10.5281/zenodo.23798795}{doi:10.5281/zenodo.23798795}} (Code and data availability).

\begin{table}[htbp]
\begin{newtext}
\centering
\small
\caption{Scripts behind Section~\ref{sec:comp}, in the order in which they run, with their main outputs; \textit{run} stands for seed1 to seed5 or nomut, the no-mutation control.}
\label{tab:appH-scripts}
\begin{tabularx}{\textwidth}{@{}>{\raggedright\arraybackslash}p{0.2\textwidth}>{\raggedright\arraybackslash}p{0.34\textwidth}>{\raggedright\arraybackslash}X@{}}
\toprule
Stage & Script & Main output \\
\midrule
Simulator, tests, replay & \texttt{mini\_avida.py}, \texttt{test\_mini\_avida.py}, \texttt{replay.py} & \texttt{run.json} and \texttt{run-withheld.json} (seed 1; the other runs carry the run name) \\ %
CSS, $k = 1$ & \texttt{css\_build\_k1.py} & \texttt{css-}\textit{run}\texttt{-k1.sqlite} \\ %
Search passes A and B & \texttt{blind\_rv.py --order ALL} & \texttt{w1-rv-all-}\textit{run}\texttt{.json} \\ %
Search pass C & \texttt{blind\_v\_path.py} & \texttt{w1-vpath-all-}\textit{run}\texttt{.json} \\ %
Search pass D, family & \texttt{blind\_t2.py --order ALL} & \texttt{w1-t2-all-}\textit{run}\texttt{.json} \\ %
Non-empty signature & \texttt{w1\_followup.py} & \texttt{w1-followup.json} \\ %
Family-level null & \texttt{family\_null.py} & \texttt{family-null-seed1.json} to \texttt{family-null-seed4.json} \\ %
Guided checks 1--3 & \texttt{blind\_rv.py}, \texttt{gmc.py} & \texttt{w1-rv-all-}\textit{run}\texttt{.json}, \texttt{w1-gmc-}\textit{run}\texttt{.json} \\ %
Guided checks 4 and 5 & \texttt{w1\_diag\_copyloop.py}, \texttt{w9\_trace.py} & \texttt{w1-diag-copyloop-seed1.json}, \texttt{w9-trace-seed1.json} \\ %
Organism level & \texttt{l2\_genotype.py}, \texttt{l2\_swap.py} & \texttt{l2g-}\textit{run}\texttt{.json}, \texttt{l2swap-}\textit{run}\texttt{.json} \\ %
Colony level & \texttt{l3\_colony.py} & \texttt{l3-}\textit{run}\texttt{.json} \\ %
Build time & \texttt{w7\_build\_time.py} & table of stage times \\ %
Comparison tables & \texttt{compare\_order\_all.py}, \texttt{compare\_l2.py}, \texttt{compare\_l3.py} & one comparison file per level \\ %
\new{Held-out test} & \new{\texttt{run\_heldout.sh}, \texttt{heldout\_run.py}, \texttt{heldout\_score.py}} & \new{\texttt{heldout-pooled.json}; per run \texttt{score-seed}\textit{N}\texttt{.json}} \\ %
\bottomrule
\end{tabularx}
\end{newtext}
\end{table}
\end{newtext}

\end{appendices}

\clearpage
\nocite{Acerbi2015} %
\bibliographystyle{plainnat}
\bibliography{refs}

\end{document}